%% file: main.tex
\documentclass[11pt,a4paper,oneside]{memoir}
\usepackage{config/rmit}

\makeindex
\setsecnumdepth{subsubsection}
\input{config/variables}

\begin{document}
\input{frontmatter/title}
\input{frontmatter/frontmatter}

\mainmatter 
\pagestyle{rmit} 
\include{chapters/0.abstract}

\include{chapters/1.introduction}

\include{chapters/2.background}

\include{chapters/3.metric}

\include{chapters/4.event}

\include{chapters/5.multisource}

\include{chapters/6.benchmark}

\include{chapters/7.evaluation}
\include{chapters/8.conclusion}

\bibliographystyle{abbrvnat}    
\bibliography{bib/strings,bib/references} 

\appendix
\include{appendices/a}

\include{appendices/b}

\include{appendices/c}

\end{document}

%% file: config/variables.tex
\lstdefinelanguage{json}{
  morestring=[b]",
  morecomment=[l]{//},
  moredelim=[l][\color{black}]{:},
  moredelim=[s][\color{red}]{[}{]},
  stringstyle=\color{teal},
  keywordstyle=\color{black},
  commentstyle=\color{gray},
  showstringspaces=false,
}
\definecolor{DarkOrange}{rgb}{0.8,0.4,0}
\definecolor{DarkRed}{rgb}{0.6,0,0}

\newcommand{\method}[1]{#1}
\newcommand{\barotool}{\method{BARO}}
\newcommand{\eventadl}{\method{EventADL}}
\newcommand{\torai}{\method{TORAI}}
\newcommand{\rcaeval}{\method{RCAEval}}
\newcommand{\nsigma}{\method{N-Sigma}}

%% file: frontmatter/title.tex

\begingroup
\thispagestyle{empty}
\begin{center}

\renewcommand{\baselinestretch}{1.5}\normalsize

\includegraphics[width=5cm]{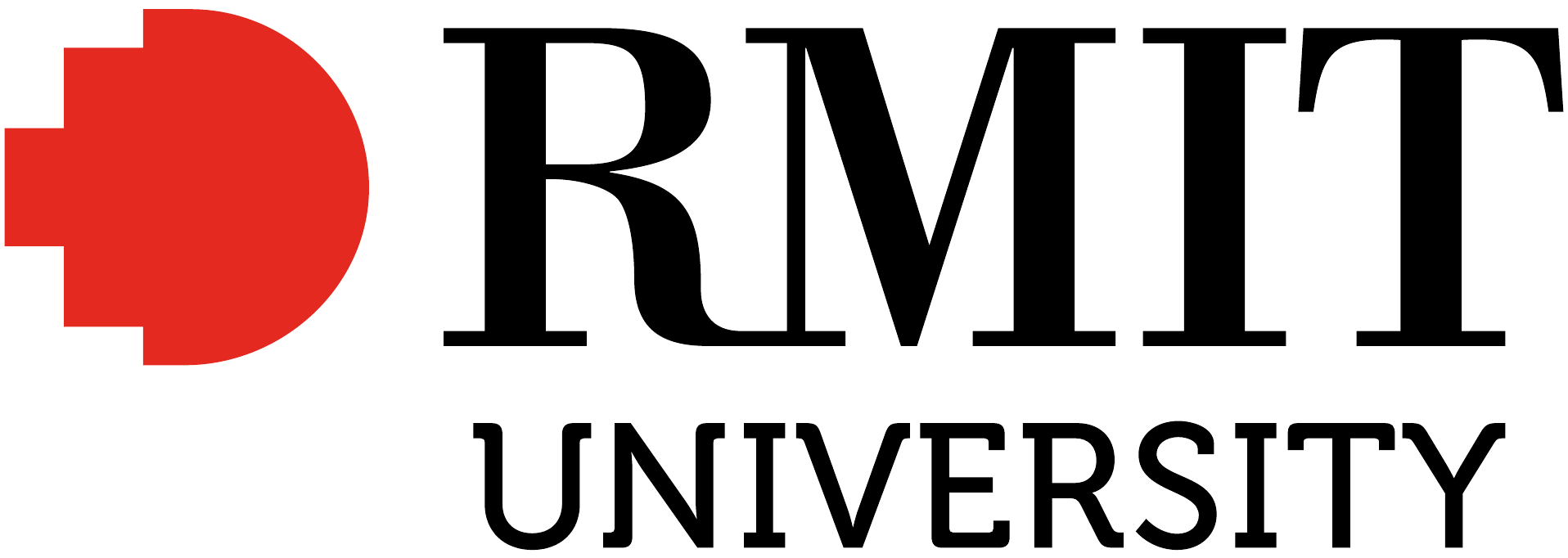}\\[1.5cm]

\vspace{1cm}

{\LARGE \textbf{Anomaly Detection and Root Cause Analysis \\ for Microservice Systems}}\\[1.5cm]

{\normalsize A thesis submitted in fulfilment of the requirements for the degree of \\[0.2cm]
Doctor of Philosophy \\[1.5cm]}

{\large Qui Luan Pham} \\[0.5cm]
{\small Bachelor of Engineering (B.Eng) in Computer Science, \\
Ho Chi Minh City University of Technology} \\[1cm]

{\normalsize ORCID: \href{https://orcid.org/0000-0001-7243-3225}{0000-0001-7243-3225}} \\[1.5cm]

\vspace*{0.8\fill}

\vspace{0.8cm}
{\normalsize School of Computing Technologies} \\[0.2cm]
{\normalsize College of Science, Technology, Engineering and Maths} \\[0.2cm]
{\normalsize RMIT University} \\[0.2cm]
{\normalsize Australia} \\[1.5cm]

\vspace{0.5cm}
{\normalsize June 2026}

\end{center}
\endgroup

%% file: frontmatter/frontmatter.tex
\pagestyle{rmitheadings}
\frontmatter

\noindent \textit{Dedicated To Those Who Came Before Me.} \\

\noindent \textit{"I did not invent the transistor, the microprocessor, object oriented programming, or most of the technology I work with. I love and admire my species, living and dead, and am totally dependent on them for my life and well being." - Steve Jobs}

\clearpage

\chapter*{Declaration}
\addcontentsline{toc}{chapter}{Declaration}

I certify that except where due acknowledgement has been made, this research is that of the author alone; the content of this research submission is the result of work which has been carried out since the official commencement date of the approved research program; any editorial work, paid or unpaid, carried out by a third party is acknowledged; and, ethics procedures and guidelines have been followed.

In addition, I certify that this submission contains no material previously submitted for award of any qualification at any other university or institution, unless approved for a joint-award with another institution, and acknowledge that no part of this work will, in the future, be used in a submission in my name, for any other qualification in any university or other tertiary institution without the prior approval of the University, and where applicable, any partner institution responsible for the joint-award of this degree.

I acknowledge that copyright of any published works contained within this thesis resides with the copyright holder(s) of those works.

I give permission for the digital version of my research submission to be made available on the web, via the University’s digital research repository, unless permission has been granted by the University to restrict access for a period of time.

\vspace{1cm}

\begin{flushright}
\textbf{Qui Luan Pham} \\
16 June 2026
\end{flushright}

\clearpage

\chapter*{Acknowledgements}
\addcontentsline{toc}{chapter}{Acknowledgements}

I started to research when I worked with Dr.~Tuan Anh Tran, and when I was at Cinnamon~AI in 2019. I was surrounded by wonderful people, including Gale, Kush, Emma, Hato, Lotus, Duc, Cat, Xing, and others, who inspired and prepared me for my own PhD journey.

In November 2022, I came to RMIT Melbourne to work on this thesis, previously titled ``Incident Management for Software-Intensive Systems'', under the supervision of Dr.~Huong Ha and Prof.~Hongyu Zhang, and later, Prof.~Xiuzhen Zhang. In July 2023, I finished a preliminary version of RCAEval, which evaluates existing RCA methods. In September 2023, I submitted the second paper proposing BARO and started working on TORAI. I am deeply grateful to my advisors at that time for their great suggestions that led me to the right paths and choices. In 2024, I visited UNSW Sydney (hosted by Prof.~Flora Salim) and the University of Newcastle (hosted by Prof.~Hongyu Zhang), continuing the development of RCAEval. In 2025, with a wonderful support from Thanh Le-Cong, Dr. Victor Nicolet and Dr. Joey Dodds, I joined Amazon as an intern, created EventADL, and was involved in benchmarking HyGLAD. After all of these, I started to write this thesis.

I would like to express my deepest gratitude to everyone who has supported me. I have been blessed with incredible people who have helped me grow as an engineer, a scientist, and a person. I am profoundly thankful to my advisors for their unwavering support. Their trust gave me the freedom to shape and pursue my own research directions. I extend my gratitude to Dr.~Maria Spichkova, Dr.~Mojtaba Shahin, Prof. Jacopo Soldani, and other reviewers for their encouragement and feedback, which have enriched my work greatly.

To my friends and family: I'd like to thank my masters and friends at the RMIT Taekwon-Do family for their companionship. To the UNSW CSE Boys including Maodong, Wilson, and Breeze for their friendship and love. Thanks to my dear friends and brothers, including Thanh Hung, Thanh Huy, Thanh Thao, Tuyet Phuong, and many more, for always replying to my calls. Thanks to my high school teachers, Mrs.~Kim Loan, Mr.~Hoan Vu, Mrs.~My Trang, Mrs.~Tuyet Nga, Mrs.~Xuan Doan, Mrs.~Quynh Tram, and others, for always believing in and encouraging me. Thanks to my family, parents, and brother. 

I also thank Thao Nguyen, Ngoc Kha, Thanh Quynh, Hai Yen, My Nhi, Phuong, and others, for the love and support they once gave me, which I still carry. Thanks for the moments shared together, the laughters and the tears. I am sorry for the promises I could not keep, and I hope that, wherever you are, life has been gentle with you.

To the mysterious universe, and my borrowed time. To all the tiny particles that have been magically assembled into me, filling me with energy, passion, and perseverance to complete this work. To unknown existences that help me without my awareness. \\

\noindent Seattle, United States, \\ 
\noindent Luan Pham.

\begin{KeepFromToc}
\maxtocdepth{subsection} 
\tableofcontents \clearpage
\end{KeepFromToc} 

\listoffigures   \clearpage
\listoftables    \clearpage

%% file: chapters/0.abstract.tex
\chapter*{Abstract} \label{chap:abstract}

\addcontentsline{toc}{chapter}{Abstract}

Microservice systems have gained significant popularity in the development of cloud applications. However, failures in microservice systems are inevitable due to their inherent complexity, leading to poor user experiences and significant economic losses. Therefore, automated anomaly detection and root cause analysis (RCA) for failures in microservice systems have been an active research area in the past few years, yielding many fruitful techniques. However, they do come with noteworthy limitations. First, most RCA techniques treat anomaly detection and RCA separately, assuming the anomalies are correctly detected. These techniques may fail when anomalies are imprecisely detected due to noise or delays from automated anomaly detectors. Second, existing techniques primarily focus on metrics, logs, and traces, while event data such as API calls and configuration changes remains underexplored for anomaly detection and RCA. Third, many RCA techniques require a given service call graph for diagnosis. Without a service call graph, these approaches fail to perform any level of diagnosis. Fourth, there is an absence of standardised evaluation datasets and framework for RCA. Existing studies use different datasets, evaluation metrics, and experimental setups, making it difficult to fairly compare methods and assess their generalisability. Fifth, while causal inference-based RCA methods have emerged as a dominant paradigm, their practical effectiveness, efficiency, and robustness across diverse scenarios remain unclear, hindering the development of more effective techniques. 

This thesis addresses the aforementioned limitations through two groups of contributions. The first group introduces novel methods for automated anomaly detection and RCA, exploring multiple sources of observability data both independently and collectively. It introduces BARO, an end-to-end anomaly detection and RCA approach for metric data. It introduces EventADL, an end-to-end anomaly detection and RCA framework for event data. It introduces TORAI, a multimodal RCA framework without requiring a service call graph. The extensive experiments demonstrate the effectiveness and robustness of these methods across multiple real microservice systems. The second group contributes benchmarking datasets, evaluation framework, and systematic evaluation efforts. It presents RCAEval, a comprehensive benchmark built upon the datasets used to evaluate the proposed methods, providing ready-to-use datasets and reproducible baselines that pave the way for future research. It provides a systematic evaluation of existing RCA methods, especially causal inference-based approaches, offering novel insights that guide future research directions. Overall, this thesis advances the field of automated anomaly detection and RCA of failures in microservice systems, enabling future research on incident mitigation and remediation.

%% file: chapters/1.introduction.tex
\chapter{Introduction}\label{chap:introduction}

\section[Motivation: Automated Detection and Analysis]{Motivation: Automated Anomaly Detection and Root Cause Analysis for Microservice Systems}

The past decade has witnessed microservices becoming the \textit{de facto} standard in the development of cloud applications, replacing traditional monolithic design. Microservice architectures offer many advantages, including a loosely coupled design that enables modular development, independent deployment, and rapid scalability. However, as microservice systems grow in size and complexity, they often become increasingly fragile~\citep{Soldani2018microservice}. A failure in one service can quickly propagate and affect other services, leading to widespread degradation in system performance and availability. This chain of failures often results in poor user experience and significant financial loss. For example, a one-hour downtime on Amazon.com could cost up to 100 million USD~\citep{yahoo_amazon_downtime_2018}. In more critical cases, system failures can even lead to fatalities~\citep{Gregory2025Optus}.

To ensure high availability and reliability of deployed systems, operators must continuously monitor these systems and analyse key runtime information to detect anomalies and diagnose their root causes as early as possible. However, the sheer scale of modern systems and the enormous volume of observability data (e.g., metrics, logs) make manual monitoring and diagnosis infeasible. Automated \textit{anomaly detection} and \textit{root cause analysis} (RCA) for failures in microservice systems have therefore become critical~\citep{Chen2020incidentmanagement}.

\begin{figure}
\centering
\includegraphics[width=\textwidth]{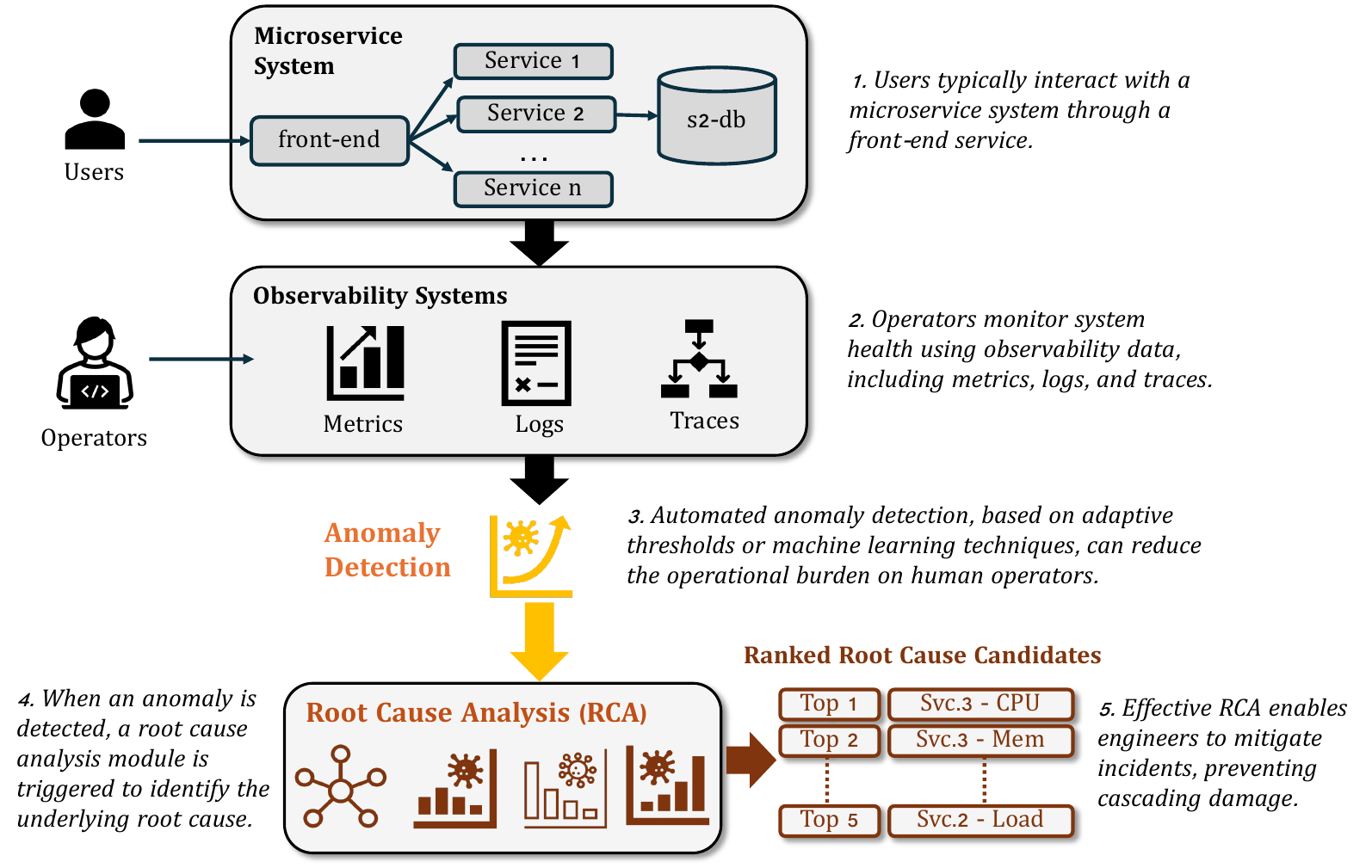}
\caption{Anomaly Detection and Root Cause Analysis for Microservice Systems.}
\label{fig:intro-ad-rca}
\end{figure}

Automated anomaly detection and RCA for failures in microservice systems have been an active research area in recent years, yielding many promising techniques~\citep{Azam2022rcd, Li2022Circa, lee2023eadro, xin2023causalrca, orchard2025root, budhathoki2022causal, zhang2025adaptive}. Their central idea is to employ different methods to analyse \textit{observability data} (including metrics, logs, and traces, see Section~\ref{chap:background} for their definition) to detect anomalies and diagnose their corresponding root causes (see Figure~\ref{fig:intro-ad-rca}). For example, the N-Sigma rule has been used and discussed in MicroRank~\citep{yu2021microrank}, CIRCA~\citep{Li2022Circa}, and Eadro~\citep{lee2023eadro} to detect anomalies in metric data. It assumes the normal metric data follow a Gaussian distribution, so any datum that deviates significantly from the mean is considered as an anomaly (e.g.,~\(x>\mu+3\cdot\sigma\)). BIRCH~\citep{zhang1996birch} is employed in MicroRCA~\citep{wu2020microrca} to detect anomalies, assuming that normal data can be grouped into one cluster represented by a centroid and radius, hence, any new datum that falls far outside that cluster is considered an anomaly. For RCA, RCD~\citep{Azam2022rcd} uses $\Psi$-PC~\citep{Jaber2020PsiFCI} to construct a causal graph from metrics data to derive the root cause, assuming that the root cause affects many metrics. Similarly, CIRCA~\citep{Li2022Circa} constructs a causal graph based on domain knowledge and applies hypothesis testing to infer the root cause from the graph. Several other methods~\citep{xin2023causalrca, liu2023pyrca, pan2021dycauserca} also explore different techniques to analyse observability data for automated anomaly detection and RCA.

However, existing studies come with several noteworthy limitations that fall into two categories. \textbf{First, there are methodological gaps in existing anomaly detection and RCA approaches.} Most studies (if not all) treat anomaly detection and RCA separately, assuming that anomalies are correctly segmented and do not affect RCA performance~\citep{orchard2025root, altenbernd2025amocrca, lee2023eadro, xin2023causalrca, Li2022Circa, Azam2022rcd, Soldani2022rcasurvey, Ma2019Msrank}. These techniques may fail when anomalies are imprecisely segmented due to noise or delays from automated detectors in real-world systems. Thus, it remains unclear whether, when combined with existing anomaly detection methods that may provide imprecise output, these RCA approaches are still effective in localising the root cause. Additionally, most existing studies focus on metrics, logs, and traces data, leaving event data underexplored~\citep{chen2022adaptive, Azam2022rcd, Li2022Circa, pham2024baro, xin2023causalrca, orchard2025root}. While event data may capture critical information about the manifestation of anomalies and support RCA, it remains unclear whether it can be effectively leveraged for automated techniques~\citep{aws_cloudtrail_events}. Furthermore, many RCA techniques require a predefined service call graph for diagnosis~\citep{lee2023eadro, Li2022Circa, rouf2024instantops, xie2024tvdiag, zhang2023diagfusion}. Without a service call graph, these approaches fail to perform any level of diagnosis.
\textbf{Second, there are evaluation gaps that hinder research progress.} The lack of standardised benchmarks with comprehensive datasets and evaluation protocols has made it difficult to fairly compare different RCA approaches and assess their generalisability~\citep{cheng2023ai}. Moreover, several RCA methods rely on causal discovery to construct a causal graph as a proxy for diagnosing failures when the service call graph is unavailable~\citep{Azam2022rcd, pham2024baro}. However, these methods are often evaluated on synthetic datasets or limited systems, raising questions about their generalisability to diverse and realistic scenarios.

\section[Research Approach: Intelligent Analysis]{Research Approach: Intelligent Analysis through Multimodal Observability Data}

Recent advances in machine learning and statistical modelling have opened up new possibilities for automating anomaly detection and RCA in microservice systems. While manual monitoring and diagnosis are infeasible, and existing automated techniques suffer from the limitations discussed above, this thesis asks: \textit{Can these failures be detected and diagnosed more effectively from multimodal observability data using machine learning tools?}

Multimodal observability data refer to the diverse yet complementary data modalities that collectively describe the operational state and behavioral dynamics of a microservice system. These modalities (including metrics, logs, traces, and events) capture system behaviour from distinct perspectives and at varying granularities. Metric data provide structured, quantitative time-series measurements such as CPU usage, latency, or request rate, offering a high-level, continuous view of system health. Log data, in contrast, contain unstructured or semi-structured textual records that describe discrete events or system activities, including warnings, errors, and application-level messages, thus enabling fine-grained contextual analysis during failure diagnosis. Trace data capture the causal relationships among distributed service calls by linking spans through trace and span identifiers, reflecting the end-to-end execution flow of requests across the system. Event data document system activities, such as API calls, infrastructure changes, and security actions. These event records offer valuable context for post-mortem analysis, explaining why anomalies are detected in other observability data modalities. The combination of these heterogeneous data types enables a richer, more comprehensive understanding of system behaviour, allowing machine learning models to characterise normal system behaviour and, therefore, detect anomalies and identify their root causes.


\begin{tcolorbox}[colback=gray!5, colframe=gray!40,
   left=5pt, right=5pt, top=5pt, bottom=5pt]
\textbf{Approach:} This thesis explores these heterogeneous sources of observability data both \textit{independently} and \textit{collectively} for automated anomaly detection and RCA. We start with the exploration of metric and event data for automated anomaly detection and RCA, ensuring that the framework can operate robustly with limited available data sources. We then explore multimodal telemetry data for RCA, ensuring high fault diagnosis coverage. For all methods developed, we employ unsupervised machine learning techniques, given the unpredictable nature of failures in microservice systems. Since failure data are difficult to collect and quickly become outdated as issues are fixed and new failures emerge, our methods are designed to work without extensive labelled data. They also emphasise speed, accuracy, and interpretability, avoiding opaque features that engineers cannot easily trust or act upon. \textbf{Together, these methods form a cohesive and generalisable framework for automated anomaly detection and root cause analysis across diverse observability modalities.}
\end{tcolorbox}

This thesis makes contributions through two groups. \textbf{The first group introduces novel methods for automated anomaly detection and RCA.} We begin with the development of \barotool{} (Chapter~\ref{chap:fse24}), an end-to-end anomaly detection and RCA framework for metric data. We show that imprecise anomaly detection results can significantly affect RCA performance, a problem overlooked by existing studies. To address this, we propose RobustScorer, which robustly diagnoses the root cause of failures even in the presence of imprecise anomaly detection results. Extending beyond metrics, we propose \eventadl{} (Chapter~\ref{chap:eventadl}), the first open-box anomaly detection and localisation framework that captures semantic and frequency patterns from event data to detect anomalies efficiently. The tool is built upon insights gathered from the analysis of 520 real-world incidents at Amazon Web Services. To enhance interpretability, \eventadl{} is equipped with a root cause localisation layer (i.e., RCA) that identifies root causes and explains why anomalies are detected. To further enhance diagnostic capability, we develop \torai{} (Chapter~\ref{chap:torai}), a multimodal RCA method capable of diagnosing failures using available observability data (metrics, logs, and traces), even with incomplete trace coverage, addressing the issue of blind spots in modern microservice systems, which has been largely overlooked by prior research.
\textbf{The second group contributes benchmarking datasets, evaluation framework, and systematic evaluation efforts.} To promote reproducibility and standardised evaluation, we introduce \rcaeval{} (Chapter~\ref{chap:www25}), an open-source benchmark providing comprehensive datasets (where widely used benchmark microservice systems are deployed to collect first-hand monitoring data) and standardised evaluation protocols for RCA research. Building on this benchmark, we conduct a systematic evaluation of existing metric-based RCA methods (Chapter~\ref{chap:ase24}) to identify their strengths, weaknesses, and practical limitations. Together, these contributions form a cohesive and comprehensive framework for automated anomaly detection and RCA that operates effectively across multiple data modalities.

\section{Contributions and Organisation}

This thesis contributes a unified set of techniques and resources that advance automated anomaly detection and root cause analysis for microservice systems. Together, these contributions deliver a robust, cohesive, and generalisable framework that operates across metric, event, and multimodal telemetry data, and provide standardised tools for evaluating RCA methods. Specifically, this thesis makes contributions through two groups. Each contribution is backed by a peer-reviewed publication at a CORE~A* ranked venue, as indicated below, and the full list of publications is provided in Appendix~A.
\textbf{The first group introduces novel methods for automated anomaly detection and RCA}, exploring multiple sources of observability data both independently and collectively:

\begin{enumerate}[font=\large\bfseries]
    \item \textbf{\barotool{}: A Metric-based Anomaly Detection and RCA Method.} We propose \barotool{}, an end-to-end framework that jointly performs anomaly detection and RCA using multivariate time-series metrics. \barotool{} employs Multivariate Bayesian Online Change Point Detection for anomaly detection together with RobustScorer, a non-parametric hypothesis testing approach for robust root cause localisation. Experiments on three benchmark systems show that \barotool{} consistently outperforms state-of-the-art baselines. This work has been published at the ACM International Conference on the Foundations of Software Engineering (FSE 2024)~\citep{pham2024baro}.

    \item \textbf{\eventadl{}: An Event-based Anomaly Detection and RCA Method.} We introduce \eventadl{}, the first open-box event-based anomaly detection and localisation framework. \eventadl{} models normal event behaviours using Event Semantic Patterns and Event Frequency Patterns, and constructs an \textit{Intervention Graph} for interpretable RCA. Evaluation on real-world systems demonstrates superior accuracy and interpretability compared with existing state-of-the-art methods. This work has been published at FSE 2026~\citep{pham2026eventadl}.

    \item \textbf{\torai{}: A Multimodal RCA Method.} We propose \torai{}, a novel unsupervised multimodal RCA approach that does not rely on full trace coverage. \torai{} integrates multiple telemetry sources, performs severity clustering, and applies causal ranking to identify fine-grained root causes, even in the presence of blind spots. This work has been published at FSE 2026~\citep{pham2026torai}.
\end{enumerate}

\noindent \textbf{The second group contributes benchmarking datasets, evaluation framework, and systematic evaluation efforts} that pave the way for future research on RCA:

\begin{enumerate}[font=\large\bfseries, resume]
    \item \textbf{\rcaeval{}: A Comprehensive Benchmark for RCA in Microservice Systems.} We develop \rcaeval{}, the first open-source benchmark with three large-scale datasets (735 failure cases) and fifteen reproducible baselines. \rcaeval{} provides a standardised evaluation environment for coarse- and fine-grained RCA methods. This work has been published at the ACM Web Conference (WWW 2025)~\citep{pham2025rcaeval}.

    \item \textbf{An Evaluation of Causal Inference-based RCA Methods.} We conduct a comprehensive empirical study of nine causal discovery methods and twenty-one failure diagnosis approaches to assess their effectiveness, efficiency, and robustness. The results reveal that no single method performs best across all settings, highlighting opportunities for further research and methodological refinement. This work has been published at the IEEE/ACM International Conference on Automated Software Engineering (ASE 2024)~\citep{pham2024root}.
\end{enumerate}

These five contributions are components of a single research programme. 
The three methods progress along two axes, namely the observability modality they consume (metrics for \barotool{}, events for \eventadl{}, and metrics, logs, and traces collectively for \torai{}) and the assumption each one removes (precise anomaly segmentation, the manifestation of anomalies in numeric telemetry, and the availability of a complete service call graph). Each method therefore widens the range of systems that can be diagnosed by relaxing a requirement that the preceding one still imposes. The methods also share a common design contract, being unsupervised, fast enough for online use, and producing a ranked list of candidate root causes together with the evidence supporting that ranking. This shared contract is what allows them to be composed, since an operator can deploy whichever subset matches the telemetry a given system actually emits and compare or combine the resulting rankings. The benchmarking contributions then close the loop around the methods. The systematic evaluation in Chapter~\ref{chap:ase24} establishes where existing techniques break down and produced the RE1 dataset, \rcaeval{} (Chapter~\ref{chap:www25}) generalises that evaluation into a reusable benchmark by adding the RE2 and RE3 datasets together with standardised protocols and reproducible baselines, and the proposed methods are released through \rcaeval{} itself, with \barotool{} distributed as one of its baselines and \torai{} integrated into the same codebase. Taken together, they form a cohesive framework that spans the diagnostic pipeline from raw multimodal telemetry to a ranked and explained root cause, and that supplies the infrastructure needed to measure any method placed inside it.

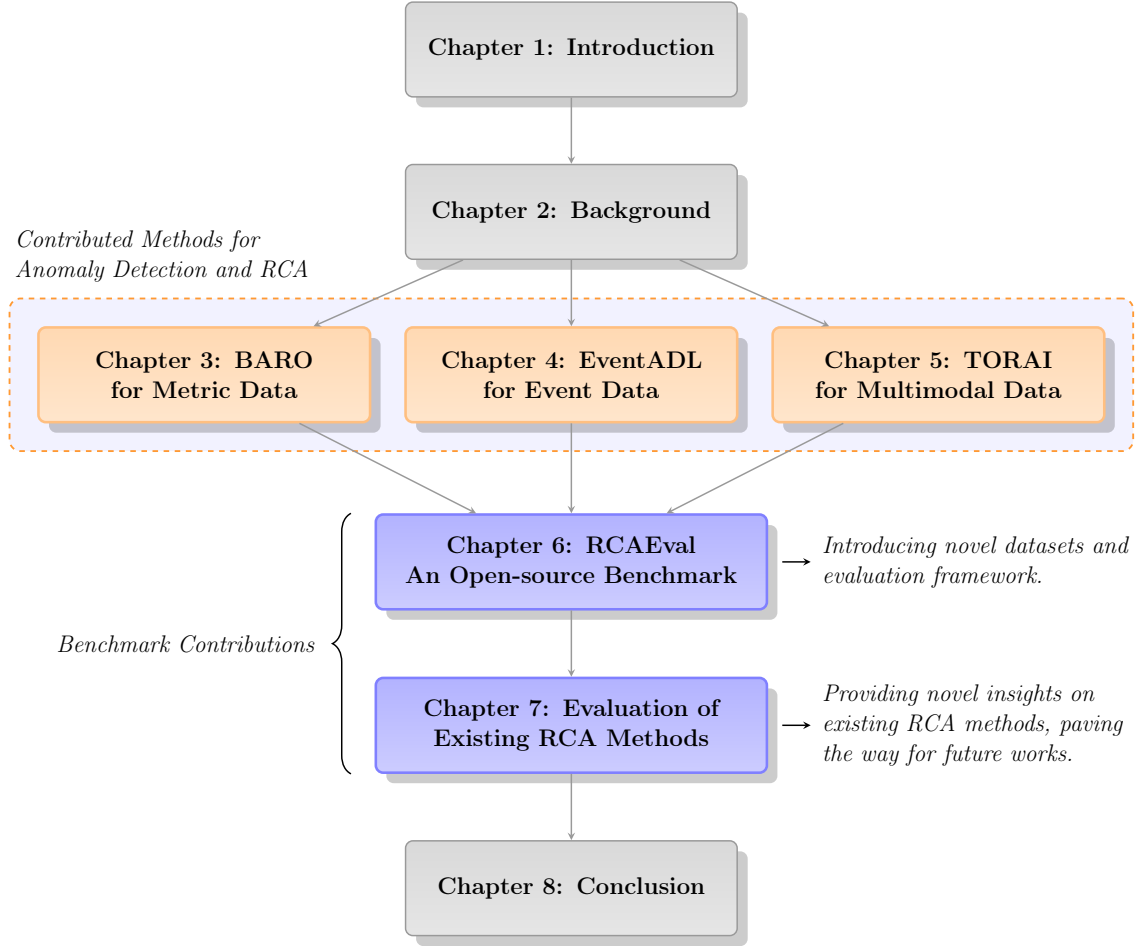
\begin{figure}[ht]
\vspace{20pt}
\centering
\resizebox{\textwidth}{!}{%
    \input{figures/common/thesis-organization.tex}%
}
\vspace{10pt}
\caption{Thesis Organisation.}
\label{fig:thesis-organisation}
\end{figure}

The remainder of this thesis is organised as follows: \textbf{Chapter~\ref{chap:background}} introduces the background on microservice architectures, multimodal observability data, and causal inference methods, and formulates the anomaly detection and RCA problems. \textbf{Chapter~\ref{chap:fse24}} introduces \barotool{}, a metric-based end-to-end anomaly detection and RCA method. \textbf{Chapter~\ref{chap:eventadl}} presents \eventadl{}, an open-box event-based anomaly detection and RCA framework that constructs interpretable intervention graphs. \textbf{Chapter~\ref{chap:torai}} introduces \torai{}, a multimodal RCA approach capable of operating under incomplete trace coverage. \textbf{Chapter~\ref{chap:www25}} presents RCAEval, a RCA benchmark for microservice systems.  \textbf{Chapter~\ref{chap:ase24}} presents a systematic evaluation of causal inference-based RCA techniques, identifying key insights that motivate subsequent research. \textbf{Chapter~\ref{chap:conclusion}} concludes the thesis and discusses potential future research directions. Figure~\ref{fig:thesis-organisation} illustrates the organisation of this thesis.

%% file: figures/common/thesis-organization.tex
\begin{tikzpicture}[
    node distance=1.4cm and 0.1cm,
    base/.style={
        rectangle,
        draw,
        thick,
        align=center,
        font=\Large\bfseries,
        minimum height=2cm,
        minimum width=7cm,
        rounded corners=4pt,
        drop shadow={
            shadow xshift=0.2cm,
            shadow yshift=-0.2cm,
            opacity=0.25,
            fill=black!80
        }
    },
    box/.style={
        base,
        text width=6.25cm,
        line width=0.8pt,
        top color=gray!30,
        bottom color=gray!20,
        draw=gray!80
    },
    methodbox/.style={
        base,
        text width=5.5cm,
        line width=1.4pt,
        top color=orange!30,
        bottom color=orange!20,
        draw=orange!50,
    },
    benchmarkbox/.style={
        base,
        text width=8cm,
        line width=1.4pt,
        top color=blue!30,
        bottom color=blue!20,
        draw=blue!50,
    },
    arrow/.style={
        ->,
        thick,
        >=stealth,
        color=gray!80
    },
    scale=0.9,
    transform shape
]

\node[box] (ch1) {Chapter 1: Introduction};
\node[box, below=of ch1] (ch2) {Chapter 2: Background};

\node[methodbox, below=1.4cm of ch2] (ch4) {Chapter 4: EventADL\\for Event Data};
\node[methodbox, left=0.7cm of ch4] (ch3) {Chapter 3: BARO\\for Metric Data};
\node[methodbox, right=0.7cm of ch4] (ch5) {Chapter 5: TORAI\\for Multimodal Data};

\node[benchmarkbox, below=of ch4, yshift=-0.5cm] (ch6) {Chapter 6: RCAEval\\An Open-source Benchmark};
\node[benchmarkbox, below=of ch6] (ch7) {Chapter 7: Evaluation of\\Existing RCA Methods};
\node[box, below=of ch7] (ch8) {Chapter 8: Conclusion};

\draw[arrow] (ch1) -- (ch2);
\draw[arrow] (ch2) -- (ch3);
\draw[arrow] (ch2) -- (ch4);
\draw[arrow] (ch2) -- (ch5);
\draw[arrow] (ch3) -- (ch6);
\draw[arrow] (ch4) -- (ch6);
\draw[arrow] (ch5) -- (ch6);
\draw[arrow] (ch6) -- (ch7);
\draw[arrow] (ch7) -- (ch8);

\begin{scope}[on background layer]
    \node[
        rectangle,
        draw=orange!80,
        dashed,
        line width=1pt,
        rounded corners=6pt,
        fill=blue!5,
        fit=(ch3)(ch4)(ch5),
        inner sep=15pt
    ] (methodgroup) {};

    \node[font=\Large\itshape, align=left, anchor=south west] 
    at ([yshift=2mm]methodgroup.north west) 
    {Contributed Methods for \\
    Anomaly Detection and RCA};
\end{scope}

\draw[decorate, decoration={brace, amplitude=10pt, mirror}, thick]
    ([xshift=-0.5cm]ch6.north west) -- ([xshift=-0.5cm]ch7.south west)
    node[left=18pt, midway, font=\Large\itshape] {Benchmark Contributions};

\draw[arrow, color=black] ($(ch6.east)+(0.3cm,0)$) -- ++(0.6cm,0) 
    node[pos=1, right=1mm, font=\Large\itshape, align=left, text=black] 
    {Introducing novel datasets and \\evaluation framework.};

\draw[arrow, color=black] ($(ch7.east)+(0.3cm,0)$) -- ++(0.6cm,0) 
    node[pos=1, right=1mm, font=\Large\itshape, align=left, text=black] 
    {Providing novel insights on \\existing RCA methods, paving \\the way for future works.};

\end{tikzpicture}

%% file: chapters/2.background.tex
\chapter{Background} \label{chap:background}

\section{Microservice System}

\subsection{Characteristics}

\begin{figure}
    \vspace{-10pt}
    \centering
    \includegraphics[width=\textwidth]{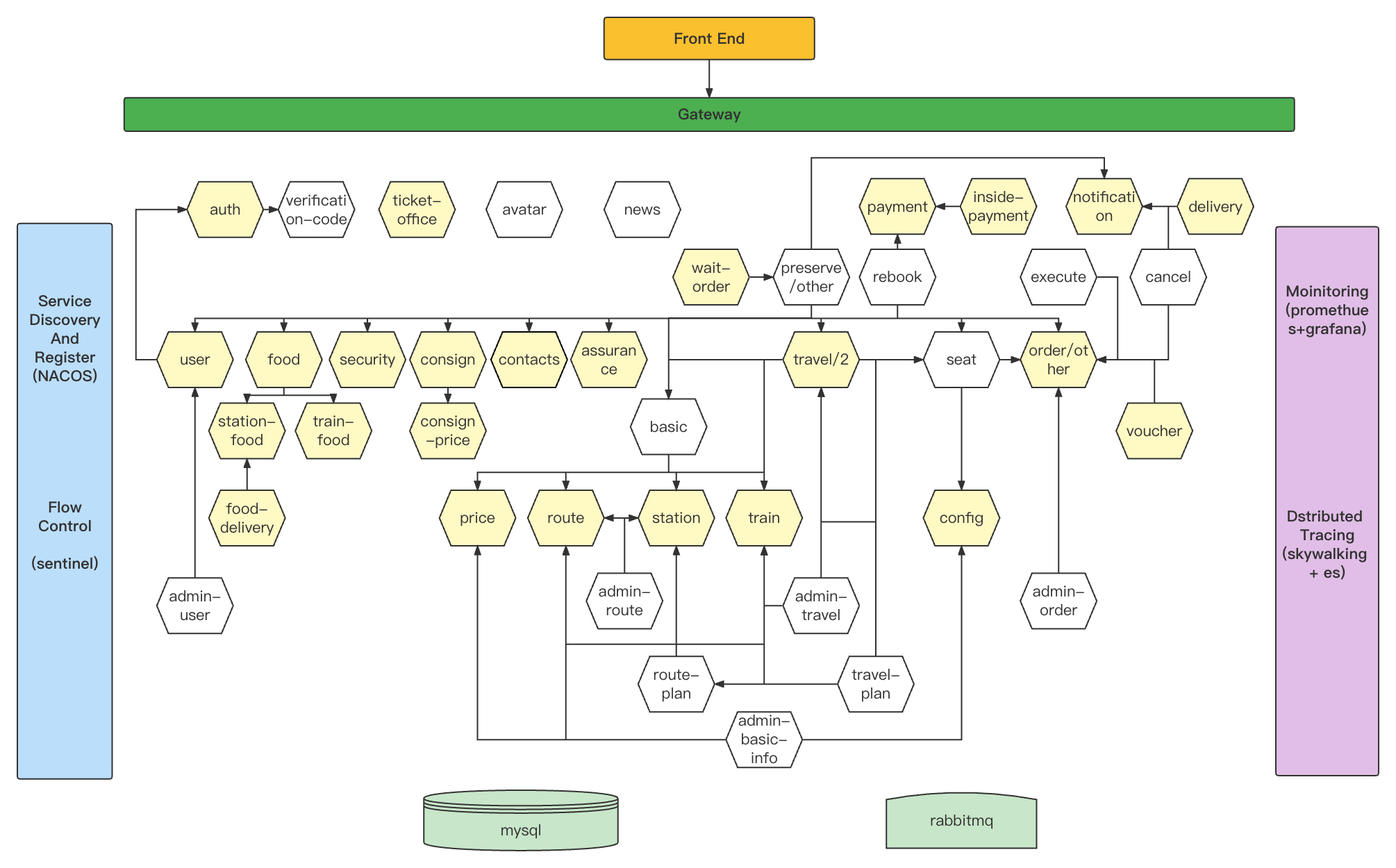}
    \vspace{-10pt}
    \caption[The architecture of the Train Ticket microservice system.]{The architecture of the Train Ticket microservice system~\citep{zhou2018trainticket}.} \label{fig:train-ticket-arch}
    \vspace{-10pt}
\end{figure}

Microservices have emerged as a dominant paradigm for developing large-scale cloud applications. Unlike monolithic architectures, which encapsulate all functionalities within a single deployable unit, microservices decompose the system into a set of small, independently deployable services, each responsible for a specific capability. These services communicate with each other through APIs (e.g., REST, gRPC) and are often managed by orchestration platforms such as Kubernetes or Docker Compose. The key characteristics of microservice systems include service autonomy, loose coupling, independent deployment, and scalability. Figure~\ref{fig:train-ticket-arch} provides an architectural overview of the Train Ticket system~\citep{zhou2018trainticket}, a well-established benchmark system for anomaly detection and RCA. 

\subsection{Telemetry Data}

Telemetry data is the cornerstone for ensuring the reliability and high availability of microservice systems. In practice, there are typically \textit{monitoring systems} deployed within or near the same infrastructure as the microservice systems to collect and analyse telemetry data. There are three major types of telemetry data: metrics, logs, and traces. Recently, events have also gained attention as a new telemetry source. These heterogeneous data sources provide complementary perspectives on the system’s runtime behaviour.

\subsubsection{Metric Data}

Metrics are numerical measurements sampled periodically to capture the quantitative state of system components. Typical examples include workload, CPU usage, memory consumption, latency, and error rate (see Figure~\ref{fig:metric_data_sample}). Metric data are generally stored as time-series and provide a high-level overview of system health over time. Metrics are widely used for anomaly detection due to their structured nature and statistical regularity. Approaches such as \nsigma, change-point detection, or clustering are often employed to identify deviations from normal behaviour~\citep{yu2021microrank, Li2022Circa, pham2024baro, wu2020microrca}. However, metrics alone provide limited context, making it difficult to infer the exact cause of failures without complementary data such as logs or traces.

\input{figures/background/metric_data_sample}

\subsubsection{Log Data}

Logs are unstructured or semi-structured textual messages produced by applications, middleware, and system components. Each log entry typically contains a timestamp, message content, and optionally severity level (e.g., INFO, WARN, ERROR), see Figure~\ref{fig:log_data_sample}. Logs record system and application states such as user request started, unexpected exceptions, and execution states, providing detailed insight into the internal execution flow of services. Log analysis plays a crucial role in anomaly detection and RCA. Techniques such as log parsing~\citep{he2017drain, le2023log}, and log-based anomaly detection~\citep{le2021neurallog, du2017deeplog} have been applied to mine useful data from logs. Nevertheless, the large volume, unstructured format, and evolving patterns of log data make automated analysis challenging in dynamic microservice environments.

\input{figures/background/log_data_sample}

\subsubsection{Trace Data}

Traces capture end-to-end request flows across services, see Figure~\ref{fig:trace_data_sample}. Each trace is composed of multiple spans, where each span records the start and end times of a single operation, along with metadata such as service name, endpoint, and status. By linking spans through trace and span identifiers, traces reveal causal relationships among distributed calls. Distributed tracing has gradually become a standard in microservice observability. However, the collection of traces usually requires an intrusive approach~\citep{shen2023deepflow}, i.e., the source code of the microservices must be modified to create traces with a distributed tracing framework and to assign the correct \texttt{trace\_id}~\citep{giamattei2023monitoring}. This process is complex and resource-demanding, causing difficulties in anomaly detection and RCA with this data source. We will discuss this further in Chapter~\ref{chap:torai}.

\input{figures/background/trace_data_sample}

\subsubsection{Event Data} \label{sec:event-data}

Events can be regarded as a specialized form of \textit{structured} logs that follow standardised schemas~\citep{opentelemetry_events_2025}. 
An event is a structured record that captures at least four key attributes: the \texttt{actor} (who performed the action), the \texttt{operation} (what action was performed), the \texttt{resources} (what was acted upon), and the \texttt{timestamp} (when the event occurred). Events may also include auxiliary attributes such as execution parameters, contextual metadata, or error indicators. Compared to unstructured logs, which require parsing for analysis~\citep{he2017drain, brianlogsurvey2025}, events have a known schema. Events are standardised in both open-source~\citep{opentelemetry_events_2025, ocsf} and commercial~\citep{aws_cloudtrail_events, azure_monitor_eventhub_ingestion, google_cloud_audit_logs} platforms. Figure~\ref{fig:event_data_sample} shows a representative event following the OCSF schema~\citep{ocsf}. Despite their diagnostic potential, event data have been underexplored in anomaly detection and RCA research.

\input{figures/background/event_data_sample}

\subsection{Faults and Failures} \label{sec:terms-fault-failure}

\textit{Failures} denote the incapacity of a service to perform its functions~\citep{Soldani2022rcasurvey}. \textit{Faults} correspond to the root causes of such failures (e.g., CPU overload, memory leaks, or network disconnections)~\citep{Soldani2022rcasurvey, avizienis2004basic}. \textit{Anomalies} are defined as observable symptoms of failures~\citep{Soldani2022rcasurvey}. \textit{Root cause analysis (RCA)} is the process of identifying why a failure has occurred~\citep{lee2023eadro}.
RCA entails a comprehensive examination of multi-sources telemetry data (i.e., metrics, logs, and traces) to derive \textit{coarse-grained root causes} (i.e., root cause services), and \textit{fine-grained root causes} (i.e., root cause indicators).
The system operators can use these suggested root cause services and indicators (e.g., specific metrics, logs, or traces) to identify the underlying root cause of the failures. The use of terminologies aligns with existing RCA works~\citep{Azam2022rcd, Li2022Circa, xin2023causalrca}.

\section{Problem Formulation} \label{sec:problem-formulation}

Let $\mathcal{S}$ denote a microservice system consisting of $n$ services $\{s^i\}_{i=1}^{n}$. Each service operates as an independent component, communicating with other services through APIs and collectively providing the system's functionality. The system continuously emits observability data in the form of metrics, events, logs, and traces, which reflect its runtime behaviour and serve as the foundation for anomaly detection.

\textbf{Metrics.} For each service $s^i$, the monitoring system collects $m$ time-series metrics at each time step $t$. These metrics capture quantitative measurements such as CPU usage, memory consumption, latency, and error rates. The metric-based formulation is primarily employed in Chapter~\ref{chap:fse24}, where we develop anomaly detection and root cause analysis methods based on time-series metric data.

\textbf{Events.} Let $\mathcal{E} = \langle e_1, e_2, \dots \rangle$ denote the stream of discrete events, where each event
\(
e_i = (\Lambda_i,\, \Omega_i,\, \mathcal{R}_i,\, \tau_i)
\)
contains the \textit{actor} $\Lambda_i$ (who performed the action), the \textit{operation} $\Omega_i$ (what action was performed), the affected \textit{resources} $\mathcal{R}_i$ (what was acted upon), and the \textit{timestamp} $\tau_i$ (when the event occurred). The event-based formulation is employed in Chapter~\ref{chap:eventadl}, where we develop an event-based anomaly detection and localisation framework.

\textbf{Logs and Traces.} Logs $\mathcal{L}_t$ capture textual records of system and application states, including execution flow, error messages, and debugging information. Distributed traces $\mathcal{T}_t$ record end-to-end request flows across services, with each trace composed of multiple spans that capture the timing and metadata of individual operations. These data sources are particularly valuable for understanding request propagation and service interactions. The log and trace formulations are employed in Chapter~\ref{chap:torai}, where we develop multimodal root cause analysis methods that leverage these complementary data sources.

\textbf{Multimodal Data.} At timestamp $t$, a window of recent multimodal observations
\[
\mathcal{W}_t = \{ \mathcal{M}_{t-\Delta:t},\, \mathcal{E}_{t-\Delta:t},\, \mathcal{L}_{t-\Delta:t},\, \mathcal{T}_{t-\Delta:t} \},
\]
is given, where $[t-\Delta:t]$ defines a window of size $\Delta$. This unified representation captures the heterogeneous nature of observability data, enabling methods that operate on individual modalities or integrate multiple data sources for comprehensive system monitoring.

\subsection{Anomaly Detection} \label{sec:problem-formulation-ad}

The task of \textbf{Anomaly Detection} is to determine whether the system exhibits abnormal behaviour within an observation window $\mathcal{W}_t$. Formally, it aims to learn a function
\[
f_{\text{AD}}: \mathcal{W}_t \rightarrow y, \quad y \in \{0, 1\},
\]
where $y = 1$ indicates the presence of anomalies and $y = 0$ indicates otherwise. Depending on the type of input data, $f_{\text{AD}}$ may operate on metrics (metric-based anomaly detection), events (event-based anomaly detection), or a combination of modalities (multimodal anomaly detection). Critically, to determine whether the current observation window $\mathcal{W}_t$ is anomalous, the anomaly detector must first establish a baseline of normal behaviour through a period of collected normal data $\mathcal{W}_{\text{normal}}$. An anomaly is then declared when the observed data deviates significantly from this learned normal pattern. If $y=1$, the system may trigger the RCA process to localise the root cause of the detected anomaly.

\subsection{Root Cause Analysis} \label{sec:problem-formulation-rca}

When an anomaly is detected ($y = 1$), the goal of \textbf{Root Cause Analysis} is to identify the underlying cause(s) responsible for the abnormal behaviour. Formally, RCA operates over a possibly extended observation window $\mathcal{W}'_t$ that may include additional temporal context before or after the detected anomaly.
RCA aims to identify a subset of causal factors
\[
\mathcal{C}_t \subseteq \mathcal{W}'_t
\]
that are responsible for the observed failure. These causal factors can correspond to faulty services, anomalous metrics, erroneous events, or abnormal log/tracing patterns. Accordingly, RCA can be instantiated in different forms:
\begin{itemize}
    \item \textbf{Metric-based RCA}, which identifies the root cause services and the corresponding root cause metrics, i.e., metrics indicating root causes;
    \item \textbf{Event-based RCA}, which identifies the root cause events or operations that triggered the observed failure; and
    \item \textbf{Multimodal RCA}, which integrates information from metrics, logs, and traces to achieve finer-grained and more robust localisation.
\end{itemize}

Formally, the RCA process can be defined as learning a mapping
\[
f_{\text{RCA}}: (\mathcal{W}'_t, y) \rightarrow \mathcal{C}_t,
\]
where $\mathcal{C}_t = \{ c_1, c_2, \dots, c_k \}$ represents the set of root cause entities. Each $c_i$ may correspond to a  \textit{coarse-grained root cause} (e.g., a specific service or entity) or a \textit{fine-grained root cause} (e.g., a specific metric or event).

The final output of the overall pipeline is the tuple $(y,\, \mathcal{C}_t)$, where $y$ denotes whether the system is anomalous and $\mathcal{C}_t$ specifies the identified root cause(s). This formulation provides a unified view that supports metric-based, event-based, and multimodal RCA.

\section{Limitations of Existing Approaches}

This section identifies critical limitations in existing approaches to anomaly detection and root cause analysis for microservice systems, organised into two categories. \textbf{The first category concerns methodological gaps} in current AD and RCA techniques. First, the decoupling of anomaly detection from RCA can lead to cascading inaccuracies when imperfect detection results propagate to the localisation phase. Second, the limited utilisation of event data represents a missed opportunity to leverage structured diagnostic information capturing actor-resource interactions. Third, the assumption of full trace coverage ignores the practical reality of blind spots in microservice deployments. \textbf{The second category concerns evaluation gaps.} First, the absence of standard benchmarks with comprehensive datasets has led to inconsistent evaluations. Second, existing causal inference-based RCA methods lack systematic assessment of their effectiveness and robustness across diverse scenarios. Collectively, these limitations restrict the effectiveness, applicability, and reliability of current RCA methods in real-world environments.

\subsection{Decoupling Anomaly Detection from RCA} \label{sec:limit-ad-decoupling}

A typical failure troubleshooting pipeline for microservice systems consists of two sequential phases: anomaly detection followed by root cause analysis~\citep{lee2023eadro}. Most existing RCA research works~\citep{Soldani2022rcasurvey, Li2022Circa, Azam2022rcd, Jinjin2018Microscope, Wang2018cloudranger, Ma2019Msrank, wu2020microrca, wu2021microdiag, xin2023causalrca} focus solely on the RCA phase while assuming the existence of an anomaly detection module that can accurately detect failures and trigger the RCA process when anomalies occur. This decoupling approach implicitly assumes that the anomaly detection component provides accurate information, such as the precise failure occurrence time and the specific services or metrics exhibiting anomalous behaviour.

However, in practice, many RCA methods either treat anomaly detection as an independent task or rely on overly simplistic anomaly detection techniques that may not provide accurate results. For instance, MicroRCA~\citep{wu2020microrca} and MicroDiag~\citep{wu2021microdiag} employ the BIRCH clustering technique~\citep{zhang1996birch} for detecting anomalies, while MicroScope~\citep{Jinjin2018Microscope} and Ms-Rank~\citep{Ma2019Msrank} use the basic N-Sigma rule (the three-sigma rule of thumb). Similarly, commercial monitoring platforms such as DataDog~\citep{datadog} and Dynatrace~\citep{dynatrace} predominantly rely on threshold-based anomaly detection techniques for univariate time series data~\citep{dynatrace1, datadog2}, which may struggle with the complexity of multivariate metrics in modern microservice systems.

More critically, some RCA approaches explicitly require specific information from the anomaly detection module. For example, CIRCA~\citep{Li2022Circa} and RCD~\citep{Azam2022rcd} specifically require the failure occurrence time as input to their RCA algorithms. These methods assume that this temporal information is provided accurately by the anomaly detection component. However, it remains unclear whether these RCA approaches can maintain their effectiveness when combined with existing anomaly detection methods that may provide imprecise or delayed failure detection, leading to inaccurate failure times.

This decoupling of anomaly detection and RCA presents a significant challenge: inaccurate anomaly detection results can cascade through the pipeline and substantially degrade root cause localisation performance. When the anomaly detection module provides incorrect failure times, misidentifies affected services, or generates false positives, the downstream RCA module operates on flawed assumptions, potentially leading to incorrect root cause identification. Despite the critical nature of this issue, most existing works do not evaluate their RCA methods under realistic conditions where anomaly detection may be imperfect. This gap motivates the need for end-to-end approaches that either integrate both phases more tightly or design RCA techniques that are robust to anomaly detection inaccuracies.

\subsection{Underexplored Event Data for Anomaly Detection and RCA} \label{sec:limit-event-data}

As discussed in Section~\ref{sec:event-data}, event data represents a valuable telemetry source in modern cloud-based systems, capturing structured records of system operations including the actor (who performed the action), the operation (what was done), the resources (what was affected), and the timestamp (when it occurred). Major cloud providers offer comprehensive event monitoring services, such as AWS CloudTrail~\citep{aws_cloudtrail_events}, Azure Event Hub~\citep{azure_monitor_eventhub_ingestion}, and Google Cloud Audit~\citep{google_cloud_audit_logs}. Despite this widespread availability and the rich diagnostic potential of event data, anomaly detection and localisation (ADL) for events remains significantly underexplored compared to metrics and logs. While metric-based ADL~\citep{xin2023causalrca, pham2024baro, Azam2022rcd, Li2022Circa, chen2022adaptive} and log-based ADL~\citep{brianlogsurvey2025, le2021neurallog, du2017deeplog} have been extensively studied, event-based ADL has received comparatively limited attention in research.

Existing metric-based and log-based approaches are insufficient for comprehensively analysing event data due to fundamental differences in data characteristics. Metric-based methods, which primarily analyse time-series numerical data, are effective at detecting frequency-based anomalies (e.g., sudden spikes or drops in event counts) but fail to detect pointwise anomalies, where a single anomalous event (such as an unauthorized resource deletion) may be the root cause of a system failure. Log-based methods~\citep{du2017deeplog, le2021neurallog}, while effective for analysing unstructured or semi-structured textual messages, often ignore the structured information readily available in event data. In particular, log-based approaches typically do not explicitly model the actor-resource interaction patterns that are central to event data, such as which users or services performed operations on which resources, thereby missing critical diagnostic information that could aid in both anomaly detection and root cause localisation.

The few existing event-based anomaly detection methods can be categorised into closed-box and open-box approaches, both of which have significant limitations. Closed-box approaches~\citep{zengy2022shadewatcher, coskun2022detecting, amin2019cadence} rely heavily on deep neural networks to identify anomalies. For instance, ShadeWatcher~\citep{zengy2022shadewatcher} employs context-aware embeddings and graph neural networks to detect anomalous interactions, while GuardDuty~\citep{coskun2022detecting} uses variational autoencoders to identify anomalies based on reconstruction error. While these methods can achieve high detection accuracy, they suffer from a critical drawback: they provide anomaly scores without interpretable explanations of why an anomaly was detected, necessitating further manual investigation by operators to understand the underlying issues. On the other hand, HyGLAD~\citep{hyglad} represents an open-box approach that provides interpretable anomaly detection through graph-based pattern matching. However, HyGLAD focuses solely on anomaly detection and lacks integrated root cause localisation capabilities, leaving operators to manually trace the causes of detected anomalies. Furthermore, HyGLAD is designed to detect only pointwise anomalies and does not handle frequency-based anomalies, which are prevalent in event data streams.

This limited utilisation of event data in ADL research presents a significant gap. There is a clear need for comprehensive, interpretable event-based ADL frameworks that can: (1) detect both pointwise and frequency-based anomalies, (2) automatically localise root causes rather than merely flagging anomalies, and (3) provide interpretable explanations that enable operators to understand both what anomalies occurred and why they occurred. Such frameworks would leverage the structured nature of event data to provide more precise and actionable diagnostic information for cloud-based systems, complementing existing metric-based and log-based approaches.

\subsection{Existing Multi-source RCA Methods Omit Blind Spots} \label{sec:limit-blind-spots}

Most existing multi-source RCA methods~\citep{yu2023nezha, lee2023eadro, zhang2023diagfusion, hou2021pdiagnose, rouf2024instantops, xie2024tvdiag} rely on distributed traces to construct a service call graph, which serves as the structural foundation for analysing causal relationships among services and localising root causes. These methods typically assume full trace coverage, meaning that all services in the microservice system are instrumented with distributed tracing~\citep{shen2023deepflow, ashok2024traceweaver}. However, this assumption is often unrealistic in practice. Services or components that lack distributed tracing instrumentation, referred to as \textit{blind spots}~\citep{shen2023deepflow}, are common in real-world microservice deployments. These blind spots may include newly introduced services, closed-source components, third-party services, or services developed using frameworks that do not yet support distributed tracing~\citep{giamattei2023monitoring}.

The prevalence of blind spots can be attributed to several practical challenges in instrumenting microservice systems with distributed tracing. First, microservice systems are highly dynamic, with new services and versions being introduced frequently as systems evolve~\citep{shen2023deepflow}. Engineers often lack sufficient time to implement distributed tracing for newly introduced services before deployment, particularly under time-sensitive release schedules. Second, distributed tracing requires heavy instrumentation involving modifications to the microservices' source code~\citep{ashok2024traceweaver}, which is both time-consuming and resource-intensive. Recent studies~\citep{shen2023deepflow} have shown that engineers can spend hours instrumenting mere tens of lines of code for a single component. Third, in large-scale microservice systems (some comprising up to 1,500 services~\citep{luo2022depth}), achieving full trace coverage is often infeasible. Industry reports~\citep{odigos2025pitfalls} further confirm that requiring complete instrumentation is ineffective and inefficient. In contrast, metrics and logs are relatively straightforward to collect, as they do not require source code modifications, making them more readily available than traces.

The presence of blind spots significantly impacts the effectiveness of existing multi-source RCA methods. When blind spots exist, the service call graph constructed from traces is incomplete, causing RCA methods to only diagnose \textit{visible} services that appear in the graph~\citep{yu2023nezha, lee2023eadro, zhang2023diagfusion, gu2023trinityrcl, xie2024tvdiag}. Consequently, key metrics and logs associated with services outside the call graph may be overlooked, and their behaviour remains unanalyzed during root cause diagnosis. This limitation is particularly problematic for methods like Eadro~\citep{lee2023eadro} and DiagFusion~\citep{zhang2023diagfusion}, which extract features from metrics and logs and integrate them into the trace-based graph. When blind spots exist, many metrics and logs cannot be mapped to any location on the graph, leading to the underutilization of valuable telemetry data. To circumvent this issue, some multi-source RCA methods~\citep{zheng2024mulan, zhang2021cloudrca} completely omit trace information from their analysis, but this approach sacrifices the rich structural information that traces provide about service dependencies. 

This limitation highlights a critical gap in existing multi-source RCA research: the need for methods that can effectively perform root cause analysis without requiring full trace coverage. The assumption that all services are instrumented with distributed tracing makes many existing approaches inflexible and challenging to deploy in real-world scenarios where blind spots are inevitable. There is a clear need for RCA techniques that can leverage available multi-source telemetry data, including metrics, logs, and partial traces, to accurately localise root causes even in the presence of blind spots. Such methods must balance the trade-off between utilising the structural information from available traces while not being entirely dependent on having a complete service call graph. Addressing this limitation would significantly improve the applicability and robustness of RCA methods in dynamic, large-scale microservice environments.

\medskip
\noindent\textbf{Evaluation Gaps.} Beyond methodological limitations, there are also significant gaps in how RCA methods are evaluated and benchmarked.

\subsection{Absence of Standard Benchmarks for RCA in Microservices} \label{sec:limit-benchmarks}

Root cause analysis for microservice systems has gained significant attention in recent years~\citep{lee2023eadro, yu2023nezha, pham2024baro}, with numerous methods proposed to address the challenges of failure diagnosis in distributed environments. However, despite this growing research interest, there remains a notable absence of standard benchmarks that include large-scale datasets and comprehensive evaluation frameworks~\citep{cheng2023ai}. This lack of standardisation has led to inconsistent evaluations across RCA studies, where different works employ different systems, datasets, fault types, and evaluation metrics. Consequently, it becomes difficult to fairly compare the effectiveness of different approaches, hindering a comprehensive understanding of their relative strengths and limitations and ultimately impeding progress in the field~\citep{cheng2023ai}.

The evaluation methodologies employed in existing RCA studies often suffer from limited scope and unrealistic experimental settings. Many studies~\citep{Azam2022rcd, Li2022Circa, run2024aaai} evaluate their proposed methods on only one or two microservice systems with a narrow range of fault types, typically covering only two to three different failure scenarios. This limited evaluation scope raises questions about the generalisability of the proposed methods to diverse real-world failure conditions. Furthermore, some studies employ unrealistic experimental configurations that may not reflect actual production environments. For instance, Eadro~\citep{lee2023eadro} conducts experiments with an extremely low request rate of only 2-3 requests per second, which does not adequately simulate the high-load conditions typical of production microservice deployments. Such limited and unrealistic evaluations make it challenging to assess whether proposed RCA methods can perform effectively across a wide range of systems, fault types, and conditions.

Beyond individual research studies, existing open-source benchmark resources also exhibit significant limitations in their scope and coverage. PyRCA~\citep{liu2023pyrca}, while providing a useful collection of RCA algorithms, focuses exclusively on metric-based methods and relies primarily on synthetic datasets rather than real-world failure data. The AIOps 2020 dataset~\citep{li2022constructing} offers multi-source data including metrics and traces, but notably omits log information, which can be crucial for diagnosing certain types of failures. Similarly, the evaluation framework presented by Pham et al.~\citep{pham2024root} concentrates solely on metric-based RCA methods without supporting trace-based or multi-source approaches. Consequently, none of the existing benchmark resources provide comprehensive coverage of all three major telemetry data sources (metrics, logs, and traces) simultaneously, limiting their utility for evaluating modern multi-source RCA methods that leverage the complementary information from heterogeneous data sources.

The absence of comprehensive standard benchmarks has several important implications for RCA research. The inconsistent evaluation methodologies make it difficult to objectively compare different RCA approaches, understand their respective strengths and weaknesses under various conditions, or validate their effectiveness across diverse fault types and system configurations. This situation creates barriers to identifying the most promising research directions and transferring research innovations into practical deployment. There is a clear and urgent need for standardised benchmarks that provide large-scale datasets covering diverse fault types observed in real-world microservice failures, include comprehensive multi-source telemetry data (metrics, logs, and traces), and offer standardised evaluation frameworks that enable fair and reproducible comparisons across different RCA methods. Such benchmarks would facilitate more rigorous evaluation, accelerate research progress, and ultimately contribute to the development of more robust and reliable RCA solutions.

\subsection{Limited Evaluation of Causal Inference-based Methods} \label{sec:limit-causal-eval}

Causal inference-based RCA methods have attracted increasing attention from researchers in recent years~\citep{Soldani2022rcasurvey, xin2023causalrca, Azam2022rcd, Li2022Circa, wu2021microdiag, Ma2019Msrank, Wang2018cloudranger, Meng2020Microcause, run2024aaai, causalai23salesforce}. The main idea is to construct a causal graph from metrics data to depict the causal relationships among services and metrics, and from this graph, infer the root cause of a failure. Notable methods include CloudRanger~\citep{Wang2018cloudranger}, Microscope~\citep{Jinjin2018Microscope}, MS-Rank~\citep{Ma2019Msrank}, MicroCause~\citep{Meng2020Microcause}, RCD~\citep{Azam2022rcd}, CIRCA~\citep{Li2022Circa}, CausalRCA~\citep{xin2023causalrca}, and RUN~\citep{run2024aaai}. Despite the proliferation of these methods, there is a notable lack of comprehensive evaluation to understand their true capabilities and limitations. 

Existing evaluations of causal inference-based RCA methods suffer from several critical limitations. First, many studies employ synthetic datasets with known ground-truth causal structures, which may not capture the complexity and noise present in real-world telemetry data~\citep{Wu2021evalcausal}. Second, evaluations are often conducted on only one or two benchmark systems with a narrow range of fault types, typically covering only two to three failure scenarios~\citep{Azam2022rcd, Li2022Circa}. Third, previous evaluation studies have significant gaps in their coverage. For example, Wu et al.~\citep{Wu2021evalcausal} evaluate six causal inference-based RCA methods but neither assess the causal graph construction step nor include recently proposed methods such as RCD, CIRCA, CausalRCA, and RUN. Similarly, evaluation studies by Arya et al.~\citep{Arya2021evalcausalai} and Wang et al.~\citep{Wang2021evalcausal} focus only on Granger-based RCA methods using time series data obtained from system logs, leaving other causal discovery approaches unexamined.

Beyond method coverage, existing evaluations also fail to investigate important factors that could significantly affect the performance of causal inference-based RCA methods. These factors include the impact of hyperparameter tuning on causal graph quality, the effect of input data length on method accuracy, the sensitivity of methods to misspecification of failure occurrence time, and the scalability of methods to large-scale microservice systems with hundreds of services and thousands of metrics. Without systematic investigation of these factors, practitioners lack guidance on how to configure and deploy causal inference-based RCA methods effectively in production environments.

This limited evaluation scope has important implications: it remains unclear how different causal discovery methods (e.g., PC, FCI, Granger, LiNGAM, GES) compare in terms of accuracy, efficiency, and robustness when applied to RCA tasks across diverse microservice systems. Furthermore, the reliance on synthetic datasets in many evaluations raises questions about whether reported performance translates to real-world deployments. There is a clear need for systematic empirical studies that evaluate causal inference-based RCA methods across multiple real-world datasets, diverse fault types, and various operational conditions to identify their strengths, weaknesses, and practical limitations. Such studies would provide valuable insights for both researchers developing new methods and practitioners deploying RCA solutions in production systems.

%% file: figures/background/metric_data_sample.tex
\begin{figure}[htbp]
    \centering
    \resizebox{0.8\textwidth}{!}{%
    \begin{tikzpicture}
        \begin{axis}[
            width=14cm, height=7cm,
            xlabel={Timestamp},
            xmin=0, xmax=10, 
            ylabel={Normalized Value},
            legend pos=north west,
            grid=both,
            thick,
            xticklabel style={rotate=45, anchor=east},
        ]
        \addplot[color=NavyBlue, mark=*, smooth] coordinates {
            (0,10) (1,12) (2,9) (3,14) (4,13) (5,15) (6,12) (7,16) (8,14) (9,17) (10,15)
        };
        \addlegendentry{serviceA-workload}

        \addplot[color=ForestGreen, mark=square*, smooth] coordinates {
            (0,30) (1,32) (2,31) (3,35) (4,33) (5,36) (6,34) (7,37) (8,35) (9,38) (10,36)
        };
        \addlegendentry{serviceA-cpu-usage}

        \addplot[color=DarkOrange, mark=triangle*, smooth] coordinates {
            (0,30) (1,35) (2,20) (3,40) (4,25) (5,35) (6,98) (7,88) (8,72) (9,95) (10,98)
        };
        \addlegendentry{serviceA-latency}

        \draw[->, thick, red!70!black] (axis cs:3.5,50) -- (axis cs:5.3,60);
        \node[red!70!black, above] at (axis cs:3.1,50) {Anomaly};
        
        \end{axis}
    \end{tikzpicture}%
    }%
    \caption[Metrics of ServiceA, including workload, CPU usage, and latency.]{Metrics of ServiceA, including workload (rps), CPU usage (\%), and latency~(ms). The \textcolor{red!70!black}{red arrow} highlights a detected anomaly in latency metric.}
    \label{fig:metric_data_sample}
\end{figure}
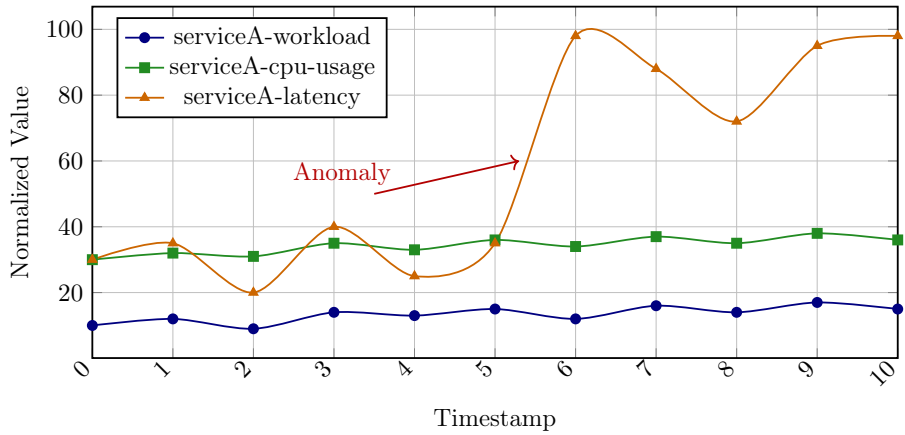

%% file: figures/background/log_data_sample.tex
\begin{figure}[htbp]
\centering
\begin{tcolorbox}[
    width=\textwidth, 
    boxrule=1pt, 
    left=1mm, right=1mm,
    top=1mm, bottom=1mm,
    boxsep=1mm
]
\small
\raggedright
\textcolor{DarkRed}{\texttt{1705734986, cartservice, GetCartAsync called with userId=d06d21b0-4d78-42e4...}}\\[1mm]
\textcolor{DarkRed}{\texttt{1705734986, cartservice, GetCartAsync called with userId=662f989d-da98-4cbf...}}\\[1mm]
\textcolor{DarkCyan}{\texttt{1705734986, currencyservice, conversion request successful}}\\[1mm]
\textcolor{DarkBlue}{\texttt{1705734986, recommendation, [RecvList] prod\_ids=['LS4PSXU', '66VCHSJ', ...]}}\\[1mm]
\textcolor{DarkBlue}{\texttt{1705734986, recommendation, [RecvList] prod\_ids=['2ZYFJ3G', '0PUK6V6', ...]}}\\[1mm]
\textcolor{DarkMagenta}{\texttt{1705734986, shippingservice, [GetQuote] received request}}\\[1mm]
\textcolor{DarkMagenta}{\texttt{1705734986, shippingservice, [CRITICAL] OOM Killed}}\tikz[remember picture, overlay] \coordinate (oom_trigger);\\[1mm]
\textcolor{DarkOrange}{\texttt{1705734986, adservice, received ad request (context\_words=[kitchen])}}\\[1mm]
\textcolor{DarkViolet}{\texttt{1705644272, user, ts=2024-01-19T06:04:32.152111575Z caller=middlewares.go:75 method=GetUsers id=65aa10d0d63d6a00014f9bf4 result=1 took=1.642652ms}}\\[1mm]
\textcolor{Teal}{\texttt{1705644272, front-end, Received: \{"firstName":"Miles Pham", ...\}}}
\end{tcolorbox}

\begin{tikzpicture}[remember picture, overlay]
    \node[red!70!black, anchor=west, align=center] (anomaly_label)
        at ([xshift=3cm, yshift=0.2cm]oom_trigger) {Anomalous\\log entry};
    \draw[->, thick, red!70!black] (anomaly_label.west) -- ([xshift=3mm,yshift=1mm]oom_trigger);
\end{tikzpicture}

\caption{Sample log entries from microservice systems.}
\label{fig:log_data_sample}
\end{figure}

%% file: figures/background/trace_data_sample.tex
\begin{figure}[htbp]
    \centering
    \resizebox{\textwidth}{!}{%
    \begin{tikzpicture}[
        span/.style={rectangle, draw=black, fill=blue!20,
                     minimum height=0.8cm, minimum width=9cm,
                     align=left, inner sep=3pt,
                     font=\ttfamily, rounded corners=2pt},
    ]
        \node[span, fill=blue!20] (n0) at (0, 0) {frontend-ui: PlaceOrder (\textcolor{red}{\textbf{389.4s}})};
        \node[span, fill=blue!20] (n1) at (2, -1) {frontendservice: PlaceOrder (\textcolor{red}{\textbf{215.5s}})};
        \node[span, fill=green!20] (n2) at (4, -2) {checkoutservice: PlaceOrder (\textcolor{red}{\textbf{213.0s}})};
        \node[span, fill=green!20] (n3) at (6, -3) {checkoutservice: GetCart (3.6s)};
        \node[span, fill=green!20] (n4) at (6, -4) {checkoutservice: GetProduct (2.7s)};
        \node[span, fill=orange!20] (n5) at (8, -5) {productcatalogservice: GetProduct (9ms)};
        \node[span, fill=green!20] (n6) at (6, -6) {checkoutservice: GetQuote (3.6s)};
        \node[span, fill=green!20] (n7) at (6, -7) {checkoutservice: Charge (\textcolor{red}{\textbf{126.5s}})};
        \node[span, fill=teal!20, minimum width=8cm] (n8) at (8, -8) {paymentservice: Charge (\textcolor{red}{\textbf{123.4s}})};
        \node[span, fill=green!20] (n9) at (6, -9) {checkoutservice: ShipOrder (6.2s)};
        \node[span, fill=green!20] (n10) at (6, -10) {checkoutservice: EmptyCart (3.7s)};
        \node[span, fill=green!20] (n11) at (6, -11) {checkoutservice: SendOrderConfirmation (3.6s)};
        \node[span, fill=magenta!20] (n12) at (8, -12) {emailservice: SendOrderConfirmation (295ms)};
        \node[span, fill=blue!20] (n13) at (2, -13) {frontendservice: ListRecommendations (95ms)};
        \node[span, fill=purple!20] (n15) at (4, -14) {recommendationservice: ListRecommendations (49ms)};
        \node[span, fill=purple!20] (n14) at (6, -15) {recommendationservice: ListProducts (43ms)};
        \node[span, fill=orange!20] (n16) at (8, -16) {productcatalogservice: ListProducts (8ms)};
        \node[span, fill=blue!20] (n17) at (2, -17) {frontendservice: GetProduct (25ms)};
        \node[span, fill=orange!20] (n18) at (4, -18) {productcatalogservice: GetProduct (8ms)};
        \node[span, fill=blue!20] (n19) at (2, -19) {frontendservice: GetProduct (31ms)};
        \node[span, fill=orange!20] (n20) at (4, -20) {productcatalogservice: GetProduct (9ms)};
        \node[span, fill=blue!20] (n21) at (2, -21) {frontendservice: GetProduct (28ms)};
        \node[span, fill=orange!20] (n22) at (4, -22) {productcatalogservice: GetProduct (11ms)};
        \node[span, fill=blue!20] (n23) at (2, -23) {frontendservice: GetProduct (32ms)};
        \node[span, fill=orange!20] (n24) at (4, -24) {productcatalogservice: GetProduct (13ms)};
        \node[span, fill=blue!20] (n25) at (2, -25) {frontendservice: GetProduct (24ms)};
        \node[span, fill=orange!20] (n26) at (4, -26) {productcatalogservice: GetProduct (7ms)};
        \node[span, fill=blue!20] (n27) at (2, -27) {frontendservice: GetSupportedCurrencies (94ms)};
        \node[span, fill=yellow!20] (n28) at (4, -28) {currencyservice: GetSupportedCurrencies (75ms)};

        \draw[->, thick] ([xshift=9mm]n0.south west) |- (n1);
        \draw[->, thick] ([xshift=9mm]n1.south west) |- (n2);
        \draw[->, thick] ([xshift=9mm]n2.south west) |- (n3);
        \draw[->, thick] ([xshift=9mm]n2.south west) |- (n4);
        \draw[->, thick] ([xshift=9mm]n4.south west) |- (n5);
        \draw[->, thick] ([xshift=9mm]n2.south west) |- (n6);
        \draw[->, thick] ([xshift=9mm]n2.south west) |- (n7);
        \draw[->, thick] ([xshift=9mm]n7.south west) |- (n8);
        \draw[->, thick] ([xshift=9mm]n2.south west) |- (n9);
        \draw[->, thick] ([xshift=9mm]n2.south west) |- (n10);
        \draw[->, thick] ([xshift=9mm]n2.south west) |- (n11);
        \draw[->, thick] ([xshift=9mm]n11.south west) |- (n12);
        \draw[->, thick] ([xshift=9mm]n0.south west) |- (n13);
        \draw[->, thick] ([xshift=9mm]n15.south west) |- (n14);
        \draw[->, thick] ([xshift=9mm]n13.south west) |- (n15);
        \draw[->, thick] ([xshift=9mm]n14.south west) |- (n16);
        \draw[->, thick] ([xshift=9mm]n0.south west) |- (n17);
        \draw[->, thick] ([xshift=9mm]n17.south west) |- (n18);
        \draw[->, thick] ([xshift=9mm]n0.south west) |- (n19);
        \draw[->, thick] ([xshift=9mm]n19.south west) |- (n20);
        \draw[->, thick] ([xshift=9mm]n0.south west) |- (n21);
        \draw[->, thick] ([xshift=9mm]n21.south west) |- (n22);
        \draw[->, thick] ([xshift=9mm]n0.south west) |- (n23);
        \draw[->, thick] ([xshift=9mm]n23.south west) |- (n24);
        \draw[->, thick] ([xshift=9mm]n0.south west) |- (n25);
        \draw[->, thick] ([xshift=9mm]n25.south west) |- (n26);
        \draw[->, thick] ([xshift=9mm]n0.south west) |- (n27);
        \draw[->, thick] ([xshift=9mm]n27.south west) |- (n28);
        
    \draw[<-, red!70!black, thick] ([xshift=2mm]n8.east) -- ++(4,0)
        node[midway, above, font=\bfseries] {Root Cause}
        node[midway, below, font=\bfseries] {Service};

    \draw[<-, red!70!black, thick] ([xshift=2mm]n0.east) -- ++(3,0)
        node[anchor=west] {Anomalous latency};
    \draw[<-, red!70!black, thick] ([xshift=2mm]n1.east) -- ++(2,0)
        node[anchor=west] {Anomalous latency};
    \draw[<-, red!70!black, thick] ([xshift=2mm]n2.east) -- ++(1,0)
        node[anchor=west] {Anomalous latency};
    \draw[<-, red!70!black, thick] ([xshift=2mm]n7.east) -- ++(4.5,0)
        node[midway, above] {Anomalous latency};
        
    \end{tikzpicture}
    }%
    \caption[Distributed trace of an anomalous PlaceOrder request.]{Distributed trace of an anomalous \texttt{PlaceOrder} request. The graph illustrates the service call topology, where each node details the service name, operation, and execution latency. The \textcolor{red!70!black}{red text} highlights the propagation of high latency from the root cause (\texttt{paymentservice}) upstream to the \texttt{frontend-ui}.}
    \label{fig:trace_data_sample}
\end{figure}
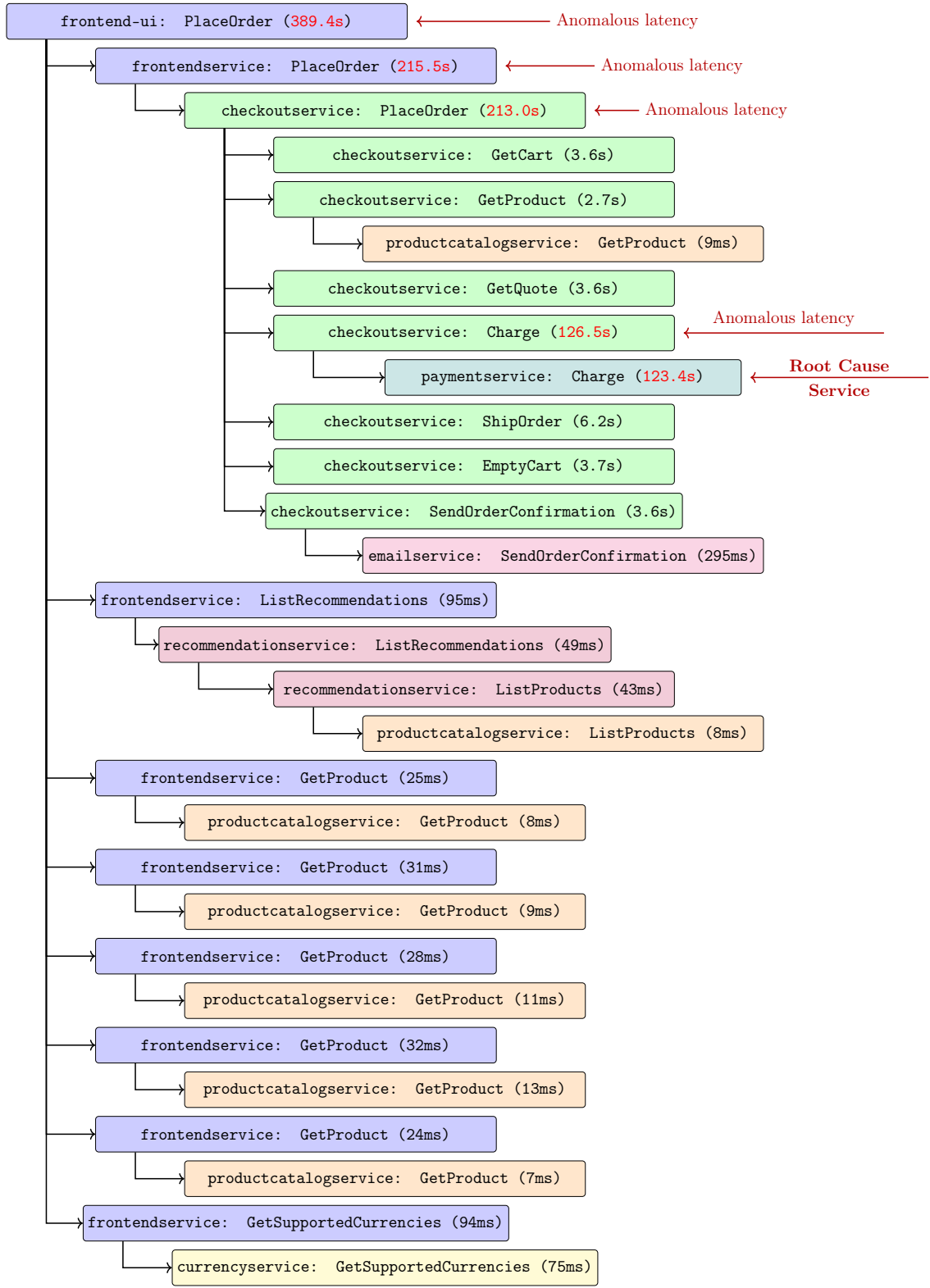

%% file: figures/background/event_data_sample.tex
\begin{figure}[htbp]
\centering
\resizebox{\textwidth}{!}{%
\begin{tikzpicture}[
    event/.style={
        rectangle,
        draw=black,
        rounded corners=2pt,
        minimum width=4.2cm,
        align=left,
        font=\ttfamily\scriptsize,
        inner sep=4pt
    },
    annotation/.style={
        font=\scriptsize\sffamily,
        text=gray!70!black
    },
    timeline/.style={
        ->,
        thick,
        gray
    }
]

\draw[timeline] (-0.5, 0) -- (14.5, 0) node[right, font=\small] {time};

\foreach \x/\t in {1.5/17:38:32, 7/17:39:15, 12.5/17:40:01} {
    \draw[gray] (\x, -0.1) -- (\x, 0.1);
    \node[below, font=\tiny\ttfamily, gray] at (\x, -0.15) {\t};
}

\node[event, fill=red!8, draw=red!70!black, line width=1pt] (e1) at (1.5, 1.5) {
    \textcolor{blue!70!black}{actor}: merlinary\\
    \textcolor{orange!70!black}{op}: DeleteSecurityGroup\\
    \textcolor{teal!70!black}{resource}: sg-0a1b2c3d\\
    \textcolor{purple!70!black}{error}: null
};
\draw[gray, dashed] (1.5, 0.1) -- (1.5, 1.3);
\node[above=2pt of e1, font=\small\bfseries, red!70!black] {$e_1$ (Root Cause)};

\node[event, fill=gray!5] (e2) at (7, 1.5) {
    \textcolor{blue!70!black}{actor}: scheduler\\
    \textcolor{orange!70!black}{op}: StartInstance\\
    \textcolor{teal!70!black}{resource}: i-98765432\\
    \textcolor{purple!70!black}{error}: null
};
\draw[gray, dashed] (7, 0.1) -- (7, 1.3);
\node[above=2pt of e2, font=\small\bfseries] {$e_2$};

\node[event, fill=orange!10, draw=orange!70!black] (e3) at (12.5, 1.5) {
    \textcolor{blue!70!black}{actor}: scheduler\\
    \textcolor{orange!70!black}{op}: AttachNetworkInterface\\
    \textcolor{teal!70!black}{resource}: i-98765432\\
    \textcolor{purple!70!black}{error}: \textbf{SG\_NOT\_FOUND}
};
\draw[gray, dashed] (12.5, 0.1) -- (12.5, 1.3);
\node[above=2pt of e3, font=\small\bfseries, orange!70!black] {$e_3$ (Failure)};

\draw[->, red!70!black, thick, dashed]
    (e1.north east) to[out=25, in=155]
    node[above, midway, font=\scriptsize\itshape, text=red!70!black] {causes}
    (e3.north west);

\node[annotation, anchor=west] at (-0.3, -0.9) {
    \textcolor{blue!70!black}{\textbullet~actor} (who) \quad
    \textcolor{orange!70!black}{\textbullet~op} (operation) \quad
    \textcolor{teal!70!black}{\textbullet~resource} (what) \quad
    \textcolor{purple!70!black}{\textbullet~error} (status)
};

\end{tikzpicture}%
}%
\caption[Event stream with causal relationship between events.]{Event stream following the OCSF schema~\citep{ocsf}. Each event captures the \textcolor{blue!70!black}{actor}, \textcolor{orange!70!black}{operation}, \textcolor{teal!70!black}{resource}, and \textcolor{purple!70!black}{error status}. Event $e_1$ (deleting a security group) is the root cause of the failure in $e_3$ (network attachment fails).}
\label{fig:event_data_sample}
\end{figure}
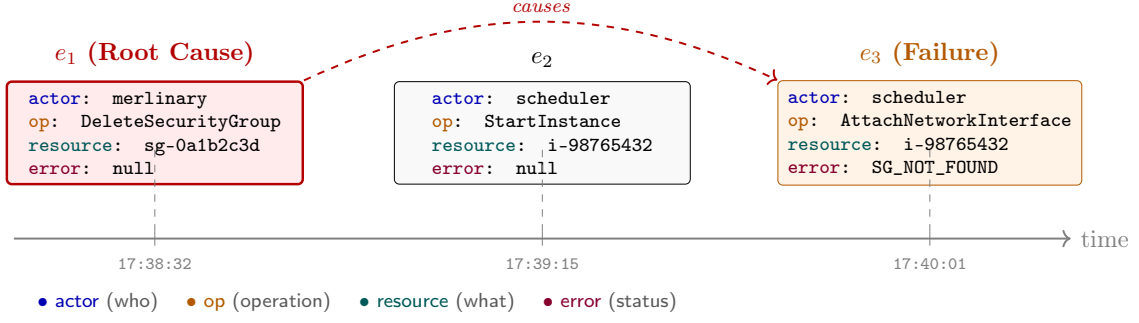

%% file: chapters/3.metric.tex
\chapter{BARO: Metric-based Anomaly Detection and Root Cause Analysis}\label{chap:fse24}

\begin{tcolorbox}[left=2pt,right=2pt,top=0pt,bottom=0pt,
  enhanced,
  drop shadow={shadow xshift=1ex, shadow yshift=-1ex, opacity=0.3}]
	\textbf{Publication:} This chapter is based on our paper titled \textbf{``BARO: Robust Root Cause Analysis for Microservices via Multivariate Bayesian Online Change Point Detection''}, Luan Pham, Huong Ha, and Hongyu Zhang, published in the Proceedings of the ACM on Software Engineering (\textit{PACMSE}), Issue \textit{FSE}, Volume 1, 2024, pp.~2214--2237 (\textbf{CORE~A*}) \citep{pham2024baro}.
\end{tcolorbox}
\vspace{10pt}

\noindent This chapter presents the first methodological contribution of this thesis: a metric-based approach to anomaly detection and root cause analysis. Building on the foundational concepts introduced in Chapter~\ref{chap:background}, we address the limitation identified in Section~\ref{sec:limit-ad-decoupling}---the decoupling of anomaly detection from RCA. In practice, automated anomaly detection is imprecise, and this imprecision may directly impact RCA performance. This chapter introduces BARO, an end-to-end framework that jointly performs anomaly detection and RCA on multivariate time-series metrics. BARO employs Multivariate Bayesian Online Change Point Detection for anomaly detection and introduces RobustScorer, a novel non-parametric hypothesis testing approach that maintains robust localisation even when anomaly detection results are imprecise.
\vspace{10pt}

\input{papers/fse24/0.abstract}

\section{Introduction}
\input{papers/fse24/1.introduction}

\section{Problem Statement and Background}
\input{papers/fse24/2.background}

\section{BARO: Proposed Approach}
\input{papers/fse24/3.method}

\section{Evaluation} \label{sec:results}

\input{papers/fse24/4.results}

\section{Threats to Validity}
\input{papers/fse24/5.threats}

\section{Summary}
\input{papers/fse24/6.conclusion}

\vspace{10pt}
\noindent This chapter demonstrated that metric-based anomaly detection and RCA can be effectively unified through BARO's end-to-end framework, achieving consistent improvements over state-of-the-art methods. However, modern microservice systems generate diverse observability data beyond metrics. Event data, which captures system activities such as API calls, infrastructure changes, and security actions, offers a complementary perspective on system behaviour that may reveal anomalies not visible in metric data alone. The next chapter extends our exploration to this alternative data modality, presenting EventADL, an open-box framework that leverages event semantic and frequency patterns for interpretable anomaly detection and localisation.
\vspace{10pt}


%% file: papers/fse24/0.abstract.tex
A typical failure troubleshooting pipeline for microservices consists of two phases: anomaly detection and root cause analysis. While various existing works on root cause analysis require accurate anomaly detection, there is no guarantee of accurate estimation with anomaly detection techniques. Inaccurate anomaly detection results can significantly affect the root cause localisation results. To address this challenge, we propose \textit{BARO}, an end-to-end approach that integrates anomaly detection and root cause analysis for effectively troubleshooting failures in microservice systems. BARO leverages the Multivariate Bayesian Online Change Point Detection technique to model the dependency within multivariate time-series metrics data, enabling it to detect anomalies more accurately. BARO also incorporates a novel nonparametric statistical hypothesis testing technique for robustly identifying root causes, which is less sensitive to the accuracy of anomaly detection compared to existing works. Our comprehensive experiments conducted on three popular benchmark microservice systems demonstrate that BARO consistently outperforms state-of-the-art approaches in both anomaly detection and root cause analysis.

%% file: papers/fse24/1.introduction.tex
In recent years, microservice systems have gained significant popularity in the development of cloud-based applications, owing to their numerous advantages such as resource flexibility, a loosely coupled architecture, and lightweight deployment. However, failures are inevitable in microservice systems due to their inherent complexity. A failure in one service can propagate across the system, affecting many other services and resulting in the degradation of the system availability. This, in turn, leads to poor user experience and incurs huge economic losses. For instance, it has been reported that a one-hour downtime on Amazon.com could potentially cost up to 100 million USD~\citep{Chen2019incidenttriage, Chen2020incidentmanagement}. Therefore, system operators must closely monitor the systems, checking key run-time information to promptly detect failures as soon as they occur, and then proceed to identify the failures' root causes and troubleshoot them. However, in practice, the complexity of microservice systems and the large volume of monitoring data make these tasks especially challenging.

Metric-based anomaly detection and root cause analysis (RCA) for microservice systems have been extensively studied in recent years \citep{Soldani2022rcasurvey, Azam2022rcd, Li2022Circa, xin2023causalrca, chen2022adaptive, siffer2017anomaly, wu2020microrca, Jinjin2018Microscope, yu2021microrank, liu2021microhecl,lee2023eadro}. Given a set of metrics data, anomaly detection techniques aim to detect whether there exist anomalies and consequently, failures within the microservice system~\citep{chen2022adaptive, lee2023eadro}. If a failure is detected, the RCA module is then triggered to locate the root cause of the failure \citep{lee2023eadro, Azam2022rcd, Li2022Circa}. The RCA module aims to address two fundamental questions: (1) which services are the root causes, and (2) what specific issues are causing that failure (e.g. high CPU utilisation, memory leak, or network congestion). Some RCA approaches use hypothesis testing or a statistical analysis method to analyse the time series metrics data to identify the candidate root causes for the detected anomalies \citep{Shan2019Ediagnosis, Li2022Circa}. Some RCA methods construct topology graphs using the information provided by the monitoring systems, such as the microservice status, the interaction between the services, and the interaction traces, to facilitate root cause analysis \citep{wu2020microrca, liu2021microhecl}. Multiple recent RCA methods \citep{xin2023causalrca, Azam2022rcd, Jinjin2018Microscope, Meng2020Microcause} use different causal discovery methods \citep{Runge2019PCMCI, Jaber2020PsiFCI, Spirtes1995fci} to derive the causal relationships among the services and metrics from the multivariate time series metrics data and employ graph centrality algorithms like random walk \citep{Jinjin2018Microscope}, PageRank \citep{xin2023causalrca,wu2021microdiag} or Depth-First Search (DFS) \citep{Chen2014Causeinfer} to infer the root causes.

Despite being closely related, existing RCA works typically either treat the anomaly detection tasks independently or rely on overly simplistic anomaly detection techniques~\citep{lee2023eadro, Soldani2022rcasurvey, Li2022Circa, Azam2022rcd}. For example, MicroRCA \citep{wu2020microrca} and MicroDiag \citep{wu2021microdiag} employ the simple BIRCH clustering technique~\citep{zhang1996birch} for detecting anomalies. MicroScope \citep{Jinjin2018Microscope} and Ms-Rank \citep{Ma2019Msrank} use a basic three-sigma rule of thumb as their anomaly detection method, known as N-Sigma. 
There are various commercial monitoring platforms, notable platforms including DataDog \citep{datadog} and Dynatrace \citep{dynatrace}, which rely on simplistic 
anomaly detection techniques \citep{dynatrace1, datadog2} for univariate time series data. 
Most RCA research works~\citep{Jinjin2018Microscope, Wang2018cloudranger, Ma2019Msrank, wu2020microrca, Ma2020Automap, wu2021microdiag, Li2022Circa, Azam2022rcd, xin2023causalrca} focus solely on identifying the root cause of the failure whilst assuming the existence of an anomaly detection module that can accurately detect failures and trigger the RCA module when failures are detected. Some of these works, such as CIRCA \citep{Li2022Circa} and RCD \citep{Azam2022rcd}, specifically require certain information from the anomaly detection module, in particular, the failure occurrence time. However, they assume this information is already known accurately. Thus, it remains unclear whether, when combined with existing anomaly detection methods that may provide imprecise information, these approaches are still effective in localising the root cause.

In this chapter, we introduce BARO, an end-to-end approach for anomaly detection and root cause analysis for microservice systems based on metrics data, which are multivariate time series data. BARO includes a Multivariate Bayesian Online Change Point Detection module for detecting anomalies. It also includes a novel RobustScorer module, which is a nonparametric statistical hypothesis testing technique and less sensitive to the accuracy of the anomaly detection, for robustly identifying the root causes. BARO offers several advantages. First, it follows an unsupervised learning approach, eliminating the need for labelled data and enabling direct application without the requirement of such labelled data. Second, it does not rely on operational knowledge (e.g., service call graphs) or causal graphs, making it suitable for large-scale evolving systems where acquiring such operational knowledge or causal graphs for numerous services is difficult \citep{Azam2022rcd, Li2022Circa, Wang2018cloudranger}. Third, BARO is nonparametric, scale-equivalent, and rotation-invariant, making it applicable to a wide range of systems. We comprehensively evaluate BARO against various state-of-the-art approaches on three popular benchmark microservice systems. Our experimental results demonstrate that BARO consistently surpasses the state-of-the-art methods. Additionally, we analyse the sensitivity of the RCA methods against their parameters to show the robustness of our method.

This chapter makes the following contributions:
\begin{itemize}
    \item We propose a new end-to-end approach for anomaly detection and root cause analysis in microservice systems based on multivariate time-series metrics data. The approach employs the Multivariate Bayesian Online Change Point Detection technique for anomaly detection and a novel nonparametric statistical hypothesis testing technique (RobustScorer) for accurately identifying root causes of microservice failures.
    \item We conduct an extensive empirical evaluation on three popular benchmark microservice systems (Online Boutique, Sock Shop, and Train Ticket), demonstrating that BARO outperforms state-of-the-art approaches in both anomaly detection and RCA.
    \item We perform a comprehensive sensitivity analysis of all studied RCA methods with respect to their parameters, showing that BARO is significantly more robust against 
    the anomaly detection time compared with baseline methods.
\end{itemize}

The remainder of this chapter is organised as follows. The following section presents the problem statement and background. The subsequent section then presents the BARO approach. Section~\ref{sec:results} reports the experimental evaluation against state-of-the-art baselines. We then discuss threats to validity, and the final section concludes the chapter.

%% file: papers/fse24/2.background.tex
\subsection{Problem Statement}

This chapter focuses on metric-based anomaly detection and root cause analysis.\footnote{The original publication~\citep{pham2024baro} contained self-contained terminology and problem definitions. In this thesis, we consolidate these in Chapter~\ref{chap:background} for consistency.}
We adopt the terminology established in Section~\ref{sec:terms-fault-failure} and introduce one additional term specific to this context: \textit{root cause metrics} are the metrics that serve as indicators of the root cause~\citep{Li2022Circa, chen2022adaptive, Azam2022rcd}. The system operators can use these suggested root cause metrics to identify the true underlying root cause of the failures.

\subsubsection{Problem Formulation} \label{sec:baro-problem}

This chapter addresses the metric-based instantiation of the general AD and RCA formulation (Sections~\ref{sec:problem-formulation-ad} and~\ref{sec:problem-formulation-rca}). We focus on the metrics component $\mathcal{M}$ collected from $n$ services from a given microservice system, over a $T$-length observation window. Our goal is to (1) predict the anomaly indicator $y \in \{0,1\}$, and (2) when $y=1$, identify the root cause services and their corresponding root cause metrics.

\subsection{Multivariate Time Series Metric Data}

As described in Chapter~\ref{chap:background}, metric-based anomaly detection and RCA are typically based on runtime information collected on the services within the microservice system. Such information includes metrics monitored on the microservices, such as workload, resource consumption, and response time. These metrics are typically represented as multivariate time series, with each time series corresponding to the data collected with a specific metric \citep{Soldani2022rcasurvey}. Microservices also generate logs to provide more detailed and meaningful information about their state. Some previous studies~\citep{Wang2021evalcausal, aggarwal2020localization, aggarwal2021causal} parse raw logs to extract log static structures (i.e., log templates \citep{le2023log}), and count the occurrences of these templates, which are subsequently transformed into time series data. These logs can also be a source of multivariate time series data, which can be used for analyse root cause. In this work, we, however, only focus on metrics as multivariate time series data, as with~\citep{Li2022Circa, Azam2022rcd, wu2020microrca, wu2021microdiag, Meng2020Microcause, Wang2018cloudranger, liu2021microhecl}.

\subsection{Metric-based Anomaly Detection} \label{sec:background-ad}

In microservice systems, once a failure occurs in a service, it is typically reflected in the metrics data of that particular service, resulting in an anomaly or a change in its data distribution. Furthermore, a failure in one service can propagate across the systems and impact other services, subsequently leading to changes to the metrics data of those services as well. To detect a failure in a microservice system, the goal is to detect any anomalies or changes in the given metrics dataset \citep{Soldani2022rcasurvey}.

A wide range of anomaly detection methods exist for time series data \citep{BlazquezGarc2021ADreview}. In this chapter, since we target the problem of identifying root causes for microservice systems, we only focus on RCA-oriented anomaly detectors employed or discussed in existing metric-based RCA studies \citep{lee2023eadro}. N-Sigma \citep{Jinjin2018Microscope} is used and discussed in MicroRank \citep{yu2021microrank}, CIRCA \citep{Li2022Circa}, Eadro \citep{lee2023eadro}, and MicroScope \citep{Jinjin2018Microscope}. SPOT \citep{siffer2017anomaly} is mentioned and evaluated in CIRCA \citep{Li2022Circa} and Eadro \citep{lee2023eadro}.  BIRCH \citep{zhang1996birch} is employed in MicroRCA \citep{wu2020microrca}  and MicroDiag \citep{wu2021microdiag}. Finally, Univariate Offline Bayesian Change Point Detection \citep{adams2007bayesian} is used in CauseInfer \citep{Chen2014Causeinfer, chen2016causeinfer}. We describe these methods in the below.

\paragraph{N-Sigma}
N-Sigma (i.e., the three-sigma rule of thumb) represents one of the simplest anomaly detection methods. It operates based on the assumption that the data points falling within three standard deviations of the mean of the data distribution are considered normal. Consequently, any data point $x$ that falls outside this range, i.e., $\mu - 3\sigma < x < \mu + 3\sigma$, is deemed abnormal. For instance, MicroScope \citep{Jinjin2018Microscope} computes the mean and standard deviation of the metrics data distribution using the most recent 10 minutes of data and then applies this method to detect anomalies.

\paragraph{SPOT}
SPOT and dSPOT \citep{siffer2017anomaly} are founded based on the principles of Extreme Value Theory \citep{Coles2001EVT} to detect anomalies and have been employed in various works~\citep{pan2021dycauserca, Li2022Circa, lee2023eadro, Meng2020Microcause}. SPOT is designed to handle data with stationary distribution, while dSPOT is developed for streaming data susceptible to concept drift. It is worth noting that different studies have used different variants of SPOT. MicroCause \citep{Meng2020Microcause} and DycauseRCA \citep{pan2021dycauserca} made use of dSPOT, whereas CIRCA~\citep{Li2022Circa} employed biSPOT. In this work, we will employ dSPOT, as in \citep{Meng2020Microcause, pan2021dycauserca}. Therefore, in the sequel, when referring to SPOT, we are specifically referring to dSPOT. 

\paragraph{BIRCH}
BIRCH (Balanced Iterative Reducing and Clustering using Hierarchies) \citep{zhang1996birch} is a well-regarded unsupervised clustering algorithm known for its efficiency in real-time data analysis and anomaly detection, particularly in large-scale datasets and time-series data. Its main idea is to generate a brief and informative summary about the original dataset, and then perform clustering on this summary. BIRCH considers a data point to be an anomaly when it is in a different cluster with other consecutive data points. MicroRCA \citep{wu2020microrca} and MicroDiag \citep{wu2021microdiag} have employed BIRCH for anomaly detection as a preliminary step before conducting RCA.

\paragraph{Univariate Offline Bayesian Change Point Detection}
Bayesian change point detection \citep{adams2007bayesian} is a statistical method for identifying change points in time series data, i.e., timesteps where the data distribution experiences significant shifts. It relies on the principle of causal predictive filtering, aiming to generate an accurate distribution of future unseen data points based solely past observations. Through Bayesian statistics, it incorporates the prior information regarding the characteristics of the change points into the modelling process, making it to be both effective and efficient. In previous RCA research works~\citep{Chen2014Causeinfer, chen2016causeinfer}, univariate offline Bayesian change point detection was used to detect change points (anomalies) within the time series metrics data. 

Besides the methods mentioned, commercial platforms like Datadog \citep{datadog} and Dynatrace \citep{dynatrace} also provide anomaly detection techniques \citep{datadog2, dynatrace1} for univariate time series. These approaches either rely on users or historical data to obtain thresholds to detect anomalies. They are similar to N-Sigma, which uses expected values along with pre-defined tolerance thresholds.

\subsection{Metric-based Root Cause Analysis}

Existing metric-based RCA algorithms can be classified into three main categories: statistical analysis, topology graph-based, and causal graph-based methods \citep{Soldani2022rcasurvey}.

\paragraph{Statistical Analysis.} These approaches pinpoint failures' root causes by identifying metrics that undergo significant changes during the anomalous period. $\epsilon$-Diagnosis \citep{Shan2019Ediagnosis} uses the two-sample test algorithm and $\epsilon$-statistics to measure the similarity among the metrics and rank the root cause based on the similarity scores. $\epsilon$-Diagnosis is evaluated against three statistical analysis methods: Pearson distance, KNN \citep{friedman1983graph, luo2014correlating} and MST \citep{friedman1979multivariate}. KNN \citep{friedman1983graph, luo2014correlating} uses nearest neighbours to model the distance between two time series, while MST \citep{friedman1979multivariate} uses a minimum spanning tree to represent the distance. In \citep{wang2020root}, a neural network is used to learn the normal behaviour and measure the similarity of monitoring data when the failure happens. They use mutual information to rank the root causes. N-Sigma \citep{Jinjin2018Microscope,Li2022Circa} is another statistical analysis technique that assesses the distance using z-score.

\paragraph{Topology Graph-based Analysis.} These approaches reconstruct a topology graph representing the microservice system using information from monitoring systems and operational knowledge. MicroRCA \citep{wu2020microrca} constructs a topology graph from monitoring data, extracts anomalous subgraphs, and uses the random walk algorithm to infer root causes. Similarly, \citep{wu2020performance} constructs a topology graph and anomalous subgraphs, followed by a neural network-based method to infer the root cause. Sieve \citep{thalheim2017sieve} uses a clustering technique to reduce the number of metrics on the constructed topology graph, then uses Granger Causality tests \citep{Granger1980causal} to determine the possible root causes. Likewise, Brandon \citep{brandon2020graph} augments the provided topology graph with monitored metrics, conducts root cause searches through extracted subgraphs, and ranks root causes based on similarity scores. DLA \citep{samir2019dla} transforms the provided topology graph and metric data into a hierarchical hidden Markov model and identifies root causes by computing the path with the highest anomalous probability. Meanwhile, CIRCA \citep{Li2022Circa} performs hypothesis testing on the structural graph to find the root causes. Commercial platforms such as Datadog \citep{datadog} and Dynatrace \citep{dynatrace} construct a topology graph from distributed traces and perform Depth First Search (DFS) to find the root cause services~\citep{datadog1, dynatrace3}.

\paragraph{Causal Graph-based Analysis.} 

Many recent RCA techniques adopt the causal graph-based approach \citep{Chen2014Causeinfer, chen2016causeinfer, Jinjin2018Microscope, Wang2018cloudranger, Chen2019airalert, Ma2019Msrank, Meng2020Microcause, Ma2020Automap, wu2021microdiag, Azam2022rcd, xin2023causalrca}. The main idea is to construct a causal graph where vertices represent services or metrics of the microservices, and edges represent the cause-effect relationships between the services/metrics. These graphs are constructed using different causal discovery methods such as PC, FCI, LiNGAM, and GES \citep{Spirtes1993Causal, Spirtes1995fci, Chickering2002Ges, Shimizu2006Lingam, Jaber2020PsiFCI}. Assuming the root cause metric would affect other services' metrics, graph centrality algorithms like Breath First Search (BFS), random walk, or PageRank are used to rank the root causes. In addition, correlation analysis can be conducted to measure the correlation among metrics \citep{wu2021microdiag}, and the graph traversal process can consider these scores to identify the root cause. Recently, RCD \citep{Azam2022rcd} employs a divide-and-conquer strategy to split the input metrics into smaller chunks and constructs a causal graph for each chunk. It then employs $\Psi$-PC \citep{Jaber2020PsiFCI} to identify the root cause for each chunk, which is later combined to yield the final rootcause. CausalRCA \citep{xin2023causalrca} introduces the use of a gradient-based causal discovery method, namely DAG-GNN, to uncover causal relationships among metrics.

%% file: papers/fse24/3.method.tex
Existing metric-based RCA methods either treat anomaly detection independently of root cause analysis or rely on imprecise detectors, leaving downstream root cause scoring sensitive to inaccurate anomaly times. To address this limitation, we propose BARO, an end-to-end approach for anomaly detection and root cause analysis that couples a more accurate anomaly detector with a robust root cause scorer. We first introduce the basic assumptions underlying our approach (Section~\ref{sec:assumption-causal}), then present BARO (Sections \ref{sec:approach-ovw}, \ref{sec:multi-bocpd}, and \ref{sec:robustscorer}). \textbf{BARO} incorporates a Multivariate \textbf{BA}yesian Online Change Point Detection technique to model the dependency and correlation structure of multivariate time series metrics data, enabling it to effectively detect anomalies and estimate the occurrence time of failures
(Section \ref{sec:multi-bocpd}). It then uses a nonparametric hypothesis testing method, referred to as \textbf{RO}bustScorer, to reliably identify and rank the potential root causes, which is less sensitive to the accuracy of the anomaly detection time (Section \ref{sec:robustscorer}).

\begin{figure}[t]
\includegraphics[width=\textwidth]{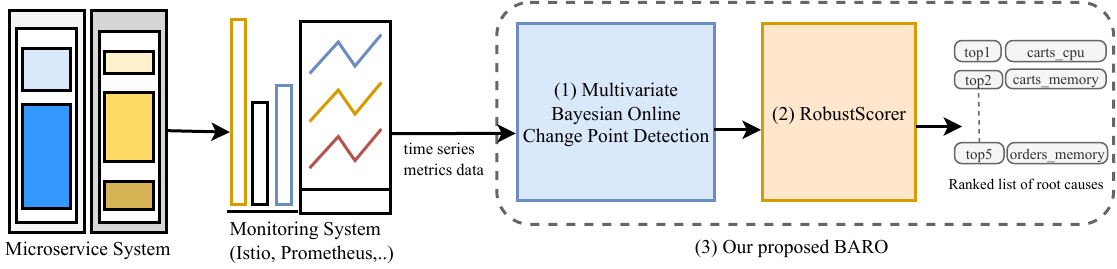}
\caption[The overview of BARO.]{The overview. The monitoring system monitors the microservice system and collects the time series data. Our BARO consists of two components, Multivariate BOCPD and RobustScorer. The Multivariate BOCPD acts as an anomaly detection module to continuously check whether there is an anomaly. If there exists an anomaly, it triggers RobustScorer to score and rank the root cause services and metrics correspondingly.} \label{fig:intro2}
\end{figure}
 
\subsection{Basic Assumptions} \label{sec:assumption-causal}

\subsubsection{Anomaly Metrics} \label{sec:anomaly-metric-selection}

As commonly recognised in previous works \citep{Li2022Circa, yu2023cmdiagnostor, liu2021microhecl}, there are generally four types of metrics: \textit{Traffic} (e.g., request count per minute), \textit{Saturation} (e.g., CPU utilisation, database records), \textit{Latency} (e.g., average response time per minute), and \textit{Errors} (e.g., the rate of failed requests). These four types are named after the four golden signals in site reliability engineering \citep{googlesre}. Following the practice in metric-based RCA studies \citep{yu2023cmdiagnostor, xin2023causalrca, Jinjin2018Microscope, Li2022Circa, Azam2022rcd}, we assume that \textit{anomalies should be visible in the metrics, subsequently affecting \textit{Latency} or \textit{Errors}.} With these assumptions, in varying conditions where there is a surge in \textit{Traffic} or \textit{Saturation} (e.g., in holiday periods) but without abnormal increases in \textit{Latency} and \textit{Errors}, our proposed method considers these situations as normal. Conversely, if a surge in \textit{Traffic} or \textit{Saturation} metrics causes abnormal increases in \textit{Latency} or \textit{Errors}, our method considers this an anomaly. Finally, for the RCA task, our method considers all the metrics to pursue the fine-grained output root cause ranking.

\subsubsection{Failure Propagation Chain} \label{sec:assumption-failure-propagation}

In practice, when a service failure occurs, it generally results in an anomaly in the data associated with the metric corresponding to that failure~\citep{Li2022Circa, Azam2022rcd}. For example, network congestion in a service typically leads to an increase in its response time. Furthermore, since a failure can propagate across the services within the system, this initial anomaly will then trigger additional anomalies in the metrics data of other services at later time~\citep{Azam2022rcd, xin2023causalrca, Jinjin2018Microscope, Li2022Circa, liu2021microhecl}. Thus, when an anomaly is detected and the RCA module is activated, the anomalous period of runtime metrics data generally consists of multiple anomalies. In this work, based on this failure propagation chain, we assume \textit{the first anomaly corresponds to the time when the failure first occurs}. Thus, in our proposed method, we use the first detected anomaly to approximate the failure occurrence time ($\hat{t}_A$ in Alg. \ref{alg:baro}) to separate the normal and abnormal metrics data. This assumption does not imply the first detected anomaly to be the root cause, and it has been implicitly used in previous RCA works \citep{Shan2019Ediagnosis, Li2022Circa, Azam2022rcd} for the same purpose as ours (i.e. to separate the abnormal and normal data). Our experimental results, along with those of previous studies, affirm the validity of this assumption.

\subsection{Approach Overview} \label{sec:approach-ovw}

We illustrate BARO, our proposed end-to-end approach for anomaly detection and root cause analysis for microservice systems based on multivariate time series metrics data, in Figure~\ref{fig:intro2}. BARO operates in two steps. First, a Multivariate Bayesian Online Change Point Detection (BOCPD) module models the dependency and correlation structure of multivariate time-series metrics data so as to detect anomalies (failures). Second, RobustScorer, a nonparametric statistical hypothesis testing technique, identifies the root cause associated with the failures. Thus, the outputs of BARO include (1) a boolean indicating whether an anomaly is presented, and (2) a ranked list of root cause metrics and their corresponding services, with the highest-ranked items having the highest probability of being the root cause of the failure. The pseudocode of BARO is described in Algorithm~\ref{alg:baro}.

\subsection{Multivariate Bayesian Online Change Point Detection} \label{sec:multi-bocpd}

Anomaly detection in microservices involves identifying anomalies, i.e., observable symptoms of failures \citep{Soldani2022rcasurvey}, while failures in microservices can be considered as interventions that change the monitoring data distribution \citep{Azam2022rcd, Li2022Circa}. Therefore, to detect anomalies (failures) within microservices via time series metrics data, we formulate this problem as a change point detection problem whose goal is to identify whether the behaviour of a time series changes significantly. We then propose to use Multivariate BOCPD, a combination of BOCPD \citep{adams2007bayesian} and MultivariateCPD \citep{xuan2007modeling}, to model the dependency and correlation among metrics to detect anomalies effectively. The motivation behind this design is twofold. First, we propose to use BOCPD \citep{adams2007bayesian} as a base technique for detecting change points as it is a simple yet effective online detection technique and it requires no user-specified thresholds to identify change points for univariate time series. It has been shown to be among the best current change point detection methods in many real-world scenarios \citep{van2020evaluation}. Second, by combining BOCPD with MultivariateCPD \citep{xuan2007modeling}, we can model the structure and dependency among the multivariate time series metrics data better. This is especially useful for detecting anomalies within microservices due to the failure propagation chain in microservices described in Section \ref{sec:assumption-failure-propagation}. Specifically, anomalies in microservices are generally propagated across the metrics data, causing correlated and dependent changes among different time series metrics. MultivariateCPD has been shown to be able to effectively detect change points when the changes occur in the correlation structure as in multivariate time series metrics data. In the following paragraphs, we describe in detail these two components of our proposed method.

\begin{figure}
\includegraphics[width=\textwidth]{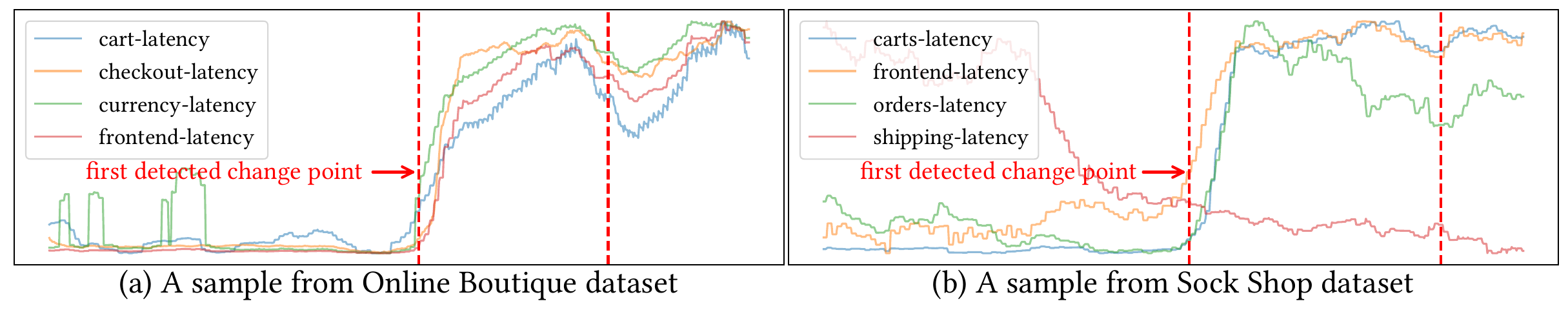}
\caption[An example of using Multivariate BOCPD to detect change points.]{An example of using Multivariate BOCPD to detect change points on multivariate time series data. \textcolor{red}{Dotted vertical red lines} indicate change points. We observe that Multivariate BOCPD can provide the anomaly detection time (the first change point) accurately that separate the normal and abnormal period. In the abnormal period there are multiple change points due to the failure propagation chain.}  \label{fig:mbocpd}
\end{figure}

The main idea of BOCPD is to model the \textit{run length}, i.e. the number of consecutive data points in the same distribution, since the last change point, given the data observed so far. Specifically, the run length $r_t$ at time $t$ is defined as $0$ if there is a change point at time $t$, and as $r_{t-1}+1$ otherwise. Given the time series metrics data $\mathcal{M}^{i,j=1:n,1:m}_{t_0:t}$, using the Bayes theorem, the posterior probability distribution of the run length $p(r_t \vert \mathcal{M}^{i=1:n,j=1:m}_{t_0:t})$ can be computed as \citep{adams2007bayesian},
\begin{equation} \label{eq:bocpd-run-length}
  \begin{aligned}
      p(r_t \vert \mathcal{M}^{i,j=1:n,1:m}_{t_0:t})
      ={} & \frac{1}{p(\mathcal{M}^{i,j=1:n,1:m}_{t_0:t})} \sum\nolimits_{r_{t-1}} p(r_t \vert r_{t-1}) \\
      & \times\, p(\mathcal{M}^{i,j=1:n,1:m}_{t} \vert r_{t-1}, (\mathcal{M}^{i,j=1:n,1:m}_{t})^{(r)}) \\
      & \times\, p(r_{t-1} \vert \mathcal{M}^{i,j=1:n,1:m}_{t_0:t-1}).
  \end{aligned}
\end{equation}
where $(\mathcal{M}^{i,j=1:n,1:m}_{t})^{(r)}$ denotes the set of observed data points associated with the run $r_t$. The formula in Eq. (\ref{eq:bocpd-run-length}) is recursive, meaning that we can compute the posterior distribution of the run length $r_t$ based on the posterior distribution of $r_{t-1}$, the conditional prior of run length $p(r_t \vert r_{t-1})$ and the distribution of the metrics data $p(\mathcal{M}^{i,j=1:n,1:m}_{t_0:t})$. As suggested in \citep{adams2007bayesian}, the marginal likelihood of the metrics data $p(\mathcal{M}^{i,j=1:n,1:m}_{t_0:t})$ can be chosen as a distribution from the exponential family and the conditional prior of run length $p(r_t \vert r_{t-1})$ can be set based on a hazard function with discrete exponential (geometric) distribution. At each time step $t$, the most probable run length is computed as the value with the highest probability $p(r_t \vert \mathcal{M}^{i,j=1:n,1:m}_{t_0:t})$. Finally, the change points are identified as the data points at the time steps whose run lengths decrease.

The main idea of MultivariateCPD, given the time series metrics data $\mathcal{M}^{i,j=1:n,1:m}_{t_0:t_0+T}$, is to model the metrics data at each data point $\mathcal{M}^{i,j=1:n,1:m}_t$ using a multivariate model \citep{xuan2007modeling}. A common choice is to use the multivariate Gaussian, i.e., $\mathcal{M}^{i,j=1:n,1:m}_t \sim \mathcal{N}(0, \Sigma)$, with $\Sigma$ is an inverse Wishart prior $\Sigma \sim IW(N_0, V_0)$ and $N_0$ is set to be $mn$ which is the number time series within the metrics dataset and $V_0$ is set to be $\hat{\sigma}^2 I$, $I$ is the identity matrix and $\hat{\sigma}$ is the mean of the empirical variance pooled across all the metrics data. With this formulation, let us denote $h=(t_2-t_1)+1$, the marginal likelihood of the multivariate time series data $\mathcal{M}^{i,j=1:n,1:m}_{t_1:t_2}$ can then be computed explicitly as~\citep{xuan2007modeling},
\begin{equation}
\begin{aligned}
p(\{\mathcal{M}^{i,j=1:n,1:m}_{t_1:t_2}\}) &= \pi^{-\frac{hmn}{2}} \frac{\vert V_0 \vert^{N_0/2}}{\vert V_h\vert^{(N_0+h)/2}}  \frac{\Gamma_{mn}(N_0/2)^{-1}}{\Gamma_{mn}((N_0+h)/2)^{-1}}, \\
V_h &= V_0 + S, S=\sum_{i=t_1}^{t_2}  \mathcal{M}^{i,j=1:n,1:m}_t {\big(\mathcal{M}^{i,j=1:n,1:m}_t\big)}^\intercal,
\end{aligned}
\end{equation}
where $\Gamma_{mn}(.)$ denotes the multivariate gamma function. This formulation of the marginal likelihood $p(\mathcal{M}^{i,j=1:n,1:m}_{t_1:t_2})$ can then incorporated into Eq. (\ref{eq:bocpd-run-length}) to replace the univariate marginal likelihood. Other steps are kept the same in order to detect change points. 

Figure~\ref{fig:mbocpd} presents two examples of using Multivariate BOCPD to detect change points within multivariate time series metrics data using two different datasets. Figure~\ref{fig:mbocpd} shows that Multivariate BOCPD accurately detects change points (data points that separate the normal and abnormal data), and thus, detects the failures.

Finally, following the assumptions in Section \ref{sec:assumption-causal}, we use \textit{Latency} and \textit{Errors} to detect anomalies, and we only output the first detected change point as the detected anomaly.

\begin{algorithm}
\caption{Pseudo-code of BARO} \label{alg:baro}
\begin{algorithmic}[1]
\REQUIRE A set of metrics data $\mathcal{M} = \{x^{i=1:n,j=1:m}_{t_0:t_0+T}\}$
\ENSURE Ranked candidate root causes $\mathcal{R}$, ($\mathcal{R} \in \mathcal{M}$) if anomaly detected

\STATE // --- Multivariate BOCPD ---
\STATE $\mathcal{M}' \gets$ select \textit{Latency} and \textit{Errors} from $\mathcal{M}$
\STATE Compute run length probability $p(r_t \vert \{\mathcal{M'}_{t_0:t}\})$, $\forall x^{(i,j)}\in \mathcal{M'}$, $\forall t\in [t_0, t_0 + T]$
\STATE $s_t \gets \arg\max p(r_t \vert \{\mathcal{M'}_{t_0:t}\})$, $\forall t\in [t_0, t_0 + T]$
\FOR{$t \in [t_0, t_0 + T]$}
    \IF{$s_t \le s_{t-1}$}
        \STATE \textbf{return} $y = 1, \hat{t}_A = t$ \COMMENT{Anomaly detected (change point)}
    \ENDIF
\ENDFOR
\STATE \textbf{return} $y = 0, \hat{t}_A = \text{null}$ \COMMENT{No change point}

\vspace{1em}
\STATE // --- Robust Scorer ---
\STATE $\mathcal{R} \gets$ empty list
\FOR{$x^{(i, j)} \in \mathcal{M}$}
    \STATE $\rho^{(i,j)} \gets 0$
    \STATE med, IQR $\gets$ learn from $\{x^{(i,j)}_{t'}\}$ where $t_0 \le t' \le \hat{t}_A$
    \FOR{$t''$ from $\hat{t}_A$ to $t_0 + T$}
        \STATE $a^{(i,j)}_{t''} =$ abs($x^{(i,j)}_{t''}$ - med) / IQR
        \STATE $\rho^{(i,j)} = \max(\rho^{(i,j)}, a^{(i,j)}_{t''})$
    \ENDFOR
    \STATE Append $(x^{(i,j)}, \rho^{(i,j)})$ to $\mathcal{R}$
\ENDFOR
\STATE $\mathcal{R} \gets$ sort $\mathcal{R}$ in descending order of $\rho^{(i,j)}$
\STATE \textbf{return} $\mathcal{R}$

\vspace{1em}
\STATE // --- BARO Main ---
\STATE $(y, \hat{t}_A) \gets$ MultivariateBOCPD($\mathcal{M}$)
\IF{$y = 1$}
    \STATE \textbf{return} RobustScorer($\mathcal{M}$, $\hat{t}_A$)
\ELSE
    \STATE \textbf{return} null
\ENDIF

\end{algorithmic}
\end{algorithm}

\subsection[RobustScorer: Robust Nonparametric Hypothesis Testing]{RobustScorer: A Robust Nonparametric Hypothesis Testing} \label{sec:robustscorer}

To identify the root cause metrics, we aim to identify the metrics that exhibit significant changes in data distribution at the anomaly detection time \citep{Shan2019Ediagnosis, liu2023pyrca, Li2022Circa}. To solve this problem, one approach is to conduct hypothesis testing and test whether the data distribution of the metrics data changes significantly after the anomaly detection time. This approach was employed in $\epsilon$-Diagnosis \citep{Shan2019Ediagnosis, liu2023pyrca} and NSigma \citep{Li2022Circa} and they have been shown to perform very well in various scenarios. Our key insight is that previous works might be extremely sensitive to the anomaly detection output (failure occurrence time $\hat{t}_A$), i.e., inaccurate specification of the failure occurrence time might yield bad root cause analysis. Therefore, we propose RobustScorer to address this problem.

Specifically, for each metric $x_{t_0:t_0+T}^{(i,j)}$ in the metrics dataset $\mathcal{M}^{i,j=1:n,1:m}_{t_0:t_0+T}= \{x^{i=1:n,j=1:m}_{t_0:t_0+T}\}$, we conduct a hypothesis test based on the following null hypothesis ($\textbf{H}_0$), namely that \textit{$x^{(i,j)}$ is not a root cause metric for the failure}. This null hypothesis means that, for $t$ after the anomaly detection time, $x^{(i,j)}_t \sim \mathcal{L}(x^{(i,j)}_{t_\text{normal}})$ with $\mathcal{L}(x^{(i,j)}_{t_\text{normal}})$ denoting the distribution of the metrics data during the normal period (when $t$ is before the anomaly detection time).

This approach generally requires the specification of the anomaly detection time. Inaccurate anomaly detection time therefore could impact the accuracy of these techniques significantly. In this section, we propose a novel nonparametric hypothesis testing technique that is robust and less sensitive to the accuracy of the anomaly detection time.

\subsubsection{A Robust Nonparametric Hypothesis Test}

RobustScorer follows a statistical approach to learn the expected distribution from the metrics data. For every time series $x^{(i,j)}_{t_0:t_0+T}$ in the metrics dataset $\mathcal{M}^{i,j=1:n,1:m}_{t_0:t_0+T}$, let us denote $\hat{t}_A$ as the anomaly detection time, which is an estimation of the time when the anomaly occurs, RobustScorer is trained using the data collected prior to the anomaly, spanning from $t_0$ to $\hat{t}_A$, to learn the median ($med$) and interquartile range ($IQR$) of this data distribution. Subsequently, for each data point $x^{(i,j)}_t$ in the anomalous period (from $\hat{t}_A$ to $t_0+T$), RobustScorer measures how significant it deviates from the expected central tendency. This deviation is denoted as $a^{(i,j)}_t$ and is computed as follows,
\begin{equation} \label{eq:robust-scoring-1}
    a^{(i,j)}_t = \big | x^{(i,j)}_t - med \big | / IQR.
\end{equation}

All the values of $a^{(i,j)}_t$ across all the metrics data during the anomalous period are then consolidated to yield $\rho^{(i, j)}$, which is an indicator measuring the changes of each metric during the anomalous period,

\begin{equation} \label{eq:robust-scoring-2}
    \rho^{(i, j)} = \max_{\hat{t}_A \leq t \leq t_0+T} a^{(i, j)}_t.
\end{equation}

A higher $\rho^{(i, j)}$ signifies a greater likelihood that $x^{(i, j)}$ serves as the root cause metric, thereby identifying $s^i$ as the root cause service. Finally, RobustScorer then generates a ranked list of root cause metrics based on the magnitudes of $\rho^{(i, j)}$, arranged in descending order, with the highest ones corresponding to the most probable root cause metrics (fine-grained root causes). The coarse-grained ranked list of root cause services can be derived from the fine-grained ranking list by extracting the services corresponding to the metrics. Note that we do not use the p-value to reject or accept a possible root cause as in standard hypothesis testing. In our method, we use hypothesis testing to rank the potential root causes as it is normal for the system operators to focus on the top candidates.

Similar to \citep{Li2022Circa, Shan2019Ediagnosis}, our proposed RobustScorer is also distribution-free, scale-equivalent (i.e. metrics data in different scales do not affect the ranked list of root causes), and rotation-invariant (i.e. timestamp shifts do not affect the ranked list of root causes).

\begin{figure}[t]
\centering
\includegraphics[width=\textwidth]{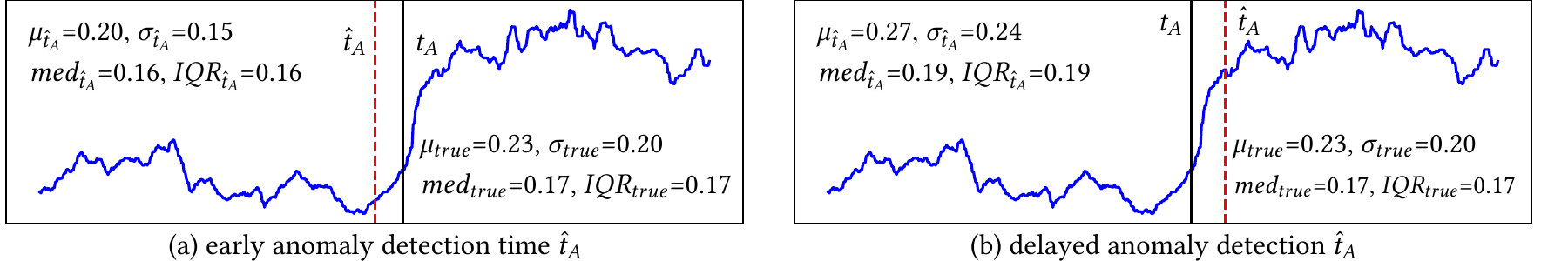}
\caption[The Robustness of RobustScorer against imprecise anomaly detection time.]{The Robustness of RobustScorer against imprecise anomaly detection time. In (a), an early anomaly detection time reduces the number of data points used to compute the distribution of the normal data in the hypothesis test. Median and IQR show greater resilience to a limited data setting compared to mean and standard deviation. In (b), a delayed anomaly detection time includes abnormal data (outliers) into the normal period. Median and IQR also show robustness to these outliers better than mean and $\sigma$.} \label{fig:robustscorer-robustness}
\end{figure}

\subsubsection{Why is RobustScorer Robust to Imprecise Anomaly Detection Time?} \label{sec:why-robust}

In contrast to previous works \citep{Li2022Circa, Shan2019Ediagnosis, Ma2019Msrank}, which compare normal and abnormal metrics data using the mean and standard deviation of the data distribution, we propose to use the median and interquartile range in our hypothesis testing module. The rationale is that mean and standard deviation are known to be sensitive to outliers, which could be introduced by inaccurate anomaly detection. On the other hand, the median and interquartile range are notably resilient to the impact of outliers \citep{modifiedzscore, robustzscore}.

In Figure~\ref{fig:robustscorer-robustness}, we illustrate the robustness to the anomaly detection time of RobustScorer corresponding to two scenarios, early anomaly detection (Figure~\ref{fig:robustscorer-robustness}a) and delayed anomaly detection (Figure~\ref{fig:robustscorer-robustness}b). An early anomaly detection reduces the number of data points used to compute the distribution of normal data in the hypothesis test. In the scenario of limited data, median and IQR are known to be more resilient than mean and standard deviation, making RobustScorer work well in this scenario. On the other hand, a delayed anomaly detection includes abnormal data (outliers) into the normal data period. Median and IQR are known to work well in the presence of outliers, making RobustScorer to also work well in this scenario. For instance, with RobustScorer, the computed median and IQR values of normal data distribution based on the anomaly detection time $\hat{t}_A$ remain close to those computed based on the true anomaly occurrence time $t_A$. In contrast, the computed mean and standard deviation based on the anomaly detection time $\hat{t}_A$ are more different compared to those computed based on $t_A$. In our evaluation in Section \ref{sec:results}, we demonstrate that RobustScorer outperforms existing RCA approaches in identifying the failure's root cause and generally is more robust to the anomaly detection time than other baselines.

\subsubsection{Handling of Correlated Failures}
For correlated failures affecting multiple services simultaneously, RobustScorer ranks affected services and metrics as the top possible root causes. This allows quicker identification of actual root causes, instead of troubleshooting all possible root causes. For example, consider a firewall misconfiguration leading to correlated failures affecting services A and B simultaneously. Although the true root cause is the misconfiguration, RobustScorer ranks services A and B as the top probable root cause services since they are the immediate successors of the true root cause. The operator can check these services and analyse the true root cause promptly.

%% file: papers/fse24/4.results.tex
This section answers the following research questions:
\begin{itemize}
    \item \textbf{\textit{RQ1: How effective is BARO in anomaly detection?}} To answer this RQ, we conduct an experiment to compare BARO with state-of-the-art anomaly detection approaches and evaluate their performance in detecting anomalies.
    \item \textbf{\textit{RQ2: How effective is BARO in root cause analysis?}} To answer this RQ, we compare BARO with state-of-the-art RCA methods and evaluate their performance in ranking both root cause services (coarse-grained) and root cause metrics (fine-grained) of the failures.
    \item \textit{\textbf{RQ3: How effective are the components of BARO?}} To answer this RQ, we evaluate the effectiveness of each main component of BARO: Multivariate BOCPD in detecting anomalies and RobustScorer in locating the failure's root cause.
    \item \textbf{\textit{RQ4: How sensitive are different RCA methods to their hyperparameters?}} To answer this RQ, we perform a sensitivity analysis to evaluate the performance of all these methods with different values of the anomaly detection time and other methods hyperparameters.
\end{itemize}

\subsection{Benchmark Microservice Systems \& Data Collection}

We deploy three well-known benchmark microservice systems, namely Online Boutique~\citep{ob}, Sock Shop~\citep{sockshop}, and Train Ticket~\citep{tt}, on a Kubernetes cluster consisting of one master node and five worker nodes. Each node has 16 CPUs and 32GB RAM, resulting in a total of 80 CPUs and 160GB RAM across five workers. We sequentially deployed the three systems with their default replicas configuration, i.e., one instance per service, to inject faults and gather metrics data under the load of 100-200 concurrent users. Online Boutique is an e-commerce application with 12 services, allowing users to view items, add them to their cart and make purchases. Each service in the Online Boutique system requires between 0.2-0.5 CPU and 64-512MB RAM for normal functioning. Sock Shop is another e-commerce system focused on selling socks, comprising 11 services that communicate via HTTP requests. Each service in the Sock Shop system requires between 0.1-1 CPU and 300MB-2GB RAM for normal functioning. On the other hand, Train Ticket is one of the largest microservice benchmark systems emulating a train ticket booking platform featuring 64 services. Compared to Sock Shop and Online Boutique, Train Ticket has longer and more complex failure propagation paths. Each service in the Train Ticket system requires between 0.2-1 CPU and 200MB-1GB RAM for normal functioning. These benchmark microservice systems have been widely recognised and employed for evaluating the performance of RCA methods \citep{Jinjin2018Microscope, Azam2022rcd, wu2021microdiag, xin2023causalrca, wu2022automatic, he2022graph, yu2021microrank, zhou2018trainticket, Wang2021evalcausal}.

\begin{figure}
\centering
\includegraphics[width=0.9\textwidth]{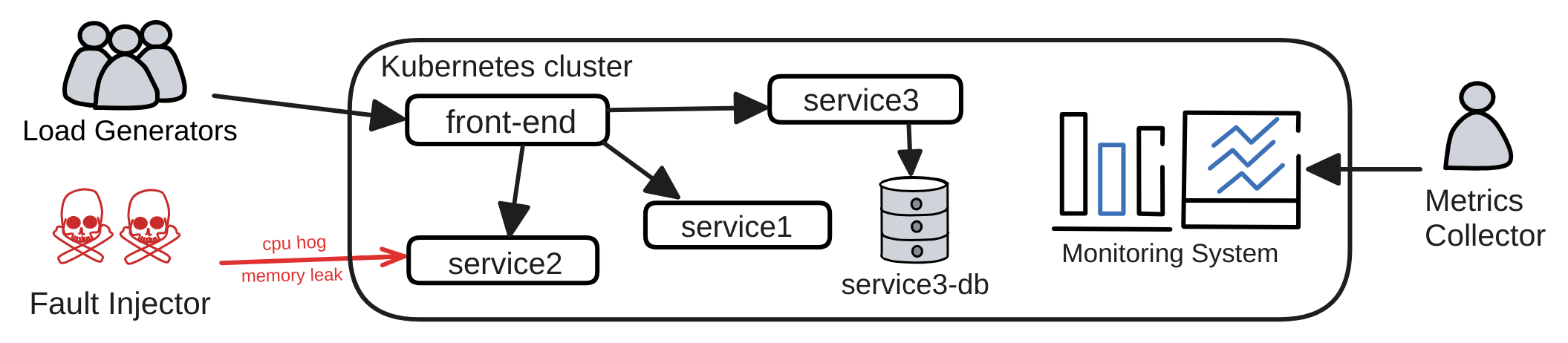}
\caption{Overview of our setup for microservice systems.} \label{fig:fse24-system_setup}
\end{figure}

To gather the metrics data, we employ the Istio service mesh \citep{istio} along with Prometheus \citep{prometheus} and cAdvisor \citep{cadvisor} to monitor and collect resource-level and service-level metrics, as previously done in~\citep{Azam2022rcd, xin2023causalrca, wu2021microdiag}. To generate traffic, we use the load generators supplied by these systems and tailor them to explore all services with the load of 40-50 requests per second. Figure~\ref{fig:fse24-system_setup} illustrates our data collection setup from the microservice systems.

We inject four common anomalies: CPU hog, memory leak, network delay, and packet loss into several key services within each benchmark microservice system. Initially, we operate the applications normally 
to gather metrics data under normal conditions. Then, we follow the existing practice \citep{Azam2022rcd, lee2023eadro, wu2021microdiag, xin2023causalrca, yu2021microrank} to inject faults into the running services. We execute into the designated container using kubectl exec. For CPU hog and memory leak, we use stress-ng \citep{stressng} to stress the container resource. For network delay and packet loss, we use tc \citep{tc} to manipulate the traffic of the container. Specifically, we inject faults into five targeted services of Sock Shop (carts, catalogue, orders, payment, and user), five targeted services of Online Boutique (adservice, cartservice, checkoutservice, currencyservice, and productcatalogue), and five targeted services of Train Ticket (ts-auth-service, ts-order-service, ts-route-service, ts-train-service, ts-travel-service). We choose these services due to their critical nature, as issues with their performance can impact other services~\citep{Azam2022rcd, xin2023causalrca, lee2023eadro}. For each combination of fault type and targeted service, we repeat the operation (i.e. fault injection and metrics data collection) five times, resulting in 100 failure cases for each microservice system. 

Furthermore, to ensure the independence of different injection experiments, we chose to restart the microservice systems after each experiment of injecting failures and collecting data, instead of waiting for a cooldown period, as described in previous studies \citep{wu2021microdiag, yu2021microrank, xin2023causalrca}. Upon visual inspections, we observed that the deviations between the fault injection time and the onset of failure symptoms in the metrics data are usually around 1-3 seconds. 
We simulate a workload of 100-200 concurrent users to interact with the systems intensively. Henceforth, the system is highly sensitive to the injected faults, and anomalies are reflected in the metrics shortly after the injection operation. Table \ref{tab:fse24-real-data} presents the statistics summarising the collected data. To the best of our knowledge, this is the first work to use all three popular benchmark microservice systems to evaluate the RCA methods.

\begin{table}
\centering
\caption[Characteristics of collected data from three benchmark microservice systems.]{Characteristics of collected data from three benchmark microservice systems (\#metrics, \#svc, \#t\_svc: number of metrics, services, and targeted services in the system. \#fault: number of fault types).} \label{tab:fse24-real-data}
\begin{tabular}{l r r r r r r}
\hline
\textbf{Name} & \textbf{\#metrics}  & \textbf{\#svc}   & \textbf{\#t\_svc}  & \textbf{\#fault} & \textbf{\#cases} \\ \hline \hline
Online Boutique & 49 & 12 & 5  & 4 & 100  \\ \hline
Sock Shop & 46 & 11 & 5 & 4 & 100   \\ \hline
Train Ticket & 212 & 64 & 5 & 4 & 100  \\ \hline
\end{tabular}%
\end{table}

\subsection{Evaluation Metrics}

\subsubsection{Anomaly Detection}
Similar to previous works \citep{Chen2014Causeinfer, chen2016causeinfer, chen2022adaptive}, we evaluate the anomaly detectors as binary classification models as the main goal is to detect whether there exist anomalies in the metrics data. We determine metrics data collected during a fault injection period as abnormal, while metrics data before that period is considered normal. Therefore, we use Precision, Recall, and F1 scores to evaluate the anomaly detectors. When an anomaly detection algorithm successfully detects an abnormal sample (i.e., a case with anomalies), it is counted as a True Positive (TP). Conversely, incorrectly classifying an abnormal sample as normal is considered False Negative (FN). Likewise, incorrectly classifying a normal sample as abnormal is considered False Positive (FP). The formulas for computing the metrics Precision, Recall, and F1-score are as follows:
\begin{equation}
    Precision = \frac{TP}{TP + FP}, \quad Recall = \frac{TP}{TP + FN}, \quad F1 = \frac{2\times Precision \times Recall}{Precision + Recall}.
\end{equation}

\subsubsection{Root Cause Analysis}

Following existing works \citep{Wang2018cloudranger, Ma2019Msrank, Ma2020Automap, Meng2020Microcause, yu2021microrank, wu2020microrca, wu2021microdiag, Azam2022rcd, Li2022Circa, xin2023causalrca}, we use two standard metrics, namely $AC@k$ and $Avg@k$ to assess the performance of the RCA methods. Herein, we set $k = 1, 3, 5$. Given a set of failure cases A, $AC@k$ and $Avg@k$ are calculated as follows,
\begin{equation}
    AC@k = \frac{1}{|A|} \sum\nolimits_{a\in A}\frac{\sum_{i<k}R^a[i]\in V^a_{rc}}{min(k, |V^a_{rc}|)}, \quad Avg@k = \frac{1}{k}\sum_{j=1}^k AC@j.
\end{equation}
where $R^a[i]$ denotes the $i$th ranking result for the failure case $a$ by an RCA method, and $V^a_{rc}$ is the true root cause set of case $a$. $AC@k$ represents the probability the top $k$ results given by a method include the real root causes. It ranges from $0$ to $1$, with higher values indicating better performance.
Meanwhile, $Avg@k$ measures the overall performance. 

\subsection{Experimental Setting}
We conduct all the experiments on Linux servers equipped with 8 CPU and 16GB RAM. To avoid any randomness, when evaluating each method, for each failure case we repeat the experiment (detecting anomalies and identifying root causes) five times, then report the average results. Our framework is implemented using Python 3.10.

\subsection{Baselines} \label{sec:baselines}

\paragraph{Anomaly Detection Baselines.} We select the following four baselines to compare against our proposed method: \textit{N-Sigma} \citep{Jinjin2018Microscope, Li2022Circa}, \textit{BIRCH} \citep{zhang1996birch, wu2020microrca, wu2021microdiag}, \textit{SPOT} \citep{siffer2017anomaly, pan2021dycauserca, Li2022Circa, lee2023eadro, Meng2020Microcause}, and \textit{Univariate Bayesian Change Point Detection} (denoted as UniBCP) \citep{adams2007bayesian, Chen2014Causeinfer, chen2016causeinfer}. These methods are commonly used in existing RCA research for anomaly detection. Their source code has been made publicly available, with the exception of UniBCP, for which we leverage an available implementation \citep{unibcp} for the evaluation. The detailed descriptions of these methods are in Section \ref{sec:background-ad}. For all methods, we use the hyperparameter settings as in previous works and in their public source code.

\paragraph{Root Cause Analysis Baselines.} We choose six representative baselines, including five state-of-the-art metric-based RCA methods: CausalRCA \citep{xin2023causalrca}, RCD \citep{Azam2022rcd}, CIRCA \citep{Li2022Circa}, $\epsilon$-Diagnosis \citep{Shan2019Ediagnosis}, and N-Sigma \citep{Li2022Circa}, for performance comparison with our proposed method, BARO. Detailed information of these methods is as follows:

\begin{itemize}
\item \textit{Dummy:} Dummy randomly selects a metric as the root cause. We use this method to assess whether our BARO and the other baselines outperform random selection.

\item \textit{CausalRCA \citep{xin2023causalrca}:} CausalRCA uses DAG-GNN \citep{Yu2019DagGNN}, a gradient-based causal structure learning method, to estimate the causal graph and uses PageRank algorithm to rank the root causes.

\item \textit{$\epsilon$-Diagnosis \citep{Shan2019Ediagnosis}:}
 $\epsilon$-Diagnosis uses the two-sample test algorithm and $\epsilon$-statistics to estimate the similarity between every pair of metrics and rank the root causes based on test scores.

\item \textit{RCD \citep{Azam2022rcd}:} RCD employs a divide-and-conquer strategy to divide the input metrics into smaller chunks. It uses the $\Psi$-PC algorithm \citep{Jaber2020PsiFCI} to construct a causal graph and identify root causes within each chunk. Subsequently, it combines these root causes and repeats this process until only one chunk remains.

\item \textit{CIRCA \citep{Li2022Circa}:} CIRCA creates a causal graph by using a provided call graph, which requires operational knowledge. Then, it performs regression-based hypothesis testing to find the root causes. Since the call graph is not available, thus following \citep{Azam2022rcd, liu2023pyrca}, we use the PC algorithm to construct this graph for CIRCA. 

\item \textit{N-Sigma \citep{Jinjin2018Microscope,Li2022Circa}:} N-Sigma is a statistical analysis technique that compares the metrics data before and after the failur time using z-score. The higher the score, the more likely that metric is the root cause of the failure. 

\end{itemize}

The source code of CausalRCA, RCD, CIRCA, and N-Sigma is publicly available. For CIRCA, since the required call graphs are unavailable, we use the PC algorithm to construct the causal graphs as done in \citep{Azam2022rcd, Ma2019Msrank, Ma2020Automap, Wang2018cloudranger, xin2023causalrca}. For $\epsilon$-Diagnosis, the original source code is unavailable thus we rely on the implementation of a related work \citep{liu2023pyrca}. We use the same hyperparameter values as reported in their papers.

\subsection{RQ1: Effectiveness in Anomaly Detection}

In this RQ, we evaluate the performance of BARO and the baseline anomaly detectors across all three datasets. We use the metrics data collected before the failure as normal data, and during the failure as abnormal data. We report the average of Precision (Pre), Recall (Rec), and F1-score (F1) over all the cases. Table \ref{tab:ad1} presents the experimental results, with the best results highlighted in \textbf{bold}. We draw the following observations:

\textbf{(1) BARO consistently outperforms all baseline methods in detecting anomalies by a large margin across all three benchmark microservice systems.} It achieves the highest Precision, Recall, and F1-score on all the datasets. BARO's performance can be attributed to three factors. First, BOCPD's ability to detect distribution changes is suitable for identifying anomalies from the failure, acting as a soft intervention \citep{Azam2022rcd, Li2022Circa}. Second, Multivariate BOCPD considers not only individual time series but also the dependencies among the time series, allowing it to detect correlation changes. Third, it does not require any user-defined thresholds, making it highly adaptive to different types of metrics data.

\begin{table}
\centering
\caption[Precision, Recall, and F1-score of five anomaly detectors on three datasets: Online Boutique, Sock Shop, and Train Ticket. The best scores are in \textbf{bold}.]{Precision, Recall, and F1-score of five anomaly detectors on three datasets: Online Boutique, Sock Shop, and Train Ticket. The best scores are in \textbf{bold}.}
\label{tab:ad1}
\begin{tabular}{l|rrr|rrr|rrr}
\hline
 & \multicolumn{3}{c|}{\textbf{Online Boutique}} & \multicolumn{3}{c|}{\textbf{Sock Shop}} & \multicolumn{3}{c}{\textbf{Train Ticket}} \\ \cline{2-10} 
Method & \multicolumn{1}{c}{\textit{Pre}} & \multicolumn{1}{c}{\textit{Rec}} & \multicolumn{1}{c|}{\textit{F1}} & \multicolumn{1}{c}{\textit{Pre}} & \multicolumn{1}{c}{\textit{Rec}} & \multicolumn{1}{c|}{\textit{F1}} & \multicolumn{1}{l}{\textit{Pre}} & \multicolumn{1}{l}{\textit{Rec}} & \multicolumn{1}{c}{\textit{F1}}  \\ \hline
N-Sigma & 0.54 & \textbf{1.00} & 0.70 & 0.56 & \textbf{1.00} & 0.72 & 0.50 & \textbf{1.00} & 0.67 \\ \hline
SPOT & 0.53 & \textbf{1.00} & 0.69 & 0.53 & \textbf{1.00} & 0.69 & 0.50 & \textbf{1.00} & 0.67 \\ \hline
BIRCH & 0.35 & 0.48 & 0.40 & 0.05 & 0.05 & 0.05 & 0.39 & 0.40 & 0.39 \\ \hline
UniBCP & 0.51 & \textbf{1.00} & 0.67 & 0.50 & 0.99 & 0.66 & 0.50 & \textbf{1.00} & 0.67 \\ \hline
\textbf{BARO (Ours)} & \textbf{0.69} & \textbf{1.00} & \textbf{0.82} & \textbf{0.60} & \textbf{1.00} & \textbf{0.75} & \textbf{0.68} & \textbf{1.00} & \textbf{0.81} \\ \hline
\end{tabular}%

{\footnotesize \textit{(*) UniBCP denotes Univariate Offline Bayesian Change Point Detection.}}

\end{table}

\textbf{(2) All baseline methods, except BIRCH, have very high recalls but low precisions across all the datasets.} This indicates that they generally can detect anomalies, however, they also frequently misclassify normal time series as anomalous. The precision of these methods generally range from 0.05 to 0.68. Based on our observations, this is due to the complexity and dynamics of the benchmark microservice systems, which include many time series metrics with complicated patterns, leading to frequent misclassifications of anomalies by these baseline methods.

In summary, BARO stands out as the best-performing method, being effective in detecting anomalies. This capability plays a crucial role in the subsequent RCA task.

\noindent\textit{Answers to RQ1: BARO consistently outperforms all four anomaly detection baselines, achieving the highest F1-scores of 0.82, 0.75, and 0.81 on Online Boutique, Sock Shop, and Train Ticket, respectively.}

\subsection{RQ2: Effectiveness in Root Cause Analysis}

In this section, we evaluate the performance of BARO and the RCA baseline methods on all three datasets. First, we assess the ideal performance of the methods when the anomaly occurrence time is known accurately. Specifically, we use the fault injection time $t_{\text{inject}}$ as a proxy for the true anomaly occurence time since we have observed that in our benchmark microservice systems, the anomaly occurence time closely aligns with the fault injection time. Then, we assess the performance of RCA methods when using the anomaly detection time provided by existing anomaly detectors described in Section \ref{sec:baselines}. Note that Dummy and CausalRCA do not require the specification of the anomaly detection time. Tables \ref{tab:q2-rca-s} and \ref{tab:q2-rca-f1} report the overall performance of all methods using the Avg@5 scores for coarse-grained root cause analysis (root cause services) and fine-grained root cause analysis (root cause metrics), respectively. We calculate the accuracy for each type of fault: CPU hog (CPU), memory leak (MEM), network delay (DELAY), and packet loss (LOSS), as well as their average (AVG) to report the overall performance across four fault types. In the tables, we denote the combination of an RCA method and an anomaly detector as \textit{RCA\_method[Anomaly\_Detector]}. For instance, RCD[N-Sigma], RCD[BIRCH], RCD[SPOT], RCD[UniBCP], and RCD[BOCPD] represent the RCA pipelines that combine RCD with the anomaly detectors N-Sigma, BIRCH, SPOT, Univariate BCPD, and Multivariate BOCPD, respectively.

\begin{table}[]
\centering
\caption[Coarse-grained performance of different RCA methods in terms of Avg@5 on three different datasets.]{Coarse-grained performance of different RCA methods in terms of Avg@5 on three different datasets. The fault types CPU, MEM, DELAY, and LOSS denote CPU overload, memory leak, network delay, and packet loss. The highest scores are in \textbf{bold}, and the second highest scores are in \underline{\mbox{underscore}}.}
\label{tab:q2-rca-s}
\resizebox{\textwidth}{!}{%
\setlength\tabcolsep{3pt}
\begin{tabular}{l|rrrrr|rrrrr|rrrrr}
\hline
 & \multicolumn{5}{c|}{\textbf{Online Boutique}} & \multicolumn{5}{c|}{\textbf{Sock Shop}} & \multicolumn{5}{c}{\textbf{Train Ticket}} \\ \cline{2-16} 
Method & \multicolumn{1}{c}{\textit{CPU}} & \multicolumn{1}{c}{\textit{MEM}} & \multicolumn{1}{c}{\textit{DELAY}} & \multicolumn{1}{c}{\textit{LOSS}} & \multicolumn{1}{c|}{\textit{AVG}} & \multicolumn{1}{c}{\textit{CPU}} & \multicolumn{1}{c}{\textit{MEM}} & \multicolumn{1}{c}{\textit{DELAY}} & \multicolumn{1}{c}{\textit{LOSS}} & \multicolumn{1}{c|}{\textit{AVG}} & \multicolumn{1}{c}{\textit{CPU}} & \multicolumn{1}{c}{\textit{MEM}} & \multicolumn{1}{c}{\textit{DELAY}} & \multicolumn{1}{c}{\textit{LOSS}} & \multicolumn{1}{c}{\textit{AVG}} \\ \hline \hline
Dummy & 0.24 & 0.26 & 0.27 & 0.24 & 0.25 & 0.37 & 0.41 & 0.38 & 0.37 & 0.38 & 0.07 & 0.08 & 0.07 & 0.07 & 0.07 \\
CausalRCA & 0.85 & 0.91 & 0.86 & 0.58 & 0.80 & 0.49 & 0.82 & 0.61 & 0.47 & 0.60 & 0.53 & 0.30 & 0.17 & 0.11 & 0.28 \\ \hline
$\epsilon$-Diagnosis {[}$t_{\text{inject}}${]} & 0.10 & 0.13 & 0.12 & 0.12 & 0.12 & 0.47 & 0.30 & 0.42 & 0.49 & 0.42 & 0.00 & 0.02 & 0.00 & 0.00 & 0.01 \\
$\epsilon$-Diagnosis {[}N-Sigma{]} & 0.19 & 0.16 & 0.23 & 0.22 & 0.20 & 0.49 & 0.39 & 0.46 & 0.50 & 0.46 & 0.00 & 0.00 & 0.00 & 0.02 & 0.01 \\
$\epsilon$-Diagnosis {[}BIRCH{]} & 0.23 & 0.33 & 0.21 & 0.14 & 0.22 & - & 0.00 & - & 1.00 & 0.80 & 0.07 & 0.00 & 0.00 & 0.00 & 0.02 \\
$\epsilon$-Diagnosis {[}SPOT{]} & 0.37 & 0.20 & 0.17 & 0.22 & 0.24 & 0.46 & 0.34 & 0.42 & 0.46 & 0.42 & 0.00 & 0.02 & 0.00 & 0.00 & 0.01 \\
$\epsilon$-Diagnosis {[}UniBCP{]} & 0.28 & 0.09 & 0.40 & 0.20 & 0.24 & 0.20 & 0.19 & 0.25 & 0.20 & 0.21 & 0.05 & 0.10 & 0.08 & 0.04 & 0.07  \\
$\epsilon$-Diagnosis {[}BOCPD{]} & 0.23 & 0.09 & 0.18 & 0.15 & 0.16 & 0.45 & 0.35 & 0.38 & 0.44 & 0.41 & 0.00 & 0.00 & 0.00 & 0.06 & 0.02 \\ \hline
RCD {[}$t_{\text{inject}}${]} & 0.69 & 0.40 & 0.27 & 0.50 & 0.47 & 0.62 & 0.46 & 0.48 & 0.37 & 0.48 & 0.08 & 0.01 & 0.09 & 0.12 & 0.08 \\
RCD {[}N-Sigma{]} & 0.67 & 0.43 & 0.31 & 0.49 & 0.48 & 0.54 & 0.42 & 0.46 & 0.37 & 0.45 & 0.07 & 0.00 & 0.05 & 0.10 & 0.06 \\
RCD {[}BIRCH{]} & 0.69 & 0.37 & 0.28 & 0.50 & 0.46 & 0.57 & 0.42 & 0.49 & 0.37 & 0.46 & 0.06 & 0.00 & 0.03 & 0.11 & 0.05 \\
RCD {[}SPOT{]} & 0.70 & 0.42 & 0.30 & 0.48 & 0.48 & 0.58 & 0.43 & 0.46 & 0.34 & 0.45 & 0.08 & 0.01 & 0.07 & 0.09 & 0.06 \\
RCD {[}UniBCP{]} & 0.66 & 0.42 & 0.29 & 0.49 & 0.47 & 0.62 & 0.44 & 0.50 & 0.36 & 0.48 & 0.05 & 0.01 & 0.05 & 0.10 & 0.05 \\
RCD {[}BOCPD{]} & 0.70 & 0.40 & 0.30 & 0.52 & 0.48 & 0.60 & 0.47 & 0.48 & 0.36 & 0.48 & 0.06 & 0.00 & 0.04 & 0.09 & 0.05 \\ \hline
CIRCA {[}$t_{\text{inject}}${]} & \underline{\mbox{0.90}} & 0.74 & 0.90 & 0.55 & 0.77 & \underline{\mbox{0.97}} & \textbf{0.98} & \underline{\mbox{0.98}} & \underline{\mbox{0.88}} & \underline{\mbox{0.95}} & 0.66 & 0.93 & \textbf{0.64} & 0.57 & 0.70 \\
CIRCA {[}N-Sigma{]} & 0.78 & 0.59 & 0.79 & 0.48 & 0.66 & 0.86 & 0.70 & 0.87 & 0.70 & 0.78 & 0.65 & 0.88 & 0.58 & 0.57 & 0.67 \\
CIRCA {[}BIRCH{]} & 0.45 & 0.38 & 0.27 & 0.48 & 0.39 & - & 0.60 & - & 0.75 & 0.72 & 0.10 & 0.00 & 0.35 & 0.13 & 0.19 \\
CIRCA {[}SPOT{]} & 0.76 & 0.70 & 0.88 & 0.66 & 0.75 & 0.83 & 0.87 & 0.95 & 0.78 & 0.86 & 0.69 & 0.96 & 0.58 & 0.53 & 0.69 \\
CIRCA {[}UniBCP{]} & 0.16 & 0.00 & 0.00 & 0.35 & 0.13 & 0.23 & 0.00 & 0.20 & 0.36 & 0.20 & 0.00 & 0.00 & 0.00 & 0.33 & 0.08 \\
CIRCA {[}BOCPD{]} & 0.68 & 0.20 & 0.28 & 0.42 & 0.40 & 0.73 & \textbf{0.98} & 0.54 & 0.66 & 0.73 & 0.37 & 0.12 & 0.30 & 0.17 & 0.24 \\ \hline
N-Sigma {[}$t_{\text{inject}}${]} & \underline{\mbox{0.90}} & 0.93 & \underline{\mbox{0.94}} & 0.66 & \textbf{0.86} & \textbf{0.98} & \textbf{0.98} & \underline{\mbox{0.98}} & \textbf{0.90} & \textbf{0.96} & 0.81 & 0.96 & 0.61 & \textbf{0.70} & 0.77 \\
N-Sigma {[}N-Sigma{]} & 0.79 & 0.69 & 0.83 & 0.63 & 0.74 & 0.85 & 0.77 & 0.89 & 0.79 & 0.83 & 0.74 & \textbf{0.98} & 0.61 & 0.65 & 0.75 \\
N-Sigma {[}BIRCH{]} & 0.83 & 0.78 & 0.64 & 0.60 & 0.71 & 		
- & 0.00 & - & 0.85 & 0.68 & 0.55 & 0.70 & 0.47 & 0.45 & 0.51 \\
N-Sigma {[}SPOT{]} & 0.79 & 0.79 & 0.93 & 0.65 & 0.79 & 0.87 & \underline{\mbox{0.91}} & \textbf{0.99} & 0.82 & 0.90 & 0.75 & \textbf{0.98} & 0.61 & 0.64 & 0.75 \\
N-Sigma {[}UniBCP{]} & 0.75 & 0.69 & 0.91 & \underline{\mbox{0.76}} & 0.78 & 0.86 & 0.79 & 0.69 & 0.81 & 0.79 & 0.38 & 0.71 & 0.38 & 0.62 & 0.52 \\
N-Sigma {[}BOCPD{]} & 0.74 & 0.32 & 0.52 & 0.58 & 0.54 & 0.80 & \textbf{0.98} & 0.63 & 0.62 & 0.76 & 0.51 & 0.16 & 0.29 & 0.22 & 0.30 \\ \hline 
RobustScorer {[}$t_{\text{inject}}${]} & \underline{\mbox{0.90}} & \underline{\mbox{0.94}} & \underline{\mbox{0.94}} & 0.65 & \textbf{0.86} & \textbf{0.98} & \textbf{0.98} & \underline{\mbox{0.98}} & 0.86 & \underline{\mbox{0.95}} & \textbf{0.90} & \underline{\mbox{0.97}} & \underline{\mbox{0.62}} & \underline{\mbox{0.67}} & \underline{\mbox{0.79}} \\
RobustScorer {[}N-Sigma{]} & \underline{\mbox{0.90}} & \textbf{0.96} & 0.88 & 0.58 & 0.83 & 0.96 & \textbf{0.98} & \underline{\mbox{0.98}} & \textbf{0.90} & \textbf{0.96} & 0.82 & \textbf{0.98} & \underline{\mbox{0.62}} & 0.64 & 0.77 \\
RobustScorer {[}BIRCH{]} & 0.92 & 0.93 & 0.81 & 0.51 & 0.79 & - & 1.00 & - & 0.85 & 0.88 & 0.76 & 0.55 & 0.57 & 0.71 & 0.65 \\
RobustScorer {[}SPOT{]} & 0.89 & \underline{\mbox{0.94}} & 0.93 & 0.61 & \underline{\mbox{0.84}} & \underline{\mbox{0.97}} & \textbf{0.98} & \textbf{0.99} & 0.86 & \underline{\mbox{0.95}} & \underline{\mbox{0.85}} & \textbf{0.98} & 0.61 & 0.66 & 0.78 \\
RobustScorer {[}UniBCP{]} & 0.73 & 0.67 & 0.89 & \textbf{0.79} & 0.77 & 0.80 & 0.65 & 0.69 & 0.84 & 0.75 & 0.37 & 0.64 & 0.38 & 0.54 & 0.48 \\ \hline
\textbf{BARO (Ours)} & \textbf{0.91} & \textbf{0.96} & \textbf{0.95} & 0.62 & \textbf{0.86} & \underline{\mbox{0.97}} & \textbf{0.98} & \underline{\mbox{0.98}} & 0.87 & \underline{\mbox{0.95}} & \textbf{0.90} & \textbf{0.98} & \textbf{0.64} & \textbf{0.70} & \textbf{0.81} \\ \hline
\end{tabular}%
}

{\footnotesize \textit{(*) BIRCH has a poor recall, so RCA results with BIRCH are obtained from a limited number of failure cases where BIRCH detects anomalies.
CIRCA results for TrainTicket are only obtained from 15/100 cases due to PC's failed causal graph construction.
}}
\end{table}

\subsubsection{Coarse-grained Root Cause Analysis}

From Table \ref{tab:q2-rca-s}, we draw the following observations.

\textbf{(1) BARO performs the best in identifying the coarse-grained root cause of the failure}, consistently achieving top accuracy scores for all types of faults across all datasets. BARO achieves the highest Avg@5 in 10 out of 15 cases and outperforms other baselines by a large margin. For example, while CausalRCA achieves 0.8, 0.6, and 0.28, BARO achieves 0.86, 0.95, and 0.81 in the overall Avg@5 for all three microservice systems, respectively. This represents 7.5\%, 58\%, and 189\% improvements compared to CausalRCA.

From the results, we can see that, when working with large-scale microservice systems, $\epsilon$-Diagnosis, RCD, CausalRCA, and CIRCA encounter huge difficulties, whereas BARO yields much better results. BARO's resistance to the imprecision of the anomaly detection time via the nonparametric RobustScorer hypothesis test demonstrates its applicability in practical scenarios, where the anomaly detection module may differ across real-world systems. Moreover, BARO leverages multivariate BOCPD to capture dependencies within multivariate metrics data, allowing for more precise localisation of the anomaly occurrence time and thus improve the RCA performance.

\textbf{(2) CausalRCA does not require separating the normal and abnormal metrics data.} It uses all the provided metrics data to construct the causal graph and then uses PageRank to rank the root causes. The experimental results indicate that it performs well on the Online Boutique system. For instance, it achieves an Avg@5 of 0.85 for the CPU overload fault, 0.91 for the memory leak fault, and an overall Avg@5 of 0.8. Nevertheless, its performance declines on the other two systems. For example, it only achieves overall Avg@5 scores of 0.6 and 0.28 on the Sock Shop and Train Ticket systems, possibly due to the challenges posed by the complex structures of these systems. Overall, we can see that the performance of CausalRCA varies significantly across different systems.

\textbf{(3) $\epsilon$-Diagnosis, a widely-known baseline RCA method, does not perform greatly superior to random selection.} For example, its overall Avg@5 scores range from 0.12 to 0.24, whereas random selection achieves 0.25 on the Online Boutique system. However, it outperforms Dummy by 20\% in the Sock Shop system, with an overall Avg@5 score of 0.46 compared to Dummy's 0.38.

\textbf{(4) RCD outperforms random selection in small-scale systems such as Online Boutique and Sock Shop.} For example, its overall Avg@5 scores range from 0.45 to 0.48 while Dummy's scores range from 0.25 to 0.38. However, for the large-scale Train Ticket system, RCD's performance drops remarkably. For instance, RCD's Avg@5 scores range from 0.05 to 0.08, similar to Dummy's. This suggests the need to include large systems like Train Ticket in future work to benchmark newly proposed RCA methods.

\textbf{(5) We observe that N-Sigma, a simple statistical analysis technique, performs better than CIRCA.} Note that, in this chapter, as discussed, for CIRCA, we use the causal graphs constructed by the PC algorithm instead of the required call graphs as these graphs are not available for these systems, which could be a cause of this observation. Overall, N-Sigma outperforms both CIRCA and RCD in coarse-grained RCA.

In summary, BARO consistently outperforms other methods for all fault types and datasets by a large margin. When tested on large 
system like Train Ticket, baselines 
like $\epsilon$-Diagnosis, RCD, CausalRCA, and CIRCA struggle, while BARO delivers superior results.

\begin{table}[!t]
\centering
\caption[Fine-grained performance of different RCA methods in terms of Avg@5 accuracy on three different datasets.]{Fine-grained performance of different RCA methods in terms of Avg@5 accuracy on three different datasets. The fault types CPU, MEM, DELAY, and LOSS denote CPU overload, memory leak, network delay, and packet loss. The highest scores are in \textbf{bold}, and the second highest scores are in \underline{\mbox{underscore}}.}
\label{tab:q2-rca-f1}
\resizebox{\textwidth}{!}{%
\setlength\tabcolsep{3pt}
\begin{tabular}{l|rrrrr|rrrrr|rrrrr}
\hline
 & \multicolumn{5}{c|}{\textbf{Online Boutique}} & \multicolumn{5}{c|}{\textbf{Sock Shop}} & \multicolumn{5}{c}{\textbf{Train Ticket}} \\ \cline{2-16} 
Method & \multicolumn{1}{c}{\textit{CPU}} & \multicolumn{1}{c}{\textit{MEM}} & \multicolumn{1}{c}{\textit{DELAY}} & \multicolumn{1}{c}{\textit{LOSS}} & \multicolumn{1}{c|}{\textit{AVG}} & \multicolumn{1}{c}{\textit{CPU}} & \multicolumn{1}{c}{\textit{MEM}} & \multicolumn{1}{c}{\textit{DELAY}} & \multicolumn{1}{c}{\textit{LOSS}} & \multicolumn{1}{c|}{\textit{AVG}} & \multicolumn{1}{c}{\textit{CPU}} & \multicolumn{1}{c}{\textit{MEM}} & \multicolumn{1}{c}{\textit{DELAY}} & \multicolumn{1}{c}{\textit{LOSS}} & \multicolumn{1}{c}{\textit{AVG}} \\ \hline \hline
Dummy & 0.08 & 0.06 & 0.07 & 0.06 & 0.07 & 0.10 & 0.07 & 0.07 & 0.06 & 0.08 & 0.02 & 0.02 & 0.02 & 0.01 & 0.02 \\
CausalRCA & \textbf{0.55} & \textbf{0.78} & 0.86 & 0.39 & \textbf{0.65} & 0.39 & 0.42 & 0.49 & 0.34 & 0.41 & 0.51 & 0.13 & 0.10 & 0.09 & 0.21 \\ \hline
$\epsilon$-Diagnosis {[}$t_{\text{inject}}${]} & 0.07 & 0.03 & 0.06 & 0.02 & 0.05 & 0.00 & 0.00 & 0.00 & 0.00 & 0.00 & 0.00 & 0.00 & 0.00 & 0.00 & 0.00 \\
$\epsilon$-Diagnosis {[}N-Sigma{]} & 0.00 & 0.06 & 0.20 & 0.13 & 0.10 & 0.02 & 0.00 & 0.00 & 0.00 & 0.01 & 0.00 & 0.00 & 0.00 & 0.00 & 0.00 \\
$\epsilon$-Diagnosis {[}BIRCH{]} & 0.00 & 0.07 & 0.21 & 0.14 & 0.11 &  - & 0.00 & - & 0.00 & 0.00 & 0.00 & 0.00 & 0.20 & 0.00 & 0.00 \\
$\epsilon$-Diagnosis {[}SPOT{]} & 0.06 & 0.02 & 0.14 & 0.06 & 0.07 & 0.00 & 0.00 & 0.00 & 0.00 & 0.00 & 0.00 & 0.00 & 0.00 & 0.00 & 0.00 \\
$\epsilon$-Diagnosis {[}UniBCP{]} & 0.00 & 0.00 & 0.30 & 0.12 & 0.11 & 0.00 & 0.00 & 0.11 & 0.09 & 0.05 & 0.05 & 0.00 & 0.05 & 0.00 & 0.03 \\
$\epsilon$-Diagnosis {[}BOCPD{]} & 0.02 & 0.04 & 0.10 & 0.12 & 0.07 & 0.00 & 0.00 & 0.00 & 0.00 & 0.00 & 0.00 & 0.00 & 0.00 & 0.00 & 0.00 \\ \hline
RCD {[}$t_{\text{inject}}${]} & 0.16 & 0.26 & 0.22 & 0.18 & 0.21 & 0.02 & 0.17 & 0.10 & 0.07 & 0.09 & 0.00 & 0.00 & 0.06 & 0.00 & 0.02 \\
RCD {[}N-Sigma{]} & 0.16 & 0.29 & 0.23 & 0.15 & 0.21 & 0.00 & 0.16 & 0.09 & 0.07 & 0.08 & 0.00 & 0.00 & 0.03 & 0.00 & 0.01 \\
RCD {[}BIRCH{]} & 0.15 & 0.27 & 0.24 & 0.19 & 0.21 & 0.01 & 0.15 & 0.11 & 0.06 & 0.08 & 0.00 & 0.00 & 0.02 & 0.00 & 0.00 \\
RCD {[}SPOT{]} & 0.14 & 0.31 & 0.23 & 0.15 & 0.21 & 0.02 & 0.16 & 0.11 & 0.06 & 0.09 & 0.00 & 0.00 & 0.04 & 0.00 & 0.01 \\
RCD {[}UniBCP{]} & 0.16 & 0.31 & 0.22 & 0.14 & 0.21 & 0.01 & 0.16 & 0.09 & 0.07 & 0.08 & 0.00 & 0.00 & 0.04 & 0.00 & 0.01 \\
RCD {[}BOCPD{]} & 0.17 & 0.28 & 0.24 & 0.17 & 0.22 & 0.04 & 0.16 & 0.14 & 0.07 & 0.10 & 0.00 & 0.00 & 0.03 & 0.00 & 0.01 \\ \hline
CIRCA {[}$t_{\text{inject}}${]} & 0.42 & 0.30 & 0.90 & 0.54 & 0.54 & 0.52 & 0.58 & \underline{\mbox{0.98}} & \underline{\mbox{0.86}} & 0.74 & 0.48 & 0.86 & 0.55 & 0.57 & 0.62 \\
CIRCA {[}N-Sigma{]} & 0.22 & 0.27 & 0.79 & 0.40 & 0.42 & 0.50 & 0.42 & 0.87 & 0.66 & 0.61 & 0.48 & 0.84 & 0.51 & 0.54 & 0.59 \\
CIRCA {[}BIRCH{]} & 0.17 & 0.00 & 0.29 & 0.38 & 0.23 & - & 0.00 & - & 0.75 & 0.60 & 0.10 & 0.00 & 0.31 & 0.13 & 0.18 \\
CIRCA {[}SPOT{]} & 0.34 & 0.18 & 0.88 & 0.60 & 0.50 & 0.50 & 0.46 & 0.94 & 0.74 & 0.66 & 0.50 & 0.88 & 0.50 & 0.50 & 0.60 \\
CIRCA {[}UniBCP{]} & 0.00 & 0.00 & 0.00 & 0.20 & 0.05 & 0.08 & 0.00 & 0.00 & 0.15 & 0.06 & 0.00 & 0.00 & 0.00 & 0.17 & 0.04 \\
CIRCA {[}BOCPD{]} & 0.29 & 0.00 & 0.18 & 0.35 & 0.21 & 0.42 & 0.49 & 0.34 & 0.47 & 0.43 & 0.30 & 0.12 & 0.18 & 0.09 & 0.17 \\ \hline
N-Sigma {[}$t_{\text{inject}}${]} & \underline{\mbox{0.54}} & 0.54 & \underline{\mbox{0.94}} & 0.50 & \underline{\mbox{0.63}} & 0.77 & 0.70 & \underline{\mbox{0.98}} & \textbf{0.88} & \underline{\mbox{0.83}} & \textbf{0.60} & 0.93 & 0.54 & \textbf{0.65} & \textbf{0.68} \\
N-Sigma {[}N-Sigma{]} & 0.38 & 0.33 & 0.83 & 0.48 & 0.51 & 0.66 & 0.46 & 0.87 & 0.66 & 0.66 & 0.55 & \underline{\mbox{0.94}} & 0.55 & 0.59 & 0.66 \\
N-Sigma {[}BIRCH{]} & 0.38 & 0.33 & 0.63 & 0.46 & 0.45 & - & 0.00 & - & 0.85 & 0.68 & 0.33 & 0.65 & 0.39 & 0.33 & 0.38 \\
N-Sigma {[}SPOT{]} & 0.41 & 0.41 & 0.92 & 0.53 & 0.57 & 0.66 & 0.58 & \textbf{0.99} & 0.77 & 0.75 & \underline{\mbox{0.56}} & \underline{\mbox{0.94}} & 0.55 & 0.58 & 0.66 \\
N-Sigma {[}UniBCP{]} & 0.37 & 0.11 & 0.91 & \textbf{0.66} & 0.51 & 0.51 & 0.08 & 0.59 & 0.74 & 0.48 & 0.20 & 0.39 & 0.38 & \underline{\mbox{0.62}} & 0.40 \\
N-Sigma {[}BOCPD{]} & 0.46 & 0.04 & 0.51 & 0.46 & 0.37 & 0.60 & 0.54 & 0.42 & 0.52 & 0.52 & 0.35 & 0.16 & 0.21 & 0.14 & 0.22 \\ \hline 
RobustScorer {[}$t_{\text{inject}}${]} & 0.51 & 0.48 & \underline{\mbox{0.94}} & 0.49 & 0.61 & \textbf{0.81} & 0.68 & \underline{\mbox{0.98}} & 0.79 & 0.82 & 0.54 & 0.90 & \underline{\mbox{0.57}} & 0.60 & 0.65 \\
RobustScorer {[}N-Sigma{]} & \underline{\mbox{0.54}} & \underline{\mbox{0.56}} & 0.88 & 0.49 & 0.62 & \underline{\mbox{0.80}} & \underline{\mbox{0.72}} & \underline{\mbox{0.98}} & 0.83 & \underline{\mbox{0.83}} & 0.46 & \textbf{0.96} & 0.52 & 0.57 & 0.63 \\
RobustScorer {[}BIRCH{]} & 0.42 & 0.51 & 0.79 & 0.43 & 0.54 & - & 0.80 & - & 0.85 & 0.84 & 0.33 & 0.45 & 0.49 & 0.67 & 0.49 \\
RobustScorer {[}SPOT{]} & \underline{\mbox{0.54}} & 0.51 & 0.93 & 0.47 & 0.61 & \textbf{0.81} & \textbf{0.78} & \textbf{0.99} & 0.78 & \textbf{0.84} & 0.50 & 0.93 & 0.52 & 0.58 & 0.63 \\
RobustScorer {[}UniBCP{]} & 0.27 & 0.04 & 0.89 & \underline{\mbox{0.61}} & 0.45 & 0.46 & 0.08 & 0.59 & 0.73 & 0.47 & 0.13 & 0.31 & 0.38 & 0.54 & 0.34 \\ \hline
\textbf{BARO (Ours)} & 0.51 & 0.51 & \textbf{0.95} & 0.47 & 0.61 & 0.79 & 0.67 & \underline{\mbox{0.98}} & 0.80 & 0.81 & 0.54 & 0.93 & \textbf{0.58} & 0.61 & \underline{\mbox{0.67}} \\ \hline
\end{tabular}%
}
\end{table}

\subsubsection{Fine-grained Root Cause Analysis}
From Table \ref{tab:q2-rca-f1}, we can see that \textbf{BARO consistently ranks among the top performers}, with overall Avg@5 scores of 0.61, 0.81, and 0.67, while the top scores among all the methods are 0.65, 0.84, and 0.68 on Online Boutique, Sock Shop, and Train Ticket systems, respectively. Notably, BARO significantly outperforms RCD and CIRCA. When provided $t_{\text{inject}}$, RCD achieves overall Avg@5 of 0.21, 0.09, and 0.02, CIRCA reaches 0.54, 0.74, and 0.62, while BARO achieves 0.61, 0.81, and 0.67 on Online Boutique, Sock Shop, and Train Ticket, respectively. BARO improves baseline RCA methods' performance from 8\% to 800\%. While BARO performs similarly to CausalRCA on Online Boutique, it outperforms CausalRCA significantly on the Sock Shop and Train Ticket systems, with improvements ranging from 97\% to 219\% in overall Avg@5 score, demonstrating its capability in working with complex systems. These consistent results establish BARO as a reliable method across different microservice systems, demonstrating BARO's superior performance compared to recent state-of-the-art approaches \citep{xin2023causalrca, Azam2022rcd, Li2022Circa}.

\noindent\textit{Answers to RQ2: BARO outperforms all RCA baselines, achieving coarse-grained Avg@5 scores of 0.86, 0.95, and 0.81 on Online Boutique, Sock Shop, and Train Ticket, with improvements of up to 800\% over baselines in fine-grained root cause localisation.}

\subsection{RQ3: Effectiveness of BARO's Components} \label{exp:q3-robustness}

In this section, we analyse the performance of major components of BARO to assess their contribution to the overall pipeline. The experimental results in Tables \ref{tab:q2-rca-s} and \ref{tab:q2-rca-f1} also support this section. We derive the following observations based on the results:

(1) \textbf{Multivariate BOCPD presents superior performance among the anomaly detectors when integrated with RobustScorer.} While different anomaly detectors can be combined with RobustScorer to provide good performance, Multivariate BOCPD stands out by consistently delivering the best results. Multivariate BOCPD enables BARO to achieve the top overall Avg@5 score (0.86 on Online Boutique, 0.95 on Sock Shop, and 0.81 in Train Ticket for coarse-grained RCA). Multivariate BOCPD is effective because it can detect distribution changes within multivariate time series, making it suitable for detecting anomalies for microservices. N-Sigma, a simple anomaly detector, can enhance the RCA performance but its best performance 
is still lower than BARO. For example, on Online Boutique, its best performance is 0.83 whilst BARO achieves 0.86. BIRCH has poor recalls (i.e. detects only 48/100 abnormal cases on Online Boutique), limiting its ability to support RCA. When it can detect anomalies, its performance is still much lower than other anomaly detection methods. SPOT's performance is similar to N-Sigma, and is lower than BARO. 
Finally, Univariate BOCPD's performance is among the worst when combined with any RCA methods.

(2) \textbf{RobustScorer demonstrates superior robustness} compared to other baselines when integrated with different anomaly detectors. When provided $t_{\text{inject}}$, all three methods (CIRCA, N-Sigma, and BARO) perform similarly across three benchmark systems. Specifically, on the Online Boutique, Sock Shop, and Train Ticket systems, CIRCA achieves scores of 0.77, 0.95, and 0.7, N-Sigma scores of 0.86, 0.96, and 0.77, and RobustScorer scores of 0.86, 0.95, and 0.79, respectively. However, when we employ different anomaly detection methods (i.e. N-Sigma, BIRCH, SPOT, Univariate BCP, and Multivariate BOCPD) to estimate the anomaly occurrence time, the RCA methods reveal varying degrees of sensitivity. For example, on the Online Boutique system, CIRCA exhibits a variation of 62\%, N-Sigma shows a variation of 33\%, and our RobustScorer displayed a variation of 25\%. 
On Online Boutique, N-Sigma achieves Avg@5 scores of 0.74, 0.46, 0.79, whilst RobustScorer achieves scores of 0.83, 0.61, 0.84, representing improvements of 12\%, 32\%, and 6\%  compared to when using N-Sigma, BIRCH, and SPOT as anomaly detectors, respectively.
Overall, our RobustScorer demonstrated less sensitivity. When combined with Multivariate BOCPD, RobustScorer achieves accuracy similar to those obtained with $t_{\text{inject}}$. This observation, once again, confirms the effectiveness of our method.
 
In summary, the two components of BARO significantly contribute to the approach's effectiveness. They both play critical roles in our proposed root cause analysis pipeline.

\noindent\textit{Answers to RQ3: Both Multivariate BOCPD and RobustScorer are essential, with Multivariate BOCPD delivering the strongest anomaly detection support and RobustScorer exhibiting the lowest sensitivity to the choice of anomaly detector (25\% variation on Online Boutique, compared to 33\% for N-Sigma and 62\% for CIRCA).}

\subsection{RQ4: Sensitivity Analysis of RCA Methods} \label{exp:q4-sensitivity}

We conduct a sensitivity analysis to assess the performance of various RCA methods with respect to different critical parameters. We use the AC@1, AC@3, and Avg@5 scores to assess the performance
. We explore two key aspects of this sensitivity analysis.

\subsubsection{Sensitivity on the Anomaly Detection Time $\hat{t}_A$}
\begin{figure}
\centering
\includegraphics[width=\textwidth]{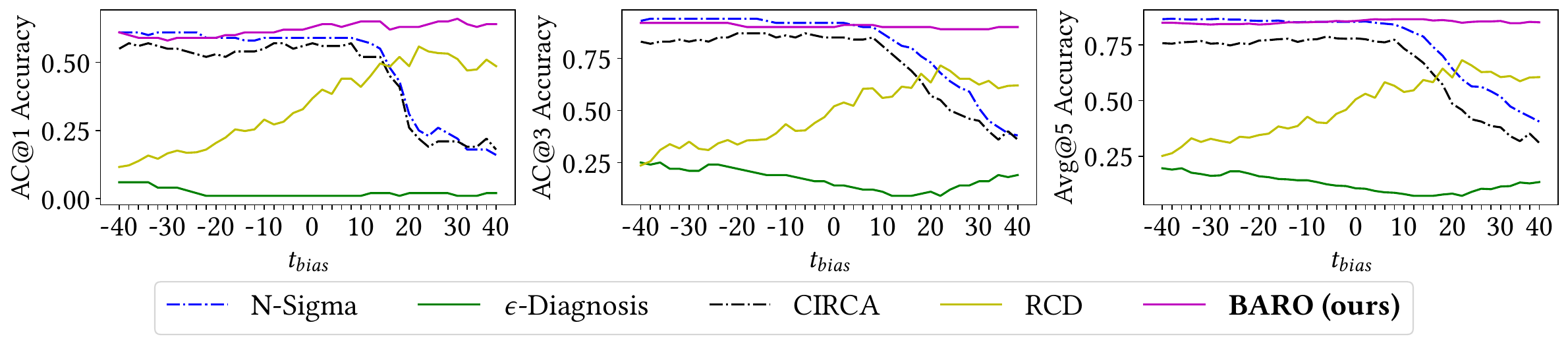}
\caption[The performance of N-Sigma, $\epsilon$-Diagnosis, CIRCA, RCD, and BARO with respect to different values of $t_{\text{bias}}$ on the Online Boutique dataset.]{The performance of N-Sigma, $\epsilon$-Diagnosis, CIRCA, RCD, and BARO with respect to different values of $t_{\text{bias}}$ on the Online Boutique dataset. The figure presents the AC@1, AC@3, and Avg@5 scores from left to right.} \label{fig:q4-sensitivity}
\end{figure}

Let us denote $t_{\text{inject}}$ as the fault injection time. Here, we aim to assess the performance of RCA methods when the anomaly detection time varies around this fault injection time. We formulate the anomaly detection time $\hat{t}_{A}$ as $\hat{t}_{A} = t_{\text{inject}} + t_{\text{bias}}$ where $t_{\text{bias}}$ ranges from -40 to 40. We then evaluate the performance of the RCA methods with different anomaly detection time within this range. We run the experiments with the Online Boutique and Sock Shop datasets. The experimental results on the Online Boutique dataset are in Figure~\ref{fig:q4-sensitivity} whilst the results on the Sock Shop dataset are in our supplementary artifact, Figure~S1 \citep{barogithubzenodo, barogithub}.

We observe that BARO and $\epsilon$-Diagnosis exhibit significant resistance to variations in the specification of $\hat{t}_{A}$. In contrast, the performance of N-Sigma and CIRCA drops significantly when the anomaly is detected late. This issue might stem from the sensitivity of the mean and standard deviation to outliers, as discussed in Section \ref{sec:why-robust}. In particular, both N-Sigma and CIRCA use z-score, $z=\frac{x-\mu}{\sigma}$, which is computed based on the mean and standard deviation of the data distribution before the anomaly detection time. When the anomaly is detected late, a number of anomalous data points is included in the normal data period, and with the use of mean and standard deviation, this leads to improper root cause ranking (see Figure~\ref{fig:robustscorer-robustness}). Finally,
RCD also has high sensitivity to $\hat{t}_A$, highlighting a weakness of the combination of $\Psi$-PC and the divide-and-conquer strategy.

\subsubsection{Hyperparameter Sensitivity}

\begin{figure}
\centering
\includegraphics[width=\textwidth]{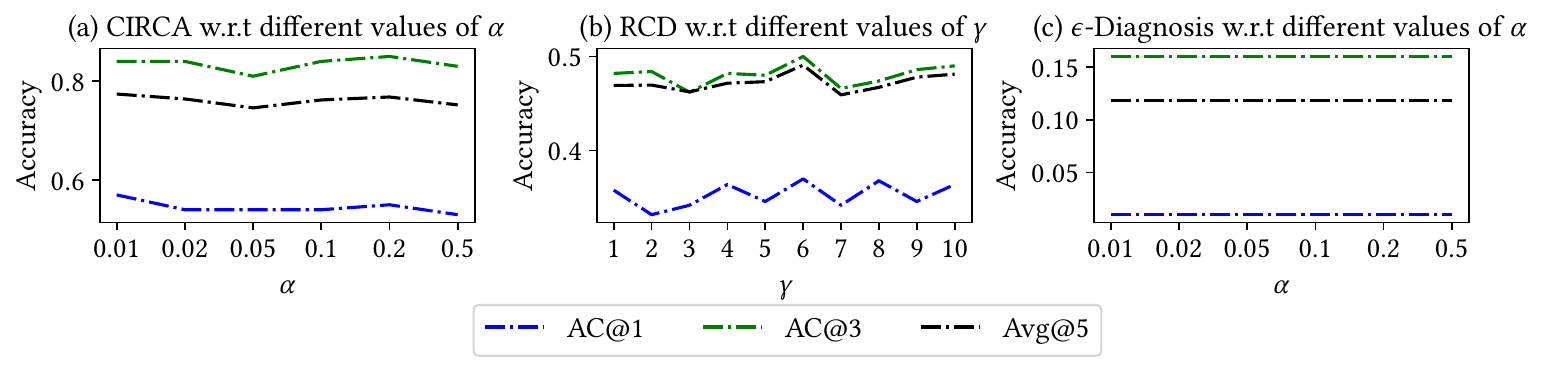}
\caption[The performance of CIRCA, RCD, and $\epsilon$-Diagnosis with respect to their different parameter values on the Online Boutique dataset.]{The performance of CIRCA (a), RCD (b), and $\epsilon$-Diagnosis (c) with respect to their different parameter values on the Online Boutique dataset. The figure presents their AC@1, AC@3, and Avg@5 scores.} \label{fig:q4-sensitivity-combine}
\end{figure}

In this section, we aim to assess the sensitivity of different RCA methods to their parameters. RCD employs a divide-and-conquer strategy, requiring the specification of a chunk size parameter denoted as $\gamma$. We vary this parameter from 1 to 10 while the default value is 5. CIRCA uses PC to construct the causal graph, requiring a threshold $\alpha$ for independence testing. We vary this parameter from 0.01 to 0.5 while the default value is 0.05. $\epsilon$-Diagnosis also requires a threshold $\alpha$ to drop the insignificant metrics. We range $\alpha$ from 0.01 to 0.5 in this case.
Note that CausalRCA's parameters are already tuned \citep{xin2023causalrca}, eliminating parameter sensitivities.
Figure~\ref{fig:q4-sensitivity-combine} presents our experimental results on the Online Boutique dataset whilst the experimental results on the Sock Shop dataset are in the supplementary material, Figure~S2 \citep{barogithub, barogithubzenodo}.

Based on these results, it is evident that RCD, CIRCA, and $\epsilon$-Diagnosis exhibit minimal sensitivity to their parameters. Consequently, we can have confidence in the reliability of the results obtained in our 
research questions. Notably, the performance of $\epsilon$-Diagnosis is identical with different values of $\alpha$. This can be attributed to the fact that in $\epsilon$-Diagnosis, the $\alpha$ parameter primarily influences the trimming of lower-ranked root causes. Given our focus on the top-k results, the length of the list becomes less critical. Similarly, for CIRCA, the main component that serves root cause analysis is its statistical scorer. The significance level $\alpha$ mainly affects the construction of the causal graph, which has minimal impact on the performance. For RCD, slight fluctuations in performance are observed with varying chunk size $\gamma$, but these variations are deemed insignificant.

\noindent\textit{Answers to RQ4: BARO exhibits strong resistance to variations in the anomaly detection time, whereas N-Sigma, CIRCA, and RCD degrade noticeably when anomalies are detected late, and the baseline RCA methods show minimal sensitivity to their own hyperparameters.}

\subsection{Running Time \& Instrumentation Cost}

\subsubsection{Running Time}

dn three datasets, BARO takes an average of 30 to 173 seconds to perform anomaly detection and root cause analysis. The highest recorded running time of BARO is 7 minutes in a case of the Train Ticket dataset. RobustScorer is very fast when it takes around 0.01 seconds to analyse the root cause, i.e., at least 3 to 1000 times faster than other methods like $\epsilon$-Diagnosis, RCD, CIRCA, and CausalRCA. The detailed running time of our proposed method and different anomaly detection and root cause analysis modules are shown in Tables \ref{tab:running-time-ad} and \ref{tab:running-time-rca}, respectively.

\begin{table}[ht]
\centering
\begin{minipage}[t]{0.48\textwidth}
\centering
\caption{Average anomaly detection run time (in seconds)}
\label{tab:running-time-ad}
\resizebox{\textwidth}{!}{%
\begin{tabular}{l|r|r|r}
\hline
Method & \multicolumn{1}{c|}{\textbf{Online Boutique}} & \multicolumn{1}{c|}{\textbf{Sock Shop}} & \multicolumn{1}{c}{\textbf{Train Ticket}} \\ \hline \hline
N-Sigma & 0.16 & 0.11 & 0.53 \\ \hline
BIRCH & 0.05 & 0.04 & 0.11 \\ \hline
SPOT & 3.17 & 1.9 & 11.4 \\ \hline
UniBCD & 819.7 & 638.19 & 3292.48 \\ \hline
\textbf{BARO (Ours)} & 44.83 & 30 & 173.37 \\ \hline
\end{tabular}%
}
\end{minipage}
\hfill
\begin{minipage}[t]{0.48\textwidth}
\centering
\caption{Average root cause analysis run time (in seconds)}
\label{tab:running-time-rca}
\resizebox{\textwidth}{!}{%
\begin{tabular}{l|r|r|r}
\hline
Method & \multicolumn{1}{l|}{\textbf{Online Boutique}} & \multicolumn{1}{l|}{\textbf{Sock Shop}} & \multicolumn{1}{l}{\textbf{Train Ticket}} \\ \hline \hline 
CausalRCA & 299.18 & 287.18 & 2638.51 \\ \hline
$\epsilon$-Diagnosis & 3.94 & 3.97 & 14.83 \\ \hline
RCD & 10.74 & 5.62 & 24.21 \\ \hline
CIRCA & 13.52 & 13.47 & 7564.88 \\ \hline
N-Sigma & 0.01 & 0.01 & 0.01 \\ \hline
\textbf{BARO (Ours)} & 0.01 & 0.01 & 0.01 \\ \hline
\end{tabular}%
}
\end{minipage}
\vspace{-10pt}
\end{table}

\subsubsection{Instrumentation Cost}

BARO requires monitoring agents (e.g., Istio, cAdvisor, Prometheus) installed alongside application services to collect system-level (e.g., CPU/Mem/Disk usage) and application-level metrics (e.g., response time, request count per minute). These metrics are used for BARO to perform anomaly detection and RCA. BARO requires Python and some standard packages (e.g. pandas, numpy). Our artifacts are available at \citep{barogithub, barogithubzenodo}.

%% file: papers/fse24/5.threats.tex
We assess potential threats to the validity of our work, considering the construct, internal, conclusion, and external factors as outlined in \citep{wohlin2012experimentation}.

\noindent\textbf{Construct validity.} The construct threat primarily concerns hyperparameter settings and evaluation metrics. To mitigate this threat, we conduct a sensitivity analysis for all methods that require specific parameters and use established evaluation metrics.

\noindent\textbf{Internal validity.} The internal threat concerns the framework implementation, where bugs may affect the reliability of our results. To mitigate this threat, we use well-maintained Python libraries, perform rigorous testing, and repeat each experiment multiple times.

\noindent\textbf{Conclusion validity.} The conclusion threat is tied to the variety of fault types, as microservices can experience various faults \citep{mariani2018localizing}. To mitigate this threat, we use four common fault types previously studied in prior work to evaluate and demonstrate the superior performance of \barotool. Furthermore, our framework relies on a set of assumptions discussed and validated in previous works on microservice systems, as described in Section~\ref{sec:assumption-causal}. If a system meets these assumptions, \barotool could be applied. Expanding \barotool to work with other types of systems, such as distributed database systems, could be a potential direction for future work.

\noindent\textbf{External validity.} The external threat is related to the deployment of microservice applications and data collection strategies. In this chapter, we employ a single deployment setting for each microservice system and a single data collection strategy, which potentially limits the generality of our work. To mitigate this threat, we deploy widely used benchmark systems, including Online Boutique, Sock Shop, and Train Ticket, on real 5-node Kubernetes clusters and repeat the experiments multiple times to obtain comprehensive datasets. These benchmark microservice applications are widely recognised in academia for testing microservices-related methods~\citep{Jinjin2018Microscope, Azam2022rcd, wu2021microdiag, xin2023causalrca, wu2022automatic, he2022graph, yu2021microrank, zhou2018trainticket, Wang2021evalcausal}. Furthermore, we adopt standard open-source tools for monitoring, and collect service-level and resource-level metrics that present the status of a running microservice application.

%% file: papers/fse24/6.conclusion.tex
In this chapter, we presented BARO, an end-to-end approach for anomaly detection and root cause analysis in microservice systems based on multivariate time series metrics data. BARO employs Multivariate Bayesian Online Change Point Detection for anomaly detection and introduces the RobustScorer, a nonparametric statistical hypothesis testing technique, for identifying root causes. We evaluated BARO on three benchmark microservice systems and found that it consistently outperforms state-of-the-art methods in both anomaly detection and root cause analysis. A comprehensive sensitivity analysis further demonstrates the robustness and broad applicability of our approach. The source code is publicly available at \url{https://github.com/phamquiluan/baro}, with an immutable artifact archived at \url{https://doi.org/10.5281/zenodo.11094092} and the experimental datasets at \url{https://zenodo.org/records/11046533}.

%% file: chapters/4.event.tex
\chapter{EventADL: Event-based Anomaly Detection and Root Cause Analysis}\label{chap:eventadl}

\input{papers/fse26-eventadl/0.symbols}

\begin{tcolorbox}[left=2pt,right=2pt,top=0pt,bottom=0pt,
  enhanced,
  drop shadow={shadow xshift=1ex, shadow yshift=-1ex, opacity=0.3}]
\textbf{Publication:} This chapter is based on our paper titled \textbf{``EventADL: Open-Box Anomaly Detection and Localization Framework for Events in Cloud-Based Service Systems''}, Luan Pham, Victor Nicolet, Joey Dodds, Hui Guan, and Daniel Kroening, published in the Proceedings of the ACM on Software Engineering (\textit{PACMSE}), Issue \textit{FSE}, Volume 3, Article FSE179, 2026 (\textbf{CORE~A*}) \citep{pham2026eventadl}\footnote{This work was conducted during the candidate's internship at Amazon Web Services in the 3rd-year of the PhD candidature. The candidate was the lead author and contributed significantly to the work.}.
\end{tcolorbox}
\vspace{10pt}

\noindent Chapter~\ref{chap:fse24} introduced BARO, an effective metric-based approach to anomaly detection and RCA. While metrics provide valuable quantitative signals about system health, they represent only one modality of observability data. This chapter addresses the limitation identified in Section~\ref{sec:limit-event-data}, the underutilization of event data in anomaly detection and RCA, by exploring events as an alternative and complementary data source. Unlike metrics, which capture continuous numerical measurements, events record discrete system activities such as API calls, configuration changes, resource modifications, and security actions. These event streams offer rich contextual information that can reveal anomalies and their root causes in ways that metrics alone cannot. This chapter presents EventADL, the first open-box event-based anomaly detection and root cause localisation framework. Informed by an analysis of 520 real-world incidents at Amazon Web Services, EventADL models normal event behaviour through Event Semantic Patterns and Event Frequency Patterns, enabling interpretable detection and diagnosis of anomalies.
\vspace{10pt}

\input{papers/fse26-eventadl/0.abstract}

\section{Introduction}
\input{papers/fse26-eventadl/1.introduction}

\section{Background and Problem Statement}
\input{papers/fse26-eventadl/2.background}

\section{Analysis of Real-world Incidents} \label{sec:analysis-real-world}

\input{papers/fse26-eventadl/3.systematic}

\section{\toolname{}: Proposed Approach} \label{sec:method}

\input{papers/fse26-eventadl/4.method}

\section{Experiments} \label{sec:experiments}

\input{papers/fse26-eventadl/5.experiments}

\input{papers/fse26-eventadl/7.conclusion}

\section*{Data Availability}
\input{papers/fse26-eventadl/data_availability}

\section{Summary}
\noindent EventADL demonstrates that event data, when properly modelled, provides valuable signals for anomaly detection and RCA that complement metric-based approaches like BARO. Together, these two chapters show the potential of analysing individual data modalities independently. However, relying on a single data source inherently limits diagnostic coverage, as each modality captures different aspects of system behaviour, and some failures may only manifest clearly when multiple signals are correlated. Furthermore, practical software systems often have incomplete observability, with ``blind spots'' where traces or other telemetry data are unavailable. The next chapter addresses these challenges by presenting TORAI, a multimodal RCA approach that leverages available metrics, logs, and traces to achieve comprehensive root cause analysis even under the presence of \textit{blindspots}.
\vspace{10pt}

%% file: papers/fse26-eventadl/0.symbols.tex
\definecolor{DarkOrange}{rgb}{0.8,0.4,0}
\definecolor{DarkRed}{rgb}{0.6,0,0}
\definecolor{ForestGreen}{rgb}{0.13,0.55,0.13}
\definecolor{Purple}{rgb}{0.70,0.0,0.70}
\definecolor{linegray}{rgb}{0.93,0.93,0.93}
\definecolor{lineblue}{rgb}{0.9,0.9,0.99}
\definecolor{linegreen}{rgb}{0.91,0.95,0.91}
\definecolor{linegreen2}{rgb}{0.98,1.0,0.98}
\newcommand{\tinycolorbox}[2]{\tikz[baseline=(a.base),inner sep=0pt]\node[fill=#1](a){#2};}

\newcommand{\anomaly}[1]{\textbf{\textcolor{DarkOrange}{#1}}}
\newcommand{\rootcause}[1]{\textbf{\textcolor{DarkRed}{#1}}}

\newcommand{\mypar}[1]{\vspace{0.15cm} \noindent {\textbf{#1.}\/}}

\newcommand{\circled}[1]{\raisebox{1pt}{\tikz[baseline=(char.base)]{%
  \node[shape=circle, fill=black, text=white, inner sep=1pt, font=\scriptsize\bfseries] (char) {#1};}}}

\def\toolname{\mbox{\textsc{EventADL}}}

\def\hyglad{\textsc{HyGLAD}}
\def\mxpEvent{AVA}
\def\zrhEvent{OUT}
\def\aioEvent{AIO}
\def\pvmEvent{PVM}
\def\losEvent{LOS}
\def\netEvent{NET}
\def\ersEvent{ERS}
\def\netflixEvent{FLIX}
\def\aws{UKW}
\def\iam{Identity}
\def\numbaselines{X}

%% file: papers/fse26-eventadl/0.abstract.tex
Anomaly detection and localisation (ADL)\footnote{The three terms: `Anomaly Localisation`, `Root Cause Localisation`, and `Root Cause Analysis' are used interchangably in this chapter.} is critical for maintaining reliability and availability in cloud-based systems. Recent ADL developments focus on metric and log data, leaving event data unexplored. To address this gap, we propose \toolname{}, the first \textit{open-box} event-based ADL framework for cloud-based service systems. To motivate the design of our framework, we conduct a systematic analysis on 520 real-world incidents, and provide insights into how anomalies and their root causes manifest through event data. \toolname{} has three phases: offline training, online anomaly detection, and root cause localisation. During the training phase, \toolname{} first learns \emph{Event Semantic Patterns (ESPs)}, which capture normal interactions between system entities using historical event data, and then learns \emph{Event Frequency Patterns (EFPs)}, which capture the normal frequency of known ESPs. In the online anomaly detection phase, any data in the event stream that deviates significantly from either pattern is identified as anomalous. For localisation, \toolname{} constructs an \textit{Intervention Graph} that models the relationships between recent system interactions and the detected anomalies for automatic root cause localisation. The framework is designed to operate efficiently with unlabeled data and to produce interpretable anomalies with their corresponding root causes. Our evaluation on three real cloud service systems and two real-world incidents demonstrates that \toolname{} outperforms existing methods, achieving F1-scores of at least 90\% for anomaly detection and 100\% top-3 accuracy in root cause localisation.

%% file: papers/fse26-eventadl/1.introduction.tex
Cloud-based service systems generate large volumes of structured event data that record who performed what operation on which resources and when. These events are captured by monitoring systems to track operations such as API calls, configuration changes, and resource updates. Leading cloud providers offer event monitoring services to support observability and auditing, including AWS CloudTrail~\citep{aws_cloudtrail_events}, Azure Event Hub~\citep{azure_monitor_eventhub_ingestion}, Google Cloud Audit~\citep{google_cloud_audit_logs}, and Alibaba ActionTrail~\citep{alibaba-actiontrail}.

A fundamental challenge faced by cloud operators is \textbf{timely anomaly detection and localisation (ADL)} from event streams. Anomaly detection identifies unexpected behaviours in the system, while root cause localisation (RCL) pinpoints the root causes of those anomalies. Notably, the root cause may not be anomalous on its own. For instance, a resource deletion may seem benign in isolation but could trigger catastrophic failures in downstream services depending on the deleted resource. Our study shows that detecting and diagnosing such incidents often involves multiple teams and significant effort to analyse observable event data. Given the operational scale and high stakes, automated and interpretable ADL for event data is a critical capability in modern cloud.

\begin{figure}[htbp]
\centering
\begin{minipage}{0.5\textwidth}
\begin{lstlisting}[language=json, basicstyle=\ttfamily\scriptsize, backgroundcolor=\color{gray!10}, frame=single]
{ "actor.user.name": "merlinary",
  "api.operation": "UpdateInstances",
  "api.request.data": {
    "force": true
  },
  "resources": [
    { "uid": "prod-04242345432" }
  ],
  "cloud.region": "us-east-1",
  "time": "2025-05-19T17:38:32Z",
  "error": null
}
\end{lstlisting}
\end{minipage}
\caption[An event in the OCSF schema.]{An event in the OCSF schema~\citep{ocsf}.} \label{fig:event-example}
\end{figure}
\mypar{Limitations of Existing Work}
While ADL has been actively studied for metrics~\citep{pham2026graph, gu2024kpiroot, pham2024baro, gu2025adamas, Li2022Circa, chen2022adaptive} and unstructured logs~\citep{brianlogsurvey2025, le2021neurallog, du2017deeplog, landauer2024critical, meng2019loganomaly, li2020swisslog}, event-based ADL remains underexplored. Metric-based methods~\citep{pham2024baro, Li2022Circa, pham2026graph} capture frequency-based anomalies but fail to detect pointwise anomalies (i.e., an individual anomalous event). Log-based methods such as DeepLog~\citep{du2017deeplog}, LogAnomaly~\citep{meng2019loganomaly}, and SwissLog~\citep{li2020swisslog} employ deep neural networks (LSTMs and BERT) to capture sequential, quantitative, and temporal patterns in log data. These methods suffer from four key limitations: \circled{1}~they are \textit{closed-box}, offering no interpretability into why an anomaly was flagged, \circled{2}~they perform detection only, without localising root causes, \circled{3}~they lack adaptive mechanisms for evolving systems, requiring retraining when system behaviour changes, and \circled{4}~they ignore structured information available, such as the interactions between actors and resources (Figure~\ref{fig:event-example}).

Most event-based anomaly detection methods~\citep{zengy2022shadewatcher, amin2019cadence, coskun2022detecting} also rely on deep neural networks and thus cannot offer interpretable decisions. ShadeWatcher~\citep{zengy2022shadewatcher} uses a context-aware embedding model and a GNN to identify anomalous interactions, with anomaly scores derived from deep GNN embeddings. GuardDuty~\citep{coskun2022detecting} employs a VAE trained on historical data to detect anomalies based on reconstruction error in the latent space. These systems provide no explanation for why an anomaly is detected, requiring further manual investigation.
HyGLAD~\citep{hyglad} is a pointwise anomaly detection technique which is unable to detect frequency-based anomalies.

Most existing RCL methods focus on metrics, logs, and traces. Nezha~\citep{yu2023nezha}, MULAN~\citep{zheng2024mulan}, CORAL~\citep{wang2023incremental}, and MRCA~\citep{wang2024mrca} construct causal graphs from metrics, logs, traces and apply centrality or causality analysis for RCL. CONAN~\citep{li2023conan} extracts contrast patterns from attribute-value pairs to diagnose batch failures. TraceContrast~\citep{zhang2024trace} mines contrast sequential patterns from distributed traces. SLIM~\citep{ren2024slim} uses supervised rule-set learning from fault data, and ReACT~\citep{roy2024exploring} leverages LLMs to reason over incident reports. While these methods advance RCL in their respective domains, none are designed for structured event data that encodes explicit actor-operation-resource relationships. Their data representations lack the semantic structure that enables open-box detection of \textit{Event Type}, \textit{Event Value}, and \textit{Event Frequency} anomalies, and interpretable RCL through intervention graphs.

\mypar{Empirical Study} Despite the importance of event data, no prior work has systematically analysed how anomalies and their root causes manifest through events. We address this gap with a novel empirical study of 520 real-world incident reports from production systems (Section~\ref{sec:analysis-real-world}). This analysis provides the empirical foundation that has been missing from prior work, offering guidance for designing effective event-based ADL techniques. Our analysis reveals two key findings. First, event-based anomalies manifest along three dimensions: \textit{Event Type (21\%)}, \textit{Event Value (68\%)}, and \textit{Event Frequency (67\%)}. This distribution indicates that effective detection requires capturing both semantic deviations (unusual types or values) and temporal deviations (abnormal frequencies). Second, root causes stem from either a single intervention (\textit{32\%}) or multiple interventions (\textit{68\%}), such as resource deletions or configuration changes. This pattern suggests that RCL must trace causal chains from observed anomalies back through potentially multiple triggering events.

\mypar{Proposed Approach} Guided by these findings, we present \toolname{}, the first \textit{open-box} ADL framework tailored for event data. \toolname{} provides both anomaly detection and RCL capabilities while maintaining \textbf{interpretability} in two ways. First, when anomalies are detected, operators receive not just alerts but also interpretable explanations that detail \textit{why} these anomalies occurred. Second, all components of \toolname{} are interpretable and transparent, enabling operators to validate the results and build trust in automated diagnostics. \toolname{} operates in three phases: offline training, online anomaly detection, and RCL. During offline training, \toolname{} learns human-interpretable patterns: \textit{Event Semantic Patterns (ESPs)} and \textit{Event Frequency Patterns (EFPs)}. ESPs capture normal event field structures (i.e., Event Types and Values), addressing the Event Type and Event Value anomalies identified in our study. EFPs represent the expected frequency patterns of events, targeting Event Frequency anomalies. During the online detection phase, \toolname{} detects anomalies by comparing incoming events against learned patterns. Violations of ESPs indicate pointwise anomalies, while deviations from EFPs signal frequency-based anomalies. For RCL, \toolname{} constructs an \textit{Intervention Graph} capturing interventions that may have triggered the anomalies. The Intervention Graph models causal relationships between actors, operations, impacted resources, and detected anomalies. \toolname{} applies a time-aware random walk over this graph to identify root causes, tracing the causal chains that our analysis found prevalent.

This chapter presents a comprehensive research effort on event-based ADL, an area that remains underexplored. Our work spans problem formulation, systematic analysis, framework design, extensive evaluation, and dataset release to facilitate future research. The contributions of this chapter are as follows:

\begin{itemize}[leftmargin=*]
\item We conduct a systematic analysis of 520 real-world incident reports, offering insights into how anomalies and their root causes manifest in event data in cloud service systems.
\item We introduce \toolname{}, the first open-box ADL framework for events in cloud service systems. \toolname{} detects anomalies along three dimensions (Event Type, Event Value, and Event Frequency) and localises their corresponding root causes. The framework is designed to operate efficiently on unlabelled data and to provide interpretable outcomes.
\item We evaluate \toolname{} on three benchmark datasets and two real-world incidents, showing that \toolname{} consistently outperforms existing state-of-the-art methods, achieving F1-scores of at least 90\% for anomaly detection and 100\% top-3 RCL accuracy.
\end{itemize}

%% file: papers/fse26-eventadl/2.background.tex
This section introduces event-specific concepts building on the foundation in Chapter~\ref{chap:background}.\footnote{The original submission contained self-contained terminology and problem definitions. In this thesis, I consolidate these in Chapter~\ref{chap:background} for consistency.}

\subsection{Terminology} \label{sec:term}

\mypar{Event} In cloud-based service systems, an event is a structured record that captures at least four key attributes: the \texttt{actor} (who performed the action), the \texttt{operation} (what action was performed), the \texttt{resources} (what was acted upon), and the \texttt{timestamp} (when the event occurred). Events may also include auxiliary attributes such as execution parameters, metadata, or error indicators. Compared to unstructured logs, which require parsing for analysis~\citep{he2017drain, brianlogsurvey2025}, events have a known schema. Events are standardised in both open-source~\citep{opentelemetry_events_2025} and commercial~\citep{aws_cloudtrail_events, azure_monitor_eventhub_ingestion, google_cloud_audit_logs} platforms. Figure~\ref{fig:event-example} shows an event following the OCSF schema~\citep{ocsf}. \textit{Event Type} refers to the categorical field that identifies the kind of interaction being recorded (e.g., the \texttt{api.operation} field in Figure~\ref{fig:event-example}, or the \texttt{eventName} field in AWS CloudTrail~\citep{aws_cloudtrail_events}). An Event Type anomaly occurs when an unseen or unexpected interaction type is observed. \textit{Event Value} refers to the attribute values within an event, such as actor identifiers, resource identifiers, or request parameters. An Event Value anomaly occurs when an unusual combination of values appears (e.g., a development user accessing a production database, or an operation performed in an unexpected region). In Figure~\ref{fig:event-example}, the Event Type is \texttt{UpdateInstances}, while Event Values include the actor \texttt{merlinary}, resource \texttt{prod-04242345432}, and region \texttt{us-east-1}.

\mypar{Anomaly} Building on the general definition of anomalies (Section~\ref{sec:terms-fault-failure}), we identify four manifestations specific to event data: (1) \textit{Event Type}, (2) \textit{Event Value}, (3) \textit{Event Frequency}, and (4) \textit{Event Order}. Although prior studies~\citep{zengy2022shadewatcher, event_ijcai16_ape, van2004workflow} have examined subsets of these anomalies, none have systematically defined or comprehensively investigated all four. In Section~\ref{sec:analysis-real-world}, we present an analysis of 520 incident reports to assess the prevalence and significance of these anomaly types.

\mypar{Root Cause of Anomalies} Following Section~\ref{sec:terms-fault-failure}, the root cause refers to the initial event(s) that trigger the observed abnormal behaviour, typically \textit{intervention events} such as configuration changes or code deployments. Notably, a root cause may not be an anomaly itself in isolation. For instance, a user deleting a resource may be consistent with past behaviour. However, if this deletion disrupts downstream services, it is causally linked to subsequent anomalies.

\subsection{Problem Formulation}
\label{sec:prob-form}

This chapter addresses the event-based instantiation of the general ADL formulation (Sections~\ref{sec:problem-formulation-ad} and~\ref{sec:problem-formulation-rca}). We focus on the event stream $\mathcal{E} = \langle e_1, e_2, \dots\rangle$ where each event $e_i = (\Lambda_i, \Omega_i, \mathcal{R}_i, \tau_i)$ contains the actor, operation, resources, and timestamp. At timestamp $t$, we observe window $\mathcal{W}_t = \{ e_i \in \mathcal{E} \mid \tau_i \in [t - \Delta, t] \}$ and aim to (1) detect anomalies ($y \in \{0,1\}$), and (2) when $y=1$, identify root cause events $\mathcal{C}_t \subseteq \mathcal{W}_t'$.

%% file: papers/fse26-eventadl/3.systematic.tex
\input{tables/fse26-eventadl/incidents-main}

To guide the design of our framework, we conducted a systematic empirical study of incident reports from a large service provider, You-Know-Where (\aws{}). Each incident report typically contains a high-level summary, detailed information on anomaly symptom(s), mitigation action(s), diagnosed root cause(s), and the incident impact (e.g., the number of affected users, the incident duration). Focusing on how engineers investigate anomalies and use event data during incident diagnosis helps us understand how anomalies and their root causes manifest in production environments.

\begin{figure*}[t]
\centering
\includegraphics[width=\linewidth]{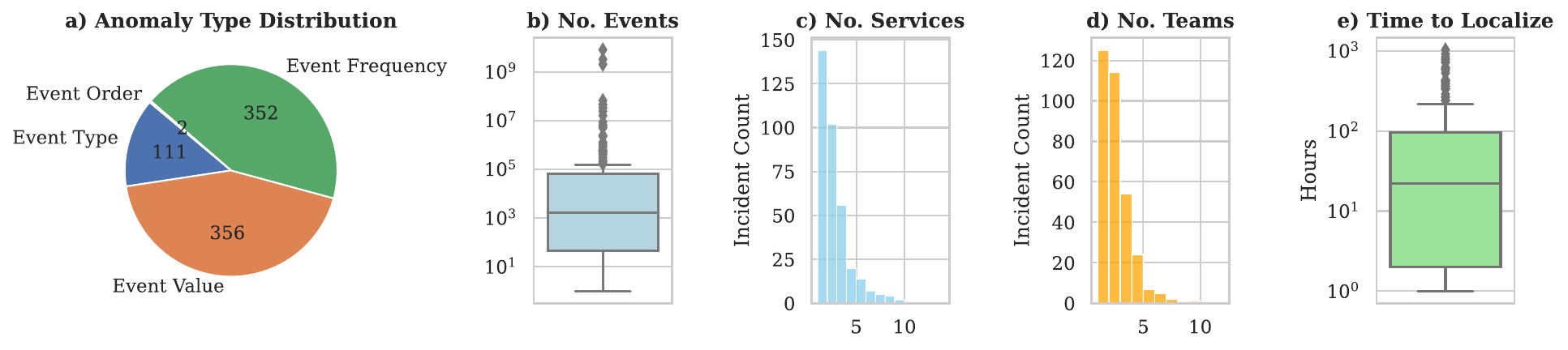}
\caption[Insights from real-world incidents.]{Insights from real-world incidents. (a) Distribution of anomaly types. (b) Number of events analysed. (c)~Number of services involved. (d) Number of teams involved. (e)~Root cause localisation time.} \label{fig:real-world-insights}
\end{figure*}

To systematically select the relevant incident reports, we first queried the incident database using the keywords \texttt{"Service X"} and \texttt{"X Data"}, where X is \aws{}'s event collection service. The search yielded 1{,}634 incidents mentioning \texttt{"Service X"} and 2{,}928 mentioning \texttt{"X Data"} among all incidents since 2013, indicating that event data is a standard part of operational diagnostics. Focusing on recent operational contexts, we restricted our analysis to incidents from the past year (June 2024, 2025). This search yielded 520 incident reports in total, of which 202 mention \texttt{"Service X"} and 354 mention \texttt{"X Data"}. Our analysis aims to answer two key questions: \textbf{(Q1)}~How do anomalies manifest in event data? and \textbf{(Q2)} How do root causes leave traces in event data?

\subsection{Findings} \label{sec:findings}

Figure~\ref{fig:real-world-insights} give our quantitative analysis, and Table~\ref{tab:incidents-main} provides representative case studies.

\mypar{Anomaly Symptoms in Event Data (Q1)} We categorised anomaly symptoms using the taxonomy introduced in Section~\ref{sec:term}. We observe that: (1) among the four anomaly types, \textit{Event Type}, \textit{Event Value}, and \textit{Event Frequency} dominate the distribution with only 2 incidents exhibiting \textit{Event Order} anomalies, and (2) most incidents exhibit more than one anomaly type. Specifically, 111 incidents (21\%) involve abnormal Event Type, 356 incidents (68\%) involve abnormal Event Value, and 352 incidents (67\%) exhibit anomalies in Event Frequency. Event Value anomalies commonly manifest as unusual actor-resource relationships, such as unauthorized actors accessing protected resources, actors operating in unexpected regions, or actors performing operations on resources outside their scope. The two incidents with abnormal Event Order are caused by secret rotation procedures (e.g., deleting the old secret before the new one becomes available). Most incidents involve multiple anomaly types (72\% or 375 incidents), and 145 incidents exhibit a single type.

The most frequent combination is Event Value and Event Frequency (29\%). Note that we report the symptoms observed in incident reports. The actual data may sometimes contain more information. For example, we found that the incidents with Event Order anomalies also exhibit anomalies in Event Frequency (e.g., a spike in error events caused by the deleted key). These findings suggest that an anomaly detection solution should monitor for anomalies in Event Type, Event Value and Event Frequency. This would allow the detection of anomalies for all incidents analysed, and provide a more complete view of anomalies for incidents that combine multiple types.

\mypar{Root Cause in Event Data (Q2)}
Root causes are often interventions, that is, explicit operations by actors that alter system resources in ways that trigger downstream issues. Those interventions are often observable in event data. These interventions vary in form: (1) In 32\% of incidents, the root cause can be attributed to a single actor performing an operation (e.g., resource deletion). In this case, a single root cause event is sufficient to explain the anomaly. (2) In 68\% of cases, root causes involve multiple actors or complex automated workflows (e.g., CI/CD pipelines), and intervention chains consist of many mutating events. Although the root cause can still be captured potentially by a single event, fully understanding it requires tracing through dependent services. (3) Common root causes include resource deletions, misconfigured deployments, and infrastructure updates.

The \zrhEvent{} incident presented in Table~\ref{tab:incidents-main} is complex, involving a series of events initiated by \rootcause{a code change}. The code change was merged into the main branch, which triggered an automated CI/CD pipeline, which in turn updated the infrastructure stack and \rootcause{improperly removed a critical \iam{} role \textit{(root cause)}}. As a result, many dependencies lost access to the removed role, leading to \anomaly{service failures in the sign-in website \textit{(anomaly)}}. It took the operators 2 hours to localise the root cause. The incident was mitigated through a manual reactivation of the deleted role. This case highlights the need for interpretable root cause localisation to provide information for detected anomalies, supporting faster incident mitigation.

\begin{tcolorbox}[left=2pt,right=2pt,top=0pt,bottom=0pt,
  enhanced,
  drop shadow={shadow xshift=1ex, shadow yshift=-1ex, opacity=0.3}]
{\small \textbf{Q2: How do root causes leave traces in event data?}}

\small \textit{\textbf{A:}
Root causes frequently manifest as one or more events representing interventions on critical resources. These may be directly attributable to a single event (e.g., a deletion action) or distributed across a causal chain of actions (e.g., deployments).}
\end{tcolorbox}

\subsection{Challenges}

Our empirical analysis reveals three major challenges in detecting anomalies and diagnosing their root causes from event data in cloud-based service systems:

\mypar{C1. Overwhelming Event Volume}
The volume of events within a short time window can be overwhelming, making manual analysis difficult. Figure~\ref{fig:real-world-insights}b shows that in over 50\% of incidents, more than one thousand events needed to be examined, and this number could reach up to one billion. This scale stems from multiple services (Figure~\ref{fig:real-world-insights}c gives the number of services involved per incident) and multiple teams (Figure~\ref{fig:real-world-insights}d presents the number of teams involved per incident) generating concurrent events, significantly increasing the effort required to localise the root cause. For 71\% of incidents, identifying the root cause took more than 10 hours (Figure~\ref{fig:real-world-insights}e). This delay prevents timely mitigation. The financial impact can be severe, as 81\% of incidents cost over \$1{,}000.

\mypar{C2. Lack of Automatic Event-Based ADL}
The detection of incidents largely depends on alerts from impacted downstream services or user reports (93\% of incidents). The detection of incidents using event data (7\%) typically relies on time-consuming manual efforts from human operators. There is a clear need for proactive, event-based mechanisms that can detect anomalies and localise root causes of the issues before they significantly affect downstream services or user experience.

\mypar{C3. Lack of Interpretability and Actionability}
Even when alarms are triggered from downstream services, they often lack actionable insights. Most systems raise alarms without explaining what went wrong (see Anomaly Symptoms in Table~\ref{tab:incidents-main}). Operators then need to search through large volumes of event data from different sources to find the root causes. There is a strong need for an ADL framework that not only detects anomalies but also provides interpretable and actionable outputs such as
root causes, so that operators can understand incidents and respond effectively.

\begin{tcolorbox}[left=2pt,right=2pt,top=0pt,bottom=0pt,
  enhanced,
  drop shadow={shadow xshift=1ex, shadow yshift=-1ex, opacity=0.3}]
\small
\textbf{Summary:} \textit{Real-world incident analysis highlights the need for an automated, scalable, and interpretable ADL framework. It motivates the design of \toolname{}, which detects anomalies by modelling event semantics and frequency, and localises root causes by tracing recent interventions from detected anomalies.}
\end{tcolorbox}

%% file: tables/fse26-eventadl/incidents-main.tex
\begin{table*}
\caption[How do anomalies and their root cause manifest through event data?]{How do \anomaly{anomalies} and their \rootcause{root cause} manifest through event data?} \label{tab:incidents-main}
\scriptsize
\begin{tabular}{p{0.02\textwidth} p{0.13\textwidth} p{0.40\textwidth} p{0.35\textwidth}}
\toprule
\textbf{ID} & \textbf{Title} & \textbf{Anomaly Symptoms} & \textbf{Root Cause Event} \\
\midrule
\mxpEvent & The Availability of Service X drops in Region A. & The events associated with session token creation in Region A \anomaly{suddenly disappeared}. There was a significant increase in API call volume (\(\approx\)14x normal) due to client retries, with \anomaly{almost all of them (99\%) resulting in errors}. No successful token creation was recorded. & The root cause event was \rootcause{the deactivation of an access key used by Service X} for encryption and signing keys. This deactivation occurred during a credentials cleanup when the credential management tool incorrectly identified used keys. \\ \midrule
\zrhEvent & Sign-in outage in Region B. &  The anomalies were observed as \anomaly{authentication errors} from AuthService with error messages "Session validation didn't return a principal". The frequency of successful authentication events suddenly stopped for Sign-in service in Region B. & The root cause event was \rootcause{the deletion of an \iam{} role} used by Sign-in service to call Service X APIs. This deletion event occurred as part of an infrastructure stack update triggered by the infrastructure pipeline deployment. \\ \midrule
\aioEvent & AIOps customers encountered 4XX errors for 3 days while creating AQuery observations. & X events show \anomaly{a sudden spike of 4XX error responses for API calls}. The errors occurred when users create investigations or observations using AQueryResult events containing newline characters. This resulted in a \anomaly{96\% failure rate in Region A} and \anomaly{100\% failure rate in Region B} for Y-related events, while other types of events continued to work normally. & The root cause event was \rootcause{the deployment of a code change that introduced an incorrect regex validation pattern} (\verb|^[\S\s]+$| and later \verb|^.*$|) for API input fields, which rejected valid XXLI queries containing newline characters. This change was to address a security requirement. \\ 
\bottomrule
\end{tabular}

{\small (*) The values in this table have been edited to hide identifiable information, but they are representative of real-world incidents.}
\end{table*}

%% file: papers/fse26-eventadl/4.method.tex
We present \toolname{}, an open-box ADL framework for event data in cloud-based service systems. \toolname{} detects three anomaly types (\textit{Event Type}, \textit{Event Value}, and \textit{Event Frequency}), and provides root causes by tracing the causal structure of recent system interventions and detected anomalies.

\subsection{Framework Overview}

\begin{figure*}
\centering
\includegraphics[width=\linewidth]{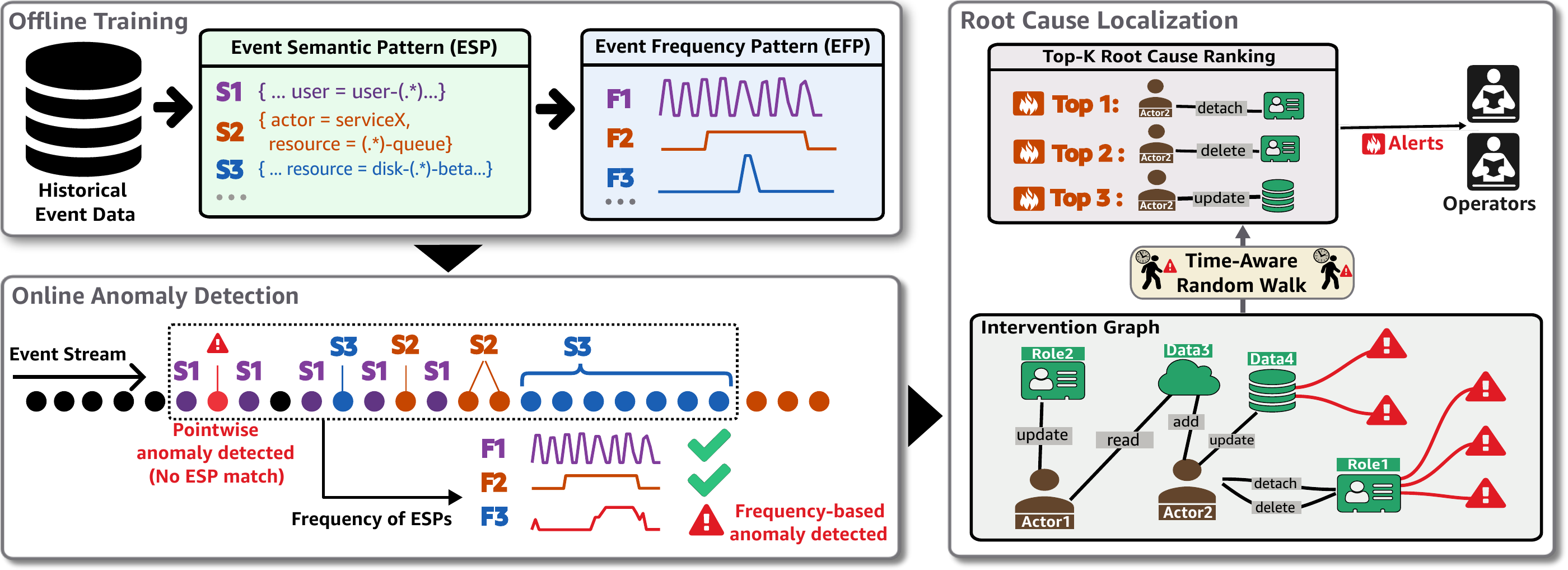}
\caption[Overview of \toolname{}.]{Overview of \toolname{}. The framework operates in three phases: offline training (upper left), online anomaly detection (lower left), and root cause localisation (right). During offline training, \toolname{} learns ESPs and EFPs that capture behaviours observed in historical event data. In the online detection phase, incoming events are continuously evaluated against these patterns: ESP identifies pointwise anomalies, while EFP detects frequency-based anomalies. When an anomaly is detected, \toolname{} constructs an Intervention Graph that encodes causal relationships between recent system interventions and the detected anomalies. Finally, it applies a time-aware random walk over the graph to rank potential root causes.}
\label{fig:overview-method}
\end{figure*}

\toolname{} operates in three phases: (1) Offline Training, (2) Online Anomaly Detection, and (3) RCL, as shown in Figure~\ref{fig:overview-method}. Each phase operates as follows:

\mypar{Offline Training}
During offline training, \toolname{} learns two types of interpretable patterns from historical event data. (1) \textit{Event Semantic Patterns (ESPs)} capture semantic relationships across event fields (e.g., \texttt{actor}, \texttt{operation}, \texttt{resource}, and \texttt{time}) to represent known behaviours (e.g., normal interactions between system entities). (2) \textit{Event Frequency Patterns (EFPs)} model the frequency of these semantic patterns over time, enabling the detection of frequency-based anomalies. Together, ESPs and EFPs form the foundation for interpretable anomaly detection.

\mypar{Online Anomaly Detection}
As new events arrive, they are continuously compared against the learned ESPs and EFPs. An event is flagged as a \textit{pointwise anomaly} if it does not match any ESP (i.e., an Event Type or Event Value anomaly).  A time-window is flagged as a \textit{frequency-based anomaly} if the frequency of ESP-matching events deviates significantly from the corresponding EFP (i.e., an Event Frequency anomaly). These mechanisms capture the three anomaly types observed in practice. Although Event Order anomalies are defined in prior work~\citep{van2004workflow}, we found that they are exceedingly rare in practice and often manifest indirectly through frequency anomalies. 

\mypar{Root Cause Localisation}
When an anomaly is detected, \toolname{} constructs an \textit{Intervention Graph}, which encodes causal relationships between recent system interventions and the detected anomalies. \toolname{} then performs a time-aware random walk over the graph to identify the root cause(s), i.e., the intervention(s) most likely to be responsible for the detected anomalies. By RCL, \toolname{} provides interpretable and actionable insights, in contrast to prior approaches~\citep{event_ijcai16_ape, zengy2022shadewatcher, coskun2022detecting}, which only produce anomaly scores and require further manual investigation.

\begin{figure}
\centering
\begin{lstlisting}[language=json, basicstyle=\ttfamily, backgroundcolor=\color{gray!10}, frame=single, linewidth=0.95\textwidth, xleftmargin=0.05\textwidth]
{"and": [
 {"==": [{"var": "actor.user.name"}, "merlinary"]},
 {"==": [{"var": "api.operation"}, "UpdateInstances"]},
 {"like": [{"var": "cloud.region"}, "^us-east-[1-4]$"]}
]}
\end{lstlisting}
\caption[An ESP in the jsonLogic schema.]{An ESP in the jsonLogic~\citep{jsonlogic} schema.}
\label{fig:event-pattern}
\end{figure} 

\subsection{Event Semantic Pattern} \label{sec:esp}

Event Semantic Pattern (ESP) is a model of the normal event types and values observed in historical event data. They serve two purposes: a set of ESPs is used to detect \emph{pointwise anomalies} (i.e., single events that are anomalous on their own), and they allow labelling normal events with a specific ESP, which is then used to compute EFPs (Section~\ref{sec:efp}).
\textit{Event Type} and \textit{Event Value} anomalies are pointwise anomalies, so we must design ESPs to capture both of those types of anomalies. As for any other component of our system, we also require the ESP model to be interpretable and scalable.

Formally, the model learned by \toolname{} on a set of events is a set \(\mathcal{S} = \{s_1, s_2, \dots, s_k\}\) of ESPs, where each ESP \(s_i\) is an event-matching rule that captures expected behaviour observed in the system. During online detection, if an incoming event \(e_j \in \mathcal{E}\) matches any pattern \(s_i \in \mathcal{S}\), it is labelled as normal with pattern \(s_i\). Otherwise, it is flagged as an anomaly. A rule-based approach provides \emph{interpretability}, \emph{scalability} (Section~\ref{sec:eval-efficiency}), and \emph{deterministic results}.

ESPs are interpretable event matching expressions that capture the events observed in historical data. Figure~\ref{fig:event-pattern} shows an ESP that captures events where actor \texttt{\textcolor{NavyBlue}{merlinary}} performs operation \texttt{\textcolor{NavyBlue}{UpdateInstances}} across regions \texttt{\textcolor{NavyBlue}{us-east-[1-4]}}. During the online anomaly detection phase, an event with actor \texttt{\textcolor{NavyBlue}{merlinary}} performing \texttt{\textcolor{NavyBlue}{UpdateInstances}} in \texttt{\textcolor{NavyBlue}{us-}\textcolor{DarkRed}{west}\textcolor{NavyBlue}{-1}} is a pointwise anomaly, as no similar action has been observed before (assuming there are no other patterns matching it). Existing pattern synthesis techniques such as HyGLAD~\citep{hyglad} or log parsing techniques such as Drain~\citep{he2017drain} can extract patterns from event data. HyGLAD has a key advantage over Drain, as it leverages the structured information in events directly to take into account entity relationships. Figure~\ref{fig:relationship-aware} gives an example where relationship-agnostic methods may mistakenly classify an abnormal event as normal, e.g., \texttt{\{user:\tinycolorbox{gray!20}{dev-1ac}, operation:\tinycolorbox{gray!20}{update}, resource:\tinycolorbox{gray!20}{prod-db-1}\}}. In our experiments, we use HyGLAD to learn ESPs in \toolname{}, and perform an ablation study using Drain (Section~\ref{sec:drain-study}).

\input{figures/fse26-eventadl/esp-comparison}

\subsection{Event Frequency Pattern} \label{sec:efp}
Event Frequency Pattern (EFP) detects anomalies in \textit{Event Frequency}, which account for 67\% of the analysed cases (Section~\ref{sec:analysis-real-world}). From our analysis, we observe that event frequencies are heterogeneous, as they may be dense (e.g., data transfers between services) or sparse (e.g., secret rotation events). We also require EFP to be interpretable and consistent with the open-box design of \toolname{}. 

Our EFP is inspired by the Matrix Profile~\citep{lu2022matrix}, which provides efficient and interpretable means to characterise frequency patterns through subsequence comparison. However, a key limitation of prior work is its emphasis on the \textit{shape} of subsequences, typically ignoring \textit{magnitude} differences~\citep{lu2022matrix}. While shape-based matching may be effective for continuously-valued signals, we argue that \textit{magnitude} is more important in the context of event-based anomaly detection. For instance, a known event pattern occurring at 1–3 requests per second (rps) may appear normal even at 5 rps, but a sudden spike to 100 rps is clearly anomalous. Shape-based anomaly detection methods~\citep{lu2022matrix} may fail to detect such deviations (Section~\ref{sec:ablation-shape-mag}). To address this, \toolname{} uses the Euclidean distance to directly compare subsequences, which is more suitable for event data. EFP is the first adaptation of the Matrix Profile for discrete time series derived from event data.

We construct a set of EFPs for a training period \([t_0, T]\) for each ESP \(s_i \in \mathcal{S}\). Let \(x^{(i)}_\tau\) denote the frequency of \(s_i\) at timestamp \(\tau\). Given a subsequence length \(M\), the set of all length-\(M\) sliding windows is defined as:
{\footnotesize \[
W^{(i)} = \left\{ w_u^{(i)} = \bigl(x_u^{(i)}, \dots, x_{u+M-1}^{(i)}\bigr) \in \mathbb{N}_0^M \;\middle|\; t_0 \le u \le T - M + 1 \right\}.
\]}

For each window \(w^{(i)}_u\), we compute the minimum Euclidean distance to a non-overlapping window \(w^{(i)}_v\), where \(\lvert u-v\rvert \ge M\):
\begin{equation}  
d^{(i)}_u \;=\; \min_{\substack{v = t_0,\dots,T-M+1 \\ |u-v| \ge M}}   \;\bigl\lVert w^{(i)}_u - w^{(i)}_v \bigr\rVert_2.
\end{equation}

We use \(f_i\) to denote the EFP with respect to the ESP \(s_i \in \mathcal{S}\) as the sequence of nearest distances computed from \(W^{(i)}\),
\begin{equation} \label{eq:fi}
f_i = \left\{ d_u^{(i)} \right\}_{u=t_0}^{T - M + 1}.
\end{equation}

\begin{figure}
\centering
\includegraphics[width=0.65\textwidth]{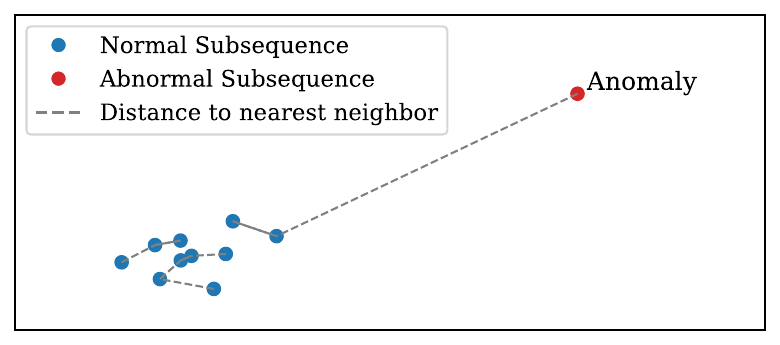}
\caption[Detecting anomalies with EFP.]{Detecting anomalies with EFP. Each subsequence \(w_u^{(i)}\) is a point in Euclidean space and linked to its nearest non-trivial neighbor \(w_v^{(i)}\). The set of distances \(\{d_u^{(i)}\}\) forms the Event Frequency Pattern (EFP) \(f_i \in \mathcal{F}\). The \textcolor{DarkRed}{abnormal subsequence} lies far from the cluster of \textcolor{NavyBlue}{normal subsequences}, as it has a \textit{statistically} large distance to its nearest neighbor, indicating a potential anomaly.}
\label{fig:efp}
\end{figure}

The set of all EFPs is denoted by \(\mathcal{F} = \{ f_1, f_2, \dots, f_{|\mathcal{S}|} \}\). Each EFP \(f_i \in \mathcal{F}\) provides an \textit{interpretable} measure of self-similarity, showing how typical each subsequence is when compared to others (see Figure~\ref{fig:efp}). An anomaly naturally arises when a subsequence has an abnormally large distance to its nearest neighbor, indicating that the observed frequency is \textit{very} different from all known patterns.

To conclude if the new observation \(w^{(i)}_{\mathrm{new}}\) is anomalous or not, we perform hypothesis testing to test whether the new observation \(d^{(i)}_{\mathrm{new}}\) is statistically consistent with the known distances in \(f_i\). Specifically, we treat \( f_i \) as samples drawn from an unknown distribution \( P_D \), representing normal frequency of \( s_i \). Then, we define the empirical cumulative distribution function over \(f_i\):
\[
\hat{F}_D(x) = \frac{1}{|f_i|} \sum_{d \in f_i} \mathbf{1}\{d \le x\}.
\]
The empirical survival function is then given by
\(
\hat{S}(d_{\mathrm{new}}) = 1 - \hat{F}_D(d_{\mathrm{new}}),
\)
which quantifies the probability of observing a distance as large as \( d_{\mathrm{new}} \). Let \( \alpha \) be a chosen significance level (e.g., \( \alpha = 0.01 \)). If \( \hat{S}(d_{\mathrm{new}}) < \alpha \), we reject the null hypothesis that \( d_{\mathrm{new}} \sim P_D \) and classify the window as anomalous. 
This design is distribution-free, making it well-suited to the heterogeneous event frequencies found in large-scale cloud systems.

\noindent \textbf{Robustness to Noise.}
A key challenge in real-world anomaly detection is the presence of noise (e.g., unknown anomalies) in training data. Many prior methods~\citep{hyglad, aydore2022detecting, event_ijcai16_ape, du2017deeplog} assume that training data is anomaly-free, which is often unrealistic. In contrast, our approach is robust to such noise because hypothesis testing treats these cases as statistically insignificant. As long as they do not dominate the empirical distribution during normal periods, they have little impact on overall performance. We empirically validate this in Section~\ref{sec:robustness}.

\noindent \textbf{Adaptation.} 
Cloud systems are constantly evolving with new behaviours, and without adaptation, new behaviours would be repeatedly flagged as anomalies. \toolname{}'s adaptive mechanism works as follows: when an event does not match any ESP, \toolname{} first flags it as an anomaly then records its count \(c_s\) per window. If \(c_s\) exceeds a  threshold \(T_s\) within \(N\) successive windows (i.e., persistent), we invoke HyGLAD to revise the ESP set, ensuring only persistent behaviours are added as new normals. For EFPs, \toolname{} continuously updates \(f_i\) with newly observed normal distances \(d_u^{(i)}\) (Equation~\ref{eq:fi}), keeping hypothesis testing (Section~\ref{sec:efp}) aligned with evolving system characteristics. We note that no online detector can instantly distinguish legitimate system changes from true anomalies without external context (e.g., release notes). Therefore, we also localise root causes to provide interpretable explanations of why the detected anomalies occur (Section~\ref{sec:anomaly-localisation}).

\begin{algorithm}
\caption{Construction of the Intervention Graph}
\label{alg:construct-intervention-graph}
\begin{algorithmic}[1]
\REQUIRE Event set $\mathcal{E} = \{e_1, e_2, \dots, e_n\}$
\ENSURE Intervention graph $\mathcal{G} = (\mathcal{V}, \mathcal{E}_G)$

\STATE Initialize graph $\mathcal{G} \gets (\mathcal{V} \gets \emptyset, \mathcal{E}_G \gets \emptyset)$

\FOR{each event $e_i \in \mathcal{E}$}
    \STATE Extract \texttt{actor}, \texttt{operation}, \texttt{resources}, \texttt{time} from $e_i$
    \STATE Add \texttt{actor} node to $\mathcal{V}$ if not already present
    \FOR{each \texttt{resource} $r_j$ in \texttt{resources}}
        \STATE Add node $r_j$ to $\mathcal{V}$ if not already present
        \STATE Add edge $(\texttt{actor} \xrightarrow{\texttt{operation}, \texttt{time}} r_j)$ to $\mathcal{E}_G$
        \IF{$e_i$ is associated with an anomaly}
            \STATE Add \texttt{Anomaly} node to $\mathcal{V}$ if not already present
            \STATE Add edge $(r_j \xrightarrow{\texttt{time}} \texttt{Anomaly})$ to $\mathcal{E}_G$
        \ENDIF
    \ENDFOR
\ENDFOR

\STATE \textbf{return} $\mathcal{G}$

\end{algorithmic}
\end{algorithm}

\subsection{Root Cause Localisation} \label{sec:anomaly-localisation}
\toolname{} localises the root cause to provide interpretable explanations of what triggered the anomaly and why it occurred. Our empirical analysis reveals that many incidents stem from improper interventions that modify one or more system resources, thus leading to anomalies observable in different parts of the system (see Section~\ref{sec:analysis-real-world}). Motivated by this observation, we introduce the concept of an \textit{Intervention Graph}, which encodes the temporal and causal relationships between system interventions and detected anomalies. This graph can be constructed directly from event data and enables automatic RCL.

\subsubsection{Intervention Graph}
Given a set of structured events $\mathcal{E} = \{e_1, e_2, \ldots, e_n\}$ within a time window $\mathcal{W}_{t}'$, the \emph{Intervention Graph} is a directed multigraph $\mathcal{G} = (\mathcal{V}, \mathcal{E}_G)$ where $\mathcal{V}$ is the set of nodes representing actors, resources, and anomalies, and $\mathcal{E}_G$ is the set of directed labelled edges representing event-level interactions. For each event $e_i$ involving actor $\Lambda_i$, operation $\Omega_i$, time $\tau_i$, and resource set $\{r_1, \ldots, r_k\}$, we add directed edges from $\Lambda_i$ to each $r_j$, labelled with $(\Omega_i, \tau_i)$. If $e_i$ is anomalous, we add an edge from each $r_j$ to an anomaly node, labelled with $\tau_i$ (see Algorithm~\ref{alg:construct-intervention-graph}).

The resulting graph provides a unified view of how interventions propagate and impact system components (see Figure~\ref{fig:overview-method}). It enables operators to visually trace anomaly paths and interpret potential root causes. Nevertheless, in practice, large-scale systems generate complex graphs that are difficult to inspect manually. To this end, we introduce a time-aware random walk algorithm that ranks root cause candidates based on traversal frequency and temporal plausibility.

\begin{algorithm}
\caption{Time-Aware Random Walk}
\label{alg:temporal-random-walk}
\begin{algorithmic}[1]
\STATE \textbf{Input:} graph $G$, anomaly nodes $A$, number of walks $N$
\STATE \textbf{Output:} node-level $visitCount$

\STATE Initialize $visitCount \gets \{\}$

\FOR{each $(a, t_a) \in A$}
    \FOR{$i \gets 1$ \TO $N$}
        \STATE $u \gets a$, $t \gets t_a$
        \WHILE{true}
            \STATE $preds \gets$ predecessors of $u$ in $G$
            \STATE $validEdges \gets \emptyset$
            \FOR{each $p \in preds$}
                \FOR{each edge $(p,u,e)$ in $G$}
                    \STATE $t_e \gets e.\text{event\_time}$
                    \IF{$t_e \le t$}
                        \STATE $validEdges \gets validEdges \cup \{(p, t_e)\}$
                    \ENDIF
                \ENDFOR
            \ENDFOR
            \IF{$validEdges = \emptyset$}
                \STATE \textbf{break}
            \ENDIF
            \STATE Sample a pair $(p', t')$ from $validEdges$
            \IF{ $p' \notin \mathrm{dom}(visitCount)$ }
                \STATE $visitCount[p'] \gets 1$
            \ELSE
                \STATE $visitCount[p'] \gets visitCount[p'] + 1$
            \ENDIF
            \STATE $u \gets p'$, $t \gets t'$
        \ENDWHILE
    \ENDFOR
\ENDFOR

\STATE \textbf{return} $visitCount$
\end{algorithmic}
\end{algorithm}

\subsubsection{Time-Aware Random Walk} \label{sec:random-walk}

To automate root cause localisation, we apply a random walk algorithm over the Intervention Graph. The walk begins at each anomaly node and traverses backward through the graph toward potential sources of the anomaly. Nodes that are visited frequently across multiple walks are identified as root causes. The intuition behind this design is that true root causes (such as problematic interventions) often have multiple causal paths leading to observed anomalies. As the random walker explores these paths, it will tend to visit the true root cause nodes more frequently than unrelated nodes. 

However, a naive traversal may violate causal time constraints. For instance, tracing from an anomaly at 11:00 AM to an operation that occurred at 2:00 PM creates an invalid path. We propose to use a \textit{time-aware random walk} that restricts movement to edges where the associated event occurred no later than the current timestamp within \(\mathcal{W}_{t}'\). This ensures that all paths are temporally valid (see Algorithm~\ref{alg:temporal-random-walk}).

This assumption is empirically grounded and used in prior work~\citep{causalai23salesforce, Azam2022rcd, pham2024root, xin2023causalrca}. Our analysis (Section~\ref{sec:analysis-real-world}) shows that root causes are often interventions that alter system resources, triggering downstream anomalies through multiple causal paths. Our Intervention Graph captures these paths directly from event data, allowing the random walk to visit the root cause frequently from the detected anomalies.

\subsubsection{Root Cause Ranking}
\toolname{} returns a ranked list of root causes (i.e., $\texttt{actor} \xrightarrow{\texttt{operation}, \texttt{time}} \texttt{resource}$), where the ranking is determined by the visit counts from the time-aware random walk. Interventions that are more frequently visited along valid temporal paths from anomalies are considered more likely to be the root causes. Given this ranked list, operators can focus on the top-ranked interventions instead of manually inspecting all events. Furthermore, the sub-Intervention Graph associated with each ranked root cause can be visualized (Figure~\ref{fig:overview-method}), reinforcing our open-box design. These visualizations provide interpretable, path-based explanations of how anomalies may have propagated from specific actors or resources through chains of operations.

%% file: figures/fse26-eventadl/esp-comparison.tex
\begin{figure}[ht]
\centering
\scriptsize
\begin{tikzpicture}[
    node distance=0cm and 0cm,
    box/.style={rectangle, draw, thick, align=left, font=\ttfamily\scriptsize, rounded corners},
    label/.style={font=\bfseries\scriptsize, align=center},
    arrow/.style={->, thick, >=stealth}
]

\node[box, fill=gray!10, text width=8cm] (input) {
    \{user:\tinycolorbox{gray!20}{admin-1}, operation:\tinycolorbox{gray!20}{update}, resource:\tinycolorbox{gray!20}{prod-db-1}\}\\
    \{user:\tinycolorbox{gray!20}{admin-2}, operation:\tinycolorbox{gray!20}{update}, resource:\tinycolorbox{gray!20}{stag-db-1}\}\\
    \{user:\tinycolorbox{gray!20}{dev-1ac}, operation:\tinycolorbox{gray!20}{update}, resource:\tinycolorbox{gray!20}{stag-db-1}\}\\
    \{user:\tinycolorbox{gray!20}{dev-2fa}, operation:\tinycolorbox{gray!20}{update}, resource:\tinycolorbox{gray!20}{stag-db-1}\}
};

\node[label, above=0.1cm of input] {Historical Events};

\node[box, fill=red!10, below right=1cm and -4cm of input, text width=6.7cm] (output1) {
    \{user:\tinycolorbox{gray!20}{*}, operation:\tinycolorbox{gray!20}{update}, resource:\tinycolorbox{gray!20}{*-db-1}\}
};

\node[label, below=0.1cm of output1, text width=6.8cm, align=center] {Relationship-agnostic~\citep{he2017drain}\\(over-generalised)};

\node[box, fill=green!10, below left=1cm and -4cm of input, text width=7.5cm] (output2) {
    \{user:\tinycolorbox{gray!20}{admin-*}, operation:\tinycolorbox{gray!20}{update}, resource:\tinycolorbox{gray!20}{*-db-1}\}\\[0.3em]
    \{user:\tinycolorbox{gray!20}{dev-*}, operation:\tinycolorbox{gray!20}{update}, resource:\tinycolorbox{gray!20}{stag-db-1}\}
};

\node[label, below=0.1cm of output2, text width=6.8cm, align=center] {Relationship-aware~\citep{hyglad}\\(more precise)};

\coordinate (splitDown) at ($(input.south)+(0,-0.5)$);   
\coordinate (splitLeft) at ($(splitDown)+(-1.9,0)$);      
\coordinate (splitRight) at ($(splitDown)+(1.9,0)$);      

\draw[thick] (input.south) -- (splitDown);
\draw[thick] (splitLeft) -- (splitRight);
\draw[arrow] (splitLeft) -- (output2.north);
\draw[arrow] (splitRight) -- (output1.north);

\end{tikzpicture}

\caption{Relationship-agnostic and relationship-aware ESP generalizations.} \label{fig:relationship-aware}
\end{figure}

%% file: papers/fse26-eventadl/5.experiments.tex
We conduct extensive experiments to answer the following questions:

\begin{itemize}
\item RQ1: How effective is \toolname{} in detecting anomalies?
\item RQ2: How effective is \toolname{} in localising root causes?
\item RQ3: How efficient is \toolname{}?
\item RQ4: What is the contribution of each component to \toolname{}?
\item RQ5: How robust is \toolname{}?
\item RQ6: How well does \toolname{} generalise to non-event data?
\end{itemize}

\subsection{Benchmark Datasets}

At the time of conducting this research, there is no publicly available benchmark for event-based ADL in cloud systems with annotated anomalies and their corresponding root causes. To address this gap, we construct \textbf{five benchmark datasets}: three by reproducing incidents on real-world cloud service systems, and two collected from \textbf{historical incidents}. We describe both the incident reproductions on benchmark systems and the collected historical incidents in detail below.

\subsubsection{Benchmark Systems}
We use three service systems, Falcon, Flask, and Live, deployed on real infrastructure to reproduce incidents and collect event data. \textbf{Falcon} is a cloud-based web application platform with 363 actors and 138{,}292 resources. \textbf{Flask} is a microservice-based music catalog and retrieval system, comprising 531 actors and 143{,}353 resources. \textbf{Live} is a real-time cricket scoring platform serving 2{,}507 actors and managing 1{,}404 resources. Each system generates events across multiple layers (e.g., API calls, configurations, and resource updates). We reproduced three common types of incidents identified in our empirical study. (1)~\textit{Secret Deactivation}: We randomly deactivate a secret (e.g., access key), causing permission errors across dependent services and triggering anomalies in event types and frequencies. The root cause is the deactivation event. (2)~\textit{Denial-of-Service (DoS)}: We execute a flood of API calls, leading to system-wide throttling. Anomalies manifest as spikes in request rates and failure events. The root cause is the actor repeatedly performing the DoS operations. (3)~\textit{Unusual Activity}: We reproduce a compromised credential scenario by creating a random resource in an unusual region. These unusual operations are simultaneously anomalies and root causes. We randomly repeat these actions across different resources and permission settings, yielding 30 one-hour test samples per system.

\subsubsection{Real-world Incidents}

We collected data from two incidents in 2024 at \aws{}: \textbf{\zrhEvent{}} and \textbf{\mxpEvent{}} (presented in Table~\ref{tab:incidents-main}). The incident \textbf{\zrhEvent{}} was reported after customers had been unable to log in to the \aws{} Management Console or switch between service consoles in Region A for 1h49m. Around 33.1\% of the requests failed due to 4xx/5xx errors. The issue stemmed from an improper deletion of a critical role during a code deployment via an infrastructure pipeline. Diagnosing the incident took approximately 100 minutes and involved multiple teams and services. The dataset includes 26{,}018 unique actors and 249 resources. \toolname{} learned 273 ESPs from the normal operation period, reflecting the complex operational patterns of this production system.
The incident \textbf{\mxpEvent{}} was reported after 3{,}355 \aws{} accounts in Region B experienced ServiceD API failures for 34 minutes, affecting 89 services in total. The root cause was an incorrect deactivation of an access key, due to a software defect and its recent deployments. The issue was resolved by reactivating the key. \toolname{} learned 308 ESPs for this dataset, capturing the diverse interaction patterns across the affected services.

\subsection{Evaluation Metrics}

\subsubsection{Anomaly Detection}

Following existing work~\citep{pham2024baro, hyglad},
we use Precision, Recall, and F1 scores to evaluate the anomaly detectors. When an anomaly detection algorithm successfully detects an abnormal sample (i.e., a case with anomalies), the detection is counted as a True Positive (TP). Conversely, incorrectly classifying an abnormal sample as normal is considered False Negative (FN). Likewise, incorrectly classifying a normal sample as abnormal is considered False Positive (FP).
\textbf{Precision} is the ratio \(\frac{TP}{TP + FP}\), \textbf{Recall} is \(\frac{TP}{TP + FN}\) and the \textbf{F1-score} is determined by \(\frac{2\times Precision \times Recall}{Precision + Recall}\).

\subsubsection{Root Cause Localisation}

We use two standard metrics, $AC@k$ and $Avg@k$, to assess the root cause localisation (RCL) performance~\citep{pham2024baro, Li2022Circa}. Herein, we set $k = 1, 3, 5$. Given a set of failure cases A, \(AC@k\) is determined by \(\frac{1}{|A|} \sum\nolimits_{a\in A}\frac{\sum_{i<k}R^a[i]\in V^a_{rc}}{min(k, |V^a_{rc}|)}\), and then \(Avg@k\) is calculated by \(\frac{1}{k}\sum_{j=1}^k AC@j\), where $R^a[i]$ denotes the $i$th ranking result for the failure case $a$ by an RCL method, and $V^a_{rc}$ is the true root cause set of case $a$. $AC@k$ represents the probability the top $k$ results given by a method include the real root causes. $Avg@k$ measures the overall performance of RCL methods.

\subsection{Baselines}

We evaluate \toolname{} against two categories of methods: (i) anomaly detection, and (ii) root cause localisation.
For anomaly detection, we include:
APE~\citep{event_ijcai16_ape},
BARO~\citep{pham2024baro},
CUSUM~\citep{cusum},
DeepSVDD~\citep{deepsvdd},
DIF~\citep{dif},
ICL~\citep{icl},
NeuralLog~\citep{le2021neurallog},
NeuTraL~\citep{neutral},
NSigma~\citep{Li2022Circa},
RCA~\citep{rca},
RDP~\citep{rdp},
ShadeWatcher~\citep{zengy2022shadewatcher},
ADAMAS~\citep{gu2025adamas}, and
KPIRoot~\citep{gu2024kpiroot}.
ADAMAS and KPIRoot are metric-based anomaly detection methods. We adapt them to event data by treating event frequency time series as input.
For root cause localisation, we compare against:
BARO~\citep{pham2024baro},
CausalAI~\citep{causalai23salesforce},
CausalRCA~\citep{xin2023causalrca},
\(\epsilon\)-Diagnosis~\citep{Shan2019Ediagnosis},
Groot~\citep{wang2021groot},
RCD~\citep{Azam2022rcd},
DeepHunt~\citep{sun2025interpretable},
TVDiag~\citep{xie2024tvdiag},
KPIRoot~\citep{gu2024kpiroot},
and a \textit{Random} baseline that selects a root cause at random. DeepHunt and TVDiag construct intervention graphs from events to localise root causes, whereas KPIRoot combines similarity analysis and causality inference on frequency time series. We use available implementations when possible. For Groot, APE, and ShadeWatcher, we implement the method based on the algorithmic details provided in the original papers. All hyperparameters follow the recommendations in the respective papers. For methods with configurable thresholds, we follow established practice by running multiple settings and reporting the best performance. Descriptions of these baselines are in our replication package.

\subsection{Experimental Setting}

We conduct all the experiments on Linux servers equipped with 12 CPU and 36\,GB RAM. To manage randomness, we repeat each experiment ten times, then report the mean and standard deviation of the results. We give each method two hours to finish their tasks, otherwise a \textit{time-out} error is thrown. Our framework is implemented using Python 3.12.

\input{tables/fse26-eventadl/exp-ad}

\subsection{RQ1: How effective is \toolname{} in detecting anomalies?}

In this RQ, we evaluate the performance of \toolname{} and the anomaly detection baselines across five datasets. We use the event data during the incident as anomalous samples and event data during normal operation as normal data. We report the average of Precision, Recall, and F1-score over all the cases. Table~\ref{tab:rq1-anomaly-detection} presents the experimental results, with the best results highlighted in \textbf{bold} and second-best results are \underline{underlined}. We draw the following observations:

\textbf{(1) \toolname{} is highly effective across all datasets, outperforming most baselines by large margins}. \toolname{} achieves the highest F1-score on 4 out of 5 datasets and consistently maintains strong performance with F1-score above 0.90 on all datasets. Because ESPs and EFPs are designed to capture any deviation from training data, \toolname{} yields 100\% recall across benchmarks. The consistent precision above 0.8 shows that ESP and EFP also generalise well over their training data, avoid false positives. Finally, \toolname{} shows deterministic performance across runs (i.e., standard deviation = 0), since ESP and EFP training is not affected by randomness.

\textbf{(2) Metric-based methods achieve high recall scores but suffer from low precision}. We observe that methods such as BARO~\citep{pham2024baro}, CUSUM~\citep{cusum}, NSigma~\citep{Li2022Circa}, ADAMAS~\citep{gu2025adamas}, and KPIRoot~\citep{gu2024kpiroot} yield very high recall. BARO employs Bayesian Online Change Point Detection to identify anomalies in multivariate time series, while CUSUM and NSigma detect deviations in univariate time series based on the mean and standard deviation. ADAMAS and KPIRoot are recent metric-based methods that we adapt to event data by treating event frequency time series as input. These methods detect many true anomalies (high recall) because anomalies in cloud systems often affect multiple entities and manifest as frequency-based deviations (e.g., spikes in error events). However, they suffer from low precision. BARO models only frequency dependencies with Bayesian statistics, whereas CUSUM and NSigma rely solely on Gaussian assumptions. Such simplifications hinder their ability to model the dynamic and heterogeneous behaviours of cloud systems, making them less suitable for complex event-based time series. In contrast, our EFP uses a magnitude-based subsequence distance approach tailored to event-frequency patterns, which combined with ESP enables \toolname{} to achieve both high recall and high precision.

\textbf{(3) Deep learning-based methods exhibit moderate performance and poor robustness.}
Deep learning baselines such as DeepSVDD~\citep{deepsvdd}, DIF~\citep{dif}, ICL~\citep{icl}, and RCA~\citep{rca} achieve considerably lower F1-score than \toolname{}. They often fail to generalise over normal event behaviour. For instance, RCA trains an ensemble of autoencoders to reconstruct normal data patterns, but the autoencoders may degenerate into learning identity mappings, thereby missing anomalies that have low reconstruction loss. More broadly, these methods assume clean training data, which is rarely realistic. Unknown anomalies in the training set can significantly degrade performance. For example, DeepSVDD~\citep{deepsvdd} is a one-class classification method that learns to enclose normal data within a hypersphere in latent space, which may also incorrectly generalise anomalies into the learnt hypersphere, leading to low recall as they cannot catch anomalies. In addition, deep learning methods are inherently stochastic, relying on random initialization and sampling, which yields slightly different performance across different runs. This can undermine engineers' trust, as repeated runs lead to inconsistent results. By contrast, \toolname{}'s open-box design and deterministic behaviour ensure both accuracy and reliability.

\input{tables/fse26-eventadl/exp-rca}

\subsection{RQ2: How effective is \toolname{} in finding the root cause of anomalies?}

In this section, we evaluate the performance of \toolname{} against the existing baselines on all five datasets. We use all events within 1 hour around the anomaly occurrence time for all methods to perform root cause localisation, as we have observed in our benchmark cloud systems and in the collected real-world incidents that the root causes are closely aligned with the anomaly occurrence time. As the cloud systems are highly dynamic, a mistaken operation propagates and affects the systems quickly. We also use the Random baseline as a proxy for a human operator that randomly generates a hypothesis and investigates the root cause, aiming to see how root cause localisation can help increase root cause analysis performance. Table~\ref{tab:rq2-rca} reports the root cause localisation performance of all methods using the AC@1, AC@3, and Avg@5 scores across all five datasets. To account for randomness, we run each experiment 10 times with different random seeds and report the mean and standard deviation of their performance. We have obtained the following insights:

\textbf{(1) \toolname{} consistently outperforms all baselines in root cause localisation}. It achieves an AC@3 score of 100\% on all systems, meaning the true root cause is always ranked among the top three candidates. In the Live, \zrhEvent{}, and \mxpEvent{} datasets, \toolname{} can rank the root causes precisely in the top-1 ranking, achieving an AC@1 of 100\%. In the Falcon and Flask datasets, there are a few cases where \toolname{} could not rank the true root cause (i.e., the actor and their interventions) as the top-1 root cause, because, right before the anomaly occurrence time, multiple actors had interacted with the same set of resources (e.g., multiple users running the same updating stack), causing confusion for the random walk backtracking. However, \toolname{} narrows down the search space by providing a small ranked list of root causes for investigation. This result validates our assumption that root causes have multiple causal paths to observed anomalies (Section~\ref{sec:random-walk}), ensuring the random walker reaches the true root cause. It is worth noting that \toolname{} is the first root cause localisation method designed to work directly with event data as defined in Section~\ref{sec:term}. Therefore, it can capture the operations performed in the systems and precisely construct the intervention graph (Algorithm~\ref{alg:construct-intervention-graph}). This graph subsequently allows our time-aware random walk to localise the root cause automatically and precisely. Meanwhile, all existing root cause localisation methods for cloud systems~\citep{brianlogsurvey2025, Soldani2022rcasurvey, cheng2023ai} are developed for metrics, logs, and traces, which undermines the power of event data and leaves it under-explored.

\textbf{(2) Metric-based root cause localisation methods show reasonable performance but fail to fully exploit event data.}
BARO and $\epsilon$-Diagnosis are metric-based root cause localisation methods. To apply them, we transform event data into time series reflecting the frequency of actor interventions, where the number of series depends on the number of actors active during the incident. BARO consistently achieves the second-best performance. Its core assumption is that the root cause exhibits a strong anomaly during the incident (i.e., the most anomalous time series reported by their scorer is considered the root cause). This assumption is surprisingly effective, but does not generalise. BARO performs poorly on the \zrhEvent{} dataset, where multiple rare interventions occurred concurrently, confusing the algorithm as BARO does not correlate interventions with detected anomalies as in \toolname{}. Moreover, BARO provides little interpretability, offering no explanation for why a candidate is flagged as the root cause, thereby shifting the burden of reasoning to operators. In contrast, \toolname{} does not rely on such assumptions. Instead, it constructs an intervention graph that explicitly encodes the causal relationships between recent interventions and detected anomalies, enabling precise and interpretable root cause localisation.

\textbf{(3) Causal inference-based methods are not well-suited for event data.}
CausalAI, CausalRCA, and RCD are causal inference-based root cause localisation techniques originally designed for metric time series. They typically apply causal discovery (e.g., RCD with $\Psi$-PC, CausalRCA with DAG-GNN) to construct a causal graph under the assumption that the root-cause time series influences many others. Root cause ranking is then performed via graph centrality algorithms such as PageRank. However, the approaches are not adequate for event data, where causal relations can be explicit in the event values. Time series may fail to capture true causal relations~\citep{pham2024root}. KPIRoot achieves AC@3 scores of 0.67 to 0.90 on the reproduced benchmarks by combining similarity and causality on event-frequency time series, but drops to 0 on the real-world ZRH incident, where the root cause is a low-connectivity intervention that does not dominate the frequency spectrum. Groot performs poorly on most datasets (AC@3 around 0.33 to 0.50 on benchmarks, 0 on real-world incidents). Its edge-weight rules assume dependencies between metrics and logs, which do not transfer directly to event-based interventions. These observations reinforce that Granger causality and metric-based causal assumptions do not capture the explicit actor-operation-resource semantics that event data provide. By contrast, the Intervention Graph in our \toolname{} leverages event information directly.

\textbf{(4) Existing graph-based RCA approaches for microservices fail due to the mismatch between microservice dependency graphs and event-based intervention graphs.}
DeepHunt~\citep{sun2025interpretable} uses a graph autoencoder over microservice dependency graphs, and TVDiag~\citep{xie2024tvdiag} applies contrastive learning over multimodal observability. Both assume a relatively stable topology where node identity matches service identity. On Falcon, DeepHunt reaches AC@3 of 0.83 because the reproduced incidents induce clear dependency anomalies. However, on the two real-world incidents, DeepHunt and TVDiag collapse to AC@3 of 0.00, since event-based intervention graphs have highly dynamic topology and do not align with a fixed service dependency. TVDiag additionally requires labelled failure samples for training, which are unavailable for the OUT and AVA incidents. This highlights a fundamental mismatch: methods trained on microservice dependency graphs cannot handle the time-varying actor-resource structure captured by event data.

\subsubsection{Failure Case Analysis}

We further analyse the failure patterns of \toolname{}. On Falcon and Flask, the AC@1 is below 1.0 because background services with high connectivity (e.g., automated health-check actors) may accumulate high visit counts during the random walk, pushing them above the true root cause in the ranking. These high-connectivity services are not truly anomalous, but they share resources with the affected paths, and the walker visits them transiently. In practice, \toolname{} still recovers the true root cause within the top-3 in all cases. Operators can then prune these background actors using recent system knowledge or by filtering nodes whose edges are not temporally close to the anomaly.

\input{tables/fse26-eventadl/exp-efficiency}

\subsection{RQ3: How Efficient is \toolname{}?}\label{sec:eval-efficiency}

Table~\ref{tab:rq3-runtime-split} presents the runtime performance of \toolname{} against the baselines. All methods are evaluated over 10 runs across all datasets. For fairness, we report only the inference time for deep learning methods (e.g., NeuralLog, APE), excluding their training overheads, which are significant.

\subsubsection{Anomaly Detection}

As shown in Table~\ref{tab:rq3-runtime-split}~(a), \textbf{\toolname{} is highly efficient}, consistently outperforming most baselines across all datasets. This efficiency can be attributed to the design of its two modules: ESP, which relies on lightweight event-matching expressions, and EFP, which uses subsequence distance comparison. NSigma is extremely efficient due to its simplicity as it only considers the deviation from mean and standard deviation of the frequency. In contrast, deep learning methods are significantly slower, even though we report only their inference time. These methods typically require substantial computational resources during both training and inference.

\subsubsection{Root Cause Localisation}

As presented in Table~\ref{tab:rq3-runtime-split}~(b), \textbf{\toolname{} can localise the root cause of anomalies within seconds.} It takes 2.4, 4.3, and 2.4 seconds to localise the root causes in the Falcon, Flask, and Live datasets, respectively. We observe that the runtime is split fairly evenly between the construction of the intervention graph and the random walk (N=100). We also observe that BARO is the fastest root cause localisation method as it is a simple statistical approach that only considers three values (median, IQR, and maximum value) across all time series, making it very efficient. On the other hand, causal inference-based methods (CausalAI, CausalRCA, RCD) are time-consuming. These methods construct causal graphs between time series, which involves computationally expensive operations such as calculating correlations between all possible pairs of time series.

\subsubsection{Scalability of \toolname{}}

\begin{figure}
\centering
\includegraphics[width=0.7\linewidth]{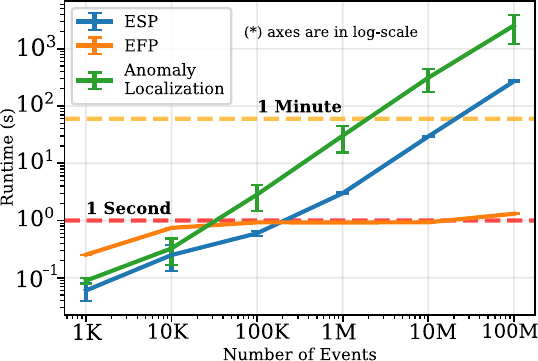}
\caption[Scalability of EventADL.]{Scalability of \toolname{}.}
\label{fig:scalability}
\end{figure}

We evaluate the scalability of three components in \toolname{} when handling large event streams. We deploy \toolname{} on a machine with 12 vCPUs and 36GB RAM, and measure its runtime when processing event streams from the deployed system at different scales. Figure~\ref{fig:scalability} presents the runtimes of \toolname{} across different scales.

We observe that \toolname{} can support real-time monitoring. Specifically, ESPs and EFPs can detect anomalies at rates of 100K events/s. Recall from our real-world analysis (Figure~\ref{fig:real-world-insights}) that the number of events per incident has a median of 1K and a \(3^{rd}\) quartile of 100K. After detection, \toolname{} takes less than 1 minute to localise the root cause with up to 1M events. In addition, ESP and root cause localisation exhibit linear growth with the number of events. This is expected, as ESP must scan the entire stream for pointwise anomaly detection and the construction of the Intervention Graph for root cause localisation also requires similar scanning. Nevertheless, both remain efficient at scale. Notably, EFP runtime remains nearly constant. Its runtime ranges from 0.25s at 1K events to only 1.3s at 100M events. The EFP module scales very favorably because EFP operates on event-based time series
, and we observe that the number of unique time series extracted does not grow proportionally with the number of events.

\subsection{RQ4: Ablation Study} \label{sec:ablation}

We conduct ablation experiments to better understand the contribution of each component in \toolname{}. First, we examine how each component contributes to the overall anomaly detection performance, then we present empirical evidence supporting our choices in EFP design (magnitude-based over shape-based) and ESP choices (HyGLAD vs Drain). Note that the RCL component is already minimal, with no parts to isolate or remove.

\subsubsection{Anomaly Detection} \label{sec:ablation-ad}

The results in Table~\ref{tab:rq1-anomaly-detection} show that \textit{ESP-only} and \textit{EFP-only} achieve high performance and positively contribute to the overall performance of \toolname{}. We can observe that EFPs outperform other statistical methods: like \textit{EFP-only}, NSigma and CUSUM also detect anomalies in frequencies of ESPs, yet \textit{EFP-only} has a higher F1-score on average across all five datasets. We also observe that \textit{ESP-only} and \textit{EFP-only} have high recall on most datasets. This is because incidents often manifest anomalies across multiple dimensions, both pointwise and frequency-based anomalies. While our analysis in Section~\ref{sec:analysis-real-world} shows that some incidents only manifest anomalies along a single dimension, there may exist undetected anomalies in other dimensions that remain unreported. To further assess how ESP and EFP complement each other, we randomly injected 20\% anomalies into the historical (training) period for \textit{ESP-only (C)} and \textit{EFP-only (C)}. Naturally, both miss the anomalies injected in training, resulting in recall scores of only $\approx$80\% each. However, their combination, \textit{\toolname{} (C)}, achieves 96\% recall, since \textit{EFP-only (C)} is able to recover 80\% of the anomalies missed by \textit{ESP-only (C)}. This experiment shows that combining ESP and EFP helps detect subtle anomalies that may occur in only one dimension. We further show that our adaptation mechanism effectively reduces false positives while maintaining high recall. In dynamic cloud systems, without adaptation to system evolution, ESP and EFP may continuously flag false anomalies. Notably, \textit{ESP-only} and \textit{EFP-only}, when run without adaptation, achieve precision scores of 0.72 and 0.79 on the Falcon dataset, already outperforming most baselines. Our \toolname{}, which combines ESP, EFP, and adaptation, further reduces false alarms, raising precision to 0.82. Importantly, when encountering system evolution for the first time, ESP and EFP alone cannot distinguish legitimate evolution from anomalies, and will report both. With root cause localisation, however, \toolname{} enables operators to identify the true sources of anomalies, discard false alarms, and trigger adaptation.

\begin{figure}
\centering
\includegraphics[width=0.7\linewidth]{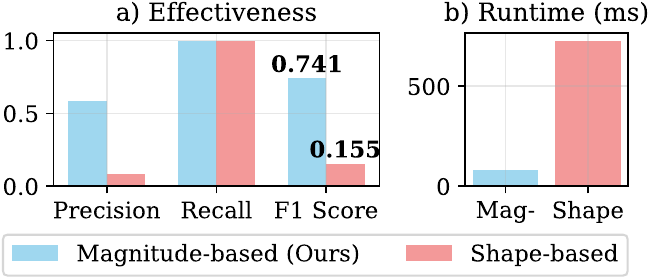}
\caption{Magnitude-based vs. shape-based EFP.}
\label{fig:ablation-mag}
\end{figure}

\subsubsection{Shape-based vs Magnitude-based EFP} \label{sec:ablation-shape-mag}
In this ablation, we compare our magnitude-based EFP with the shape-based variant~\citep{lu2022matrix, lee2024explainable} to assess their impact on the \zrhEvent{} dataset. As shown in Figure~\ref{fig:ablation-mag}, our magnitude-based EFP outperforms the shape-based approach with a substantially higher F1-score (0.741 vs. 0.155), demonstrating its effectiveness in detecting anomalies in event frequency (Section~\ref{sec:efp}). Moreover, it achieves a 9$\times$ speedup in runtime (74.98ms vs. 664.08ms) because our method does not require preprocessing the time series to match shapes. These results indicate that our magnitude-based subsequence comparison is better suited for event data.

\subsubsection{HyGLAD-based vs Drain-based ESPs}\label{sec:drain-study}

As discussed in Section~\ref{sec:esp}, \toolname{} can use different methods to learn ESPs. In this section, we replace HyGLAD~\citep{hyglad} with Drain~\citep{he2017drain} to examine how the performance varies. Since Drain operates on unstructured logs, we implement two variants of event-to-log conversion. In the first variant, \textit{\toolname{}(D1)}, we directly flatten each structured event into a log string. In the second variant, \textit{\toolname{}(D2)}, we extract key fields (time, actor, operation, resource) from the events and construct log entries in the format: \texttt{"time=<time> actor=<actor> operation=<ops> resources=<res>"}. Drain then learns the corresponding log templates, which we use as ESPs.

The experimental results presented in Table~\ref{tab:rq1-anomaly-detection} show that both variants of \toolname{} using Drain for learning ESPs still outperform most baselines consistently across all datasets, demonstrating robustness when using different methods to learn ESPs. However, as discussed in Section~\ref{sec:esp}, Drain may over-generalise some specific patterns on complex datasets (see Table~\ref{fig:relationship-aware}), as it does not take into account the relationships between system entities, resulting in missed anomalies. Experimentally, in the Live dataset, \toolname{}'s recall drops from 100\% to 93\% when using Drain to learn ESPs instead of HyGLAD, showing that HyGLAD is more suitable for event data.

\subsection{RQ5: Robustness of \toolname{}}\label{sec:robustness}

\begin{sidewaysfigure}[p]
\centering
\includegraphics[width=\textheight]{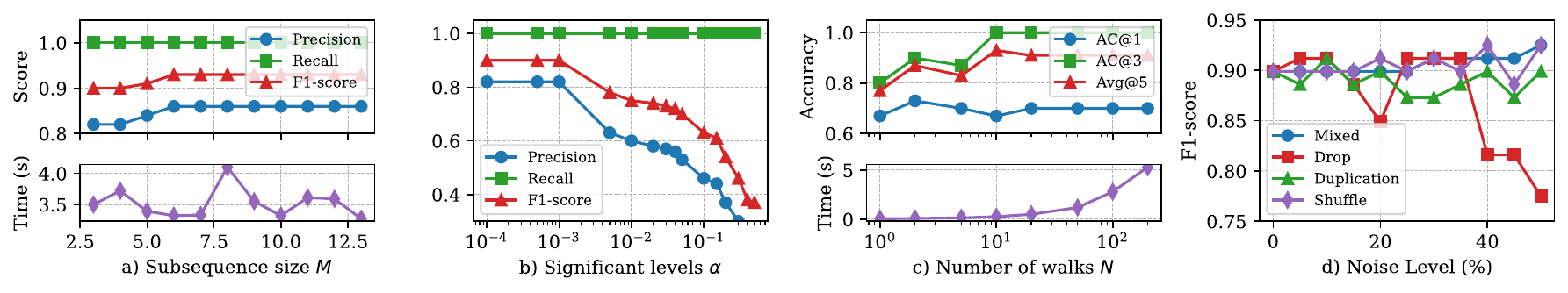}
\caption[Robustness analysis of \toolname{}.]{Robustness analysis of \toolname{} w.r.t. different parameters and noise levels on the Falcon dataset.}
\label{fig:sensitivity-m}
\end{sidewaysfigure}
\subsubsection{Robustness to Parameter Settings}

In this section, we conduct a robustness analysis to understand how the performance of \toolname{} varies under different parameter settings. We refer readers to~\citep{hyglad} for information about parameters for ESP. EFP has two parameters: (1) the subsequence size $M$, and (2) the significance level $\alpha$. Root Cause Localisation has one parameter: the number of walks $N$ used in the time-aware random walk (Algorithm~\ref{alg:temporal-random-walk}).

We first vary the subsequence size $M$ from 3 to 13 to see how \toolname{} performs with both smaller and larger subsequences. Second, we vary the significance level $\alpha$ from $10^{-4}$ to 0.05 to examine the sensitivity of \toolname{}. Third, we vary the number of walks $N$ used in root cause localisation from 1 to 200 to measure changes in performance. Due to space constraints, we only conduct these experiments on the Falcon dataset. The results are presented in Figure~\ref{fig:sensitivity-m}a, b, and c.

We observe that \toolname{} maintains stable performance across all values of $M$. However, shorter subsequences (e.g., $M < 6$) cause a slight drop in precision (from 0.86 to 0.82), as they may fail to capture more complex frequency patterns. Second, we find that \toolname{} becomes more sensitive as $\alpha$ increases (i.e., detecting more anomalies), resulting in lower precision and F1-scores. For example, precision drops from 0.82 to 0.23 as $\alpha$ increases from $10^{-4}$ to 0.05, reflecting a higher number of false positives. This finding suggests that a smaller $\alpha$ (e.g., $10^{-3}$) is preferred to avoid over-triggering alarms. Third, we observe that very small values of $N$ (e.g., 1 or 2) yield suboptimal localisation performance due to limited exploration on the Intervention Graph. Accuracy improves significantly and stabilizes once $N \geq 10$, although runtime increases linearly. For instance, Avg@5 improves from 0.77 (at $N=1$) to 0.91 (at $N=20$), but runtime also increases from 0.05s to 0.5s. In practice, \toolname{} can execute root cause localisation continuously, therefore, operators receive an initial ranked list of root causes that becomes progressively refined and more accurate as additional nodes are traversed.

\subsubsection{Robustness to Noise}

We add noise to training event data by randomly (1) dropping, (2) duplicating, (3) shuffling, and (4) applying a mixture of these three types of noise, with intensity ranging from 1\% to 50\% of the total events. The results presented in Figure~\ref{fig:sensitivity-m}d show that \toolname{} mostly maintains stable performance under these noise conditions. However, it degrades under event dropping, as fewer events reduce the quality of learned patterns (e.g., F1 = 0.775 at 50\% drop). These results demonstrate that our \toolname{} is resilient to imperfect training data, a critical property for deployment in noisy, evolving cloud environments.

\input{tables/fse26-eventadl/public_benchmarks}

\subsection{RQ6: Generalisability to Non-Event Data} \label{sec:generalisability-non-event}

To examine whether the design of \toolname{} transfers beyond event data, we evaluate the EFP module on three public benchmarks: Eadro~\citep{lee2023eadro} (with the Train Ticket and Social Network sub-datasets), GAIA~\citep{gaia}, and AIOps21~\citep{aiops21}. These benchmarks contain metric, log, and trace time series from microservice systems. We apply EFP directly to the available frequency-based time series, using the same magnitude-based subsequence comparison and hypothesis testing described in Section~\ref{sec:efp}. Table~\ref{tab:public-benchmarks} summarises the results.

\textbf{(1) EFP achieves competitive F1-scores on non-event benchmarks.} On Eadro TT, Eadro SN, GAIA, and AIOps21, \toolname{} reaches F1-scores of 84.5\%, 92.3\%, 90.4\%, and 89.9\%, respectively, comparable to or surpassing baselines such as TraceAnomaly, MultimodalTrace, MS-LSTM, MS-DCC, LogFormer, ART~\citep{sun2024art}, and Hades~\citep{lee2023heterogeneous}. The strong performance on Eadro SN (92.3\% F1) and GAIA (90.4\% F1) shows that magnitude-based subsequence analysis captures frequency-based anomalies across different data modalities.

\textbf{(2) EFP generalises without retraining or parameter tuning.} We use the same parameter settings as on event data ($M = 7$, $\alpha = 10^{-3}$). This demonstrates that the design principles behind EFP (distribution-free hypothesis testing, magnitude-based comparison) apply to any frequency-based time series, not just event frequencies. The generalisation result suggests that EFP can serve as a general-purpose anomaly detection primitive for systems with heterogeneous observability data.

%% file: tables/fse26-eventadl/exp-ad.tex
\begin{sidewaystable}[p]
\caption{The anomaly detection performance of \toolname{} and fourteen baselines on five datasets (Falcon, Flask, Live, \zrhEvent{}, \mxpEvent{}) in terms of Precision, Recall, and F1-score. We report the mean and standard deviation over ten runs with different random seeds. We \textbf{bold} the best values and \underline{underline} the second-best.}
\label{tab:rq1-anomaly-detection}
\resizebox{\textheight}{!}{%
\setlength\tabcolsep{1pt}
\begin{tabular}{lccccccccccccccc}
\toprule
\multirow{2}{*}{\textbf{Method}} 
 & \multicolumn{3}{c}{\textbf{Falcon}} 
 & \multicolumn{3}{c}{\textbf{Flask}} 
 & \multicolumn{3}{c}{\textbf{Live}} 
 & \multicolumn{3}{c}{\textbf{\zrhEvent{}}} 
 & \multicolumn{3}{c}{\textbf{\mxpEvent{}}} \\ 
 \cmidrule(lr){2-4} \cmidrule(lr){5-7} \cmidrule(lr){8-10} \cmidrule(lr){11-13} \cmidrule(lr){14-16}
& Precision & Recall & F1-Score & Precision & Recall & F1-Score & Precision & Recall & F1-Score & Precision & Recall & F1-Score & Precision & Recall & F1-Score \\ \midrule
\rowcolor{linegray} APE~\citep{event_ijcai16_ape} & 0.46±0.15 & 0.35±0.16 & 0.38±0.12 & 0.47±0.18 & 0.68±0.28 & 0.50±0.09 & 0.36±0.21 & 0.55±0.28 & 0.37±0.15 & 0.16±0.07 & \textbf{1.00±0.00} & 0.27±0.10  & \underline{0.25±0.09} & 0.98±0.04 & \underline{0.39±0.11} \\ 
BARO~\citep{pham2024baro} & 0.18±0.00 & \textbf{1.00±0.00} & 0.30±0.00 & 0.17±0.00 & \textbf{1.00±0.00} & 0.29±0.00 & 0.17±0.00 & \textbf{1.00±0.00} & 0.29±0.00 & 0.06±0.00 & \textbf{1.00±0.00} & 0.12±0.00 & 0.12±0.00 & \textbf{1.00±0.00} & 0.22±0.00 \\ 
\rowcolor{linegray} CUSUM~\citep{cusum} & 0.53±0.00 & \underline{0.84±0.00} & 0.65±0.00 & 0.71±0.00 & \textbf{1.00±0.00} & \underline{0.83±0.00} & \underline{0.77±0.00} & \textbf{1.00±0.00} & \underline{0.87±0.00} & 0.06±0.00 & \textbf{1.00±0.00} & 0.12±0.00 & 0.12±0.00 & \textbf{1.00±0.00} & 0.22±0.00 \\ 
DeepSVDD~\citep{deepsvdd} & 0.62±0.45 & 0.39±0.38 & 0.44±0.40 & 0.47±0.46 & 0.34±0.34 & 0.34±0.36 & 0.70±0.44 & 0.37±0.34 & 0.43±0.36 & 0.82±0.10 & 0.81±0.24 & 0.79±0.12 & 0.15±0.04 & 0.80±0.21 & 0.25±0.05 \\
\rowcolor{linegray} DIF~\citep{dif} & 0.15±0.17 & 0.27±0.41 & 0.10±0.15 & 0.21±0.26 & 0.37±0.38 & 0.19±0.20 & 0.18±0.28 & 0.24±0.37 & 0.15±0.23 & 0.43±0.12 & 0.84±0.24 & 0.53±0.03 & 0.10±0.01 & \underline{0.99±0.01} & 0.18±0.01 \\
ICL~\citep{icl} & \underline{0.66±0.44} & 0.39±0.38 & 0.43±0.40 & 0.54±0.49 & 0.36±0.38 & 0.32±0.35 & 0.61±0.49 & 0.22±0.27 & 0.29±0.32 & \underline{0.87±0.05} & \underline{0.99±0.01} & \textbf{0.92±0.03} & - & - & - \\
\rowcolor{linegray} NeuralLog~\citep{le2021neurallog} & 0.60±0.03 & \textbf{1.00±0.00} & \underline{0.75±0.02} & 0.42±0.18 & 0.39±0.21 & 0.33±0.09 & 0.53±0.01 & \textbf{1.00±0.00} & 0.69±0.01 & 0.06±0.00 & \textbf{1.00±0.00} & 0.12±0.00 &  0.15±0.00 & \textbf{1.00±0.00} & 0.26±0.00  \\ 
NeuTraL~\citep{neutral} & \textbf{0.82±0.34} & 0.55±0.39 & 0.60±0.38 & 0.68±0.45 & 0.45±0.38 & 0.49±0.39 & 0.68±0.46 & 0.35±0.36 & 0.41±0.37 & \textbf{0.89±0.03} & 0.91±0.21 & 0.89±0.14 & - & - & -\\
\rowcolor{linegray} NSigma~\citep{Li2022Circa} & 0.37±0.00 & \textbf{1.00±0.00} & 0.54±0.00 & 0.17±0.00 & \textbf{1.00±0.00} & 0.29±0.00 & 0.17±0.00 & \textbf{1.00±0.00} & 0.29±0.00 & 0.07±0.00 & \textbf{1.00±0.00} & 0.13±0.00 & 0.16±0.00 & \textbf{1.00±0.00} & 0.28±0.00 \\ 
RCA~\citep{rca} & 0.40±0.39 & 0.37±0.41 & 0.26±0.30 & 0.30±0.37 & 0.46±0.41 & 0.28±0.31 & 0.27±0.37 & 0.32±0.40 & 0.22±0.30 & 0.55±0.21 & 0.62±0.22 & 0.53±0.08 & 0.14±0.02 & 0.45±0.12 & 0.21±0.04 \\
\rowcolor{linegray} RDP~\citep{rdp} & 0.14±0.16 & 0.80±0.40 & 0.21±0.23 & 0.12±0.26 & 0.80±0.40 & 0.15±0.28 & 0.12±0.27 & 0.50±0.50 & 0.14±0.28 & 0.11±0.00 & \textbf{1.00±0.00} & 0.20±0.00 & - & - & -\\
ShadeWat~\citep{zengy2022shadewatcher} & 0.17±0.02& 0.76±0.25 & 0.27±0.03 & 0.17±0.01 & \underline{0.82±0.20} & 0.28±0.03 & 0.17±0.02 & \underline{0.90±0.12} & 0.30±0.03 & 0.06±0.01 & 0.90±0.14 & 0.11±0.01 & 0.11±0.03 & 0.83±0.35 & 0.19±0.06 \\ 
\rowcolor{linegray} ADAMAS~\citep{gu2025adamas} & 0.19±0.00 & \textbf{1.00±0.00} & 0.31±0.00 & 0.18±0.00 & \textbf{1.00±0.00} & 0.30±0.00 & 0.17±0.00 & \textbf{1.00±0.00} & 0.29±0.00 & 0.06±0.00 & \textbf{1.00±0.00} & 0.12±0.00 & 0.12±0.00 & \textbf{1.00±0.00} & 0.22±0.00 \\
KPIRoot~\citep{gu2024kpiroot} & 0.61±0.00 & 0.71±0.00 & 0.66±0.00 & \underline{0.80±0.00} & 0.53±0.00 & 0.64±0.00 & 0.70±0.00 & 0.53±0.00 & 0.60±0.00 & 0.21±0.00 & 0.40±0.00 & 0.28±0.00 & 0.00±0.00 & 0.00±0.00 & 0.00±0.00 \\
\midrule
\rowcolor{lineblue} \textbf{\toolname{}} & \textbf{0.82±0.00} & \textbf{1.00±0.00} & \textbf{0.90±0.00} & \textbf{0.81±0.00} & \textbf{1.00±0.00} & \textbf{0.90±0.00} & \textbf{0.91±0.00} & \textbf{1.00±0.00} & \textbf{0.95±0.00} & 0.81±0.00 & \textbf{1.00±0.00} & \underline{0.90±0.00} & \textbf{0.91±0.00} & \textbf{1.00±0.00} & \textbf{0.95±0.00} \\ \midrule
\rowcolor{linegreen} \textit{ESP-only} & 0.72±0.00 & 1.00±0.00 & 0.84±0.00 & 0.37±0.00 & 1.00±0.00 & 0.54±0.00 & 0.31±0.00 & 1.00±0.00 & 0.47±0.00 & 0.72±0.00 & 1.00±0.00 & 0.84±0.00 & 0.91±0.00 & 1.00±0.00 & 0.95±0.00\\
\rowcolor{linegreen2} \textit{EFP-only} & 0.79±0.00 & 1.00±0.00 & 0.88±0.00 & 0.52±0.00 & 0.83±0.00 & 0.64±0.00 & 0.18±0.00 & 0.50±0.00 & 0.26±0.00 & 0.59±0.00 & 1.00±0.00 & 0.74±0.00 & 1.00±0.00 & 1.00±0.00 & 1.00±0.00 \\ 
\midrule
\rowcolor{linegreen} \textit{EventADL (C)} & 0.83±0.01 & 0.96±0.05 & 0.89±0.03 & 0.82±0.00 & 0.93±0.00 & 0.87±0.00 & 0.92±0.02 & 0.90±0.03 & 0.91±0.01 & 0.62±0.08 & 0.93±0.12 & 0.75±0.09 & 0.94±0.06 & 0.93±0.06 & 0.85±0.18 \\ 
\rowcolor{linegreen2} \textit{ESP-only (C)} & 0.73±0.03 & 0.78±0.05 & 0.76±0.04 & 0.37±0.06 & 0.81±0.13 & 0.51±0.08 & 0.34±0.03 & 0.88±0.04 & 0.49±0.03 & 0.93±0.06 & 0.70±0.17 & 0.78±0.10 & 0.97±0.06 & 0.87±0.06 & 0.83±0.16 \\
\rowcolor{linegreen} \textit{EFP-only (C)} & 0.65±0.05 & 0.83±0.05 & 0.72±0.05 & 0.59±0.07 & 0.78±0.09 & 0.67±0.08 & 0.17±0.01 & 0.40±0.03 & 0.24±0.01 & 0.67±0.05 & 0.87±0.06 & 0.75±0.01 & 1.00±0.00 & 0.83±0.21 & 0.86±0.20 \\ \midrule
\rowcolor{linegreen2}\textit{\toolname{} (D1)} & 0.53±0.00 & 1.00±0.00 & 0.70±0.00 & 0.51±0.00 & 1.00±0.00 & 0.67±0.00 & 0.88±0.00 & 0.93±0.00 & 0.90±0.00 & 0.62±0.00 & 1.00±0.00 & 0.77±0.00 & 0.91±0.00 & 1.00±0.00 & 0.95±0.00 \\ 
\rowcolor{linegreen} \textit{\toolname{} (D2)} & 0.72±0.00 & 0.94±0.00 & 0.82±0.00 & 0.75±0.00 & 1.00±0.00 & 0.86±0.00 & 0.97±0.00 & 0.93±0.00 & 0.95±0.00 & 0.43±0.00 & 1.00±0.00 & 0.61±0.00 & 0.83±0.00 & 1.00±0.00 & 0.91±0.00 \\ 
\bottomrule
\end{tabular}
}

{\tiny \textit{Note}: Methods shown in \tinycolorbox{linegreen}{\textit{italic}} are used in the ablation study discussed in Section~\ref{sec:ablation}. (-) ICL, NeuTraL, and RDP encounter OOM/time-out errors in the \mxpEvent{} dataset.}
\end{sidewaystable}

%% file: tables/fse26-eventadl/exp-rca.tex
\begin{sidewaystable}[p]
\caption{The anomaly localisation performance of \toolname{} and ten baselines on five datasets (Falcon, Flask, Live, \zrhEvent{}, and \mxpEvent{}) in terms of AC@1, AC@3, and Avg@5. We report the mean and standard deviation over ten runs with different random seeds. We \textbf{bold} the best values and \underline{underline} the second-best.}
\label{tab:rq2-rca}
\resizebox{\textheight}{!}{%
\setlength\tabcolsep{1pt}
\begin{tabular}{lccccccccccccccc}
\toprule
\multirow{2}{*}{\textbf{Method}} & \multicolumn{3}{c}{\textbf{Falcon}} & \multicolumn{3}{c}{\textbf{Flask}} & \multicolumn{3}{c}{\textbf{Live}} & \multicolumn{3}{c}{\textbf{\zrhEvent{}}} & \multicolumn{3}{c}{\textbf{\mxpEvent{}}} \\ \cmidrule(lr){2-4} \cmidrule(lr){5-7} \cmidrule(lr){8-10} \cmidrule(lr){11-13} \cmidrule(lr){14-16}
 & AC@1 & AC@3 & Avg@5 & AC@1 & AC@3 & Avg@5 & AC@1 & AC@3 & Avg@5 & AC@1 & AC@3 & Avg@5 & AC@1 & AC@3 & Avg@5 \\ \midrule
\rowcolor{linegray} Random & 0.05±0.03 & 0.22±0.07 & 0.22±0.04 & 0.04±0.04 & 0.20±0.08 & 0.19±0.06 & 0.08±0.06 & 0.24±0.05 & 0.23±0.05 & 0.05±0.08 & 0.20±0.13 & 0.20±0.11 & 0.09±0.09 & 0.24±0.10 & 0.23±0.09 \\
BARO~\citep{pham2024baro} & 0.50±0.00 & 0.83±0.00 & \underline{0.73±0.00} & \underline{0.67±0.00} & 0.83±0.00 & \underline{0.81±0.00} & 0.80±0.00 & \textbf{1.00±0.00} & \underline{0.92±0.00} & \underline{0.20±0.00} & 0.20±0.00 & 0.28±0.00 & 0.70±0.00 & \textbf{1.00±0.00} & \underline{0.90±0.00} \\ 
\rowcolor{linegray} CausalAI~\citep{causalai23salesforce} & 0.18±0.09 & 0.40±0.10 & 0.37±0.07 & 0.14±0.05 & 0.40±0.06 & 0.40±0.04 & 0.24±0.07 & 0.55±0.07 & 0.49±0.04 & 0.00±0.00 & 0.00±0.00 & 0.00±0.00 & 0.07±0.26 & \underline{0.27±0.45} & 0.26±0.35 \\ 
CausalRCA~\citep{xin2023causalrca} & 0.00±0.00 & 0.67±0.00 & 0.44±0.00 & 0.00±0.00 & 0.00±0.00 & 0.23±0.00 & 0.00±0.00 & 0.60±0.00 & 0.44±0.00 & 0.00±0.00 & 0.00±0.00 & 0.00±0.00 & 0.00±0.00 & \textbf{1.00±0.00} & 0.60±0.00 \\ 
\rowcolor{linegray} \(\epsilon\)-Diagnosis~\citep{Shan2019Ediagnosis} & 0.03±0.02 & 0.12±0.06 & 0.13±0.05 & 0.03±0.03 & 0.10±0.05 & 0.08±0.04 & 0.00±0.00 & 0.00±0.00 & 0.00±0.00 & 0.04±0.05 & \underline{0.23±0.12} & 0.22±0.09 & 0.00±0.00 & 0.00±0.00 & 0.12±0.00\\ 
Groot~\citep{wang2021groot} & 0.33±0.00 & 0.33±0.00 & 0.33±0.00 & 0.50±0.00 & 0.50±0.00 & 0.50±0.00 & 0.40±0.00 & 0.40±0.00 & 0.40±0.00 & 0.00±0.00 & 0.00±0.00 & 0.00±0.00 & 0.00±0.00 & 0.00±0.00 & 0.00±0.00 \\ 
\rowcolor{linegray} RCD~\citep{Azam2022rcd} & \underline{0.57±0.00} & 0.69±0.04 & 0.65±0.03 & 0.41±0.03 & 0.66±0.05 & 0.60±0.04 & 0.00±0.00 & 0.09±0.02 & 0.11±0.01 & 0.02±0.04 & 0.07±0.09 & 0.05±0.07 & 0.00±0.00 & 0.20±0.40 & 0.14±0.29 \\ 
DeepHunt~\citep{sun2025interpretable} & 0.53±0.10 & 0.83±0.00 & 0.65±0.06 & 0.53±0.12 & \underline{0.89±0.07} & 0.73±0.05 & \underline{0.82±0.09} & \textbf{1.00±0.00} & 0.90±0.05 & 0.00±0.00 & 0.00±0.00 & 0.00±0.00 & 0.00±0.00 & 0.00±0.00 & 0.04±0.00 \\
\rowcolor{linegray} TVDiag~\citep{xie2024tvdiag} & 0.00±0.00 & 0.67±0.00 & 0.40±0.00 & 0.00±0.00 & 0.00±0.00 & 0.00±0.00 & 0.00±0.00 & 0.00±0.00 & 0.00±0.00 & - & - & - & - & - & -\\ 
KPIRoot~\citep{gu2024kpiroot} & 0.43±0.00 & \underline{0.90±0.00} & 0.63±0.00 & 0.13±0.00 & 0.67±0.00 & 0.41±0.00 & 0.60±0.00 & \underline{0.88±0.00} & 0.74±0.00 & 0.00±0.00 & \textbf{1.00±0.00} & \underline{0.50±0.00} & \underline{0.80±0.00} & \textbf{1.00±0.00} & 0.87±0.00\\
\midrule
\rowcolor{lineblue} \textbf{\toolname{}} & \textbf{0.70±0.02} & \textbf{1.00±0.00} & \textbf{0.91±0.00} & \textbf{0.68±0.01} & \textbf{1.00±0.00} & \textbf{0.91±0.01} & \textbf{1.00±0.00} & \textbf{1.00±0.00} & \textbf{1.00±0.00} & \textbf{1.00±0.00} & \textbf{1.00±0.00} & \textbf{1.00±0.00} & \textbf{1.00±0.00} & \textbf{1.00±0.00} & \textbf{1.00±0.00} \\ \bottomrule
\end{tabular}
}

{\tiny (-) TVDiag~\citep{xie2024tvdiag} requires failure training samples. The OUT and AVA incidents are test data without corresponding failure training samples.}
\end{sidewaystable}

%% file: tables/fse26-eventadl/exp-efficiency.tex
\begin{table}
\caption{Runtime comparison.} \label{tab:rq3-runtime-split}
\centering
{\small \textbf{(a) Anomaly Detection Runtime (in seconds).}} \\
\vspace{2pt}
\resizebox{\linewidth}{!}{%
\setlength{\tabcolsep}{1pt}
\begin{tabular}{lccccc}
\toprule
\textbf{Method} & \textbf{Falcon} & \textbf{Flask} & \textbf{Live} & \textbf{\zrhEvent{}} & \textbf{\mxpEvent{}} \\
\midrule
APE~\citep{event_ijcai16_ape} & 0.060±0.08 & 0.075±0.08 & 0.141±0.14 & 0.093±0.02 & 1.290±0.19\\
\rowcolor{linegray} BARO~\citep{pham2024baro} & 0.886±0.08 & 0.913±0.06 & 0.890±0.06 & 0.883±0.06 & 0.830±0.06 \\
CUSUM~\citep{cusum} & 0.011±0.04 & 0.010±0.04 & 0.003±0.00 & 0.010±0.04 & 0.016±0.05 \\
\rowcolor{linegray} DeepSVDD~\citep{deepsvdd} & 7.780±3.59 & 10.25±7.01 & 9.500±7.74 & 5.560±2.12 & 5.416±1.75\\ 
DIF~\citep{dif} & 46.22±46.5 & 95.03±130 & 86.54±117 & 12.68±1.39 & 17.31±1.38 \\ 
\rowcolor{linegray} ICL~\citep{icl} & 110.2±144 & 245.9±404 & 229.3±329 & 27.34±3.56 & - \\ 
NeuralLog~\citep{le2021neurallog} & 4.642±2.83 & 1.375±0.11 & 0.971±0.27 & 0.338±0.02 & 3.820±0.21\\ 
\rowcolor{linegray} NeuTraL~\citep{neutral} & 11.32±6.88 & 16.79±15.6 & 18.09±20.3 & 9.290±0.56 & - \\ 
NSigma~\citep{Li2022Circa} & 0.003±0.00 & 0.003±0.00 & 0.003±0.00 & 0.012±0.01 & 0.006±0.01 \\
\rowcolor{linegray} RCA~\citep{rca} & 18.91±15.1 & 32.75±39.6 & 31.25±39.2 & 12.38±5.30 & 143.0±1.91 \\ 
RDP~\citep{rdp} & 8.440±3.94 & 11.37±8.33 & 10.64±8.99 & 9.860±5.17 & - \\ 
\rowcolor{linegray} ShadeWat~\citep{zengy2022shadewatcher} & 0.147±0.75 & 0.131±0.71 & 0.047±0.22 & 0.010±0.01 & 0.570±0.18 \\ 
ADAMAS~\citep{gu2025adamas} & 4.847±0.98 & 5.547±0.99 & 4.599±0.70 & 2.839±0.26 & 92.89±1.52 \\
\rowcolor{linegray} KPIRoot~\citep{gu2024kpiroot} & 0.040±0.01 & 0.022±0.01 & 0.041±0.02 & 0.030±0.01 & 0.353±0.01\\
\midrule
\rowcolor{lineblue} \textbf{\toolname{}}  & \textbf{0.031±0.01} & \textbf{0.104±0.01} & \textbf{0.017±0.01} & \textbf{0.096±0.03} & \textbf{0.022±0.01} \\
\textit{ESP-only} & 0.001±0.00 & 0.004±0.00 & 0.007±0.00 & 0.006±0.02 & 0.006±0.02 \\ 
\textit{EFP-only} & 0.030±0.01 & 0.100±0.01 & 0.010±0.01 & 0.090±0.02 & 0.016±0.01 \\ 
\bottomrule
\end{tabular}
}

\vspace{6pt}

{\small \textbf{(b) RCL Runtime (in seconds).}} \\
\vspace{2pt}
\resizebox{\linewidth}{!}{%
\setlength{\tabcolsep}{1pt}
\begin{tabular}{lccccc}
\toprule
\textbf{Method} & \textbf{Falcon} & \textbf{Flask} & \textbf{Live} & \textbf{\zrhEvent{}} & \textbf{\mxpEvent{}} \\
\midrule

BARO~\citep{pham2024baro} & 0.120±0.01 & 0.135±0.01 & 0.290±0.03 & 0.009±0.00 & 0.010±0.00 \\
\rowcolor{linegray} \(\epsilon\)-Diagnosis~\citep{Shan2019Ediagnosis} & 2.320±3.25 & 2.470±3.15 & 1.620±2.02 & 1.130±0.12 & 0.160±0.05  \\
Groot~\citep{wang2021groot} & 9.797±0.11 & 13.81±0.22 & 8.236±0.03 & 0.357±0.01 & 0.760±0.09 \\
\rowcolor{linegray} RCD~\citep{Azam2022rcd} & 15.78±36.1 & 10.79±24.1 & 9.530±17.9 & 0.190±0.02 & 81.21±0.67 \\ 
 CausalAI~\citep{causalai23salesforce} & 81.06±52.7 & 63.35±34.9 & 73.73±44.5 & 2.580±0.01 & 1.460±0.98 \\
\rowcolor{linegray} CausalRCA~\citep{xin2023causalrca} & 185.7±296 & 189.8±299 & 232.1±341 & 55.95±4.67 & 54.66±4.47\\
DeepHunt~\citep{sun2025interpretable} & 30.10±0.31 & 33.88±0.48 & 14.24±0.31 & 4.100±0.14 & 243.0±1.90 \\
\rowcolor{linegray} TVDiag~\citep{xie2024tvdiag} & 66.90±0.22 & 73.10±0.00 & 85.11±0.00 & - & -\\
KPIRoot~\citep{gu2024kpiroot} & 554.9±0.56 & 907.4±27.2 & 445.3±13.5 & 11.37±0.29 & 600.21±0.84 \\
\midrule
\rowcolor{lineblue} \textbf{\toolname{}} & \textbf{2.366±0.06} & \textbf{4.319±0.04} & \textbf{2.420±0.04} & \textbf{0.130±0.00} & \textbf{0.034±0.00} \\
\bottomrule
\end{tabular}
}
\end{table}

%% file: tables/fse26-eventadl/public_benchmarks.tex
\begin{sidewaystable}[p]
\centering
\caption{Performance comparison on public benchmarks: Eadro, GAIA, and AIOps21.} \label{tab:public-benchmarks}
\resizebox{\textheight}{!}{%
\setlength\tabcolsep{3pt}
\begin{tabular}{ccccccccccccccc}
\toprule
\multirow{2}{*}{\textbf{Method}} & \multicolumn{3}{c}{\textbf{Eadro TT}} & \multicolumn{3}{c}{\textbf{Eadro SN}} & \multirow{2}{*}{\textbf{Method}} & \multicolumn{3}{c}{\textbf{GAIA}} & \multirow{2}{*}{\textbf{Method}} & \multicolumn{3}{c}{\textbf{AIOps21}} \\
\cmidrule(lr){2-4} \cmidrule(lr){5-7} \cmidrule(lr){9-11} \cmidrule(lr){13-15}
& Precision & Recall & F1-Score & Precision & Recall & F1-Score & & Precision & Recall & F1-Score & & Precision & Recall & F1-Score \\
\midrule
\rowcolor{linegray} TraceAnomaly~\citep{liu2020unsupervised} & 0.486 & 0.414 & 0.589 & 0.539 & 0.468 & 0.636 & SVM~\citep{guo2024logformer} & 0.210 & 0.540 & 0.300 & ART~\citep{sun2024art} & 0.877 & 0.960 & 0.917 \\
MultimodalTrace~\citep{nedelkoski2019anomaly} & 0.608 & 0.576 & 0.644 & 0.676 & 0.632 & 0.726 & DeepLog~\citep{du2017deeplog} & 0.180 & 0.820 & 0.310 & Eadro~\citep{lee2023eadro} & 0.767 & 0.935 & 0.842 \\
\rowcolor{linegray} MS-RF-AD~\citep{lee2023eadro} & 0.817 & 0.705 & 0.971 & 0.773 & 0.866 & 0.700 & LogAnomaly~\citep{meng2019loganomaly} & 0.230 & 0.800 & 0.360 & Hades~\citep{lee2023heterogeneous} & 0.867 & 0.868 & 0.868 \\
MS-SVM-AD~\citep{lee2023eadro} & 0.787 & 0.678 & 0.938 & 0.789 & 0.770 & 0.808 & PLELog~\citep{yang2021semi} & 0.810 & 0.860 & 0.840 & - & - & - & - \\
\rowcolor{linegray} MS-LSTM~\citep{lee2023eadro} & 0.967 & 0.997 & 0.940 & 0.948 & 0.959 & 0.937 & LogRobust~\citep{zhang2019robust} & 0.830 & 0.940 & 0.880 & - & - & - & -\\
MS-DCC~\citep{lee2023eadro} & 0.965 & 0.993 & 0.938 & 0.948 & 0.962 & 0.934 & LogFormer~\citep{guo2024logformer} & 0.890 & 0.980 & 0.930 & - & - & - & -\\
\midrule
\rowcolor{lineblue} \textbf{\toolname{}} & \textbf{0.745} & \textbf{0.975} & \textbf{0.845} & \textbf{0.857} & \textbf{1.000} & \textbf{0.923} & \textbf{\toolname{}} & \textbf{1.000} & \textbf{0.825} & \textbf{0.904} & \textbf{\toolname{}} & \textbf{0.870} & \textbf{0.930} & \textbf{0.899} \\
\bottomrule
\end{tabular}%
}
\end{sidewaystable}

%% file: papers/fse26-eventadl/7.conclusion.tex
\section{Threats to Validity}

We assess potential threats to the validity of our work, following the guidelines outlined by Wohlin et al.~\citep{wohlin2012experimentation}. The \textbf{construct validity} primarily concerns the hyperparameter settings and evaluation metrics. To mitigate this, we use established evaluation metrics and adopt recommended configurations from previous works~\citep{chen2022adaptive, pham2024baro, pham2026torai, pham2025rcaeval}. Another threat lies in the use of ESPs, which may be susceptible to adversarial evasion (e.g., attackers may mimic normal patterns to bypass detection). However, such evasive behaviour likely triggers other consequences, which will eventually be detected by our framework. The \textbf{internal validity} stems from potential implementation bugs that could affect result reliability. We mitigate this by using well-maintained Python libraries, extensive testing, and repeating each experiment multiple times to ensure consistency. The \textbf{conclusion validity} stems from our benchmark datasets not covering the full range of anomaly types. While we base our incident reproduction on a systematic analysis of 520 real-world incident reports, certain scenarios require manual intervention because they fall outside \toolname{}'s design scope. For anomaly detection, issues that do not manifest through event data cannot be detected. For example, a hardware fault (e.g., disk full) causing a node crash may not generate events. Similarly, performance degradation captured only in metrics (e.g., increased latency) without corresponding event signatures would be missed. For RCL, interventions with low connectivity in the Intervention Graph may receive fewer random walk visits than unrelated high-activity nodes, as discussed in Section~\ref{sec:experiments}. Nevertheless, \toolname{} still narrows the search space by identifying affected resources and recent interventions, enabling operators to extend their investigation. The \textbf{external validity} concerns the generalisability of our findings. In this study, we deployed our method in real cloud systems and evaluated against real incident data, grounding evaluation in realistic settings.

\section{Conclusion}

We present \toolname{}, the first open-box anomaly detection and RCL framework designed for event data in cloud systems. Our real-world incident analysis provides the empirical foundation for this work, revealing that event-based anomalies manifest through Event Type, Event Value, and Event Frequency, and their root causes require tracing intervention chains. Guided by these findings, \toolname{} detects pointwise anomalies through \textit{Event Semantic Patterns (ESPs)} and frequency-based anomalies through \textit{Event Frequency Patterns (EFPs)}, and localises root causes by constructing an \textit{Intervention Graph} and performing a time-aware random walk. Our evaluation on three benchmark systems and two real-world incidents demonstrates that \toolname{} achieves F1-scores of at least 90\% for anomaly detection and 100\% top-3 accuracy for RCL. We further show that EFP generalises to non-event data, achieving competitive performance on three public benchmarks. We release our event datasets to facilitate future research on event-based ADL.

%% file: papers/fse26-eventadl/data_availability.tex
The implementation of \toolname{}, the three experimental datasets, and the supplementary materials are available on Zenodo at \url{https://zenodo.org/records/19433493}~\citep{eventadl_artifact}.

%% file: chapters/5.multisource.tex
\chapter{TORAI: Multimodal Root Cause Analysis for Microservice Systems}\label{chap:torai}

\input{papers/fse26-torai/0.symbols}

\begin{tcolorbox}[left=2pt,right=2pt,top=0pt,bottom=0pt,
  enhanced,
  drop shadow={shadow xshift=1ex, shadow yshift=-1ex, opacity=0.3}]
\textbf{Publication:} This chapter is based on our paper titled \textbf{``TORAI: Multi-Source Root Cause Analysis for Blind Spots in Microservice Service Call Graph''}, Luan Pham, Huong Ha, Xiuzhen Zhang, and Hongyu Zhang, published in the Proceedings of the ACM on Software Engineering (\textit{PACMSE}), Issue \textit{FSE}, Volume 3, Article FSE130, 2026 (\textbf{CORE~A*}) \citep{pham2026torai}.
\end{tcolorbox}
\vspace{10pt}

\noindent The previous chapters addressed anomaly detection and RCA using single data modalities: Chapter~\ref{chap:fse24} focused on metrics, while Chapter~\ref{chap:eventadl} explored events. Each approach demonstrated strong performance within its respective domain, but single-modality methods face limitations: (1) anomalies may manifest through one modality but not the other, and (2) root cause may only become apparent when multiple signals are correlated. This chapter targets multimodal observability data and addresses the limitation identified in Section~\ref{sec:limit-blind-spots}, the assumption of complete trace coverage in existing multimodal RCA methods. Many existing multimodal RCA approaches require every service call to be instrumented, but in practice, this assumption rarely holds due to sampling, third-party integrations, or legacy components, creating ``blind spots'' that render such methods ineffective. This chapter presents TORAI, an unsupervised RCA approach that combines multimodal data to achieve effective diagnosis without requiring full trace coverage. 
\vspace{10pt}

\input{papers/fse26-torai/0.abstract}

\section{Introduction}

\input{papers/fse26-torai/1.introduction}

\section{Background}
\input{papers/fse26-torai/2.background}

\section{TORAI: The Proposed Method}

\input{papers/fse26-torai/4.method}

\section{Results}
\input{papers/fse26-torai/5.results}

\section{Discussion}
\subsection{Threats to Validity}

\input{papers/fse26-torai/6.discussion}

\section{Summary}
\input{papers/fse26-torai/7.conclusion}

\section*{Data Availability}
\input{papers/fse26-torai/data_availability}

\vspace{10pt}
\noindent Together, BARO (Chapter~\ref{chap:fse24}), EventADL (Chapter~\ref{chap:eventadl}), and TORAI form a comprehensive suite of methods for anomaly detection and RCA across different data modalities and scenarios. These contributed methods address key limitations identified in Chapter~\ref{chap:background}: the coupling between anomaly detection and RCA (Section~\ref{sec:limit-ad-decoupling}), the underutilization of event data (Section~\ref{sec:limit-event-data}), and the assumption of complete trace coverage (Section~\ref{sec:limit-blind-spots}). However, the lack of standardised benchmarks (Section~\ref{sec:limit-benchmarks}) with comprehensive datasets and evaluation protocols has made it difficult to fairly compare different RCA approaches and assess their generalisability across diverse failure scenarios. The next chapter addresses this gap by introducing RCAEval, an open-source benchmark that provides large-scale datasets and reproducible baselines for systematic evaluation of RCA methods.
\vspace{10pt}

%% file: papers/fse26-torai/0.symbols.tex
\definecolor{ballblue}{rgb}{0.13, 0.67, 0.8}

\newcommand{\rectangle}{{%
\ooalign{$\sqsubset\mkern3mu$\cr$\mkern4mu\sqsupset$\cr}%
}}

\definecolor{codegreen}{rgb}{0,0.6,0}
\definecolor{codegray}{rgb}{0.5,0.5,0.5}
\definecolor{codepurple}{rgb}{0.58,0,0.82}
\definecolor{backcolour}{rgb}{0.95,0.95,0.92}

\lstdefinelanguage{json}{
  morestring=[b]",
  morecomment=[l]{//},
  moredelim=[l][\color{black}]{:},
  moredelim=[s][\color{red}]{[}{]},
  stringstyle=\color{teal},
  keywordstyle=\color{black},
  commentstyle=\color{gray},
  showstringspaces=false,
}
\lstset{
  aboveskip=0pt,
  belowskip=0pt,
  lineskip=-1pt
}

%% file: papers/fse26-torai/0.abstract.tex
Existing multi-source root cause analysis (RCA) methods for microservice systems assume all services have traces to construct a service call graph. However, this assumption is not practical as microservice systems evolve rapidly and may contain black-box services without traces, such as compiled software or unsupported services. We refer to these services as \textit{blind spots}. In the presence of blind spots, the performance of existing multi-source RCA methods may be affected, as they only diagnose \textit{visible} services on the call graph. To overcome this limitation, we propose TORAI, a novel unsupervised approach that effectively pinpoints fine-grained root causes without relying on the service call graph. Instead, TORAI first measures anomaly severity using available multi-source telemetry data. It then performs clustering to group services based on their severity symptoms and conducts causal analysis to rank services within each severity cluster. Finally, TORAI aggregates the cluster rankings and uses hypothesis testing to identify fine-grained root causes. TORAI provides an unsupervised approach that leverages available multi-source telemetry data for RCA without requiring a constructed service call graph or further intrusive actions, thus addressing the limitations of existing methods. Our experiments on three benchmark systems demonstrate that TORAI outperforms state-of-the-art baselines remarkably in the presence of blind spots. Performance on real-world failures further shows that TORAI can accurately pinpoint the root causes in top-3 recommendations.

%% file: papers/fse26-torai/1.introduction.tex
Microservices have emerged as a popular form of software systems because of their advantages such as loose coupling, resource flexibility, and simplified deployment processes. However, the inherent complexity of microservice systems often leads to failures, affecting user experience, and causing substantial economic losses. For example, a one-hour downtime on Amazon may cost up to 100 million dollars~\citep{pham2024baro, pham2026eventadl, pham2026graph}. Therefore, effectively and efficiently diagnosing the root causes of failures in microservice systems is crucial to quickly mitigate the failure and minimize its impact.

To effectively conduct root cause analysis (RCA), system operators typically gather three primary sources of telemetry data: metrics, logs, and traces~\citep{lee2023eadro, yu2023nezha, zhang2023diagfusion}. Metrics, typically presented as time series, provide insights into service status, including system-level and application-level metrics like CPU usage and average response time. Logs, comprising semi-structured text, record events that occur at runtime, including request handling events, state changes. Traces, structured in a tree format, record the sequence of user request invocations. Unlike metrics and logs, traces usually require heavy instrumentation with distributed tracing, often involving modifications to microservices' source code~\citep{janes2023open, shen2023deepflow, ashok2024traceweaver, yu2023nezha}. Figure~\ref{fig:main}A illustrates the multi-source telemetry data.

\begin{figure}
\centering
\includegraphics[width=0.7\textwidth]{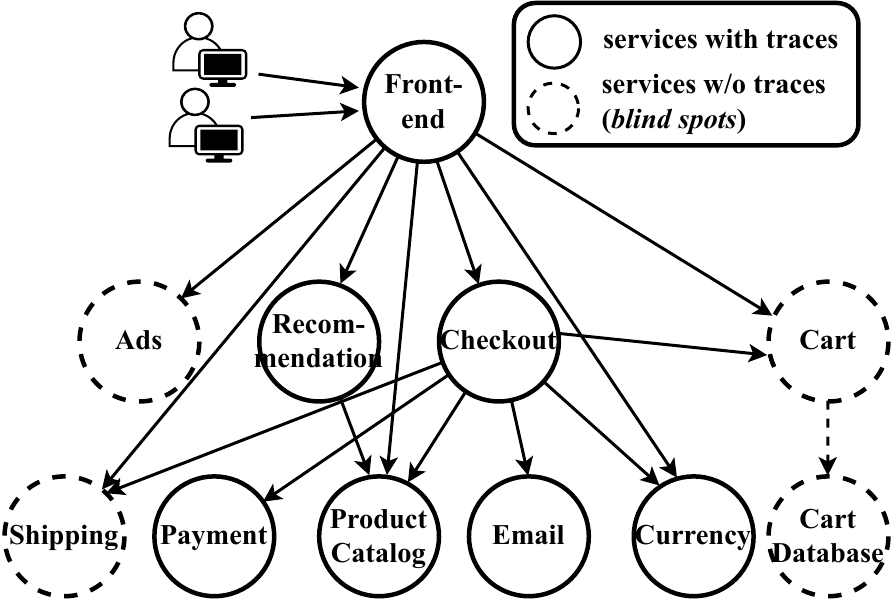}
\caption[Service Call Graph of the Online Boutique system containing \textit{blind spots}.]{Service Call Graph of the Online Boutique microservice system containing \textit{blind spots} (i.e., services without distributed tracing instrumentation).}\label{fig:missing-traces-1}
\end{figure} 

We have identified three key limitations of existing multi-source RCA methods~\citep{lee2023eadro, zhang2023diagfusion, rouf2024instantops, xie2024tvdiag, yu2023nezha}: they are either \textit{(1) approaches that assume the system has full trace coverage, allowing them to construct a full service call graph from traces, or (2) supervised approaches requiring a large volume of labelled data, or (3) intrusive approaches requiring heavy instrumentation of the microservices' source code}.
First, most multi-source RCA methods~\citep{lee2023eadro, zhang2023diagfusion, rouf2024instantops, xie2024tvdiag, yu2023nezha}, except those omitting trace information~\citep{zheng2024mulan, zhang2021cloudrca}, assume the system has full trace coverage, i.e., all services in the system are instrumented with distributed tracing, because they rely on traces to construct the service call graph required in their RCA methods. This requirement limits their applicability, especially in large and evolving microservice systems where new services or versions are introduced all the time, the engineers may not have enough effort to implement the distributed tracing for newly introduced services~\citep{shen2023deepflow, ashok2024traceweaver}, resulting in many \textit{blind spots} (i.e., services are not in the service call graph), that directly impact RCA performance, as shown in Figure~\ref{fig:missing-traces-1}.  Second, many multi-source RCA approaches ~\citep{lee2023eadro, zhang2023diagfusion, liu2022microcbr, xie2024tvdiag} are supervised approaches, requiring a large volume of training data to train their models. It is generally impractical to expect such a large volume of training data, as microservice systems are typically large and rapidly developed with many updates~\citep{shen2023deepflow, uber}. Third, a recent multi-source RCA method~\citep{yu2023nezha} requires a tight integration between traces and logs (i.e., each log line must contain the \texttt{trace\_id}), which demands extensive
effort to implement the necessary changes in the microservices' source code and may become infeasible when there are closed-source components. Consequently, these three limitations make existing multi-source RCA methods inflexible and challenging to implement in real-world scenarios.

\begin{tcolorbox}[left=2pt,right=2pt,top=0pt,bottom=0pt]
\textbf{Q: How do we address the mentioned limitations?}

\textit{We aim to propose a novel multi-source RCA method that satisfies three criteria: (1) avoiding using the call graph constructed from traces, (2) relying on unsupervised techniques, and (3) leveraging available telemetry data without requiring further updates to the source code.}
\end{tcolorbox}

In this chapter, we introduce TORAI, a novel multi-source RCA method for identifying fine-grained root causes of microservice failures. TORAI employs unsupervised techniques intuitively. First, it measures the anomaly severity from multi-source telemetry data. Second, it clusters services with similar severity symptoms. Next, it applies causal analysis to identify coarse-grained root causes (i.e., root cause services). Finally, it uses hypothesis testing to derive fine-grained root causes indicators (e.g., metrics or logs indicating the underlying root cause). TORAI offers several advantages over existing multi-source RCA methods. First, it does not rely on the service call graph (see Figure~\ref{fig:missing-traces-1}), eliminating the need for full trace coverage. Second, it uses unsupervised techniques such as clustering, and causal analysis, removing the need for labelled data and enabling direct application to microservice systems without requiring training. Third, it effectively leverages multi-source telemetry data, allowing for the root cause diagnosis in services not present in the call graph, without the need for additional intrusive instrumentation to achieve full trace coverage.

To evaluate TORAI, we conducted extensive experiments on 270 failure cases collected from three benchmark microservice systems and 10 real-world incidents, comparing TORAI against nine state-of-the-art RCA methods. The experimental results demonstrate that TORAI outperforms all baselines in localising both coarse-grained and fine-grained root causes in terms of effectiveness and efficiency. We further evaluate TORAI on real-world failures from a major internet provider. The results show that TORAI outperforms state-of-the-art baselines and achieves 100\% accuracy in ranking the root cause within the top three recommendations. 

\noindent This chapter makes the following contributions:
\begin{itemize}[leftmargin=*]
    \item We identify and characterise three key limitations of existing multi-source RCA methods, namely the assumption of full trace coverage (i.e., assuming the constructed service call graph has no \textit{blind spots}), the reliance on labelled data, and the requirement for intrusive instrumentation. These limitations motivate the design of \torai{}.
    \item We introduce \torai{}, a novel unsupervised multi-source RCA method that pinpoints fine-grained root causes by diagnosing available multi-source telemetry without requiring a constructed call graph or labelled data, making it adaptable to a wide range of microservice systems.
    \item We conduct an extensive empirical evaluation on 270 failure cases collected from three benchmark microservice systems and 10 real-world incidents, demonstrating that \torai{} outperforms nine state-of-the-art baselines in both effectiveness and efficiency.
\end{itemize}

%% file: papers/fse26-torai/2.background.tex
\subsection{Problem Formulation}

This chapter addresses multi-source root cause analysis, building on the terminology and general formulation established in Chapter~\ref{chap:background} (Sections~\ref{sec:terms-fault-failure}, \ref{sec:problem-formulation-ad}, and~\ref{sec:problem-formulation-rca}).\footnote{The original submission contained self-contained terminology and problem definitions. In this thesis, we consolidate these in Chapter~\ref{chap:background} for consistency.}

We focus on the full multimodal case where a microservice system $\mathcal{S}$ comprising $N$ services emits multi-source telemetry data $\mathcal{D}_t^i=\{\mathcal{M}_t^{(i,m)}, \mathcal{L}_t^{(i,l)}, \mathcal{TC}_t^{(i,s)}\}$ for each service $s^i$, where $\mathcal{M}^{(i,m)}$ denotes metrics, $\mathcal{L}_t^{(i,l)}$ represents logs, and $\mathcal{TC}_t^{(i,s)}$ denotes traces. Given the anomaly detection time $\hat{t}_A$, our goal is to identify root cause services and their corresponding fine-grained indicators (specific metrics, logs, or traces).

\subsection{Related Work and Motivation} \label{sec:related-world-and-motivation}

\subsubsection{The Blind Spots of Existing Multi-Source RCA}
Existing multi-source RCA methods~\citep{yu2023nezha, lee2023eadro, zhang2023diagfusion, hou2021pdiagnose, rouf2024instantops, xie2024tvdiag} typically construct a service call graph from traces, assuming the system has full trace coverage, i.e., 100\% of services are instrumented with distributed tracing \citep{shen2023deepflow, ashok2024traceweaver}. They then integrate metrics and logs into the service call graph to perform RCA. This approach is only effective when the system has no \textit{blind spots}, i.e., services without traces or closed-source components \citep{shen2023deepflow, lee2023eadro, yu2023nezha, ashok2024traceweaver}. In this chapter, we refer to services or components that should be detected by traces but are not as blind spots. In the presence of blind spots, existing multi-source RCA methods \citep{yu2023nezha, lee2023eadro, zhang2023diagfusion, gu2023trinityrcl, xie2024tvdiag} may not perform RCA effectively, as the constructed service call graph is incomplete. Consequently, key metrics and logs may be overlooked, and their behaviour may not be analysed when diagnosing the root cause of failure in microservice systems. Recent studies~\citep{shen2023deepflow, ashok2024traceweaver} and industry reports~\citep{odigos2025pitfalls} show that requiring the microservice systems to be fully instrumented is ineffective and inefficient. In reality, microservice systems are often developed in many different languages, and the call graph can become exceedingly intricate, with some systems comprising up to 1,500 services \citep{luo2022depth}. Shen et al.~\citep{shen2023deepflow} show that engineers spend hours to instrument mere tens of lines of code for a single component. Hence, they may not have sufficient time to instrument every component before deployment, leading to numerous blind spots in the constructed call graph. TraceWeaver~\citep{ashok2024traceweaver} explicitly acknowledges that eBPF is ``insufficient to solve the tracing challenge'' as ``there is no guarantee that the application will propagate these headers to related outgoing backend requests.'' Industry reports~\citep{odigos2025pitfalls} confirm that blind spots remain prevalent in production systems despite advances in eBPF-based tracing. In this chapter, we aim to bypass the assumption of full trace coverage and develop a multi-source RCA method to localise root causes effectively, even when some or all traces are missing. It is important to note that metrics and logs are relatively easy to collect, as they do not require source code modification, unlike traces. Metrics such as response time and error rates can be collected without traces by monitoring agents or service mesh without requiring any source code modification~\citep{pham2024baro, shen2023deepflow}. This is standard practice in industry~\citep{googlesre} and is how existing metric-based RCA works collect data~\citep{pham2024baro, Azam2022rcd, Li2022Circa}.

\subsubsection{Supervised RCA Approaches In A Dynamic World} \label{sec:motivation-unsupervise}
Supervised multi-source RCA methods~\citep{lee2023eadro, zhang2023diagfusion, zhang2021cloudrca, xie2024tvdiag} rely on large volumes of manually labelled training data to train their models, which often incorporate graph neural networks or convolutional layers. However, requiring such large amounts of labelled data is often impractical or challenging in real-world scenarios. First, microservice systems are highly dynamic, with old services being updated and new services being released frequently. Second, historical faults are often resolved, while new faults are introduced over time~\citep{shen2023deepflow}. Additionally, future edge-case faults, which could cause significant losses, are unlikely to be present in historical failure datasets~\citep{zhang2023benefit}. Third, microservice systems are typically large in scale~\citep{uber}, making it costly and impractical to manually label data to cover all services and fault types. Therefore, our aim is to develop an RCA method that leverages unsupervised techniques to reduce the dependency on labelled data.

\subsubsection{Tight Integration between Multi-source Data} \label{sec:motivation-2-independent}
Recent multi-source RCA studies~\citep{yu2023nezha, zhang2023diagfusion, lee2023eadro, zheng2024mulan, zhang2021cloudrca} require multi-source telemetry data to be tightly integrated. For example, Nezha~\citep{yu2023nezha} requires every log line in the microservices to be modified to contain a \texttt{trace\_id}. To achieve this, engineers must manually modify every log or print statement in the microservices' source code, which is time-consuming and costly. This task becomes impossible when dealing with black-box components, third-party services, or new frameworks that do not yet support distributed tracing~\citep{giamattei2023monitoring, shen2023deepflow}. In addition, Eadro \citep{lee2023eadro} and DiagFusion \citep{zhang2023diagfusion} extract features from metrics and logs, integrate these features into a graph constructed using traces. These methods struggle in the presence of blind spots, as many metrics and logs may not correspond to any location on the graph, leading to the failure to use these valuable data for failure diagnosis. In summary, most existing multi-source RCA approaches may be impractical, as their assumptions are often not fully met in real-world scenarios. Some other multi-source RCA methods do not require this tight integration, but interestingly, they completely omit trace information \citep{zhang2021cloudrca, zheng2024mulan, wang2020root} or log information \citep{zhu2024hemirca}. In other words, they fail to fully take advantage of all available telemetry data. 

\subsubsection{Summary} These motivations drive the design of TORAI, a novel multi-source RCA method that overcomes the existing limitations. Firstly, our method leverages multi-source telemetry data without requiring full trace coverage \citep{yu2023nezha, lee2023eadro, zhang2023diagfusion}, allowing it to diagnose the fine-grained root causes of failures with high performance even in the absence of some or all traces. We do not attempt to reconstruct the call graph, instead, we design TORAI to perform RCA effectively without requiring a complete call graph. This design is inspired by recent theoretical work~\citep{orchard2025root}, which proves that identifying root causes based on anomaly scores is causally justified. Recent studies~\citep{pham2024root} also show that call graph reconstruction can be ineffective for RCA due to limitations of causal discovery algorithms. Secondly, our method can be applied without modifying the source code of microservices for tight integration between traces and logs~\citep{yu2023nezha}. Thirdly, our method does not rely on labelled training data \citep{lee2023eadro, zhang2023diagfusion, zhang2021cloudrca, xie2024tvdiag}, making it adaptable to a wide range of systems. The details of our proposed RCA method, TORAI, are presented in the next section.

%% file: papers/fse26-torai/4.method.tex
\begin{figure}
\centering
\includegraphics[width=\textwidth]{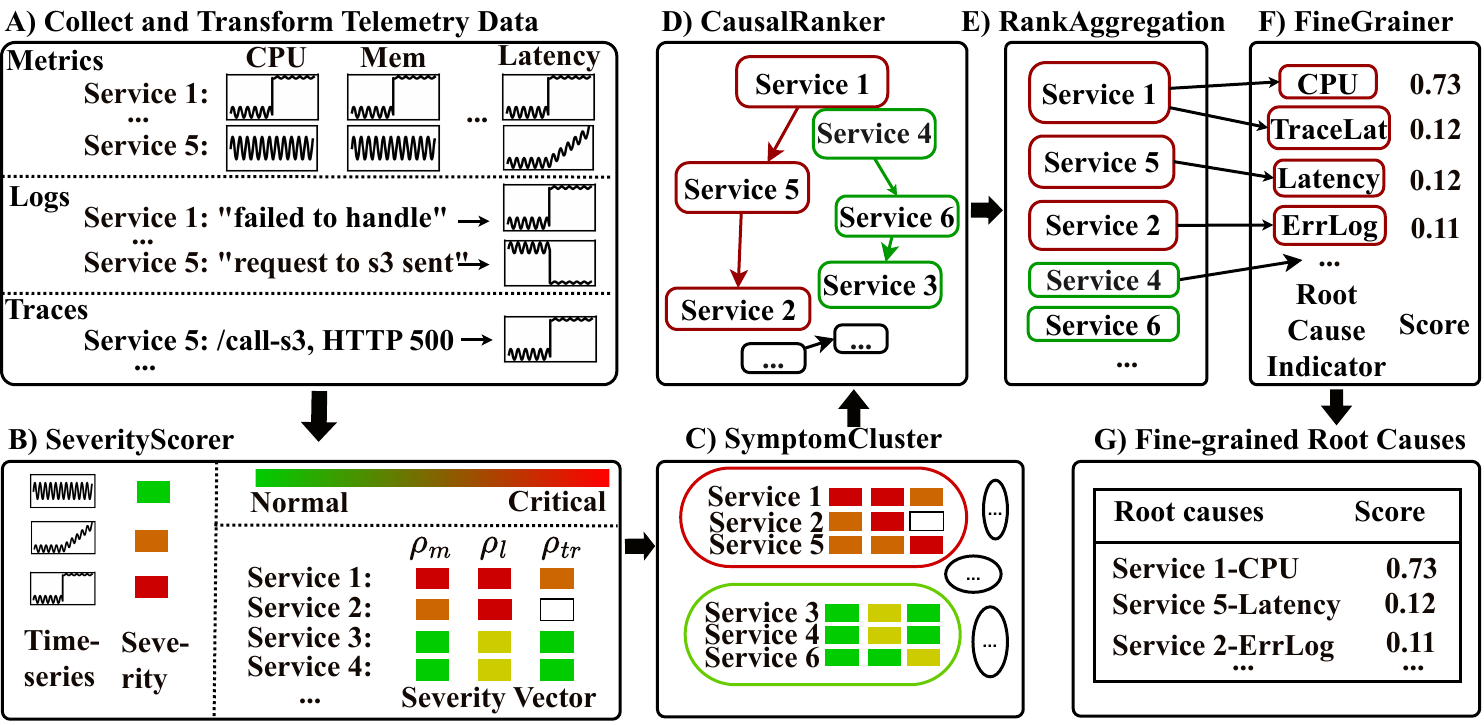}
\caption[Overview of TORAI.]{Overview of TORAI. (A) TORAI transforms telemetry data into time series. (B) It computes anomaly severity scores, producing vectors [$\rho_m^i, \rho_l^i, \rho_{tc}^i$] for each service $s^i$ (with missing data sources denoted by $\protect\rectangle$, as in Service 2, which is a \textbf{blind spot}). (C) TORAI clusters services based on their severity symptoms. (D) Within each cluster, it applies causal inference-based RCA to rank potential root causes. (E) TORAI aggregates rankings from steps C and D to generate a coarse-grained root cause list. (F) It then conducts fine-grained RCA using hypothesis testing to produce (G) a fine-grained ranked list of root causes.} \label{fig:main}
\end{figure}

When an anomaly is detected, TORAI is triggered to perform RCA as follows. First, TORAI collects the multi-source telemetry data and transforms them into time series (Sec. \ref{sec:method-collect}). Second, it measures the anomaly severity of each data source for all services using \textit{SeverityScorer} (Sec.~\ref{sec:severity-scorer}). Third, it uses \textit{SymptomCluster} to group services exhibiting similar severity symptoms (Sec.~\ref{sec:symptom-cluster}). Fourth, it conducts causal analysis to rank the root causes within each severity group using \textit{CausalRanker} (Sec. \ref{sec:causal-ranker}). Then, it performs \textit{RankAggregation} to aggregate the results from the SymptomCluster and CausalRanker steps, yielding a root cause service ranked list (Sec.~\ref{sec:rank-aggregation}). Finally, TORAI uses \textit{FineGrainer} to perform hypothesis testing and derive fine-grained root cause indicators for the corresponding root cause services (Sec. \ref{sec:fine-grainer}). The overview of TORAI is shown in Figure~\ref{fig:main}.

\subsection{Transform Multi-source Telemetry Data} \label{sec:method-collect}

TORAI transforms multi-source telemetry data into time series for RCA. It collects data during normal periods to learn expected behaviours and during abnormal periods to analyse and identify root causes. The processing steps for each data source are as follows:

\subsubsection{Metrics}
TORAI collects four metrics types: \textit{Traffic} (e.g., requests per minute), \textit{Saturation} (e.g., CPU/memory utilisation), \textit{Latency} (e.g., average response time), and \textit{Errors} (e.g., the rate of failed requests)~\citep{pham2024baro}, known as the four golden signals in site reliability engineering~\citep{googlesre}, during both normal and abnormal periods. For each service $s^i$, TORAI compiles a set of time series $\mathcal{X}_\mathcal{M}^i$ to represent the metrics data.

\subsubsection{Logs}
We focus on log occurrences rather than log semantics, as empirical studies have shown that the quality of log semantics cannot be guaranteed~\citep{he2021survey, lee2023eadro} and semantic extraction requires significant computational resources. First, TORAI parses logs into log templates using Drain \citep{he2017drain} by removing variables from the log messages. Next, TORAI bins the occurrence of log templates based on their timestamps to obtain time series. This process generates a set of time series $\mathcal{X}_\mathcal{L}^i$, representing the log occurrences of each service $s^i$. 

\subsubsection{Traces.}
We extract available information from traces, but do not construct the call graph as it may be incomplete (see Sec. ~\ref{sec:related-world-and-motivation}).
TORAI transforms latency and status code in traces into time series. Each time series presents the frequency of latency and status code for each trace operation.

We refer to time series data derived from multi-source telemetry data as \textbf{multi-source time series data}. Each service $s^i$ has a set of multi-source time series $\mathcal{X}^i = \{ \mathcal{X}^i_\mathcal{M}, \mathcal{X}^i_\mathcal{L}, \mathcal{X}^i_\mathcal{TC} \}_{t_0 \le t < \hat{t}_A +T}$ where $\mathcal{X}^i_\mathcal{M}$, $\mathcal{X}^i_\mathcal{L}$, and $\mathcal{X}^i_\mathcal{TC}$ denote the set of time series for metrics, logs, and traces. The data is split at the anomaly detection time $\hat{t}_A$ into normal ($\mathcal{X}^i_{t_0\le t<\hat{t}_A}$) and abnormal ($\mathcal{X}^i_{\hat{t}_A\le t< \hat{t}_A+T}$) sets.

\subsection{SeverityScorer} \label{sec:severity-scorer}

SeverityScorer measures anomaly severity scores for each time series and generates a severity vector for each service. These severity vectors effectively assist SymptomCluster in grouping services based on their severity symptoms, thereby quickly separating abnormal services from normal ones (see Sec.~\ref{sec:symptom-cluster}). This also reduces overhead for CausalRanker, which further analyses the multi-source time series to derive the root causes (see Sec.~\ref{sec:causal-ranker}). It is important to note that SeverityScorer does not aim to identify the root causes directly but instead focuses on quickly assessing the severity of each service based on its multi-source telemetry data after a failure occurs, thereby helping subsequent components of TORAI concentrate their focus on analysing the abnormal services.

In particular, for each time series $x^{(i, j)}$, SeverityScorer learns the mean $\mu^{(i, j)}$ and standard deviation $\smash{\sigma^{(i, j)}}$ during normal period. Then during the abnormal period, for each data point $\smash{x_t^{(i, j)}}$ of the time series $x^{(i, j)}$, SeverityScorer measures how far it diverges from the expected value. This deviation is denoted as $a^{(i, j)}_t$ and is computed as $\smash{a^{(i, j)}_t = \big | x^{(i, j)}_t - \mu^{(i, j)} \big | / \sigma^{(i, j)}}$.  SeverityScorer then aggregates $a^{(i, j)}_{t}$ for all the available data during the abnormal period, yielding the anomaly severity score $\smash{\rho^{(i, j)} = \max_{\hat{t}_A \leq t < \hat{t}_A + T} a^{(i, j)}_t}$.

Using the anomaly severity scores ${\rho^{(i, j)}}$ across all time series and data sources, SeverityScorer assigns each service $s^i$ a severity vector $[\rho_m^i$,~$\rho_l^i$,~$\rho_{tc}^i]$, which represents the severity status of each service as observed through its multi-source telemetry data. If a data source is missing, SeverityScorer imputes the corresponding $\rho$ value as 0, indicating no anomalies were detected from that source. This flexible design enables TORAI to use available data sources, such as metrics and logs, without requiring the presence of traces. Specifically, $\rho_m^i$ and $\rho_{tc}^i$ are obtained by integrating the severity scores of the time series derived from metrics and traces associated with service $s^i$, i.e.,  $\rho_m^i = \sum_{x^{(i,j)}\in \mathcal{X}_\mathcal{M}^i} \rho_m^{(i,j)}$ and $\rho_{tc}^i = \sum_{x^{(i,j)}\in \mathcal{X}_\mathcal{TC}^i} \rho_{tc}^{(i,j)}$. For logs, $\rho_l^i$ is determined as the score of the most anomalous log templates for $s^i$, i.e., $\smash{\rho_{l}^i = \max_{x^{(i,j)}\in \mathcal{X}_\mathcal{L}^i} \rho_{l}^{(i,j)}}$, since the number of log templates between services varies significantly.

\subsection{SymptomCluster} \label{sec:symptom-cluster}

SymptomCluster performs clustering to group services with similar severity symptoms. The goal is to effectively separate abnormal services from normal ones, reducing the overhead for CausalRanker when conducting causal analysis, inspired by recent research~\citep{pham2024root} that highlights how including all services in the causal analysis can be inefficient and ineffective.

Specifically, SymptomCluster proposes using Gaussian Mixture Model (GMM) \citep{gmm} for this clustering task. GMM is a probabilistic model that assumes the data come from a mixture of several Gaussian distributions, each representing a cluster. GMM is known for its efficiency and has been successfully applied in various domains~\citep{patel2020clustering}. We first use the Bayesian Information Criterion (BIC)~\citep{Schwarz1978bic} to identify the optimal number of clusters (i.e., the number of Gaussian distributions). Specifically, we iterate through all possible values for the number of clusters. For each iteration, SymptomCluster runs GMM on the set of severity vectors $\{[\rho_m^i, \rho_l^i, \rho_{tc}^i]\}_{i=1}^N$ ($N$ is the number of services) and calculates the corresponding BIC score. The optimal number of clusters is chosen as the number that yields the lowest BIC score \citep{gmm, pham2024root}.

Once the optimal number of clusters is determined, SymptomCluster performs clustering again using this value. For each cluster, SymptomCluster measures the cluster severity score as the mean severity score of all services within that cluster (i.e., the mean of the Gaussian distribution). Finally, SymptomCluster generates a ranked list of clusters (a cluster-level ranked list), where the highest-ranked clusters are the most likely to contain the root cause. This approach aligns with the intuition that operators should prioritize higher severity services for failure diagnosis, while setting lower priority to services with less severity and normal services. In addition, GMM allows soft clustering, meaning a service may be probabilistically assigned to multiple clusters. Consequently, a ``weak'' anomalous service can be included in multiple clusters for causal analysis, enabling the discovery of cross-cluster causal relationships.

\subsection{CausalRanker: Causal-based Root Cause Analysis} \label{sec:causal-ranker}

CausalRanker is designed to identify the root causes within each severity group obtained from the previous step. It operates on the belief that the root cause services may not exhibit the strongest anomaly severity but will have cause-effect relationships with affected services. These causal relationships are reflected in their time series data. CausalRanker performs causal inference-based RCA, analysing the multi-source time series of all services in each severity group by examining their anomalous patterns (e.g., periodicity, causality) to rank the root causes in each group.

\begin{figure}
\centering
\includegraphics[width=0.5\linewidth]{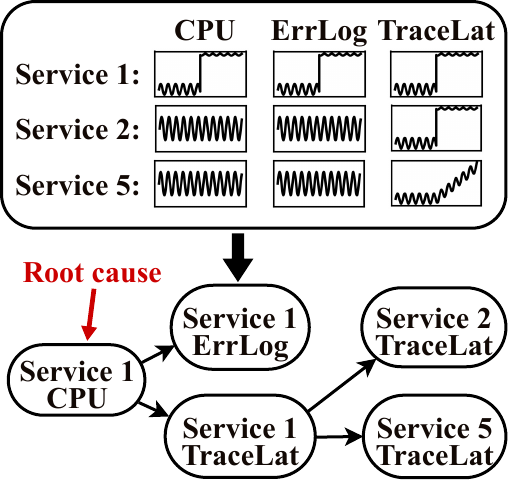}
\caption[CausalRanker analyses the multi-source time series data of all services within each severity group to construct a causal graph and identify the root causes.]{CausalRanker analyses the multi-source time series data of all services within each severity group to construct a causal graph and identify the root causes. (ErrLog: Error Logs, TraceLat: Latency extracted from Traces).}
\label{fig:causal-ranker}
\end{figure}

Our CausalRanker ranks services within each cluster via divide-and-conquer causal inference using multi-source time series data. A causal graph is a directed acyclic graph (DAG) where nodes represent time series (e.g., CpuUsage, ErrLog) and directed edges represent causal relationships (e.g., CpuUsage $\rightarrow$ ErrLog indicates that CPU anomaly causes error log anomaly). This is a standard representation in causal discovery literature~\citep{Spirtes1995fci, Jaber2020PsiFCI}. A root cause is identified as a node that has no parents in the causal graph (i.e., an interventional target).

Specifically, CausalRanker randomly partitions the given set of time series into smaller groups called chunks. The time series are randomly partitioned into smaller chunks for efficient causal analysis, as analysing all time series simultaneously is computationally expensive and has been shown to degrade performance~\citep{pham2024root}. It then applies the $\Psi$-PC \citep{Jaber2020PsiFCI}, an advanced causal discovery method in the presence of \textit{interventions} (i.e., failures), to construct causal graphs and identify root causes, referred to as interventional targets, within each chunk. Recursively, CausalRanker groups these root causes and repeats the process until only one chunk remains. Finally, the root causes are ranked based on their (conditional) independence test scores. Lower independence test scores indicate stronger causal influence, suggesting higher likelihood of being the root cause. This provides a ranked list of root cause services for each cluster.

Consider the example in Figure~\ref{fig:main}C and Figure~\ref{fig:causal-ranker}, ``Service~1", ``Service~2", and ``Service~5" are grouped in the same severity group. In this example, the root cause is ``Service~1~-~CPU", which causes an increase in error logs (Service 1-ErrLog) and latency (Service~1~-~TraceLat). Therefore, we design CausalRanker to construct the causal graph and identify ``Service~1" as the root cause service within this severity group. 

Our CausalRanker offers two key advancements over existing methods~\citep{Azam2022rcd, xin2023causalrca}. First, prior studies~\citep{Azam2022rcd, xin2023causalrca} use all available time series for causal analysis, which has recently been shown to degrade the performance of causal inference~\citep{pham2024root}. In contrast, CausalRanker leverages SymptomCluster, which groups services with similar severity levels, ensuring that only relevant time series are analysed for causal relationships to derive the root cause. Second, previous studies rely solely on causal discovery algorithms to identify fine-grained root causes (e.g., RCD~\citep{Azam2022rcd} employs the \(\Psi\)-PC algorithm, while CausalRCA~\citep{xin2023causalrca} uses DAG-GNN~\citep{Yu2019DagGNN}), an approach shown to be ineffective~\citep{pham2024root}. In TORAI, we introduce FineGrainer, which applies a robust hypothesis testing method to produce more accurate fine-grained root cause rankings within our framework (see Sec.~\ref{sec:fine-grainer}).

\subsection{RankAggregation} \label{sec:rank-aggregation}

In this phase, we combine the ranked lists from SymptomCluster and CausalRanker to produce an aggregated ranked list of coarse-grained root cause services.  We first prioritize the cluster-level ranking provided by SymptomCluster (Sec. \ref{sec:symptom-cluster}), meaning that services in clusters with higher severity are ranked higher. Second, within each cluster, we follow the ranking provided by CausalRanker (Sec. \ref{sec:causal-ranker}), meaning that if Service A is ranked higher than Service B within a cluster, then Service A will also be ranked higher than Service B in the final aggregated list. For instance, given a cluster-level ranking [Group A, Group B], where the internal ranking for Group A is [a,c] and for Group B is [e,p,t], our RankAggregation produces the corresponding aggregated ranking [a,c,e,p,t]. The highest-ranked items in the aggregated list have the highest probability of being the root cause service of the failure. The pseudo code for RankAggregation is presented in Algorithm \ref{alg:rank-aggregation}.

\begin{algorithm}[h]
\caption{Pseudo-code of RankAggregation} \label{alg:rank-aggregation}
\begin{algorithmic}[1]
\REQUIRE Ranked list $R_{cluster}$ from SymptomCluster (Sec. \ref{sec:symptom-cluster}), function CausalRanker (Sec. \ref{sec:causal-ranker})
\ENSURE Ranked list of coarse-grained root cause services $R_s$

\STATE Initialize $R_s \gets \emptyset$

\FOR{each $cluster \in R_{cluster}$ (from highest to lowest rank)}
    \STATE Obtain the corresponding ranked list $R_i \gets$ \textit{CausalRanker}($cluster$)
    \FOR{each service $s \in R_i$}
        \STATE $R_s \gets R_s \cup \{s\}$
    \ENDFOR
\ENDFOR

\STATE \textbf{return} $R_s$

\end{algorithmic}
\end{algorithm}

From our empirical observations, the top severity cluster may contain a single or multiple services. When the top severity cluster contains only one service, it is possible that the root cause service belongs to the second-highest severity cluster. Therefore, we apply CausalRanker to all clusters, not just the top severity cluster, and aggregate the results through this Rank Aggregation step.

In cases where the root cause does not exhibit strong anomalies (e.g., a code defect that only manifests in downstream services), TORAI ranks the affected services at the top. This allows quicker identification of actual root causes, instead of troubleshooting all possible services. For example, if a code defect causes correlated failures in services A and B, TORAI ranks A and B as top candidates, enabling operators to inspect these services and trace back to the true root cause promptly.

\subsection{FineGrainer} \label{sec:fine-grainer}

After identifying the coarse-grained root cause services, we propose FineGrainer to determine fine-grained root cause indicators (e.g., specific metrics or logs indicating the root cause of failure) for each service, which better assists operators in diagnosing underlying issues~\citep{Li2022Circa, pham2024baro, liu2023pyrca}.

FineGrainer performs hypothesis testing on time series indicators to detect significant distribution changes after the anomaly detection time~\citep{Li2022Circa, pham2024baro}. Our key insight is that root cause indicators exhibit significant distributional changes after a failure occurs.
We perform hypothesis testing where the null hypothesis $H_0$ states that the time series distribution remains unchanged after $\hat{t}_A$.

Specifically, for the time series $x$ of each indicator (i.e., the time series data corresponding to each metric, log template, or trace), FineGrainer learns the median ($med$) and interquartile range ($IQR$) of the time series $x$ prior to the anomaly, spanning from $t_0$ to $\hat{t}_A$. It then measures how significantly each time series $x$ deviates from its expected central tendency during the failure period. This deviation is quantified as $a_t = \big | x_t - med \big | / IQR$. All values of $a_t$ across the time step $t$ during the failure period are consolidated to yield $\gamma = \max_{\hat{t}_A \leq t < \hat{t}_A + T} a_t$. Higher $\gamma$ values provide stronger evidence against $H_0$, indicating greater likelihood that the indicator is the root cause.

\begin{figure}
\centering
\includegraphics[width=0.6\textwidth]{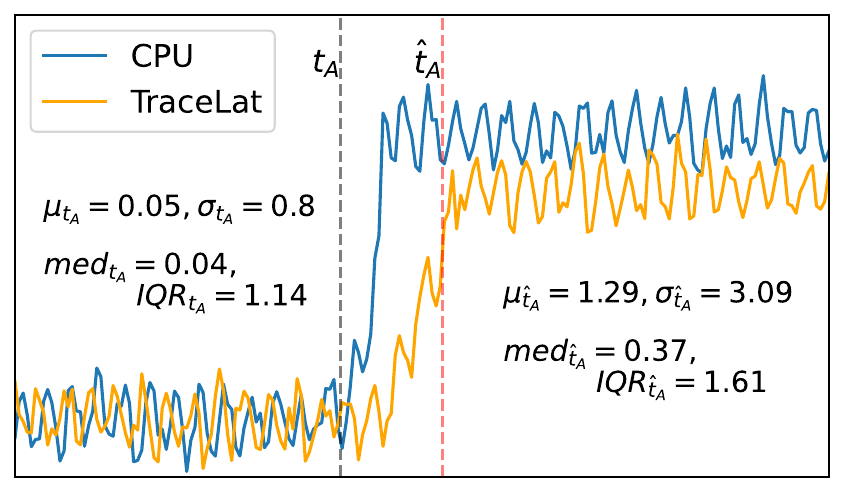}
\caption[The Robustness of FineGrainer to Imprecise Anomaly Detection.]{\textbf{The Robustness of FineGrainer to Imprecise Anomaly Detection.} At time $t_A$, a failure occurs, causing a spike in CPU that eventually leads to increased latency (TraceLat). At $\hat{t}_A$, TraceLat surpasses the anomaly detection threshold, triggering an anomaly detection. This delayed detection introduces abnormal data (outliers) into the normal period of the CPU time series. The median and interquartile range (IQR) demonstrate greater robustness to these outliers compared to the $\mu$ and $\sigma$.}\label{fig:fine-grainer}
\end{figure}

We prefer median and IQR over mean and standard deviation for analysing fine-grained root cause indicators. Prior work using mean and standard deviation~\citep{Li2022Circa} can degrade when anomaly detection times $\hat{t}_A$ are imprecise, as outliers from the actual failure period contaminate the normal period statistics. Median and IQR are more robust to such outliers (see Figure~\ref{fig:fine-grainer}). 

%% file: papers/fse26-torai/5.results.tex
This section addresses the following research questions:
\begin{itemize}[leftmargin=*]
    \item RQ1: How effective is TORAI in coarse-grained RCA? 
    \item RQ2: How effective is TORAI in fine-grained RCA? 
    \item RQ3: How efficient is TORAI in performing RCA? 
    \item RQ4: How do TORAI's core components contribute to its overall performance?  
    \item RQ5: How does TORAI perform in real-world scenarios?
\end{itemize}

\subsection{Datasets} \label{sec:dataset}

We deploy three widely used microservice benchmark systems, namely Online Boutique \citep{ob}, Sock Shop \citep{sockshop}, and Train Ticket \citep{tt}, on a Kubernetes cluster featuring five worker nodes, all configured with their default settings. Online Boutique is an e-commerce platform comprising 11 services that facilitate tasks such as browsing items, adding items to a user's cart, and making orders. Sock Shop, another online shopping system, consists of 11 services communicating via HTTP. Train Ticket is a ticket booking system that has 64 services, making it one of the largest benchmark microservice systems. Compared to Online Boutique and Sock Shop, Train Ticket has a complex design, various types of invocations and many log templates. The three benchmark microservice systems are widely used to evaluate RCA performance in existing literature~\citep{Azam2022rcd, wu2021microdiag, xin2023causalrca, he2022graph, dan2021tracerca, yu2021microrank, zhou2018trainticket, Wang2021evalcausal}.

\begin{figure}[h]
\centering
\includegraphics[width=0.8\linewidth]{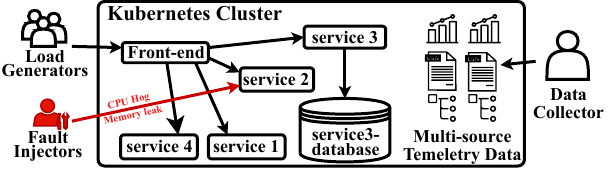}
\caption[Our fault injection setup for data collection from the benchmark systems.]{Illustration of our setup for the microservice systems and the multi-source telemetry data collection.} \label{fig:experiment-setup}
\end{figure}

To simulate user interactions, we customize the provided load generators of these systems to emulate a load of 40-50 requests per second across all services. To gather metrics, we employ Prometheus \citep{prometheus}, along with cAdvisor \citep{cadvisor} and the Istio service mesh \citep{istio}, to monitor and collect application-level and resource-level metrics. For log collection, we deploy Datadog Vector \citep{vector} and Loki \citep{loki} to gather and aggregate logs from all service instances, storing them in Elasticsearch \citep{elasticsearch}. Traces are gathered using Jaeger \citep{jaeger}, with data sent to Elasticsearch for storage. Our telemetry data collection setup is depicted in Figure~\ref{fig:experiment-setup},
similar to existing works~\citep{Azam2022rcd, xin2023causalrca, wu2021microdiag, wu2022automatic, yu2023nezha, lee2023eadro, Jinjin2018Microscope, he2022graph, dan2021tracerca, yu2021microrank, zhou2018trainticket, Wang2021evalcausal}.

We inject six common faults: CPU hog (CPU), memory leak (MEM), disk IO stress (DISK), socket stress (SOCKET), network delay (DELAY), and packet loss (LOSS) into key services of each benchmark system. In particular, we inject faults into five Sock Shop services (user, catalogue, orders, payment, and carts), five Online Boutique services (email, currency, recommendation, product, checkout), and five Train Ticket services (order, route, auth, train, travel). These services play an important role in their respective systems, as issues with their performance quickly impact the overall health of the system \citep{Azam2022rcd, xin2023causalrca, lee2023eadro}. Firstly,  we let the systems run normally for ten minutes to gather normal metrics, logs, and traces. Then, we follow the existing practice \citep{Azam2022rcd, lee2023eadro, wu2021microdiag, xin2023causalrca, yu2021microrank} to inject faults into the running services. For CPU, MEM, DISK, and SOCKET, we use stress-ng \citep{stressng} to stress the container resource. For DELAY and LOSS, we use tc \citep{tc} to manipulate the container traffic. For each combination of fault type and targeted service, we repeat the fault injections and data collections three times, resulting in a total of 90 collected failure cases for each benchmark microservice system. Table \ref{tab:torai-real-data} presents the statistics summarising the collected data.

\begin{table}
\centering
\caption[Properties of collected datasets.]{Properties of collected datasets (\#service, \#metric, \#log, \#trace: number of services, metrics, log templates, and trace operations, per case. \#fault: no. fault types).} \label{tab:torai-real-data}
\begin{tabular}{l c c c c c c c c}
\hline
\textbf{Name} & \textbf{\#service} & \textbf{\#metric} & \textbf{\#log} & \textbf{\#trace} & \textbf{\#fault} & \textbf{\#cases} \\ \hline \hline
Online Boutique & 11 & 77 & 33$\pm$9 & 17  & 6 & 90  \\ \hline
Sock Shop  & 11 & 74 & 67$\pm$53 & - & 6 & 90   \\ \hline
Train Ticket & 64 & 376 & 163$\pm$44 & 148.3$\pm$26 & 6 & 90  \\ \hline
\end{tabular}%
\end{table}

\textbf{Blind spots in the benchmark systems.} Despite their widespread use for benchmarking, these systems have several blind spots (i.e., services without traces) by default. In the Online Boutique system, 7 out of 11 services are instrumented with tracing, leaving \textbf{4 blind spots}. The Sock Shop system is not instrumented at all, meaning \textbf{no traces are available} to construct a service call graph. In the Train Ticket system, 27 out of 64 services are instrumented with tracing, resulting in \textbf{37 blind spots}. Existing RCA methods that rely on call graphs~\citep{lee2023eadro, yu2023nezha, zhang2023diagfusion} will be unable to diagnose root causes in these blind spots, as they do not appear in the call graph.

\subsection{Evaluation Metrics}

Following existing works \citep{Azam2022rcd, pham2024baro, pham2024root},
we use two standard evaluation metrics: $AC@k$ and $Avg@k$ to measure the RCA performance. Given a set of failure cases A, $AC@k$ is calculated as follows,
\begin{equation}
    AC@k = \frac{1}{|A|} \sum\nolimits_{a\in A}\frac{\sum_{i<k}R^a[i]\in V^a_{rc}}{min(k, |V^a_{rc}|)},
\end{equation}
where $R^a[i]$ is the $i$th ranking result for the failure case $a$ by an RCA method, and $V^a_{rc}$ is the true root cause set of case $a$. $AC@k$ represents the probability the top $k$ results of the given method include the true root causes. Its values range from $0$ to $1$, with higher values indicating better performance. $Avg@k$, which shows the overall RCA performance, is measured as $Avg@k = \frac{1}{k}\sum_{j=1}^k AC@j$. For brevity, we also refer to $AC@1$, $AC@3$, and $Avg@5$ as T1, T3, and A5, respectively.

We run all experiments on Linux servers each with 8 CPUs and 16GB RAM. In addition, we repeat each experiment five times and report the average results to minimize the impact of randomness. We use one-way ANOVA to assess overall differences among methods and pairwise t-tests for comparing individual method pairs. We report results as statistically significant when $p < 0.05$.

\subsection{Baselines} \label{sec:torai-baselines}

We select nine RCA baselines from previous studies for performance comparison with our proposed multi-source RCA method, namely: PDiagnose \citep{hou2021pdiagnose}, HeMiRCA \citep{zhu2024hemirca}, CausalRCA \citep{xin2023causalrca}, MicroCause \citep{Meng2020Microcause}, RCD \citep{Azam2022rcd}, CIRCA \citep{Li2022Circa}, BARO \citep{pham2024baro}, MicroRank \citep{yu2021microrank}, and TraceRCA \citep{dan2021tracerca}. Detailed information of these methods is as follows:

\begin{itemize}[leftmargin=*]
\item \textit{PDiagnose \citep{hou2021pdiagnose}:} PDiagnose is a multi-source RCA method that transforms metrics, logs, and traces into time series and determines root causes through voting. It relies on traces to determine the root cause services, and metrics to derive fine-grained root causes.

\item \textit{HeMiRCA \citep{zhu2024hemirca}}: HeMiRCA is a multi-source RCA method, which relies on the monotonic correlation between metrics and trace-based anomaly scores. HeMiRCA first measures trace-based anomaly scores and then exploits the correlations between metrics and the trace anomaly scores to rank the suspicious metrics and microservices. However, HeMiRCA does not use logs.

\item \textit{CausalRCA \citep{xin2023causalrca}:} CausalRCA constructs the causal graph from time series derived from metrics data using DAG-GNN \citep{Yu2019DagGNN}, a gradient-based causal structure learning method. Then it employs PageRank to rank the root causes from the estimated graph.

\item \textit{MicroCause \citep{Meng2020Microcause}:} MicroCause uses PCMCI \citep{Runge2019PCMCI} to construct the causal graph. 
Then, it applies temporal cause-oriented random walk to rank the root causes from the estimated causal graph.

\item \textit{RCD \citep{Azam2022rcd}:} RCD adopts a divide-and-conquer strategy to partition time series into chunks. Then, it uses the $\Psi$-PC  \citep{Jaber2020PsiFCI} to build causal graphs and identify root causes within each chunk. Recursively, it combines these root causes and iterates until only one chunk remains.

\item \textit{CIRCA \citep{Li2022Circa}:} CIRCA relies on a provided call graph to construct a causal graph. It then uses regression-based hypothesis testing analysis to identify root causes.

\item \textit{BARO \citep{pham2024baro}:} 
BARO uses a nonparametric hypothesis testing technique based on median and IQR to measure the change of metrics time series after the failure time and rank the root causes.

\item \textit{MicroRank \citep{yu2021microrank}:} MicroRank is a trace-based RCA approach that combines personalized PageRank and Spectrum method to identify suspicious root causes from the collected trace data.

\item \textit{TraceRCA \citep{dan2021tracerca}:} TraceRCA uses spectrum analysis to identify the root cause services, based on the insight that a service with more abnormal and fewer normal traces passing through it is more likely to be the root cause.
\end{itemize}

For the baselines HemiRCA, CausalRCA, MicroCause, RCD, CIRCA, BARO, MicroRank, and TraceRCA, we use their publicly available implementation and default hyperparameter settings suggested in their respective papers. We verified their correctness of the obtained source code  by reproducing the presented results in the original and related papers. For PDiagnose, we follow previous works \citep{yu2023nezha, zhang2023diagfusion, hou2021pdiagnose} to implement it since its source code is unavailable. Furthermore, recent multi-source RCA methods \citep{lee2023eadro, yu2023nezha, zhang2023diagfusion, zhang2021cloudrca, li2022actionable} exhibit limitations that prevent us from adopting them as baselines. Specifically,  some methods~\citep{lee2023eadro, zhang2023diagfusion, li2022actionable, zhang2021cloudrca} require labelled training data, which is unavailable to us, while~\citep{yu2023nezha} requires manual effort to integrate \texttt{trace\_id} into every log line. In contrast, our method does not require labelled data or heavy instrumentation into the systems.

It is important to note that four of our selected baselines (CausalRCA, MicroCause, RCD, BARO) are metric-based methods that do not require call graphs or trace data. They take time series as input and perform RCA using causal discovery or statistical analysis. These methods represent state-of-the-art metric-based RCA approaches in the literature.

\begin{sidewaystable}[p]
\centering
\caption[RCA performance of TORAI and baselines on the Online Boutique dataset.]{RCA performance of TORAI and baselines on the Online Boutique dataset, across six fault types. The best results are in \textbf{bold} iff the t-test reported a significant difference compared to other baselines ($p<0.05$).}
\label{tab:rq1-ob}
\resizebox{0.85\textheight}{!}{
\setlength\tabcolsep{2pt}
\begin{tabular}{c|l|rrr|rrr|rrr|rrr|rrr|rrr|rrr}
\hline
\multirow{2}{*}{\begin{tabular}[c]{@{}c@{}}Data\\ Source\end{tabular}} & \multicolumn{1}{c|}{\multirow{2}{*}{Method}} & \multicolumn{3}{c|}{\textbf{CPU}} & \multicolumn{3}{c|}{\textbf{MEM}} & \multicolumn{3}{c|}{\textbf{DISK}} & \multicolumn{3}{c|}{\textbf{SOCKET}} & \multicolumn{3}{c|}{\textbf{DELAY}} & \multicolumn{3}{c|}{\textbf{LOSS}} & \multicolumn{3}{c}{\textbf{AVERAGE}} \\ \cline{3-23} 
 & \multicolumn{1}{c|}{} & \multicolumn{1}{c}{\textit{T1}} & \multicolumn{1}{c}{\textit{T3}} & \multicolumn{1}{c|}{\textit{A5}} & \multicolumn{1}{c}{\textit{T1}} & \multicolumn{1}{c}{\textit{T3}} & \multicolumn{1}{c|}{\textit{A5}} & \multicolumn{1}{c}{\textit{T1}} & \multicolumn{1}{c}{\textit{T3}} & \multicolumn{1}{c|}{\textit{A5}} & \multicolumn{1}{c}{\textit{T1}} & \multicolumn{1}{c}{\textit{T3}} & \multicolumn{1}{c|}{\textit{A5}} & \multicolumn{1}{c}{\textit{T1}} & \multicolumn{1}{c}{\textit{T3}} & \multicolumn{1}{c|}{\textit{A5}} & \multicolumn{1}{c}{\textit{T1}} & \multicolumn{1}{c}{\textit{T3}} & \multicolumn{1}{c|}{\textit{A5}} & \multicolumn{1}{c}{\textit{T1}} & \multicolumn{1}{c}{\textit{T3}} & \multicolumn{1}{c}{\textit{A5}} \\ \hline \hline
\multirow{6}{*}{Metric}
 & BARO & 0.47 & 0.80 & 0.72 & \textbf{0.93} & \textbf{1.00} & \textbf{0.99} & \textbf{1.00} & \textbf{1.00} & \textbf{1.00} & 0.60 & 0.87 & 0.83 & 0.47 & 0.67 & 0.63 & 0.53 & 0.60 & 0.64 & 0.67 & 0.82 & 0.80 \\ 
 & CausalRCA & 0.20 & 0.60 & 0.53 & 0.33 & 0.87 & 0.77 & 0.07 & 0.67 & 0.52 & 0.27 & 0.67 & 0.61 & 0.20 & 0.73 & 0.61 & 0.20 & 0.53 & 0.53 & 0.21 & 0.68 & 0.60 \\
 & CIRCA & 0.73 & 0.93 & 0.88 & 0.67 & \underline{0.93} & 0.89 & 0.80 & 0.87 & 0.85 & 0.67 & 0.87 & 0.83 & 0.47 & 0.80 & 0.75 & 0.67 & \underline{0.93} & 0.88 & 0.67 & 0.89 & 0.85 \\
 & MicroCause & 0.20 & 0.33 & 0.33 & 0.07 & 0.20 & 0.23 & 0.27 & 0.40 & 0.37 & 0.27 & 0.40 & 0.37 & 0.00 & 0.07 & 0.07 & 0.00 & 0.07 & 0.11 & 0.14 & 0.25 & 0.25 \\
 & RCD & \underline{0.87} & \textbf{1.00} & \underline{0.94} & 0.67 & 0.87 & 0.81 & 0.73 & 0.87 & 0.81 & 0.80 & \underline{0.93} & \underline{0.91} & 0.27 & 0.60 & 0.52 & 0.20 & 0.53 & 0.44 & 0.59 & 0.80 & 0.74 \\
 \hline
\multirow{6}{*}{Log}
  & BARO & 0.00 & 0.00 & 0.00 & 0.00 & 0.07 & 0.07 & 0.07 & 0.13 & 0.12 & 0.00 & 0.07 & 0.09 & 0.00 & 0.00 & 0.00 & 0.07 & 0.13 & 0.11 & 0.02 & 0.07 & 0.06 \\
  & CausalRCA & 0.00 & 0.13 & 0.15 & 0.07 & 0.13 & 0.16 & 0.00 & 0.00 & 0.12 & 0.00 & 0.13 & 0.13 & 0.00 & 0.27 & 0.25 & 0.07 & 0.67 & 0.52 & 0.02 & 0.22 & 0.22 \\
  & CIRCA & 0.00 & 0.20 & 0.24 & 0.07 & 0.27 & 0.27 & 0.13 & 0.27 & 0.32 & 0.13 & 0.27 & 0.32 & 0.00 & 0.07 & 0.13 & 0.07 & 0.13 & 0.19 & 0.07 & 0.20 & 0.25 \\
  & MicroCause & 0.27 & 0.67 & 0.57 & 0.13 & 0.40 & 0.39 & 0.31 & 0.69 & 0.68 & 0.13 & 0.47 & 0.47 & 0.00 & 0.07 & 0.11 & 0.33 & 0.47 & 0.51 & 0.20 & 0.46 & 0.46 \\
  & RCD & 0.00 & 0.07 & 0.05 & 0.00 & 0.07 & 0.05 & 0.00 & 0.07 & 0.04 & 0.00 & 0.13 & 0.11 & 0.07 & 0.13 & 0.12 & 0.27 & 0.33 & 0.32 & 0.06 & 0.13 & 0.12 \\

  \hline
\multirow{8}{*}{Trace}
 & BARO & 0.40 & 0.80 & 0.72 & 0.07 & 0.33 & 0.28 & 0.27 & 0.47 & 0.49 & 0.53 & 0.60 & 0.61 & 0.00 & 0.07 & 0.07 & 0.27 & 0.60 & 0.51 & 0.26 & 0.48 & 0.45 \\
 & CausalRCA & 0.30 & 0.37 & 0.36 & 0.23 & 0.33 & 0.36 & 0.07 & 0.20 & 0.19 & 0.20 & 0.27 & 0.27 & 0.00 & 0.03 & 0.03 & 0.03 & 0.30 & 0.30 & 0.14 & 0.25 & 0.25 \\
 & CIRCA & 0.40 & 0.60 & 0.55 & 0.27 & 0.53 & 0.53 & 0.33 & 0.67 & 0.59 & 0.33 & 0.53 & 0.52 & 0.07 & 0.13 & 0.12 & 0.73 & 0.80 & 0.83 & 0.36 & 0.54 & 0.52 \\
 & MicroCause & 0.17 & 0.33 & 0.27 & 0.00 & 0.00 & 0.04 & 0.10 & 0.40 & 0.40 & 0.00 & 0.20 & 0.32 & 0.00 & 0.00 & 0.00 & 0.00 & 0.00 & 0.00 & 0.05 & 0.16 & 0.17 \\
 & RCD & 0.47 & 0.49 & 0.49 & 0.31 & 0.38 & 0.38 & 0.17 & 0.43 & 0.37 & 0.24 & 0.52 & 0.46 & 0.00 & 0.04 & 0.03 & 0.04 & 0.06 & 0.06 & 0.21 & 0.32 & 0.30 \\
 & MicroRank & 0.00 & 0.40 & 0.36 & 0.00 & 0.40 & 0.36 & 0.00 & 0.40 & 0.36 & 0.00 & 0.40 & 0.36 & 0.00 & 0.40 & 0.36 & 0.00 & 0.27 & 0.19 & 0.00 & 0.38 & 0.33 \\
 & TraceRCA & 0.07 & 0.80 & 0.69 & 0.00 & 0.53 & 0.57 & 0.20 & 0.80 & 0.72 & 0.07 & 0.73 & 0.67 & 0.20 & 0.53 & 0.57 & 0.07 & 0.53 & 0.49 & 0.10 & 0.65 & 0.62 \\
 \hline
\multirow{7}{*}{\begin{tabular}[c]{@{}c@{}}Metric\\ +\\ Log\end{tabular}} 
  & BARO & 0.47 & 0.80 & 0.75 & \textbf{0.93} & \textbf{1.00} & \textbf{0.99} & \textbf{1.00} & \textbf{1.00} & \textbf{1.00} & 0.60 & 0.80 & 0.79 & 0.47 & 0.67 & 0.61 & 0.67 & 0.67 & 0.71 & 0.69 & 0.82 & 0.81 \\
 & CausalRCA & 0.13 & 0.30 & 0.30 & 0.12 & 0.24 & 0.25 & 0.00 & 0.07 & 0.10 & 0.00 & 0.13 & 0.15 & 0.00 & 0.30 & 0.28 & 0.07 & 0.50 & 0.41 & 0.09 & 0.24 & 0.22 \\
 & CIRCA & 0.00 & 0.07 & 0.04 & 0.00 & 0.00 & 0.00 & 0.00 & 0.00 & 0.00 & 0.00 & 0.00 & 0.00 & 0.00 & 0.00 & 0.00 & 0.00 & 0.07 & 0.07 & 0.00 & 0.02 & 0.02 \\
 & MicroCause & 0.29 & 0.50 & 0.46 & 0.07 & 0.33 & 0.31 & 0.21 & 0.43 & 0.43 & 0.27 & 0.67 & 0.61 & 0.07 & 0.20 & 0.19 & 0.13 & 0.50 & 0.48 & 0.17 & 0.44 & 0.41 \\
 & RCD & 0.79 & \underline{0.95} & 0.91 & 0.45 & 0.79 & 0.71 & 0.77 & 0.84 & 0.84 & \underline{0.91} & \underline{0.96} & \underline{0.95} & 0.36 & 0.67 & 0.60 & 0.29 & 0.67 & 0.59 & 0.60 & 0.81 & 0.77 \\
 \cline{2-23} 
 & \textbf{TORAI} & 0.67 & 0.80 & 0.81 & \underline{0.87} & \textbf{1.00} & \underline{0.97} & \underline{0.93} & \textbf{1.00} & \underline{0.95} & 0.80 & 0.87 & 0.84 & \textbf{0.73} & \textbf{1.00} & \textbf{0.92} & \underline{0.77} & \textbf{1.00} & \underline{0.93} & \underline{0.80} & \textbf{0.95} & \underline{0.91} \\ \hline
\multirow{8}{*}{\begin{tabular}[c]{@{}c@{}}Metric\\ +\\ Log\\ +\\ Trace\end{tabular}} 
 & BARO & 0.47 & 0.80 & 0.75 & \textbf{0.93} & \textbf{1.00} & \textbf{0.99} & \textbf{1.00} & \textbf{1.00} & \textbf{1.00} & 0.60 & 0.80 & 0.79 & 0.47 & 0.67 & 0.61 & 0.67 & 0.67 & 0.71 & 0.69 & 0.82 & 0.81 \\
 & CausalRCA & 0.20 & 0.27 & 0.27 & 0.13 & 0.27 & 0.29 & 0.07 & 0.07 & 0.09 & 0.07 & 0.20 & 0.23 & 0.07 & 0.33 & 0.31 & 0.07 & 0.27 & 0.28 & 0.10 & 0.24 & 0.25 \\
 & CIRCA & 0.00 & 0.07 & 0.04 & 0.00 & 0.00 & 0.00 & 0.00 & 0.00 & 0.00 & 0.00 & 0.00 & 0.00 & 0.00 & 0.00 & 0.00 & 0.00 & 0.00 & 0.03 & 0.00 & 0.01 & 0.01 \\
 & MicroCause & 0.50 & 0.67 & 0.60 & 0.25 & 0.25 & 0.25 & 0.33 & 0.78 & 0.76 & 0.50 & 0.50 & 0.50 & 0.00 & 0.33 & 0.30 & 0.14 & 0.29 & 0.37 & 0.29 & 0.47 & 0.46 \\
 & PDiagnose & 0.00 & 0.80 & 0.60 & 0.00 & 0.40 & 0.51 & 0.00 & 0.73 & 0.59 & 0.00 & 0.60 & 0.52 & 0.20 & 0.40 & 0.55 & 0.00 & 0.47 & 0.47 & 0.03 & 0.57 & 0.54 \\ 
 & HeMiRCA & 0.43 & 0.77 & 0.72 & 0.23 & 0.36 & 0.37 & 0.89 & \underline{0.95} & 0.94 & 0.20 & 0.40 & 0.43 & 0.33 & 0.53 & 0.55 & 0.40 & 0.40 & 0.40 & 0.41 & 0.57 & 0.57 \\
 & RCD & 0.80 & \textbf{1.00} & \underline{0.94} & 0.47 & 0.80 & 0.71 & 0.80 & \textbf{1.00} & \underline{0.95} & \textbf{1.00} & \textbf{1.00} & \textbf{1.00} & 0.33 & 0.40 & 0.51 & 0.40 & 0.60 & 0.60 & 0.63 & 0.80 & 0.79 \\
 \cline{2-23} 
 & \textbf{TORAI} & \textbf{0.92} & \textbf{1.00} & \textbf{0.98} & 0.67 & \textbf{1.00} & 0.93 & \textbf{1.00} & \textbf{1.00} & \textbf{1.00} & 0.87 & 0.93 & \underline{0.95} & \underline{0.67} & 0.73 & \underline{0.76} & \textbf{0.87} & \textbf{1.00} & \textbf{0.96} & \textbf{0.83} & \textbf{0.95} & \textbf{0.93} \\ \hline
\end{tabular}%
}

\vspace{0.05cm}
{\footnotesize (*) T1, T3, and A5 denote AC@1, AC@3, and Avg@5, respectively.}
\end{sidewaystable}

\begin{sidewaystable}[p]
\centering
\caption[RCA performance of TORAI and baselines on the Sock Shop dataset.]{RCA performance of TORAI and baselines on the Sock Shop dataset, across six fault types. The best results are in \textbf{bold} iff the t-test reported a significant difference compared to other baselines ($p<0.05$).}
\label{tab:rq1-ss}
\resizebox{0.85\textheight}{!}{%
\setlength\tabcolsep{2pt}
\begin{tabular}{c|l|rrr|rrr|rrr|rrr|rrr|rrr|rrr}
\hline
\multirow{2}{*}{\begin{tabular}[c]{@{}c@{}}Data\\ Source\end{tabular}} & \multicolumn{1}{c|}{\multirow{2}{*}{Method}} & \multicolumn{3}{c|}{\textbf{CPU}} & \multicolumn{3}{c|}{\textbf{MEM}} & \multicolumn{3}{c|}{\textbf{DISK}} & \multicolumn{3}{c|}{\textbf{SOCKET}} & \multicolumn{3}{c|}{\textbf{DELAY}} & \multicolumn{3}{c|}{\textbf{LOSS}} & \multicolumn{3}{c}{\textbf{AVERAGE}} \\ \cline{3-23} 
 & \multicolumn{1}{c|}{} & \multicolumn{1}{c}{\textit{T1}} & \multicolumn{1}{c}{\textit{T3}} & \multicolumn{1}{c|}{\textit{A5}} & \multicolumn{1}{c}{\textit{T1}} & \multicolumn{1}{c}{\textit{T3}} & \multicolumn{1}{c|}{\textit{A5}} & \multicolumn{1}{c}{\textit{T1}} & \multicolumn{1}{c}{\textit{T3}} & \multicolumn{1}{c|}{\textit{A5}} & \multicolumn{1}{c}{\textit{T1}} & \multicolumn{1}{c}{\textit{T3}} & \multicolumn{1}{c|}{\textit{A5}} & \multicolumn{1}{c}{\textit{T1}} & \multicolumn{1}{c}{\textit{T3}} & \multicolumn{1}{c|}{\textit{A5}} & \multicolumn{1}{c}{\textit{T1}} & \multicolumn{1}{c}{\textit{T3}} & \multicolumn{1}{c|}{\textit{A5}} & \multicolumn{1}{c}{\textit{T1}} & \multicolumn{1}{c}{\textit{T3}} & \multicolumn{1}{c}{\textit{A5}} \\ \hline \hline
\multirow{5}{*}{Metric}
  & BARO & 0.00 & \textbf{1.00} & 0.80 & 0.20 & \textbf{1.00} & 0.83 & 0.00 & \underline{0.93} & 0.77 & 0.00 & \underline{0.93} & 0.71 & 0.00 & \underline{0.87} & 0.68 & 0.20 & \textbf{1.00} & 0.80 & 0.07 & \textbf{0.96} & 0.76 \\
 & CausalRCA & 0.20 & 0.60 & 0.55 & 0.40 & 0.80 & 0.75 & 0.20 & 0.60 & 0.55 & 0.33 & 0.60 & 0.60 & 0.27 & 0.40 & 0.43 & 0.00 & 0.33 & 0.32 & 0.23 & 0.56 & 0.53 \\
 & CIRCA & \underline{0.87} & \textbf{1.00} & \underline{0.97} & \underline{0.87} & \underline{0.93} & \underline{0.95} & \underline{0.87} & 0.87 & \underline{0.89} & \underline{0.67} & \textbf{1.00} & \underline{0.92} & \underline{0.67} & \underline{0.87} & \underline{0.85} & \underline{0.47} & 0.87 & \underline{0.81} & \underline{0.74} & \underline{0.92} & \underline{0.90} \\
 & MicroCause & 0.07 & 0.13 & 0.16 & 0.00 & 0.20 & 0.23 & 0.00 & 0.13 & 0.12 & 0.33 & 0.40 & 0.44 & 0.13 & 0.27 & 0.28 & 0.07 & 0.27 & 0.24 & 0.10 & 0.23 & 0.25 \\ 
 & RCD & 0.47 & 0.73 & 0.68 & 0.27 & 0.40 & 0.36 & 0.47 & 0.67 & 0.61 & 0.47 & 0.87 & 0.77 & 0.40 & 0.73 & 0.64 & 0.20 & 0.40 & 0.35 & 0.38 & 0.63 & 0.57 \\ \hline
\multirow{5}{*}{Log}
 & BARO & 0.20 & 0.47 & 0.51 & 0.20 & 0.47 & 0.48 & 0.13 & 0.33 & 0.39 & 0.13 & 0.33 & 0.39 & 0.20 & 0.33 & 0.40 & 0.13 & 0.60 & 0.52 & 0.17 & 0.42 & 0.45 \\
 & CIRCA & 0.13 & 0.53 & 0.48 & 0.00 & 0.33 & 0.29 & 0.07 & 0.47 & 0.41 & 0.00 & 0.40 & 0.36 & 0.07 & 0.33 & 0.36 & 0.20 & 0.47 & 0.52 & 0.08 & 0.42 & 0.40 \\
 & CausalRCA & 0.10 & 0.33 & 0.37 & 0.10 & 0.53 & 0.47 & 0.23 & 0.43 & 0.42 & 0.20 & 0.50 & 0.47 & 0.10 & 0.20 & 0.21 & \underline{0.47} & 0.80 & 0.76 & 0.20 & 0.47 & 0.45 \\ 
 & MicroCause & 0.50 & 0.75 & 0.70 & 0.29 & 0.71 & 0.60 & 0.00 & 0.33 & 0.27 & 0.13 & 0.38 & 0.38 & 0.00 & 0.17 & 0.13 & 0.00 & 0.43 & 0.34 & 0.15 & 0.46 & 0.40 \\ 
 & RCD & 0.07 & 0.21 & 0.18 & 0.09 & 0.16 & 0.14 & 0.12 & 0.33 & 0.28 & 0.09 & 0.19 & 0.17 & 0.08 & 0.15 & 0.13 & 0.37 & 0.48 & 0.46 & 0.14 & 0.25 & 0.23 \\ \hline
\multirow{8}{*}{\begin{tabular}[c]{@{}c@{}}Metric\\ +\\ Log\end{tabular}} 
  & BARO & 0.00 & \textbf{1.00} & 0.80 & 0.20 & \textbf{1.00} & 0.83 & 0.00 & \textbf{1.00} & 0.79 & 0.00 & \underline{0.93} & 0.71 & 0.00 & 0.80 & 0.65 & 0.20 & \textbf{1.00} & 0.80 & 0.07 & \textbf{0.96} & 0.76 \\
 & CausalRCA & 0.20 & 0.33 & 0.33 & 0.47 & 0.73 & 0.73 & 0.20 & 0.40 & 0.44 & 0.07 & 0.40 & 0.35 & 0.20 & 0.53 & 0.48 & 0.20 & 0.47 & 0.51 & 0.22 & 0.48 & 0.47 \\
 & CIRCA & 0.00 & 0.00 & 0.00 & 0.07 & 0.07 & 0.08 & 0.00 & 0.00 & 0.05 & 0.00 & 0.00 & 0.00 & 0.00 & 0.07 & 0.07 & 0.00 & 0.20 & 0.19 & 0.01 & 0.06 & 0.07 \\
 & MicroCause & 0.33 & 0.33 & 0.40 & 0.17 & 0.33 & 0.37 & 0.00 & 0.67 & 0.53 & 0.25 & 0.75 & 0.63 & 0.20 & 0.80 & 0.72 & 0.00 & 0.40 & 0.36 & 0.16 & 0.55 & 0.50 \\
 & RCD & 0.61 & \underline{0.77} & 0.74 & 0.20 & 0.49 & 0.42 & 0.36 & 0.64 & 0.59 & \underline{0.67} & 0.89 & 0.85 & 0.28 & 0.67 & 0.58 & 0.19 & \underline{0.92} & 0.76 & 0.39 & 0.73 & 0.66 \\
 \cline{2-23} 
 & \textbf{TORAI} & \textbf{0.93} & \textbf{1.00} & \textbf{0.98} & \textbf{1.00} & \textbf{1.00} & \textbf{1.00} & \textbf{0.93} & \underline{0.93} & \textbf{0.95} & \textbf{0.84} & \underline{0.93} & \textbf{0.94} & \textbf{0.75} & \textbf{0.88} & \textbf{0.86} & \textbf{0.60} & \textbf{1.00} & \textbf{0.92} & \textbf{0.84} & \textbf{0.96} & \textbf{0.94} \\ \hline
\end{tabular}%
}
\end{sidewaystable}

\begin{sidewaystable}[p]
\centering
\caption[RCA performance of TORAI and baselines on the Train Ticket dataset.]{RCA performance of TORAI and baselines on the Train Ticket dataset, across six fault types. The best results are in \textbf{bold} iff the t-test reported a significant difference compared to other baselines ($p<0.05$).}
\label{tab:rq1-tt}
\resizebox{0.85\textheight}{!}{%
\setlength\tabcolsep{2pt}
\begin{tabular}{c|l|rrr|rrr|rrr|rrr|rrr|rrr|rrr}
\hline
\multirow{2}{*}{\begin{tabular}[c]{@{}c@{}}Data\\ Source\end{tabular}} & \multicolumn{1}{c|}{\multirow{2}{*}{Method}} & \multicolumn{3}{c|}{\textbf{CPU}} & \multicolumn{3}{c|}{\textbf{MEM}} & \multicolumn{3}{c|}{\textbf{DISK}} & \multicolumn{3}{c|}{\textbf{SOCKET}} & \multicolumn{3}{c|}{\textbf{DELAY}} & \multicolumn{3}{c|}{\textbf{LOSS}} & \multicolumn{3}{c}{\textbf{AVERAGE}} \\ \cline{3-23} 
 & \multicolumn{1}{c|}{} & \multicolumn{1}{c}{\textit{T1}} & \multicolumn{1}{c}{\textit{T3}} & \multicolumn{1}{c|}{\textit{A5}} & \multicolumn{1}{c}{\textit{T1}} & \multicolumn{1}{c}{\textit{T3}} & \multicolumn{1}{c|}{\textit{A5}} & \multicolumn{1}{c}{\textit{T1}} & \multicolumn{1}{c}{\textit{T3}} & \multicolumn{1}{c|}{\textit{A5}} & \multicolumn{1}{c}{\textit{T1}} & \multicolumn{1}{c}{\textit{T3}} & \multicolumn{1}{c|}{\textit{A5}} & \multicolumn{1}{c}{\textit{T1}} & \multicolumn{1}{c}{\textit{T3}} & \multicolumn{1}{c|}{\textit{A5}} & \multicolumn{1}{c}{\textit{T1}} & \multicolumn{1}{c}{\textit{T3}} & \multicolumn{1}{c|}{\textit{A5}} & \multicolumn{1}{c}{\textit{T1}} & \multicolumn{1}{c}{\textit{T3}} & \multicolumn{1}{c}{\textit{A5}} \\ \hline
\multirow{4}{*}{Metric} 
 & BARO & 0.47 & 0.80 & 0.72 & \underline{0.93} & \textbf{1.00} & \underline{0.99} & \textbf{1.00} & \textbf{1.00} & \textbf{1.00} & 0.60 & \underline{0.87} & 0.83 & 0.47 & 0.67 & 0.63 & 0.53 & 0.60 & 0.64 & 0.67 & 0.82 & 0.80 \\
 & CausalRCA & 0.40 & 0.63 & 0.59 & 0.10 & 0.27 & 0.24 & 0.43 & \underline{0.83} & \underline{0.75} & 0.23 & 0.50 & 0.45 & 0.13 & 0.23 & 0.21 & 0.03 & 0.37 & 0.33 & 0.22 & 0.47 & 0.43 \\
 & CIRCA & 0.27 & 0.27 & 0.28 & 0.47 & 0.73 & 0.68 & 0.53 & 0.67 & 0.64 & 0.27 & 0.53 & 0.52 & 0.20 & 0.27 & 0.28 & 0.20 & 0.33 & 0.35 & 0.32 & 0.47 & 0.46 \\
 & MicroCause & 0.19 & 0.44 & 0.40 & 0.00 & 0.09 & 0.07 & 0.40 & 0.40 & 0.40 & 0.00 & 0.17 & 0.15 & 0.00 & 0.22 & 0.13 & 0.00 & 0.00 & 0.07 & 0.10 & 0.22 & 0.20 \\ 
 & RCD & 0.13 & 0.13 & 0.16 & 0.07 & 0.07 & 0.07 & 0.00 & 0.07 & 0.05 & 0.13 & 0.33 & 0.29 & 0.13 & 0.13 & 0.15 & 0.07 & 0.07 & 0.07 & 0.09 & 0.13 & 0.13 \\ \hline
\multirow{5}{*}{Log} 
 & BARO & 0.00 & 0.00 & 0.00 & 0.00 & 0.07 & 0.07 & 0.07 & 0.13 & 0.12 & 0.00 & 0.07 & 0.09 & 0.00 & 0.00 & 0.00 & 0.07 & 0.13 & 0.11 & 0.02 & 0.07 & 0.06 \\
 & CausalRCA & 0.07 & 0.20 & 0.22 & 0.07 & 0.10 & 0.09 & 0.07 & 0.10 & 0.12 & 0.00 & 0.07 & 0.08 & 0.03 & 0.20 & 0.15 & 0.10 & 0.20 & 0.20 & 0.06 & 0.15 & 0.14 \\
 & CIRCA & 0.13 & 0.20 & 0.27 & 0.07 & 0.33 & 0.27 & 0.07 & 0.07 & 0.12 & 0.07 & 0.20 & 0.23 & 0.13 & 0.33 & 0.33 & 0.13 & 0.33 & 0.33 & 0.10 & 0.24 & 0.26 \\
 & MicroCause & 0.00 & 0.00 & 0.10 & 0.20 & 0.40 & 0.40 & 0.14 & 0.14 & 0.14 & 0.13 & 0.38 & 0.28 & 0.11 & 0.11 & 0.16 & 0.11 & 0.22 & 0.20 & 0.12 & 0.21 & 0.21 \\ 
 & RCD & 0.07 & 0.16 & 0.14 & 0.12 & 0.15 & 0.14 & 0.07 & 0.15 & 0.13 & 0.06 & 0.19 & 0.17 & 0.11 & 0.24 & 0.21 & 0.12 & 0.16 & 0.15 & 0.09 & 0.18 & 0.16 \\ \hline
\multirow{8}{*}{Trace}
 & BARO & 0.40 & 0.80 & 0.72 & 0.07 & 0.33 & 0.28 & 0.27 & 0.47 & 0.49 & 0.53 & 0.60 & 0.61 & 0.00 & 0.07 & 0.07 & 0.27 & 0.60 & 0.51 & 0.26 & 0.48 & 0.45 \\
 & CausalRCA & 0.07 & 0.33 & 0.29 & 0.07 & 0.20 & 0.19 & 0.20 & 0.27 & 0.25 & 0.00 & 0.00 & 0.03 & 0.00 & 0.07 & 0.05 & 0.27 & 0.47 & 0.44 & 0.10 & 0.22 & 0.21 \\
 & CIRCA & 0.13 & 0.20 & 0.23 & 0.20 & 0.27 & 0.25 & 0.20 & 0.20 & 0.21 & 0.13 & 0.13 & 0.15 & 0.20 & 0.33 & 0.35 & 0.13 & 0.27 & 0.27 & 0.17 & 0.23 & 0.24 \\
 & MicroCause & 0.00 & 0.00 & 0.00 & 0.00 & 0.00 & 0.00 & 0.00 & 0.00 & 0.00 & 0.00 & 0.00 & 0.00 & 0.00 & 0.00 & 0.00 & 0.00 & 0.00 & 0.00 & 0.00 & 0.00 & 0.00 \\
 & MicroRank & 0.21 & 0.43 & 0.34 & 0.25 & 0.38 & 0.33 & 0.00 & 0.36 & 0.27 & 0.30 & 0.40 & 0.36 & 0.08 & 0.31 & 0.23 & 0.14 & 0.36 & 0.30 & 0.16 & 0.37 & 0.31 \\
 & RCD & 0.53 & \textbf{0.87} & 0.79 & 0.53 & 0.67 & 0.63 & \underline{0.67} & 0.67 & 0.69 & \underline{0.73} & 0.80 & 0.79 & 0.13 & 0.27 & 0.21 & 0.60 & 0.73 & \underline{0.71} & 0.53 & 0.67 & 0.64 \\
 & TraceRCA & \textbf{0.64} & 0.79 & 0.74 & 0.63 & \underline{0.88} & 0.83 & 0.64 & 0.71 & 0.74 & 0.60 & 0.80 & 0.76 & \underline{0.85} & \underline{0.85} & \textbf{0.88} & 0.57 & 0.71 & 0.67 & 0.66 & 0.79 & 0.77 \\ \hline
\multirow{5}{*}{\begin{tabular}[c]{@{}c@{}}Metric\\ +\\ Log\end{tabular}}
 & BARO & 0.47 & 0.80 & 0.75 & \underline{0.93} & \textbf{1.00} & \underline{0.99} & \textbf{1.00} & \textbf{1.00} & \textbf{1.00} & 0.60 & 0.80 & 0.79 & 0.47 & 0.67 & 0.61 & \underline{0.67} & 0.67 & \underline{0.71} & 0.69 & 0.82 & 0.81 \\
 & CIRCA & 0.00 & 0.13 & 0.09 & 0.00 & 0.07 & 0.08 & 0.07 & 0.07 & 0.09 & 0.00 & 0.13 & 0.15 & 0.00 & 0.07 & 0.05 & 0.07 & 0.07 & 0.09 & 0.02 & 0.09 & 0.09 \\
 & MicroCause & 0.00 & 0.00 & 0.00 & 0.07 & 0.07 & 0.07 & 0.00 & 0.00 & 0.00 & 0.07 & 0.07 & 0.08 & 0.00 & 0.00 & 0.00 & 0.07 & 0.13 & 0.13 & 0.04 & 0.05 & 0.05 \\ 
 & RCD & 0.13 & 0.27 & 0.23 & 0.00 & 0.07 & 0.08 & 0.13 & 0.27 & 0.21 & 0.20 & 0.33 & 0.29 & 0.00 & 0.13 & 0.12 & 0.07 & 0.13 & 0.11 & 0.09 & 0.20 & 0.17 \\
 \cline{2-23} 
 & \textbf{TORAI} & 0.53 & \textbf{0.87} & 0.80 & \textbf{1.00} & \textbf{1.00} & \textbf{1.00} & \textbf{1.00} & \textbf{1.00} & \textbf{1.00} & \textbf{0.80} & \underline{0.87} & \underline{0.87} & 0.53 & 0.67 & 0.64 & 0.57 & 0.63 & 0.67 & \underline{0.74} & \underline{0.84} & \underline{0.83} \\ \hline
\multirow{6}{*}{\begin{tabular}[c]{@{}c@{}}Metric\\ +\\ Log\\ +\\ Trace\end{tabular}} 
 & BARO & 0.47 & 0.80 & 0.75 & \underline{0.93} & \textbf{1.00} & \underline{0.99} & \textbf{1.00} & \textbf{1.00} & \textbf{1.00} & 0.60 & 0.80 & 0.79 & 0.47 & 0.67 & 0.61 & \underline{0.67} & 0.67 & \underline{0.71} & 0.69 & 0.82 & 0.81 \\
 & CIRCA & 0.00 & 0.07 & 0.09 & 0.07 & 0.13 & 0.21 & 0.00 & 0.07 & 0.09 & 0.07 & 0.13 & 0.16 & 0.07 & 0.07 & 0.07 & 0.13 & 0.20 & 0.17 & 0.06 & 0.11 & 0.13 \\
 & HeMiRCA & 0.00 & 0.07 & 0.15 & 0.13 & 0.27 & 0.27 & 0.13 & 0.20 & 0.19 & 0.20 & 0.47 & 0.43 & 0.07 & 0.33 & 0.32 & 0.07 & 0.13 & 0.15 & 0.10 & 0.25 & 0.25 \\
 & MicroCause & 0.07 & 0.13 & 0.11 & 0.07 & 0.07 & 0.07 & 0.00 & 0.07 & 0.11 & 0.00 & 0.00 & 0.03 & 0.00 & 0.00 & 0.00 & 0.00 & 0.13 & 0.16 & 0.02 & 0.07 & 0.08 \\
 & PDiagnose & \underline{0.60} & \textbf{0.87} & \underline{0.81} & 0.40 & 0.47 & 0.48 & 0.33 & 0.73 & 0.69 & 0.33 & 0.67 & 0.60 & \textbf{0.87} & \textbf{0.87} & \underline{0.87} & 0.33 & 0.60 & 0.57 & 0.48 & 0.70 & 0.67 \\ 
 & RCD & 0.17 & \underline{0.84} & 0.72 & 0.00 & 0.44 & 0.39 & 0.07 & 0.76 & 0.62 & 0.21 & 0.77 & 0.66 & 0.05 & 0.28 & 0.25 & 0.07 & \underline{0.75} & 0.60 & 0.10 & 0.64 & 0.54 \\
 \cline{2-23} 
 & \textbf{TORAI} & \underline{0.60} & \textbf{0.87} & \textbf{0.83} & \textbf{1.00} & \textbf{1.00} & \textbf{1.00} & \textbf{1.00} & \textbf{1.00} & \textbf{1.00} & \underline{0.73} & \textbf{1.00} & \textbf{0.93} & 0.56 & 0.67 & 0.66 & \textbf{0.73} & \textbf{1.00} & \textbf{0.92} & \textbf{0.77} & \textbf{0.92} & \textbf{0.89} \\ \hline
\end{tabular}%
}

{\footnotesize (*) CausalRCA exceeds the limit of 2 hours per case. T1, T3, and A5 denote AC@1, AC@3, and Avg@5, respectively.}
\end{sidewaystable}

\subsection{RQ1. Effectiveness in Coarse-grained Root Cause Analysis} \label{sec:experimental-rq1}

In this section, we evaluate the coarse-grained RCA performance of our proposed TORAI and the RCA baseline methods on all three datasets. 
Tables \ref{tab:rq1-ob}, \ref{tab:rq1-ss}, and \ref{tab:rq1-tt} report the overall performance of all methods on the Online Boutique, Sock Shop, and Train Ticket datasets at coarse-grained level, respectively.  
We calculate the accuracy for each type of fault: CPU hog (CPU), memory leak (MEM), disk I/O stress (DISK), socket stress (SOCKET), network delay (DELAY), and packet loss (LOSS). Additionally, we report the AVERAGE scores to present the overall performance across fault types and data sources. We perform statistical analysis on the results using the t-tests to check the pairwise differences among all RCA methods. We \textbf{bold} the best result iff the statistical tests report a significant difference ($p<0.05$) compared to others. In the tables, the first column indicates the data sources used for the methods in the second column. We draw the following observations:

\textbf{(1) TORAI performs the best on all three datasets.} For example, on the Online Boutique dataset, TORAI achieves an average T1, T3, and A5 score of 0.83, 0.95, and 0.93 when diagnosing root cause service, while metric-based CausalRCA achieves 0.21, 0.68, and 0.6, respectively. On the Sock Shop dataset, TORAI achieves the averages of 0.84, 0.96, and 0.94 for T1, T3, and A5, respectively, while the multi-source version of RCD reaches 0.39, 0.73, and 0.66. On the Train Ticket dataset, the best average A5 scores of CausalRCA, RCD, and CIRCA are 0.43, 0.64, and 0.46, respectively. Our TORAI beats them with a large margin, achieving an A5 of 0.89.

\textbf{(2) TORAI effectively integrates multi-source data, surpassing most baselines.} On the Train Ticket dataset, when incorporating metrics, logs, and traces, TORAI achieves the T1, T3, and A5 scores of 0.77, 0.92, and 0.89, respectively. In contrast, other RCA methods like CausalRCA, RCD, CIRCA, and MicroCause perform worse because their designs are less effective. These methods typically attempt to build causal graphs from all time series data and rely on scoring techniques like PageRank, random walk, or hypothesis testing, without leveraging clustering or severity scoring as TORAI does.

\textbf{(3) TORAI can diagnose failures in blind spots.} PDiagnose and HeMiRCA are multi-source RCA methods but they rely heavily on trace data to construct the service call graph. Consequently, they cannot perform RCA for the Sock Shop dataset, where there is no trace data to construct the call graph (i.e., all services are blind spots). Meanwhile, TORAI can still perform RCA effectively in the presence of blind spots using metrics and logs without requiring traces to construct a call graph (see Table~\ref{tab:rq1-ss}).
It is worth noting that metrics and logs are easier to collect as developers do not need to spend much effort to obtain them. Metrics can be automatically obtained by a monitoring system, and logs are naturally produced by the systems for troubleshooting purposes. In contrast, developers need to spend tremendous effort to instrument the system with distributed tracing. 

\textbf{(4) On the large Train Ticket system, TORAI outperforms other methods by a significant margin.} For example, our method achieves an Avg@5 score of 0.89, while CausalRCA, CIRCA, and RCD achieve their best scores of 0.43, 0.46, and 0.64, respectively.  Across all faults, TORAI achieves Avg@5 scores of 0.83, 1, 1, 0.93, 0.66, and 0.92 for CPU, MEM, DISK, SOCKET, DELAY, and LOSS, respectively. This demonstrates the effectiveness of our proposed TORAI, which works not only on small demo systems but also on large systems with many services.

\subsection{RQ2. Effectiveness in Fine-grained Root Cause Analysis}

\begin{table}[t]
\centering
\caption[Fine-grained RCA performance of TORAI on the Online Boutique dataset.]{Fine-grained RCA performance of TORAI and baselines on the Online Boutique dataset, across six fault types. The best results are in \textbf{bold} iff the t-test reported a significant difference compared to other baselines ($p<0.05$). For all baselines, we select their best setup when taking different data sources.}
\label{tab:rq1-ob-fine}
\resizebox{\textwidth}{!}{%
\setlength\tabcolsep{3pt}
\begin{tabular}{l|rrr|rrr|rrr|rrr|rrr|rrr|rrr}
\hline
\multicolumn{1}{c|}{\multirow{2}{*}{Method}} & \multicolumn{3}{c|}{\textbf{CPU}} & \multicolumn{3}{c|}{\textbf{MEM}} & \multicolumn{3}{c|}{\textbf{DISK}} & \multicolumn{3}{c|}{\textbf{SOCKET}} & \multicolumn{3}{c|}{\textbf{DELAY}} & \multicolumn{3}{c|}{\textbf{LOSS}} & \multicolumn{3}{c}{\textbf{AVERAGE}} \\ \cline{2-22} 
\multicolumn{1}{c|}{} & \multicolumn{1}{c}{\textit{T1}} & \multicolumn{1}{c}{\textit{T3}} & \multicolumn{1}{c|}{\textit{A5}} & \multicolumn{1}{c}{\textit{T1}} & \multicolumn{1}{c}{\textit{T3}} & \multicolumn{1}{c|}{\textit{A5}} & \multicolumn{1}{c}{\textit{T1}} & \multicolumn{1}{c}{\textit{T3}} & \multicolumn{1}{c|}{\textit{A5}} & \multicolumn{1}{c}{\textit{T1}} & \multicolumn{1}{c}{\textit{T3}} & \multicolumn{1}{c|}{\textit{A5}} & \multicolumn{1}{c}{\textit{T1}} & \multicolumn{1}{c}{\textit{T3}} & \multicolumn{1}{c|}{\textit{A5}} & \multicolumn{1}{c}{\textit{T1}} & \multicolumn{1}{c}{\textit{T3}} & \multicolumn{1}{c|}{\textit{A5}} & \multicolumn{1}{c}{\textit{T1}} & \multicolumn{1}{c}{\textit{T3}} & \multicolumn{1}{c}{\textit{A5}} \\ \hline
BARO & 0 & 0.33 & 0.43 & 0.2 & \textbf{0.8} & 0.67 & 0 & 0 & 0 & 0 & 0 & 0 & 0 & \textbf{0.73} & 0.6 & 0.33 & 0.73 & 0.65 & 0.09 & 0.43 & 0.39 \\ 
CausalRCA & \textbf{0.2} & 0.53 & 0.45 & \textbf{0.47} & 0.67 & 0.65 & 0.07 & 0.27 & 0.32 & 0 & 0.4 & \textbf{0.35} & 0.27 & 0.67 & 0.61 & 0.13 & 0.27 & 0.23 & 0.19 & 0.47 & 0.44 \\
CIRCA & 0.13 & 0.47 & 0.44 & \textbf{0.47} & 0.67 & 0.64 & 0.87 & 0.87 & 0.87 & 0 & 0.07 & 0.08 & 0.47 & \textbf{0.73} & 0.67 & 0.33 & 0.6 & 0.61 & 0.38 & 0.57 & 0.55 \\
MicroCause & 0.13 & 0.4 & 0.32 & 0 & 0 & 0.04 & 0.04 & 0.31 & 0.29 & 0 & 0.07 & 0.09 & 0 & 0 & 0 & 0 & 0 & 0 & 0.03 & 0.13 & 0.12 \\
HeMiRCA & 0.25 & 0.27 & 0.29 & 0.1 & 0.15 & 0.15 & 0.77 & 0.82 & 0.82 & 0.07 & 0.07 & 0.1 & 0.13 & 0.23 & 0.25 & 0.4 & 0.4 & 0.4 & 0.29 & 0.32 & 0.34 \\
RCD & 0.07 & 0.07 & 0.07 & 0.27 & 0.33 & 0.31 & 0 & 0 & 0 & 0 & 0 & 0 & 0.33 & 0.33 & 0.4 & 0.2 & 0.2 & 0.2 & 0.15 & 0.16 & 0.16 \\
\hline
TORAI & 0.13 & \textbf{0.73} & \textbf{0.63} & 0.4 & 0.67 & 0.6 & \textbf{1} & \textbf{1} & \textbf{1} & 0 & 0.07 & \textbf{0.35} & \textbf{0.6} & 0.67 & 0.65 & \textbf{0.8} & \textbf{0.87} & \textbf{0.85} & \textbf{0.49} & \textbf{0.67} & \textbf{0.68} \\ \hline
\end{tabular}%
}
\end{table}

\begin{table}[t]
\caption[Fine-grained RCA performance of TORAI on the Sock Shop dataset.]{Fine-grained RCA performance of TORAI and baselines on the Sock Shop dataset, across six fault types. The best results are in \textbf{bold} iff the t-test reported a significant difference compared to others ($p<0.05$). For all baselines, we select their best setup when taking different data sources.}
\label{tab:rq1-ss-fine}
\resizebox{\columnwidth}{!}{%
\setlength\tabcolsep{3pt}
\begin{tabular}{l|rrr|rrr|rrr|rrr|rrr|rrr|rrr}
\hline
\multicolumn{1}{c|}{\multirow{2}{*}{Method}} & \multicolumn{3}{c|}{\textbf{CPU}} & \multicolumn{3}{c|}{\textbf{MEM}} & \multicolumn{3}{c|}{\textbf{DISK}} & \multicolumn{3}{c|}{\textbf{SOCKET}} & \multicolumn{3}{c|}{\textbf{DELAY}} & \multicolumn{3}{c|}{\textbf{LOSS}} & \multicolumn{3}{c}{\textbf{AVERAGE}} \\ \cline{2-22} 
\multicolumn{1}{c|}{} & \multicolumn{1}{c}{\textit{T1}} & \multicolumn{1}{c}{\textit{T3}} & \multicolumn{1}{c|}{\textit{A5}} & \multicolumn{1}{c}{\textit{T1}} & \multicolumn{1}{c}{\textit{T3}} & \multicolumn{1}{c|}{\textit{A5}} & \multicolumn{1}{c}{\textit{T1}} & \multicolumn{1}{c}{\textit{T3}} & \multicolumn{1}{c|}{\textit{A5}} & \multicolumn{1}{c}{\textit{T1}} & \multicolumn{1}{c}{\textit{T3}} & \multicolumn{1}{c|}{\textit{A5}} & \multicolumn{1}{c}{\textit{T1}} & \multicolumn{1}{c}{\textit{T3}} & \multicolumn{1}{c|}{\textit{A5}} & \multicolumn{1}{c}{\textit{T1}} & \multicolumn{1}{c}{\textit{T3}} & \multicolumn{1}{c|}{\textit{A5}} & \multicolumn{1}{c}{\textit{T1}} & \multicolumn{1}{c}{\textit{T3}} & \multicolumn{1}{c}{\textit{A5}} \\ \hline
BARO & 0 & 0.87 & 0.68 & 0.2 & \textbf{1} & 0.8 & 0 & 0 & 0 & 0 & 0 & 0.01 & 0 & \textbf{0.87} & 0.68 & 0.2 & \textbf{1} & 0.8 & 0.07 & 0.62 & 0.50 \\
CausalRCA & 0.27 & 0.63 & 0.58 & 0.3 & 0.67 & 0.59 & 0.07 & 0.17 & 0.17 & 0 & 0 & 0.03 & 0.33 & 0.57 & 0.56 & 0.23 & 0.37 & 0.37 & 0.2 & 0.37 & 0.36 \\
CIRCA & 0.4 & 0.8 & 0.72 & 0.8 & \textbf{1} & 0.95 & \textbf{0.67} & \textbf{0.67} & \textbf{0.67} & 0 & 0.2 & 0.16 & 0.53 & 0.8 & \textbf{0.79} & 0.6 & 0.87 & \textbf{0.85} & 0.5 & \textbf{0.69} & \textbf{0.67} \\
MicroCause & 0 & 0.09 & 0.07 & 0 & 0.08 & 0.05 & 0.09 & 0.18 & 0.18 & 0.07 & 0.2 & 0.23 & 0 & 0 & 0 & 0 & 0 & 0 & 0.03 & 0.08 & 0.08 \\ 
RCD & 0.11 & 0.15 & 0.15 & 0.05 & 0.16 & 0.15 & 0 & 0 & 0 & 0.13 & 0.19 & 0.2 & 0.24 & 0.27 & 0.27 & 0.03 & 0.03 & 0.03 & 0.09 & 0.13 & 0.13 \\
\hline
\textbf{TORAI} & \textbf{0.6} & \textbf{0.93} & \textbf{0.85} & \textbf{0.93} & \textbf{1} & \textbf{0.99} & \textbf{0.67} & \textbf{0.67} & \textbf{0.67} & 0 & 0.13 & \textbf{0.32} & \textbf{0.73} & 0.73 & 0.73 & 0.6 & 0.6 & 0.6 & \textbf{0.59} & \textbf{0.68} & \textbf{0.69} \\ \hline
\end{tabular}%
}
\end{table}

\begin{table}[t]
\caption[Fine-grained RCA performance of TORAI on the Train Ticket dataset.]{Fine-grained RCA performance of TORAI and baselines on the Train Ticket dataset, across six fault types. The best results are in \textbf{bold} iff the t-test reported a significant difference compared to other baselines ($p<0.05$). For all baselines, we select their best setup when taking different data sources.}
\label{tab:rq1-tt-fine}
\resizebox{\columnwidth}{!}{%
\setlength\tabcolsep{3pt}
\begin{tabular}{l|rrr|rrr|rrr|rrr|rrr|rrr|rrr}
\hline
\multicolumn{1}{c|}{\multirow{2}{*}{Method}} & \multicolumn{3}{c|}{\textbf{CPU}} & \multicolumn{3}{c|}{\textbf{MEM}} & \multicolumn{3}{c|}{\textbf{DISK}} & \multicolumn{3}{c|}{\textbf{SOCKET}} & \multicolumn{3}{c|}{\textbf{DELAY}} & \multicolumn{3}{c|}{\textbf{LOSS}} & \multicolumn{3}{c}{\textbf{AVERAGE}} \\ \cline{2-22} 
\multicolumn{1}{c|}{} & \multicolumn{1}{c}{\textit{T1}} & \multicolumn{1}{c}{\textit{T3}} & \multicolumn{1}{c|}{\textit{A5}} & \multicolumn{1}{c}{\textit{T1}} & \multicolumn{1}{c}{\textit{T3}} & \multicolumn{1}{c|}{\textit{A5}} & \multicolumn{1}{c}{\textit{T1}} & \multicolumn{1}{c}{\textit{T3}} & \multicolumn{1}{c|}{\textit{A5}} & \multicolumn{1}{c}{\textit{T1}} & \multicolumn{1}{c}{\textit{T3}} & \multicolumn{1}{c|}{\textit{A5}} & \multicolumn{1}{c}{\textit{T1}} & \multicolumn{1}{c}{\textit{T3}} & \multicolumn{1}{c|}{\textit{A5}} & \multicolumn{1}{c}{\textit{T1}} & \multicolumn{1}{c}{\textit{T3}} & \multicolumn{1}{c|}{\textit{A5}} & \multicolumn{1}{c}{\textit{T1}} & \multicolumn{1}{c}{\textit{T3}} & \multicolumn{1}{c}{\textit{A5}} \\ \hline
BARO & 0.07 & 0.33 & 0.32 & \textbf{0.33} & 0.8 & 0.75 & \textbf{1} & \textbf{1} & \textbf{1} & 0 & 0 & 0.01 & 0.47 & \textbf{0.67} & \textbf{0.63} & 0.53 & 0.53 & 0.53 & 0.4 & 0.56 & 0.54 \\
CausalRCA & \textbf{0.33} & 0.47 & \textbf{0.49} & 0 & 0.03 & 0.02 & 0.43 & 0.73 & 0.68 & 0 & 0 & 0 & 0.13 & 0.13 & 0.13 & 0.03 & 0.27 & 0.22 & 0.15 & 0.27 & 0.26 \\
CIRCA & 0.01 & 0.05 & 0.05 & 0 & 0.23 & 0.15 & 0.05 & 0.16 & 0.12 & 0.02 & 0.12 & 0.07 & 0 & 0 & 0.09 & 0 & 0.09 & 0.11 & 0.01 & 0.11 & 0.10 \\
HeMiRCA & 0 & 0.07 & 0.05 & 0.07 & 0.07 & 0.07 & 0 & 0 & 0 & 0 & 0 & 0 & 0.07 & 0.13 & 0.12 & 0 & 0 & 0 & 0.02 & 0.05 & 0.04 \\
MicroCause & 0.06 & 0.11 & 0.11 & 0 & 0.02 & 0.02 & 0.12 & 0.12 & 0.12 & 0 & 0.01 & 0.04 & 0 & 0 & 0.03 & 0 & 0 & 0 & 0.03 & 0.04 & 0.05 \\ 
RCD & 0.03 & 0.07 & 0.05 & 0 & 0 & 0 & 0 & 0 & 0 & 0.2 & 0.23 & 0.23 & 0 & 0 & 0 & 0 & 0 & 0 & 0.04 & 0.05 & 0.05 \\
\hline
\textbf{TORAI} & 0.2 & \textbf{0.6} & \textbf{0.51} & 0.27 & \textbf{1} & \textbf{0.85} & \textbf{1} & \textbf{1} & \textbf{1} & 0.07 & \textbf{0.47} & \textbf{0.43} & \textbf{0.6} & 0.6 & 0.6 & \textbf{0.6} & \textbf{0.73} & \textbf{0.68} & \textbf{0.46} & \textbf{0.73} & \textbf{0.68} \\ \hline
\end{tabular}%
}
\end{table}

In this section, we evaluate the fine-grained RCA performance of our proposed TORAI alongside baseline RCA methods across all three datasets. Tables \ref{tab:rq1-ob-fine}, \ref{tab:rq1-ss-fine}, and \ref{tab:rq1-tt-fine} report the fine-grained RCA performance of all methods on the Online Boutique, Sock Shop, and Train Ticket datasets. The fine-grained ground truths are derived from the indicators linked to the fault injection operations. For instance, when a CPU hog fault is injected into the order service, the fine-grained ground truth indicator is the "order-cpu" metric. We measure and report the accuracy for each fault type. Additionally, we perform statistical analysis using t-tests to check for pairwise differences among all RCA methods.

\textbf{(1) TORAI outperforms all baselines in fine-grained RCA}. In the Online Boutique dataset (Table \ref{tab:rq1-ob-fine}), TORAI achieves an average accuracy of 0.68 when diagnosing the root cause indicators. Meanwhile, BARO, the second-best method, achieves the average accuracy of 0.39. This is because BARO only uses hypothesis testing to infer the root cause while our TORAI also uses causal information and cluster the anomaly severity behaviour.

\textbf{(2) Hypothesis testing-based and causal inference-based methods deliver competitive fine-grained RCA performance.} BARO, CausalRCA, and CIRCA demonstrate strong fine-grained RCA capabilities. On the Train Ticket dataset, BARO achieves an average A5 score of 0.54 (the second highest), while our TORAI achieves a score of 0.68. Notably, BARO relies solely on hypothesis testing without considering causal relationships between time series and their anomaly severity. CausalRCA consistently performs well in fine-grained RCA, ranking among the top methods (TORAI, CIRCA, BARO). It constructs a causal graph using DAG-GNN and applies PageRank to identify root causes. CIRCA achieves an average A5 score of 0.67 on the Sock Shop dataset, while TORAI achieves a slightly higher score of 0.69. CIRCA combines causal inference with regression-based hypothesis testing. Notably, TORAI integrates causal inference in its CausalRanker, hypothesis testing in FineGrainer, and clustering in SymptomCluster, enabling it to achieve the best overall performance.

\begin{table}[h]
\centering
\caption{Efficiency comparison.}
\label{tab:r2-efficiency-1}
\begin{tabular}{l|r|r|r}
\hline
\multicolumn{1}{c|}{Method} & \multicolumn{1}{c|}{\textbf{\begin{tabular}[c]{@{}c@{}}Online\\ Boutique\end{tabular}}} & \multicolumn{1}{c|}{\textbf{\begin{tabular}[c]{@{}c@{}}Sock\\ Shop\end{tabular}}} & \multicolumn{1}{c}{\textbf{\begin{tabular}[c]{@{}c@{}}Train\\ Ticket\end{tabular}}} \\ \hline \hline 
BARO & 0.01 & 0.01 & 0.01 \\
TraceRCA & 2.55 & - & 14.75 \\
RCD & 3.91 & 2.96 & 20.87 \\
PDiagnose & 3.24 & 1.26 & 61.53 \\
MicroRank & 77.56 & - & 91.04 \\
HeMiRCA & 62.08 & 55.38 & 206.02 \\
CIRCA & 4.65 & 3.5 & 312.77 \\
MicroCause & 177.09 & 129.13 & 3935.85 \\
CausalRCA & 179.45 & 397.12 & (*) \\ 
\hline \hline
\textbf{TORAI} & 12.9 & 15.63 & 20.59 \\ \hline 
\end{tabular}%

\vspace{5pt}
{\footnotesize (*) Exceeding the limit of 2 hours per case. (-) Sock Shop's trace data is unavailable.
}\
\end{table}

\subsection{RQ3. Efficiency in Root Cause Analysis} \label{sec:experimental-rq2}

\quad \textbf{(1) TORAI analyses the root cause of failures in seconds.} TORAI takes an average of 12.9, 15.63, and 20.59 seconds to perform RCA on the Online Boutique, Sock Shop, and Train Ticket datasets, respectively. In comparison, MicroCause takes 177.09, 129.13, and 3935.85 seconds while RCD, completes RCA in 3.91, 2.96, and 20.87 seconds.

\textbf{(2) TORAI's efficiency remains stable regardless of the system's scale.} TORAI takes an average of 12.9 seconds to perform RCA on the Online Boutique system (11 services), and 20.59 seconds on the Train Ticket system (64 services). In contrast, most baselines tend to slow down considerably when applied to larger systems. For instance, CausalRCA takes 180 seconds to handle Online Boutique but 20200 seconds to handle a case Train Ticket (100 times slower). This is because CausalRCA uses deep neural networks to perform causal discovery on all input time series, leading to a huge number of possible edges to analyse. BARO is the fastest method since it uses a simple statistical method applied to time series data individually.

In summary, our TORAI can efficiently analyse root causes using multi-source telemetry data with minimal overhead, providing results in seconds even for microservice systems with a large number of services.

\subsection{RQ4. Ablation Study} \label{sec:experimental-rq3}

In this section, we conduct an ablation analysis to assess the contribution of each constituent component to TORAI's overall performance. Table \ref{tab:r3-ablation} presents the effectiveness comparison of TORAI and its three components: SeverityScorer, CausalRanker, and FineGrainer, across three datasets. Specifically, SeverityScorer ranks the root cause services based on the average value of the severity vector [$\rho_m$,~$\rho_l$,~$\rho_{tc}$]. CausalRanker uses RCD \citep{Azam2022rcd} to rank the root cause services from multi-source time  
series data,~(Sec. \ref{sec:torai-baselines}). Meanwhile, FineGrainer performs hypothesis testing on multi-source time series data and ranks root causes based on the magnitude of $\gamma$.

\begin{table}[h]
\centering
\caption{Ablation Study of TORAI}
\label{tab:r3-ablation}
\begin{tabular}{c|l|rrr|ccc|rrr}
\hline
\multirow{2}{*}{\begin{tabular}[c]{@{}c@{}}Data\\ Source\end{tabular}} & \multicolumn{1}{c|}{\multirow{2}{*}{Method}} & \multicolumn{3}{c|}{\textbf{Online Boutiq}} & \multicolumn{3}{c|}{\textbf{Sock Shop}} & \multicolumn{3}{c}{\textbf{Train Ticket}} \\ \cline{3-11} 
 &  & \multicolumn{1}{c}{\textit{T1}} & \multicolumn{1}{c}{\textit{T3}} & \multicolumn{1}{c|}{\textit{A5}} & \textit{T1} & \textit{T3} & \textit{A5} & \multicolumn{1}{c}{\textit{T1}} & \multicolumn{1}{c}{\textit{T3}} & \multicolumn{1}{c}{\textit{A5}} \\ \hline \hline
\multirow{3}{*}{\begin{tabular}[c]{@{}c@{}}Metric,\\Log\end{tabular}} & CausalRanker & 0.6 & 0.81 & 0.77 & \multicolumn{1}{r}{0.39} & \multicolumn{1}{r}{0.73} & \multicolumn{1}{r|}{0.66} & 0.09 & 0.2 & 0.17 \\
 & SeverityScorer & 0.8 & 0.93 & 0.91 & \multicolumn{1}{r}{0.8} & \multicolumn{1}{r}{0.95} & \multicolumn{1}{r|}{0.93} & 0.72 & 0.83 & 0.81 \\
 & FineGrainer & 0.69 & 0.82 & 0.81 & 0.07 & 0.96 & 0.76 & 0.69 & 0.82 & 0.81 \\
 & \textbf{TORAI} & 0.8 & 0.95 & 0.91 & \multicolumn{1}{r}{0.84} & \multicolumn{1}{r}{0.96} & \multicolumn{1}{r|}{0.94} & 0.74 & 0.84 & 0.83 \\ \hline
\multirow{3}{*}{\begin{tabular}[c]{@{}c@{}}Metric,\\Log,\\Trace\end{tabular}} & CausalRanker & 0.63 & 0.8 & 0.79 & - & - & - & 0.1 & 0.64 & 0.54 \\
 & SeverityScorer & 0.82 & 0.94 & 0.92 & - & - & - & 0.71 & 0.87 & 0.85 \\
 & FineGrainer & 0.69 & 0.82 & 0.81 & - & - & - & 0.69 & 0.82 & 0.81 \\
 & \textbf{TORAI} & 0.83 & 0.95 & 0.93 & - & - & - & 0.77 & 0.92 & 0.89 \\ \hline
\end{tabular}%

\vspace{10pt}
{\footnotesize (-) Sock Shop's trace data is unavailable.}
\end{table}

The ablation results from Table \ref{tab:r3-ablation} demonstrate that all constituent components have positive effects on TORAI's overall performance. For instance, on the Train Ticket dataset, TORAI achieves T1, T3, and A5 accuracies of 0.77, 0.92, and 0.89, respectively, while SeverityScorer achieves 0.71, 0.87, and 0.85, and FineGrainer achieves 0.69, 0.82, and 0.81. SeverityScorer relies on the anomaly severity score to rank the root causes without considering the causal structure between services, hence its performance is lower than TORAI. Similarly, FineGrainer also relies purely on the hypothesis testing results of the time series individually without taking into account the causal relationship among them, resulting in poorer performance compared to our TORAI. Meanwhile, in the same scenario, CausalRanker achieves scores of 0.1, 0.64, and 0.54, respectively. It has relatively poor performance by itself because it solely focuses on causal learning without considering the anomaly severity score.


\subsection{RQ5: How does TORAI perform in real-world scenarios?}

In this section, we further evaluate TORAI on 10 real-world failures collected from a production system. We also demonstrate the ability of TORAI in diagnose code-level failures in Section~\ref{sec:code-level}.

\begin{table}[t]
\centering
\caption{Example of collected data.} \label{tab:metrics}
\vspace{3pt}

\begin{tabular}{@{}c c@{}}
\begin{minipage}{0.45\linewidth}
\centering
{\footnotesize (a) Metrics.}\\[3pt]
\resizebox{\linewidth}{!}{%
\setlength\tabcolsep{7pt}
\begin{tabular}{c|c|c}
\hline
\textbf{time} & \textbf{docker1\_cpu} & \textbf{docker1\_mem} \\ \hline \hline
17336 & 0.216 & 0.352 \\ 
17337 & 0.115 & 0.401 \\ 
17338 & 0.116 & 0.386 \\  
17339 & 0.118 & 0.398 \\  \hline
\end{tabular}%
}
\end{minipage}
&
\begin{minipage}{0.5\linewidth}
\centering
{\footnotesize (b) Traces.}\\[3pt]
\resizebox{\linewidth}{!}{%
\setlength\tabcolsep{2pt}
\begin{tabular}{l|l|l|l|c}
\hline
\textbf{time} & \textbf{id} & \textbf{service} & \textbf{callType} & \textbf{elapsedTime} \\ \hline \hline
17336 & cf8b.. & osb\_001 & OSB & 497 \\ 
17337 & 60cf.. & os\_021 & CSF & 102 \\ 
17338 & 4a93.. & docker\_003 & Remote.. & 1310 \\
17338 & fe23.. & product.. & ListProd.. & 56 \\ \hline
\end{tabular}%
}
\end{minipage}
\\
\end{tabular}
\end{table}

\subsubsection{System and Data Description.}
The real-world failures are collected from a production microservice system of a major Internet service provider. The system consists of multiple components (e.g., load balancers, web/app servers, databases) categorised into five classes: OSB (Oracle Service Bus), service, DB, Docker, and OS. The system serves more than 50 million users. To monitor the system, engineers collect multi-source telemetry data. Due to confidentiality concerns, only metrics and traces are available, whereas logs are not collected. The dataset contains 10 failures with ground-truth labels for root cause components (e.g., services, databases). Each failure originates from a single component and propagates to others due to complex dependencies, causing their telemetry data to become abnormal. 

Regarding the collected telemetry data, metrics are recorded at multiple layers. For service instances (i.e., Docker containers), 10 metrics are collected (e.g., thread\_total, fgct, session\_used, cpu\_used, mem\_used, etc.). For physical Linux host machines, 50 metric types are collected (e.g., Agent\_ping, Buffers\_used, Zombie\_Process, ss\_total, etc.). For Oracle databases, 47 database-specific metrics are collected (e.g., UndoTbs\_Pct, Used\_Tbs\_Size, User\_Commit, tnsping\_result\_time). The dataset contains over 169,000 traces in total. The data was collected since May 31, 2020, spanning 15 days. Each fault lasts for 5 minutes, and the root cause is confirmed by engineers. There are five fault types: CPU exhaustion, network delay, packet loss, container network error, and database failures. Fault locations are in databases, hosts, or service instances (containers). Table~\ref{tab:metrics} presents examples of metrics and trace data. Each column in the metric data represents a time series, and each row contains a value for a specific timestamp. Each trace has a unique ID, timestamp, corresponding service, call type, and elapsed time.

\subsubsection{Experimental setup}
In the collected dataset, logs are not available. We only have metrics and traces. We transform metrics and traces into time series as described in Sec.~\ref{sec:method-collect}. To ensure fairness, we feed the processed time series into TORAI along with four state-of-the-art RCA methods, namely BARO, RCD, CIRCA, and PDiagnose. We use AC@1, AC@3, AC@5, and Avg@5 as evaluation metrics to assess the ability to identify the root cause service of the failures.

\begin{table}
\centering
\caption{Real-world RCA Performance.} \label{tab:real-world-performance}
\begin{tabular}{l|l|l|l|l}
\hline
Method & \textit{AC@1} & \textit{AC@3} & \textit{AC@5} & \textit{Avg@5} \\ \hline \hline
BARO & \textbf{0.70} & \underline{0.90} & \underline{0.90} & \underline{0.82} \\
CIRCA & 0.20 & 0.50 & 0.70 & 0.46 \\
PDiagnose & \underline{0.60} & 0.80 & 0.80 & 0.74 \\
RCD & 0.10 & 0.20 & 0.20 & 0.16 \\
\hline
\textbf{TORAI} & \underline{0.60} & \textbf{1.00} & \textbf{1.00} & \textbf{0.88} \\ \hline
\end{tabular}%
\end{table}

\subsubsection{Results}
Table~\ref{tab:real-world-performance} presents the experimental results, demonstrating that TORAI outperforms state-of-the-art baselines. Notably, TORAI achieves 100\% in AC@3, meaning it can correctly rank the root cause within the top three recommendations with perfect accuracy. On average, TORAI attains an Avg@5 score of 0.88, outperforming BARO (0.82), PDiagnose (0.74), RCD (0.16), and CIRCA (0.46). RCD performs poorly because it processes all time series data indiscriminately, a limitation previously highlighted in~\citep{pham2024root}. Similarly, CIRCA constructs a causal graph using the PC algorithm on all time series data, leading to suboptimal performance. PDiagnose, which relies on traces to identify the root causes without incorporating metrics, achieves relatively strong results. Meanwhile, BARO employs hypothesis testing and delivers competitive performance. However, it does not account for causal relationships or severity symptoms. TORAI achieves the highest overall performance by effectively grouping abnormal services and conducting precise causal analysis, resulting in superior accuracy compared to the baselines. A replicable notebook is available in our replication package.

\subsection{How Robust Is TORAI Under Varying Blind Spots?} \label{sec:sensitivity-blindspots}

\begin{figure}
\centering
\includegraphics[width=\linewidth]{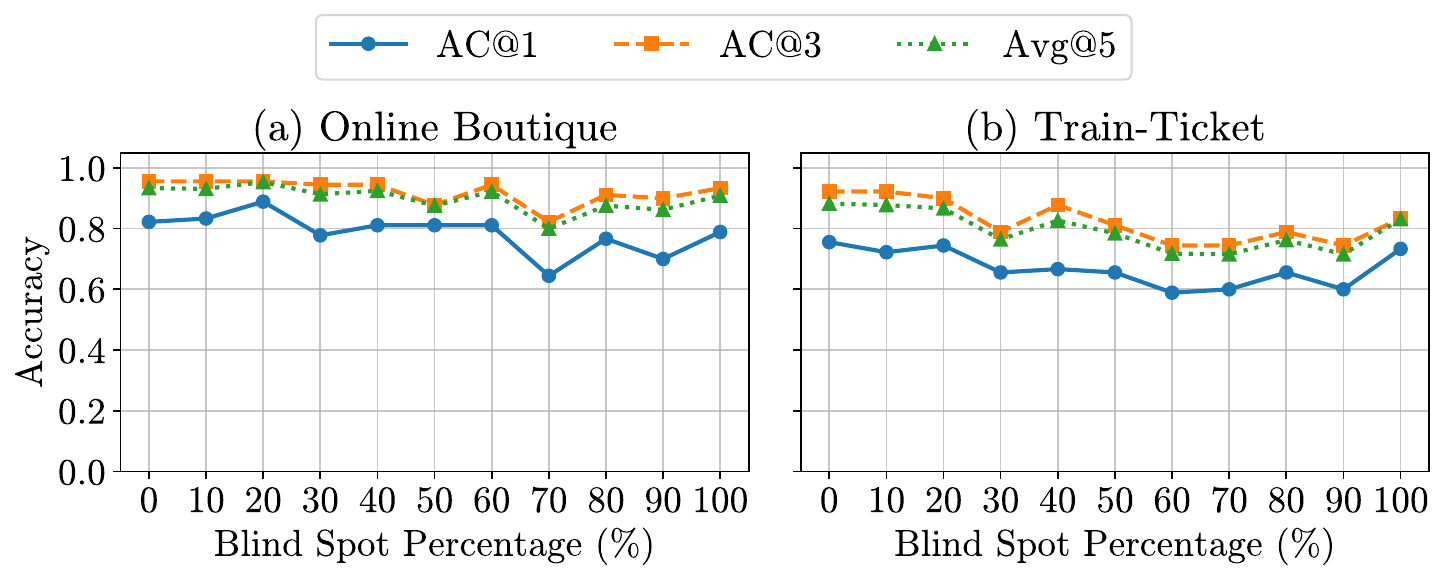}
\caption[Blind spot sensitivity analysis of TORAI.]{Sensitivity of TORAI to varying levels of blind spots on (a) the Online Boutique system and (b) the Train Ticket system.}
\label{fig:blindspot_sensitivity}
\end{figure}

To evaluate TORAI's sensitivity to varying levels of blind spots (i.e., trace unavailability), we conduct a sensitivity analysis by randomly removing traces from 0\% to 100\% of the instrumented services in 10\% increments on the Online Boutique and Train Ticket systems\footnote{The Sock Shop system has no traces by default (i.e., 100\% blind spots).}. Figure~\ref{fig:blindspot_sensitivity} presents TORAI's performance across these blind spot levels. We draw the following observations:

\textbf{(1) TORAI remains robust on the Online Boutique system.} As shown in Figure~\ref{fig:blindspot_sensitivity}(a), AC@1 ranges from 64.4\% to 88.9\% across all blind spot levels, with AC@3 remaining consistently above 82\%. Notably, even at 100\% blind spots (i.e., no trace data available), TORAI achieves 80\% AC@1 and 95\% AC@3, demonstrating that metrics and logs alone provide sufficient diagnostic signals for this system.

\textbf{(2) TORAI shows graceful degradation on the larger Train Ticket system.} Figure~\ref{fig:blindspot_sensitivity}(b) reveals that the 64-service Train Ticket system exhibits more sensitivity to blind spots, with Avg@5 dropping from 89\% (at 0\%) to a minimum of 71.6\% (at 70\%). However, the degradation is gradual rather than catastrophic. Interestingly, performance recovers at 100\% blind spots (74\% AC@1, 84\% AC@3). We found that, when traces are entirely unavailable, the method relies purely on metrics and logs without the potential noise introduced by incomplete or partial trace information.

\textbf{(3) AC@3 is more stable than AC@1 across both systems.} While AC@1 fluctuates more noticeably, AC@3 remains relatively high (above 74\% for Train Ticket and above 82\% for Online Boutique) across all blind spot levels. This indicates that the true root cause is consistently ranked within the top three candidates, which is practical for engineers who can efficiently investigate a small set of services.

These findings are further supported by TORAI's strong performance on the Sock Shop system (Table~\ref{tab:rq1-ss}), which has 100\% blind spots by default and where TORAI achieves 0.84 AC@1 and 0.94 Avg@5. In summary, TORAI's multi-source design enables reliable RCA even under the presence of blind spots, addressing a key practical concern for systems where full distributed tracing instrumentation is infeasible.

\subsection{Diagnosing code-level faults} \label{sec:code-level}

While we follow established practice to benchmark on resource and network-related faults, TORAI can diagnose other fault types (e.g., code-level faults~\citep{cotroneo2019bad}) as long as the issues manifest symptoms in telemetry data. For example, if a code-level fault produces observable symptoms, such as exception logs or erroneous traces, our method can detect the root cause services with these indicators to assist engineers in diagnosing the actual fault more efficiently.

\begin{figure}[t]
\vspace{10pt}
\centering
\begin{minipage}{\textwidth}
\begin{lstlisting}[basicstyle=\ttfamily\scriptsize, backgroundcolor=\color{gray!10}, frame=single, linewidth=0.59\textwidth]
info: Grpc.AspNetCore.Server.CallHandler[7]
      Error status code 'FailedPrecondition'
      with detail 'Can't access cart storage.
      System.OverflowException: Value was either
      too large or too small for an Int32.
    at System.Number.ThrowOverflowException()
    at cartservice.cartstore.RedisCartStore
       .AddItemAsync(String, String, String)
       in /RedisCartStore.cs:line 54'
raised.
\end{lstlisting}
\end{minipage}
\caption[A stack trace indicating that line 54 in \texttt{RedisCartStore.cs} is the root cause.]{Example of a stack trace showing a fine-grained root cause of a code-level fault, indicating that line 54 in \texttt{RedisCartStore.cs} is the source of the error.}
\label{fig:stack-traces}
\end{figure}

To demonstrate this capability, we modified the \textit{cartservice} in the Online Boutique system to inject a code-level fault, namely \textit{Incorrect parameter values}, as described in~\citep{cotroneo2019bad}. Empirical studies show that incorrect parameter values are among the five most common faults in real-world projects and that injected faults can realistically simulate actual software faults~\citep{cotroneo2019bad}. After injecting this fault, the cartservice became unstable, resulting in failures in all its callers. The front-end services reported error codes, and the resource usage of the cartservice spiked. Other services, such as recommendation and shipping, were also affected. The root cause was traced back to the cartservice (RedisCartStore.cs:54), with the fine-grained root cause indicator being the stack trace presented in Figure~\ref{fig:stack-traces}.

\begin{figure}
\centering
\includegraphics[width=0.7\linewidth]{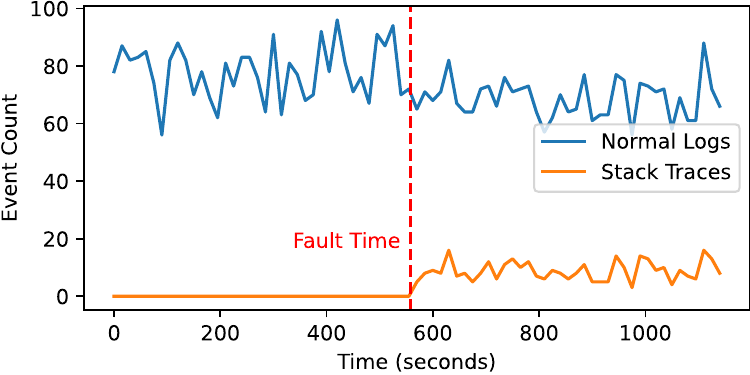}
\caption{The frequency of normal logs versus stack traces of cartservice.} \label{fig:cart-stack-trace}
\end{figure}

Our FineGrainer precisely identifies the stack trace in the cartservice as a fine-grained root cause indicator, as its frequency during the failure period deviates significantly from the expected median, which is 0. The frequency of normal logs and stack traces for the cartservice is shown in Figure~\ref{fig:cart-stack-trace}. This demonstration is included in our replication package.

%% file: papers/fse26-torai/6.discussion.tex
We now discuss threats to the validity of our study and the means we undertook to mitigate these threats. The \textit{internal threat} concerns the implementation, where bugs may affect the reliability of the results. To address this, we reused the public code for the baselines and performed experiments to replicate their results, ensuring their correctness. To avoid the influence of randomness, we repeated the experiments five times and reported the average results together with the statistical analysis. The \textit{construct threat} concerns the evaluation metrics. To address this, we used standard evaluation metrics extensively employed in the literature to evaluate the performance of RCA methods~\citep{Azam2022rcd, pham2024baro, pham2024root}. The \textit{external threat} concerns the deployment of microservice applications and data collection strategies. To address this, we follow established practice to deploy these systems and collect data, as described in Section \ref{sec:dataset}. These systems are widely recognised in academia for testing microservices-related methods~\citep{Jinjin2018Microscope, Azam2022rcd, wu2021microdiag, xin2023causalrca, wu2022automatic, he2022graph, yu2021microrank, zhou2018trainticket, Wang2021evalcausal}. Another potential threat concerns the assumptions underlying our method. We assume that root cause indicators exhibit significant distributional changes, which is supported by prior works~\citep{pham2024baro, Shan2019Ediagnosis} and recent theoretical analysis~\citep{orchard2025root}. We do not assume the root cause service always exhibits the highest raw anomaly score, instead, severity-based clustering serves as a coarse candidate selection mechanism, with final decisions made by CausalRanker and FineGrainer.

The \textit{conclusion threat} is tied to the fault types as microservices can experience various faults \citep{mariani2018localizing}. We acknowledge that different software applications and faults could have different properties and failure propagation mechanisms, which could impact the conclusions in this paper. However, we believe that these fault types are representative since they have been used in many previous studies~\citep{Jinjin2018Microscope, Azam2022rcd, wu2021microdiag, xin2023causalrca, he2022graph, dan2021tracerca, zhou2018trainticket}. Furthermore, our method is specifically designed for microservices systems. However, if a software system has multiple components interacting with each other, our method can be adapted to identify the root cause of failures in these systems. Expanding TORAI to work with other types of systems, such as distributed database systems, could be a potential future work. There may be other threats related to the underlying tools, our extracted data, that we have not considered here. To enable exploration of these potential threats and to facilitate replication and extension of our work, we make available our tools and data.

%% file: papers/fse26-torai/7.conclusion.tex
In this work, we propose TORAI, a novel unsupervised fine-grained RCA method that leverages multi-source telemetry data to accurately identify both coarse-grained root cause services and fine-grained root cause indicators of failures in microservice systems. The novelty of our proposed TORAI lies in the effective combination of unsupervised techniques, including clustering, causal analysis, and hypothesis testing, which enables the integration of multi-source telemetry data for RCA without requiring full trace coverage or labelled data, addressing limitations of existing RCA approaches. Extensive experiments on three benchmark systems and real-world failures demonstrate TORAI's superiority in both effectiveness and efficiency compared to existing methods.

%% file: papers/fse26-torai/data_availability.tex
We have integrated TORAI into our open-source benchmark RCAEval~\citep{pham2025rcaeval}, which can be accessed on GitHub at \url{https://github.com/phamquiluan/rcaeval}. Additionally, an immutable artifact for TORAI is available on Figshare~\citep{torai_sourcecode}, together with the experimental datasets~\citep{torai_datasets}.

%% file: chapters/6.benchmark.tex
\chapter[RCAEval: A Benchmark for Root Cause Analysis]{RCAEval: A Benchmark for Root Cause Analysis of Microservice Systems with Telemetry Data}\label{chap:www25}

\begin{tcolorbox}[left=2pt,right=2pt,top=0pt,bottom=0pt,
  enhanced,
  drop shadow={shadow xshift=1ex, shadow yshift=-1ex, opacity=0.3}]
\textbf{Publication:} This chapter is based on our paper titled \textbf{``RCAEval: A Benchmark for Root Cause Analysis of Microservice Systems with Telemetry Data''}, Luan Pham, Hongyu Zhang, Huong Ha, Flora Salim, and Xiuzhen Zhang, published in the Companion Proceedings of the ACM Web Conference (\textit{WWW}), 2025, pp.~777--780 (\textbf{CORE~A*}) \citep{pham2025rcaeval}.
\end{tcolorbox}
\vspace{10pt}

\noindent Chapters~\ref{chap:fse24}--\ref{chap:torai} introduced novel methods for anomaly detection and RCA, each demonstrating strong performance on their respective evaluation benchmarks. This chapter addresses the limitation identified in Section~\ref{sec:limit-benchmarks}: the absence of standard benchmarks for RCA. A persistent challenge in RCA research is the lack of standardised evaluation infrastructure. Existing studies often use different datasets, evaluation metrics, and experimental setups, making it difficult to fairly compare methods and assess their generalisability. Furthermore, many evaluations rely on limited failure scenarios or synthetic data that may not reflect the complexity of real-world microservice systems. This chapter addresses these challenges by presenting RCAEval, the first comprehensive open-source benchmark for RCA in microservice systems. RCAEval provides 735 failure cases collected from three widely-used benchmark microservice systems, along with fifteen reproducible baselines and standardised evaluation protocols for both coarse-grained and fine-grained RCA.
\vspace{10pt}

\input{papers/www25/0.abstract}

\section{Introduction}
\input{papers/www25/1.introduction}

\section{Motivation} 
\input{papers/www25/2.datasets}

\section{Evaluation Framework} 
\input{papers/www25/3.framework}

\section{Preliminary Experiments} \label{sec:experiment}

\input{papers/www25/4.results}

\section{Summary}

\input{papers/www25/6.conclusion}

\vspace{10pt}
\noindent RCAEval establishes the infrastructure for systematic, reproducible evaluation of RCA methods. With this benchmark in place, we can now conduct a comprehensive assessment of existing approaches to understand their strengths, weaknesses, and practical limitations. The next chapter presents such a study, evaluating nine causal discovery methods and twenty-one causal inference-based RCA approaches using the RCAEval framework. This systematic evaluation reveals key insights about the current state of the art and identifies opportunities for methodological improvement.
\vspace{10pt}

%% file: papers/www25/0.abstract.tex
Root cause analysis (RCA) for microservice systems has gained significant attention in recent years. However, there is still no standard benchmark that includes large-scale datasets and supports comprehensive evaluation environments. In this paper, we introduce RCAEval, an open-source benchmark that provides datasets and an evaluation environment for RCA in microservice systems. First, we introduce three comprehensive datasets comprising 735 failure cases collected from three microservice systems, covering various fault types observed in real-world failures. Second, we present a comprehensive evaluation framework that includes fifteen reproducible  baselines covering a wide range of RCA approaches, with the ability to evaluate both coarse-grained and fine-grained RCA. We hope that this ready-to-use benchmark will enable researchers and practitioners to conduct extensive analysis and pave the way for robust new solutions for RCA of microservice systems.

%% file: papers/www25/1.introduction.tex
Root cause analysis (RCA) for microservice systems is an important problem that has been studied recently, as failures are inevitable, and ensuring the reliability of microservice systems is critical. RCA aims to analyse the available telemetry data (i.e., metrics, logs, and traces) of the system during failure periods to identify the root cause service and root cause indicators (e.g., specific metrics or logs pointing to the root cause). This field has gained significant attention recently~\citep{lee2023eadro, yu2023nezha, pham2024baro, pham2024root}.
However, there is still no standard benchmark that includes large-scale datasets and a comprehensive evaluation framework~\citep{cheng2023ai}. This limitation leads to inconsistent evaluations in RCA studies, hindering understanding and impeding progress in the field~\citep{cheng2023ai}. For example, existing studies~\citep{Azam2022rcd, Li2022Circa, run2024aaai} typically evaluate their methods on only 1-2 systems with 2-3 fault types. Eadro~\citep{lee2023eadro} uses a dataset with an unrealistic load (2-3 requests per second). Existing open-source RCA resources also suffer from several limitations. For example, PyRCA~\citep{liu2023pyrca} offers only a limited set of metric-based RCA methods and relies on synthetic datasets. The AIOps 2020 dataset~\citep{li2022constructing} contains failures with metrics and traces but omits valuable log information. Pham et al.~\citep{pham2024root} evaluate only on metric-based RCA methods. As a result, existing resources are often inadequate for benchmarking purposes, hampering the development of new RCA approaches, see Table \ref{tab:intro}.

\begin{table}[ht]
\caption{Comparison of studies.}
\label{tab:intro}
\resizebox{\columnwidth}{!}{%
\begin{tabular}{l|l|c|c|c}
\hline
\textbf{Study} & \textbf{Fault Types} & \textbf{Metric} & \textbf{Log} & \textbf{Trace} \\ \hline \hline
PyRCA~\citep{liu2023pyrca} & Synthetic & \checkmark & - & - \\ \hline
AIOps 2020~\citep{li2022constructing} & Resource, Network & \checkmark & - & \checkmark \\ \hline
Pham et al.~\citep{pham2024root} & Resource, Network & \checkmark      & - & - \\ \hline 
\textbf{RCAEval (ours)} & Resource, Network, Code-level  & \checkmark & \checkmark   & \checkmark \\ \hline
\end{tabular}%
}
\end{table}

In this work, we introduce RCAEval, a benchmark including three datasets and a comprehensive evaluation environment. First, our datasets include~735 failure cases collected from three systems, covering 11 fault types observed in real-world failures. We collected multi-source telemetry data (i.e., metrics, logs, and traces), supporting a variety of RCA approaches (e.g., metric-based, multi-source RCA). Second, we open-source our evaluation framework, which includes several reproducible baselines. The prior version of this framework was used to evaluate metric-based RCA~\citep{pham2024root}. In this work, we have upgraded it to support trace-based and multi-source RCA. We also provide preliminary experiments, highlighting the need for further investigation on a robust RCA approach. We have open-sourced our benchmark, including the datasets and the evaluation framework at~\href{https://github.com/phamquiluan/RCAEval}{\textbf{https://github.com/phamquiluan/RCAEval}}.

%% file: papers/www25/2.datasets.tex

A major limitation in this field is the absence of a reproducible and open-source public benchmark for evaluating RCA in practical scenarios. Most RCA studies evaluate their methods using limited faults on limited systems~\citep{pham2024root, cheng2023ai}. For example, some works~\citep{lee2023eadro, yu2023nezha, Azam2022rcd} inject 2-3 faults into 1-2 systems, resulting in limited datasets. Others assess their solutions using private data, such as AWS~\citep{Azam2022rcd} or Oracle~\citep{Li2022Circa}. Reproducibility and fair comparison remain open challenges in RCA research~\citep{cheng2023ai}, 
hindering progress and preventing fair evaluation of new RCA approaches.

There have been some related works that introduce datasets or evaluation frameworks, but all of them suffer from several limitations, see Table~\ref{tab:intro}. PyRCA~\citep{liu2023pyrca} from Salesforce supports only metric-based RCA and relies on synthetic datasets. Our prior work~\citep{pham2024root} demonstrates that performance on synthetic datasets often fails to reflect RCA performance on real systems. Li et al.~\citep{li2022constructing} introduced datasets with metrics and traces on private systems but omitting logs and did not provide a benchmarking framework. To address these limitations, in this study, we provide a benchmark consisting of three RCA datasets and an open-source evaluation environment.

\begin{figure}
\centering
\includegraphics[width=\textwidth]{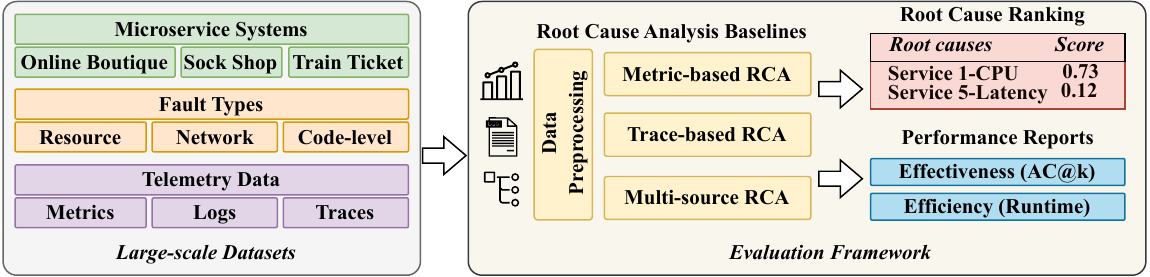}
\caption{Overview of the RCAEval benchmark.}
\end{figure}

\section{Datasets} 

RCAEval benchmark includes \textbf{three datasets}: RE1, RE2, and RE3, designed to comprehensively support benchmarking RCA in microservice systems. Together, our three datasets feature 735 failure cases collected from three microservice systems (described in Section~\ref{sec:microservice-system}) and including 11 fault types (described in Section~\ref{sec:fault-type}). Each failure case also includes annotated root cause service and root cause indicator (e.g., specific metric or log indicating the root cause). The statistics of the datasets are presented in Table~\ref{tab:dataset}.

\noindent \textbf{RE1 Dataset.} The RE1 dataset, introduced in our prior work on metric-based RCA~\citep{pham2024root}, contains 375 failure cases collected from three microservice systems (125 cases per system). These cases combine five fault types across five services, and five repetitions per fault-service pair. The RE1 dataset exclusively contains metrics data, supporting the development of metric-based RCA methods. The fault types in RE1 include CPU, MEM, DISK, DELAY, LOSS (see Section~\ref{sec:fault-type}). The number of metrics ranges from 49 to 212, depending on the system size, with smaller systems (e.g., Online Boutique, Sock Shop) having fewer metrics compared to larger system (Train Ticket). This dataset does not include logs or traces. 

\noindent \textbf{RE2 Dataset.} The RE2 dataset, newly collected for this study, supports the development of multi-source RCA methods. It includes 270 failure cases (90 cases per system), combining six fault types across five services, and three repetitions per fault-service pair. RE2 provides  multi-source telemetry data, including metrics, logs, and traces. The number of metrics ranges from 77 to 327 per failure case. Each system generates a substantial volume of logs from (8.6 to 26.9 million lines), and traces (39.6 to 76.7 million traces). The fault types include those in RE1 and an additional SOCKET fault.

\noindent \textbf{RE3 Dataset.} The RE3 dataset, also newly collected, focuses on supporting multi-source RCA methods with the ability to diagnose code-level faults. It has 90 failure cases (30 per system), involving code-level faults. The fault types in RE3 are F1, F2, F3, F4, F5 (see Section~\ref{sec:fault-type}). Like RE2, RE3 includes multi-source telemetry data (metrics, logs, and traces). This dataset emphasises diagnosing code-level faults through telemetry data, e.g., leveraging stack traces in logs or response code in traces to pinpoint root causes, making it invaluable for advancing multi-source RCA methods.

\begin{table}[t] 
\caption{Statistics of the RCAEval datasets.}
\label{tab:dataset}
\resizebox{\textwidth}{!}{%
\setlength{\tabcolsep}{3pt}
\begin{tabular}{ccccccc}
\toprule
\textbf{Dataset} & \textbf{Systems} & \textbf{Fault Types} & \textbf{Cases} & \textbf{Metrics} & \textbf{Logs (millions)} & \textbf{Traces (millions)} \\ \midrule \midrule
RE1 & 3 & 3 Resource, 2 Network & 375 & 49–212 & N/A & N/A \\ 
RE2 & 3 & 4 Resource, 2 Network & 270 & 77–376 & 8.6–26.9 & 39.6–76.7 \\
RE3 & 3 & 5 Code-level & 90 & 68–322 & 1.7–2.7 & 4.5–4.7 \\ \bottomrule
\end{tabular}%
}
\end{table}

\subsection{Microservice Systems} \label{sec:microservice-system}

We collect our three datasets from three microservice systems, ranging from 12 to 64 services. These systems are used in our previous works for evaluating RCA methods~\citep{pham2024root, pham2024baro, pham2026graph, pham2026torai}.

\noindent \textbf{1) Online Boutique.} The Online Boutique system~\citep{ob}, developed by Google, consists of 12 services forming an e-commerce application where users can browse, add, and purchase items. The services are written in five different programming languages and communicate with each other using the gRPC protocol.
 
\noindent \textbf{2) Sock Shop.} The Sock Shop system~\citep{sockshop}, developed by Weaveworks, is a sock-selling e-commerce application comprising 15 services. The services are written in three different programming languages and communicate with each other via HTTP. 
 
\noindent \textbf{3) Train Ticket.} Train Ticket~\citep{tt} is a ticket booking system with~64 services, featuring both synchronous and asynchronous communication. Compared to Sock Shop and Online Boutique, Train Ticket has more complex call chains. To the best of our knowledge, Train Ticket is the largest benchmark system for RCA evaluation.

While no single system can fully capture the diversity of real-world environments, the developers of these systems have intentionally included diverse features, such as multiple programming languages (e.g., Java, Go, Python, C\#) and communication protocols 
(e.g., HTTP, gRPC), to emulate real-world complexity.

\subsection{Fault Types} \label{sec:fault-type}

Our three datasets consist of 11 fault types (4~resource faults, 2~network faults, and 5~code-level faults). In this section, we describe these faults and the way we introduce them into the microservice systems. The RE1 dataset, which includes 3 resource faults and 2 network faults, was used in our previous metric-based RCA work~\citep{pham2024root}. In this study, we introduce two additional datasets, RE2 and RE3, which include one additional resource fault and 5 new code-level faults, covering a broader range of faults commonly found in open-source projects~\citep{cotroneo2019bad}. To the best of our knowledge, our datasets are the first to cover code-level faults for RCA in microservice systems.

\noindent \textbf{1) Resource Faults.} We introduce four resource faults into the running container (i.e., service instance) using stress-ng: CPU hog~(\textbf{CPU}), Memory leak (\textbf{MEM}), Disk stress (\textbf{DISK}), and Socket stress (\textbf{SOCK}). Symptoms of resource faults may include observable changes in resource usage of co-located containers, increased latency, and time-out requests. The system may crash when resources are severely constrained. The root cause indicator for these faults is the metric specifying resource usage (e.g., for a CPU hog, the root cause indicator is the container's CPU usage metric).

\noindent \textbf{2) Network Faults.} We use traffic control (tc) to intercept the network packets of the running container, introducing delay variations~(\textbf{DELAY}) or randomly dropping packets (\textbf{LOSS}). Symptoms of network faults may include increased latency metrics and error response codes in traces/metrics of the affected service. The root cause indicator for a DELAY fault is the latency metric, while for a LOSS fault, it is the metric showing failed requests and/or error response codes in the traces of the corresponding container.

\noindent \textbf{3) Code-Level Faults.} We modify the source code of random services to introduce five bugs commonly found in open-source projects~\citep{cotroneo2019bad}: Incorrect parameter values (\textbf{F1}), Missing parameters~(\textbf{F2}), Missing Function Call (\textbf{F3}), Incorrect Return Values (\textbf{F4}), and Missing Exception Handlers (\textbf{F5}). These were the most frequent bug types in real OpenStack. Symptoms of code-level faults may include increased failed requests, error response codes in traces, higher latency, and stack traces emitted in logs. The root cause of code-level faults is determined using the stack traces in logs of the corresponding service, which indicate the faulty line of code. If stack traces are unavailable, the root cause indicator may be derived from error logs or response codes of the affected service.

\subsection{Telemetry Data Collection Process}

\begin{figure}
\centering
\includegraphics[width=0.8\linewidth]{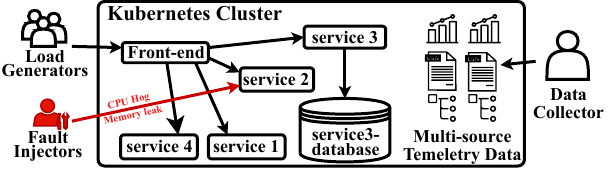}
\caption{Illustration of our data collection setup.}
\label{fig:www25-experiment-setup}
\end{figure}

We deploy three microservice systems to Kubernetes clusters and generate a random load of 10–200 requests per second across all services. We use standard, well-known open-source tools to monitor and collect telemetry data. To gather metrics, we use Prometheus, cAdvisor, and Istio to monitor and collect both application-level and resource-level metrics. For logs, we use Vector and Loki to gather logs from all service instances and store them in Elasticsearch. Traces are collected using Jaeger and sent to Elasticsearch for storage, see Figure~\ref{fig:www25-experiment-setup}. These telemetry data span the three pillars of observability~\citep{googlesre}. We allow the microservice systems to run normally to collect normal telemetry data. Then, we inject a fault into a randomly selected running service and collect the abnormal telemetry data. To ensure data quality, we engaged a DevOps engineer with five years of experience in microservices to assist with system deployment, data collection, and data verification. 

\begin{figure}[t]
    \centering
    \begin{subfigure}[t]{0.30\textwidth}
        \centering
        \resizebox{\textwidth}{!}{
        \begin{tabular}{|c|c|c|}
        \hline
        \textbf{time} & \textbf{cart\_cpu} & \textbf{cart\_mem} \\ \hline
        17336 & 0.216 & 0.352 \\ 
        17337 & 0.115 & 0.401 \\ 
        17338 & 0.116 & 0.386 \\  
        17339 & 0.118 & 0.398 \\  \hline
        \end{tabular}
        }
        \caption{Metrics}
        \label{fig:data-metrics}
    \end{subfigure}
    \hspace{-11pt}
    \begin{subfigure}[t]{0.32\textwidth}
        \centering
        \resizebox{\textwidth}{!}{
        \begin{tabular}{|c|l|l|}
        \hline
        \textbf{time} & \textbf{service} & \textbf{message} \\ \hline
        17336 & cart & GetCart called... \\ 
        17337 & currency & Getting values...  \\ 
        17338 & frontend & request complete.  \\ 
        17339 & frontend & request started.  \\ \hline
        \end{tabular}}
        \caption{Logs}
        \label{fig:data-logs}
    \end{subfigure}
    \hspace{-11pt}
    \begin{subfigure}[t]{0.399\textwidth}
        \centering
        \resizebox{\textwidth}{!}{
        \setlength\tabcolsep{2pt}
        \begin{tabular}{|l|l|l|l|l|}
        \hline
        \textbf{time} & \textbf{id} & \textbf{service} & \textbf{operation} & \textbf{duration} \\ \hline
        17336 & cf8b.. & frontend & GetCurrencies & 497 \\ 
        17337 & 60cf.. & currency & Convert & 102 \\ 
        17338 & 4a93.. & frontend & GetProduct & 1310 \\
        17338 & fe23.. & product.. & ListProducts & 56 \\ \hline
        \end{tabular}}
        \caption{Traces}
        \label{fig:data-traces}
    \end{subfigure}

    \caption{Examples of heterogeneous telemetry data in microservice systems.}
    \label{fig:heterogeneous-data}
\end{figure}

\subsection{Data Format} \label{sec:data-statistic}

The raw telemetry data collected is stored as CSV files. The key structures are presented in Figure~\ref{fig:heterogeneous-data}. A complete dataset with full structure can be downloaded from our GitHub repository. Metrics are stored as time series, with each row corresponding to a timestamp at which the metrics were collected. Logs for a failure case are stored in a single CSV file, with each row containing the timestamp, the service name, and the corresponding log message. Similarly, traces for a failure case are stored in a CSV file, where each row includes the timestamp, trace\_id, span\_id, service name, operation, duration, and response code if available. This follows the standard span model of distributed tracing~\citep{janes2023open}.

%% file: papers/www25/3.framework.tex
\begin{sidewaystable}[p]
\caption{RCA performance of eight baselines on the Train Ticket system of the RE2 dataset, across six fault types.}
\label{tab:prelim-1}
\resizebox{\textheight}{!}{%
\setlength\tabcolsep{1pt}
\begin{tabular}{c|l|rrr|rrr|rrr|rrr|rrr|rrr|rrr}
\hline
\multirow{2}{*}{\begin{tabular}[c]{@{}c@{}}Data\\ Source\end{tabular}} & \multicolumn{1}{c|}{\multirow{2}{*}{Method}} & \multicolumn{3}{c|}{\textbf{CPU}} & \multicolumn{3}{c|}{\textbf{MEM}} & \multicolumn{3}{c|}{\textbf{DISK}} & \multicolumn{3}{c|}{\textbf{SOCKET}} & \multicolumn{3}{c|}{\textbf{DELAY}} & \multicolumn{3}{c|}{\textbf{LOSS}} & \multicolumn{3}{c}{\textbf{AVERAGE}} \\ \cline{3-23} 
 & \multicolumn{1}{c|}{} & \multicolumn{1}{c}{\textit{AC@1}} & \multicolumn{1}{c}{\textit{AC@3}} & \multicolumn{1}{c|}{\textit{Avg@5}} & \multicolumn{1}{c}{\textit{AC@1}} & \multicolumn{1}{c}{\textit{AC@3}} & \multicolumn{1}{c|}{\textit{Avg@5}} & \multicolumn{1}{c}{\textit{AC@1}} & \multicolumn{1}{c}{\textit{AC@3}} & \multicolumn{1}{c|}{\textit{Avg@5}} & \multicolumn{1}{c}{\textit{AC@1}} & \multicolumn{1}{c}{\textit{AC@3}} & \multicolumn{1}{c|}{\textit{Avg@5}} & \multicolumn{1}{c}{\textit{AC@1}} & \multicolumn{1}{c}{\textit{AC@3}} & \multicolumn{1}{c|}{\textit{Avg@5}} & \multicolumn{1}{c}{\textit{AC@1}} & \multicolumn{1}{c}{\textit{AC@3}} & \multicolumn{1}{c|}{\textit{Ag@5}} & \multicolumn{1}{c}{\textit{AC@1}} & \multicolumn{1}{c}{\textit{AC@3}} & \multicolumn{1}{c}{\textit{Avg@5}} \\ \hline \hline
\multirow{5}{*}{Metric} & BARO & 0.47 & 0.8 & 0.72 & 0.93 & 1 & 0.99 & 1 & 1 & 1 & 0.6 & 0.87 & 0.83 & 0.47 & 0.67 & 0.63 & 0.53 & 0.6 & 0.64 & 0.67 & 0.82 & 0.8 \\
 & CausalRCA & 0.4 & 0.63 & 0.59 & 0.1 & 0.27 & 0.24 & 0.43 & 0.83 & 0.75 & 0.23 & 0.5 & 0.45 & 0.13 & 0.23 & 0.21 & 0.03 & 0.37 & 0.33 & 0.22 & 0.47 & 0.43 \\
 & CIRCA & 0.27 & 0.27 & 0.28 & 0.47 & 0.73 & 0.68 & 0.53 & 0.67 & 0.64 & 0.27 & 0.53 & 0.52 & 0.2 & 0.27 & 0.28 & 0.2 & 0.33 & 0.35 & 0.32 & 0.47 & 0.46 \\
 & MicroCause & 0.19 & 0.44 & 0.4 & 0 & 0.09 & 0.07 & 0.4 & 0.4 & 0.4 & 0 & 0.17 & 0.15 & 0 & 0.22 & 0.13 & 0 & 0 & 0.07 & 0.1 & 0.22 & 0.2 \\ 
 & RCD & 0.13 & 0.13 & 0.16 & 0.07 & 0.07 & 0.07 & 0 & 0.07 & 0.05 & 0.13 & 0.33 & 0.29 & 0.13 & 0.13 & 0.15 & 0.07 & 0.07 & 0.07 & 0.09 & 0.13 & 0.13 \\ \hline
\multirow{2}{*}{Trace} & MicroRank & 0.21 & 0.43 & 0.34 & 0.25 & 0.38 & 0.33 & 0 & 0.36 & 0.27 & 0.3 & 0.4 & 0.36 & 0.08 & 0.31 & 0.23 & 0.14 & 0.36 & 0.3 & 0.16 & 0.37 & 0.31 \\
 & TraceRCA & 0.64 & 0.79 & 0.74 & 0.63 & 0.88 & 0.83 & 0.64 & 0.71 & 0.74 & 0.6 & 0.8 & 0.76 & 0.85 & 0.85 & 0.88 & 0.57 & 0.71 & 0.67 & 0.66 & 0.79 & 0.77 \\ \hline
\multirow{4}{*}{\begin{tabular}[c]{@{}c@{}}Multi-\\ Source\end{tabular}} 
 & BARO & 0.47 & 0.8 & 0.75 & 0.93 & 1 & 0.99 & 1 & 1 & 1 & 0.6 & 0.8 & 0.79 & 0.47 & 0.67 & 0.61 & 0.67 & 0.67 & 0.71 & 0.69 & 0.82 & 0.81 \\ 
 & CIRCA & 0 & 0.07 & 0.09 & 0.07 & 0.13 & 0.21 & 0 & 0.07 & 0.09 & 0.07 & 0.13 & 0.16 & 0.07 & 0.07 & 0.07 & 0.13 & 0.2 & 0.17 & 0.06 & 0.11 & 0.13 \\
 & PDiagnose & 0.6 & 0.87 & 0.81 & 0.4 & 0.47 & 0.48 & 0.33 & 0.73 & 0.69 & 0.33 & 0.67 & 0.6 & 0.87 & 0.87 & 0.87 & 0.33 & 0.6 & 0.57 & 0.48 & 0.7 & 0.67 \\ 
 & RCD & 0.17 & 0.84 & 0.72 & 0 & 0.44 & 0.39 & 0.07 & 0.76 & 0.62 & 0.21 & 0.77 & 0.66 & 0.05 & 0.28 & 0.25 & 0.07 & 0.75 & 0.6 & 0.1 & 0.64 & 0.54 \\
 \hline
\end{tabular}%
}
\end{sidewaystable}

To ensure the comprehensiveness of RCAEval, we also provide an evaluation framework as an open-source library alongside our datasets. Our evaluation framework includes fifteen baselines covering a wide range of state-of-the-art RCA approaches and offers functionalities for data processing and benchmark evaluation at both coarse-grained and fine-grained levels. The RCAEval evaluation framework is an extension of our previous work \citep{pham2024root}, which focused on metric-based RCA and coarse-grained RCA. In this work, we expand it by incorporating trace-based and multi-source RCA baselines. RCAEval is released as an open-source library and can be installed via PyPI. Comprehensive documentation on installing and using the framework with our datasets, as well as guidance on extending it with new methods and datasets, is available on our GitHub repository. The documentation also includes basic usage examples and detailed instructions for reproducibility.

\subsection{Evaluation Baselines}

Our evaluation framework features 15 baselines covering a variety of state-of-the-art RCA methods. \textbf{Metric-based RCA baselines} include causal inference-based methods such as RUN, CausalRCA, CIRCA, RCD, MicroCause, EasyRCA, MSCRED, as well as non-causal methods such as BARO, and $\epsilon$-Diagnosis~\citep{pham2024root, liu2023pyrca, Li2022Circa, Azam2022rcd}. \textbf{Trace-based RCA baselines} include TraceRCA and MicroRank~\citep{yu2021microrank, dan2021tracerca}. \textbf{Multi-source RCA baselines} include PDiagnose, multi-source BARO, multi-source RCD, multi-source CIRCA~\citep{pham2024baro, Li2022Circa, Azam2022rcd}. For baselines like RUN, CausalRCA, CIRCA, RCD, MicroCause, EasyRCA, MSCRED, BARO, $\epsilon$-Diagnosis, MicroRank, and TraceRCA, we adapt their available implementations and use the default hyperparameter settings recommended in their respective papers. We verified their correctness by reproducing the results presented in the original and related studies. For multi-source BARO, multi-source RCD, and multi-source CIRCA, we updated their source code to handle time series data from logs and traces. For PDiagnose, we follow previous works \citep{yu2023nezha, zhang2023diagfusion, hou2021pdiagnose} to implement it since its source code is unavailable. 
Previous works such as PyRCA~\citep{liu2023pyrca} and Pham et al.~\citep{pham2024root} offer only a limited set of metric-based RCA methods, while our framework provides a more comprehensive set of baselines by also including trace-based and multi-source RCA methods.

\subsection{Evaluation Metrics}

We support evaluation at both the coarse-grained level (i.e., root cause service) and the fine-grained level (i.e., root cause indicator). The evaluation script executes the analysis and stores the results in a report file. We currently support two standard metrics~\citep{pham2024root, pham2024baro}: $AC@k$ and $Avg@k$ to measure the RCA performance. Given a set of failure cases A, $AC@k$ is calculated as $AC@k = \frac{1}{|A|} \sum\nolimits_{a\in A}\frac{\sum_{i<k}R^a[i]\in V^a_{rc}}{min(k, |V^a_{rc}|)}$, where $R^a[i]$ is the $i$th ranking result for the failure case $a$ by an RCA method, and $V^a_{rc}$ is the true root cause set of case $a$. $AC@k$ represents the probability the top $k$ results of the given method include the true root causes. Its values range from $0$ to $1$, with higher values indicating better performance. $Avg@k$, which shows the overall RCA performance, is measured as $Avg@k = \frac{1}{k}\sum_{j=1}^k AC@j$.

%% file: papers/www25/4.results.tex
We conduct preliminary experiments on our benchmark to evaluate the performance of existing baselines on the collected datasets, highlighting both the potential and challenges in the field. Due to space constraints, we select 11 baselines: 5 metric-based RCA methods (BARO, CausalRCA, CIRCA, MicroCause), 2 trace-based RCA methods (MicroRank, TraceRCA), and 4 multi-source RCA methods (PDiagnose, multi-source CIRCA, multi-source RCD, multi-source BARO). These methods are used to diagnose 4 resource faults (CPU, MEM, DISK, SOCK) and 2 network faults (DELAY, LOSS) using data collected from the Train Ticket system in the RE2 dataset. The RCA performance is evaluated using the AC@1, AC@3, and Avg@5 metrics, with coarse-grained results presented in Table~\ref{tab:prelim-1}. A demonstration of diagnosing root causes for code-level faults (e.g., F1 to F5) is available on our GitHub repository.

Our preliminary results show that there is still ample room for further improvement. Existing methods mostly obtain moderate results. For example, CIRCA and RCD obtain the best average Avg@5 score of 0.46 and 0.54, respectively. Notably, BARO shows encouraging results when obtaining high accuracy in diagnosing the resource fault (e.g. DISK), however, it shows limitations when dealing with network faults (e.g. DELAY, LOSS). Hence, we believe further research is needed to develop a holistic RCA solution.

%% file: papers/www25/6.conclusion.tex
In this chapter, we presented RCAEval, the first comprehensive open-source benchmark for RCA of microservice systems. RCAEval provides three datasets totalling 735 failure cases across 11 fault types collected from three microservice systems, together with an evaluation framework featuring fifteen reproducible RCA baselines covering metric-based, trace-based, and multi-source paradigms. We hope that this benchmark will be useful for industry practitioners and academic researchers in the field. The source code, datasets, and documentation are publicly available at \url{https://github.com/phamquiluan/rcaeval}.

%% file: chapters/7.evaluation.tex
\chapter[An Evaluation on Causal Inference-based Root Cause Analysis]{Causal Inference-based Root Cause Analysis: How Far Are We?}\label{chap:ase24}


\begin{tcolorbox}[left=2pt,right=2pt,top=0pt,bottom=0pt,
  enhanced,
  drop shadow={shadow xshift=1ex, shadow yshift=-1ex, opacity=0.3}]
\textbf{Publication:} This chapter is based on our paper titled \textbf{``Root Cause Analysis for Microservice System Based on Causal Inference: How Far Are We?''}, Luan Pham, Huong Ha, and Hongyu Zhang, published in the 39th IEEE/ACM International Conference on Automated Software Engineering (\textit{ASE}), 2024, pp.~706--715 (\textbf{CORE~A*}) \citep{pham2024root}.
\end{tcolorbox}
\vspace{10pt}

\noindent The previous chapters introduced novel RCA methods (Chapters~\ref{chap:fse24}--\ref{chap:torai}) and established a standardised benchmark for their evaluation (Chapter~\ref{chap:www25}). This chapter complements these contributions by conducting a systematic evaluation of existing causal inference-based RCA methods---a dominant paradigm in the field. While causal inference offers principled approaches to identifying root causes through causal reasoning, the practical effectiveness of these methods across diverse scenarios remains unclear. Using the RCAEval benchmark, this chapter evaluates nine causal discovery methods and twenty-one failure diagnosis approaches to assess their effectiveness, efficiency, and robustness. The results provide valuable insights into when different methods excel, where they fall short, and what opportunities exist for future research.
\vspace{10pt}

\input{papers/ase24/0.abstract}

\section{Introduction}
\input{papers/ase24/1.introduction}

\section{Background}
\input{papers/ase24/2.background}

\section{Study Design}
\input{papers/ase24/3.study_design}

\section{Results}
\input{papers/ase24/4.results}

\section{Discussion}
\input{papers/ase24/5.discussion}

\section{Threats to Validity}
\input{papers/ase24/6.threats_to_validity}

\section{Related Work}
\input{papers/ase24/7.related_work}

\section{Summary}
\input{papers/ase24/8.conclusion.tex}

\vspace{10pt}
\noindent This evaluation reveals that no single causal inference-based method performs best across all settings, highlighting both the complexity of RCA in microservice systems and the need for continued methodological innovation. The findings complement the novel methods introduced in earlier chapters by contextualizing their contributions relative to existing approaches and identifying directions for future improvement. The next chapter concludes this thesis by synthesising our contributions, discussing their broader implications, and outlining promising avenues for future research in automated anomaly detection and RCA.
\vspace{10pt}


%% file: papers/ase24/0.abstract.tex
Microservice architecture has become a popular architecture adopted by many cloud applications. However, identifying the root cause of a failure in microservice systems is still a challenging and time-consuming task. In recent years, researchers have introduced various causal inference-based root cause analysis methods to assist engineers in identifying the root causes. To gain a better understanding of the current status of causal inference-based root cause analysis techniques for microservice systems, we conduct a comprehensive evaluation of nine causal discovery methods and twenty-one root cause analysis methods. Our evaluation aims to understand both the effectiveness and efficiency of causal inference-based root cause analysis methods, as well as other factors that affect their performance. Our experimental results and analyses indicate that no method stands out in all situations; each method tends to either fall short in effectiveness, efficiency, or shows sensitivity to specific parameters. Notably, the performance of root cause analysis methods on synthetic datasets may not accurately reflect their performance in real systems. Indeed, there is still a large room for further improvement. Furthermore, we also suggest possible future work based on our findings.

%% file: papers/ase24/1.introduction.tex
\begin{figure}
\centering
\includegraphics[width=0.65\textwidth]{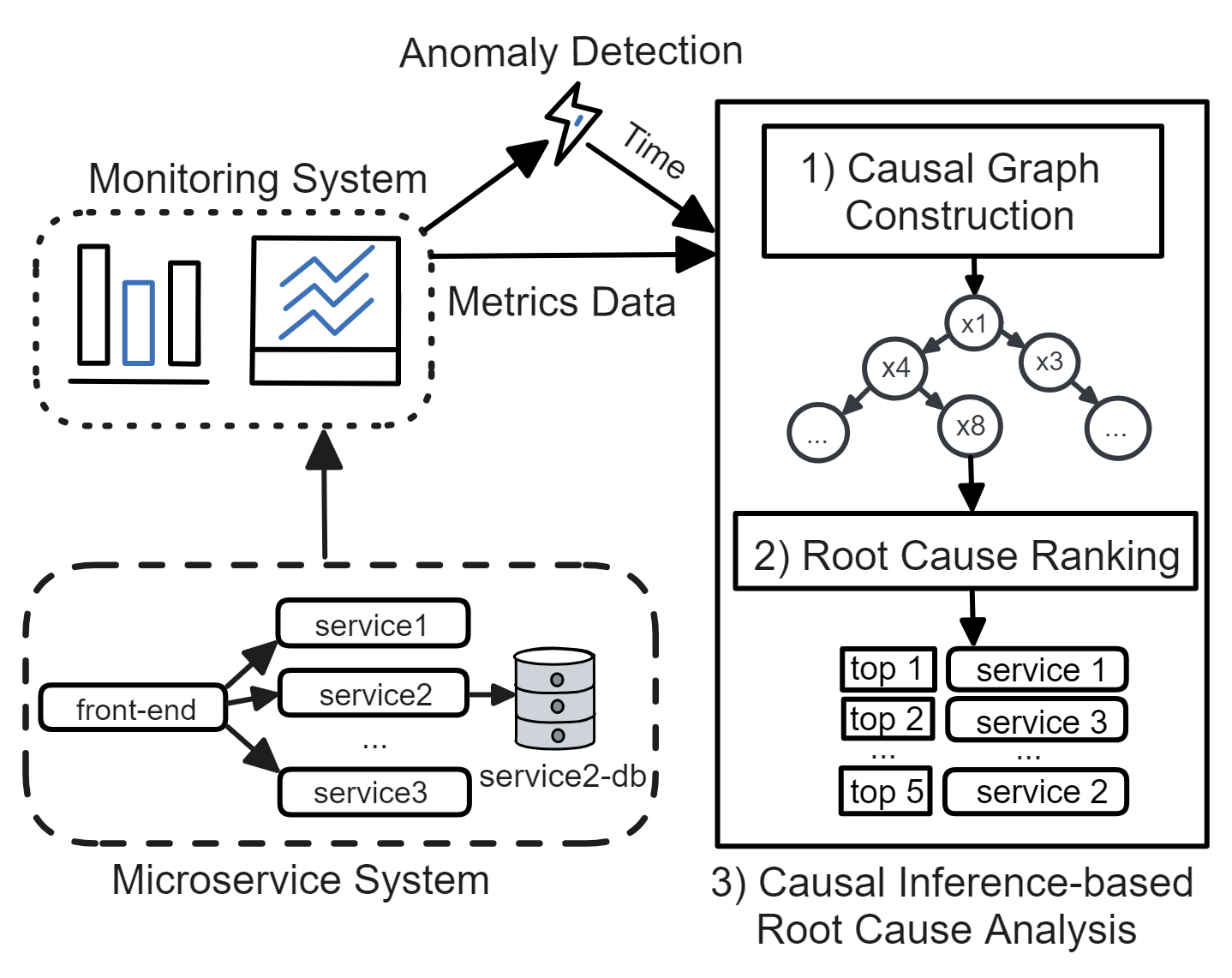}
\caption[Overview of the causal inference-based root cause analysis.]{Overview of the causal inference-based root cause analysis for microservice systems using metrics data.} \label{fig:intro}
\end{figure}

In recent years, microservice architecture has become a popular paradigm in the development of large-scale cloud-based systems (e.g., social networks, online shopping, video streaming services) owing to its scalability, resiliency, and elasticity. A microservice system consists of multiple loosely coupled services where each service can be developed and updated without requiring too much knowledge of the rest of the system. This property makes microservice systems highly adaptable for cloud environments and easy to be deployed, scaled, and maintained by different engineering groups. A large number of prominent enterprises, including Amazon, Netflix, Twitter, and Spotify, have extensively employed a wide range of microservice systems as their core business solutions~\citep{Soldani2022rcasurvey}.

Although microservice systems offer various significant benefits, they come with several drawbacks, one of the most notable ones being the challenge of analysing the root causes of system failures. A typical microservice system can consist of a dozen to hundreds of services, with each service having a large number of metrics to be continuously monitored. Once a failure occurs, it can propagate across the services and affect a large number of metrics, making it especially challenging for engineers to identify the failure's root cause promptly. It was reported that without using an automated tool, it could take engineers at least several hours to identify a failure's root cause~\citep{Azam2022rcd, Wang2018cloudranger}. This delay can affect a large number of users, incurring substantial economic losses and other unintended consequences. It has been reported that a one-hour downtime on Amazon.com could potentially cost up to $100$ million USD~\citep{Chen2020incidentmanagement, Chen2019incidenttriage}.

Causal inference-based root cause analysis (RCA) methods for microservice systems via metrics data have attracted increasing attention from researchers in recent years~\citep{Soldani2022rcasurvey, xin2023causalrca, Azam2022rcd, Li2022Circa, wu2021microdiag, Ma2019Msrank, Wang2018cloudranger, Meng2020Microcause, run2024aaai, causalai23salesforce}. The main idea is to construct a graph from metrics data to depict the causal relationships among the services and metrics (\textit{causal graph}) and, from this graph, infer the root cause of a failure. A number of methods including CloudRanger~\citep{Wang2018cloudranger}, Microscope~\citep{Jinjin2018Microscope}, MS-Rank~\citep{Ma2019Msrank}, AutoMap~\citep{Ma2020Automap}, MicroCause~\citep{Meng2020Microcause}, and CausalAI~\citep{causalai23salesforce} rely on the Peter-Clark (PC) algorithm~\citep{Spirtes1993Causal} or its variants~\citep{Runge2019PCMCI} to construct the causal graph. Then a scoring method such as random walk~\citep{Spitzer1976randomwalk}, PageRank~\citep{Brin1998Pagerank} or Depth-First Search (DFS) is used to traverse the causal graph to locate the root cause. CIRCA~\citep{Li2022Circa} constructs a causal graph based on domain knowledge and causal assumptions. Then, it uses a regression-based hypothesis testing method to infer the root cause. RCD~\citep{Azam2022rcd} uses the $\Psi$-PC algorithm~\citep{Jaber2020PsiFCI, Spirtes1993Causal} and a divide-and-conquer strategy to infer the failure's root cause. More recently, CausalRCA~\citep{xin2023causalrca} introduces a gradient-based structure learning method to generate a weighted causal graph and combines it with the PageRank algorithm to locate the root cause. Meanwhile, RUN~\citep{run2024aaai} uses neural Granger causal discovery with contrastive learning to construct the causal graph and the PageRank algorithm to infer the root cause.

Despite significant progress, we notice there is a lack of comprehensive evaluation of causal inference-based RCA methods. Existing research works only assess the methods from a limited number of aspects on a restricted set of datasets, and thus, do not provide adequate insights into the capability of these methods. For example, Wu et al.~\citep{Wu2021evalcausal} evaluate six causal inference-based RCA methods. However, it neither evaluates the causal graph construction step nor includes recent proposed methods (e.g., RCD, CIRCA, CausalRCA, RUN), and other important aspects such as the impact of hyperparameter tuning, input data length, among others. Both the work in~\citep{Arya2021evalcausalai} and~\citep{Wang2021evalcausal} only evaluate Granger-based RCA methods for AIOps using time series data obtained via the system log data. \textit{In this work, we aim to understand the current state of causal inference-based RCA methods by thoroughly assessing their performance and the factors that could affect their performance}.

We conduct a comprehensive evaluation of causal inference-based RCA methods on six synthetic datasets and four datasets from three benchmark microservice systems with different types of failures. First, we assess the performance of various common causal discovery methods in constructing causal graphs for microservice systems from metrics data. Second, we evaluate the performance of existing state-of-the-art causal inference-based RCA methods to understand whether they can accurately locate a failure's root cause. Third, we analyse the runtime of these methods to understand their efficiency. Finally, we investigate different factors that could affect the performance of causal inference-based RCA methods, such as the input data length, the hyperparameter tuning process, and the misspecification of the failure occurrence time. Through extensive experiments, we obtain the following major findings about the causal inference-based RCA methods for microservice systems:

\begin{itemize}[leftmargin=.2in]
    \item All common causal discovery methods have difficulties in working with large graphs and estimating edge directions, suggesting that directly applying causal discovery methods for RCA in large-scale microservices might not be effective. Hyperparameter tuning could help improve the performance of small-scale graphs but fail on large-scale ones.
    \item Some causal inference-based RCA methods are better than others in identifying the root causes of microservice systems' failures but at the cost of 
    losing efficiency or being highly sensitive to some parameters. Large-scale microservice systems remain challenging for causal inference-based RCA methods.
    \item Synthetic datasets used in previous research may not accurately reflect the performance of causal discovery and RCA methods for real-world microservice systems, highlighting the need to reconsider the process of generating synthetic data.
    \item The running time of most causal inference-based RCA methods increases significantly with the size of microservices and the number of metrics. Some methods are always faster than others.
    \item Long input data lengths could 
    improve the performance of causal discovery and causal inference-based RCA methods. Some methods are still effective with shorter data.
\end{itemize}

Based on our study, we identify several challenges of causal inference-based RCA methods in the context of microservice systems and propose possible future research work.

\noindent This chapter makes the following contributions:

\begin{itemize} [leftmargin=.22in]
    \item We conduct a comprehensive evaluation of causal inference-based RCA methods using metrics for microservice systems, spanning nine causal discovery methods and twenty-one causal inference-based RCA methods across six synthetic datasets and four real-world microservice benchmark datasets.
    \item We obtain a set of empirical findings showing that most existing causal inference-based RCA methods do not perform well across all scenarios, particularly on large-scale microservice systems, and identify the factors (data length, hyperparameter tuning, anomaly time misspecification) that most affect performance.
    \item We suggest a set of future research directions for advancing causal inference-based RCA for microservice systems using metrics data.
\end{itemize}

%% file: papers/ase24/2.background.tex
\subsection{Problem Statement}

This chapter focuses on evaluating causal inference-based RCA.\footnote{The published version of this work~\citep{pham2024root} contained self-contained terminology. To avoid duplication, this thesis consolidates the shared terminology in Chapter~\ref{chap:background}.}
We adopt the terminology established in Section~\ref{sec:terms-fault-failure} and the metric-based RCA formulation from Section~\ref{sec:problem-formulation-rca}. In metric-based RCA, \textit{causal graphs} depict cause-and-effect connections, with each node representing a metric of a service and each edge indicating a causal relationship~\citep{Soldani2022rcasurvey}. Causal graphs can capture relationships even among non-communicating services, such as those collocated on the same virtual machine~\citep{Jinjin2018Microscope}.

The main idea to solve the RCA problem via causal inference is to first construct a causal graph from the collected metrics data and then use a scoring method to identify the service that is likely to be the root cause of the failure (see Figure~\ref{fig:intro}). In the sections below, we first describe the existing causal discovery techniques to construct causal graphs from time series data (Sec. \ref{sec:causal-discovery}). Then, we outline the scoring methods to locate the failure's root cause based on the causal graph (Sec. \ref{sec:scoring}), and the state-of-the-art causal inference-based RCA methods for microservice systems (Sec. \ref{sec:causal-rca}).

\subsection{Causal Discovery Methods for Time Series} \label{sec:causal-discovery}

In recent years, causal discovery methods have gained attention for their ability to infer causal relationships from time series data~\citep{Vowels2022causaldiscoverysurvey}. In this section, we briefly summarise representative causal discovery methods commonly used in causal inference-based RCA methods. More information on these causal discovery methods is in our supplementary material (in \href{https://github.com/phamquiluan/rcaeval}{our GitHub page}), Sec. A. 

\textbf{Peter-Clark (PC) Algorithm~\citep{Spirtes1993Causal}.} PC is arguably the most popular causal discovery algorithm. It uses conditional independence testing and a series of rules~\citep{Spirtes1993Causal} to construct the causal graph. PCMCI~\citep{Runge2019PCMCI}, a PC variant, can handle time-lagged causal relations.

\textbf{Fast Causal Inference (FCI) Algorithm~\citep{Spirtes1993Causal}.} Similar to PC, FCI also uses conditional independence testing and a series of rules to construct the causal graph. However, FCI can deal with the presence of confounders, which is an advantage over the PC algorithm. 

\textbf{Granger Algorithm~\citep{Granger1980causal}.} It relies on the concept of \textit{Granger causality}, where a time series causes another if the former provides statistically significant information about future values of the latter. An advanced nonlinear variant, Neural Granger Causal Discovery (NGCD)~\citep{run2024aaai}, can leverage contextual information in temporal data.

\textbf{LiNGAM Algorithm~\citep{Shimizu2006Lingam}.} LiNGAM uses a linear, acyclic structural equation model to construct the causal graph. Similar to PC, it assumes no hidden confounders that affect the time series. 

\textbf{Greedy Equivalence Search (GES) Algorithm~\citep{Chickering2002Ges}.} GES uses a greedy strategy and the Bayesian Information Criterion (BIC)~\citep{Schwarz1978bic} to build the causal graph. Besides, fGES (Fast GES)~\citep{Ramsey2017fges} constructs the causal graph relying on the collider causal structure when orienting edges, making it more 
efficient.

\textbf{NOTEARS-Low-Rank (NTLR) Algorithm~\citep{fang2023low}.} NTLR is a gradient-based causal discovery method which adapts NOTEARS~\citep{zheng2018dags} with low-rank causal graphs. Note that NLTR has not yet been explored in the RCA literature but we include it to evaluate how a new causal discovery method performs in RCA. 

\subsection{Scoring Methods for Root Cause Analysis} \label{sec:scoring}

After obtaining a causal graph, a scoring method is used to locate the root cause. We 
outline below the scoring methods that have been 
used in causal inference-based RCA.

\textbf{Random Walk.} The main idea of random walk is to walk through all the nodes in the causal graph and randomly choose the next nodes to visit~\citep{Wang2018cloudranger, Jinjin2018Microscope, Meng2020Microcause}. With this strategy, the nodes that are visited most often are considered the root cause of the failure.

\textbf{PageRank.} PageRank assesses the importance of each node in the causal graph based on the number and quality of the incoming edges and identifies nodes with more incoming edges from influential nodes as potential root causes~\citep{xin2023causalrca, wu2021microdiag}.

\textbf{Depth First Search (DFS).} DFS traverses all nodes in the causal graph and determines whether they are abnormal via an anomaly detection technique. It then identifies the abnormal sub-graphs, ranks their roots via anomaly scores and determines root causes as the root nodes of the abnormal sub-graphs~\citep{Chen2014Causeinfer}.

\textbf{Hypothesis Testing.} This method formulates a failure as an intervention that alters the pre-failure data distribution. It conducts hypothesis testing to assess if a node's data, after a failure, follows the pre-failure data distribution. Nodes with the most deviation from this distribution are considered root causes~\citep{Li2022Circa, pham2024baro}. 

\subsection{Causal Inference-based Root Cause Analysis Methods} \label{sec:causal-rca}

In recent years, many causal inference-based RCA methods have been proposed to analyse and identify the root cause of failures in microservice systems~\citep{Soldani2022rcasurvey, xin2023causalrca, Azam2022rcd, Li2022Circa, wu2021microdiag, Ma2019Msrank, Meng2020Microcause, run2024aaai, causalai23salesforce}. We briefly describe recent causal inference-based RCA methods as follows.

\textbf{PC-based~\citep{Wang2018cloudranger, Jinjin2018Microscope, Meng2020Microcause, Chen2014Causeinfer, Ma2020Automap, Ma2019Msrank}.} These methods use PC or its variant to construct a causal graph from time series metrics data and use a scoring method to locate the root cause. Notable methods include CloudRanger~\citep{Wang2018cloudranger}, 
Microscope~\citep{Jinjin2018Microscope}, CauseInfer~\citep{Chen2014Causeinfer}, AutoMap~\citep{Ma2020Automap}, MicroCause~\citep{Meng2020Microcause} and MS-Rank~\citep{Ma2019Msrank}. 

\textbf{FCI-based~\citep{Chen2019airalert}.} AirAlert proposes to use the FCI algorithm to infer the causal relationships among the metrics data and serves as a diagnosis tool to help engineers identify the root cause easier.

\textbf{Granger-based~\citep{thalheim2017sieve, Wang2021evalcausal, Arya2021evalcausalai}.} These methods employ the Granger algorithm to derive causal graphs from time series logs data and then identify the root cause using a scoring technique.

\textbf{LiNGAM-based~\citep{wu2021microdiag}.} MicroDiag is a popular method following this approach. It uses the DirectLiNGAM algorithm to construct the causal graph from the metrics data and the PageRank method to determine the location of the root cause from the causal graph.

\textbf{MicroCause~\citep{Meng2020Microcause}}. MicroCause uses PCMCI~\citep{Runge2019PCMCI} to construct the causal graph. Then, it applies temporal cause oriented random walk to rank the root causes from the estimated causal graph.

\textbf{CIRCA~\citep{Li2022Circa}.} CIRCA constructs the causal graph from the call graph using operators' knowledge about the system and a mapping of metrics into some defined categories. It formulates a failure as an intervention that alters the metrics data distribution and performs a regression-based hypothesis test to identify the root cause.

\textbf{RCD~\citep{Azam2022rcd}.} RCD uses a divide-and-conquer approach that splits 
metrics into smaller chunks and learns a causal graph for each chunk. It employs $\Psi$-PC algorithm~\citep{Jaber2020PsiFCI} to find the root cause within each chunk and then merges 
the potential root causes together and runs $\Psi$-PC recursively to identify the final root cause. 

\textbf{CausalRCA~\citep{xin2023causalrca}.} CausalRCA uses a gradient-based variational autoencoder causal structure learning method called DAG-GNN~\citep{Yu2019DagGNN}, to generate the 
graph. To infer the root cause, CausalRCA uses the PageRank algorithm.

\textbf{RUN~\citep{run2024aaai}.} RUN uses NGCD with contrastive learning to construct the causal graph. Then, it applies PageRank with a personalized vector to recommend the top-k root causes.

\textbf{NSigma~\citep{Jinjin2018Microscope}.} NSigma is a hypothesis testing method that uses the z-score to compare the distributions of pre- and post-failure data. The higher the score, the more likely that metric is the root cause. NSigma does not construct 
causal graph.

\textbf{$\epsilon$-Diagnosis~\citep{Shan2019Ediagnosis}.} $\epsilon$-Diagnosis uses a two-sample test algorithm and $\epsilon$-statistics to estimate the similarity between every pair of metrics and rank the root causes based on test scores. Similar to NSigma, $\epsilon$-Diagnosis does not construct causal graphs.

\textbf{BARO~\citep{pham2024baro}.} BARO uses a variant of hypothesis testing technique based on median and interquartile range (IQR), to analyse pre-failure and post-failure data distributions. This makes BARO more resistant to noise compared to NSigma. Similar to NSigma and $\epsilon$-Diagnosis, BARO also does not construct causal graphs.

\textbf{CausalAI~\citep{causalai23salesforce}.} 
CausalAI is an open-source industrial library for causal analysis. To conduct RCA, CausalAI uses PC to build the causal graph using time series metrics data. Subsequently, it derives root nodes from the causal graph as potential root causes.

%% file: papers/ase24/3.study_design.tex
To understand the current state of causal inference-based RCA methods, we study the following four RQs to thoroughly assess their performance and the factors that could affect their performance:

\begin{itemize}
\item \textbf{RQ1:} How effective are causal discovery algorithms in constructing causal graphs from time series metrics data? (Sec. \ref{sec:rq1-results})

\item \textbf{RQ2:} How effective are causal inference-based RCA methods in locating the failure's root cause? (Sec. \ref{sec:rca-effectiveness})

\item \textbf{RQ3:} How efficient are causal discovery methods and causal inference-based RCA methods? (Sec. \ref{sec:efficency})

\item \textbf{RQ4:} How do causal discovery and causal inference-based RCA methods perform w.r.t. different input data lengths? (Sec. \ref{sec:eval-input-data})
\end{itemize}

\subsection{Datasets}

\begin{table}[t]
\centering
\caption[Characteristics of synthetic datasets.]{Characteristics of synthetic datasets (
\#nodes, \#edges: number of nodes and edges in the graph, \#cases: number of cases in the dataset, \#type: time series data type)
}
\label{tab:synthetic-data}
\resizebox{0.55\textwidth}{!}{%
\begin{tabular}{l r r r c}
\hline
\textbf{Name} & \textbf{\#nodes}  & \textbf{\#edges} & \textbf{\#cases}  & \textbf{\#type} \\ \hline \hline
CIRCA10 & 10 & 20  & 200 &  cts \\ \hline
CIRCA50 & 50 & 100  & 200 & cts \\ \hline
RCD10 & 10  & 13-19 & 200 & dct \\ \hline
RCD50 & 50  & 85-104 & 200 & dct \\ \hline
CausIL10  & 10  & 19 & 10 & cts \\ \hline
CausIL50  & 50  & 125 & 10 & cts \\ \hline
\end{tabular}%
}

{\footnotesize (*) 'cts' stands for 'continuous', 'dct' stands for 'discrete'.}
\end{table}

\begin{table}[t]
\centering
\caption[Characteristics of collected data from benchmark microservice systems.]{Characteristics of collected data from benchmark microservice systems (\#metrics, \#svc, \#t\_svc, \#fault: number of metrics, services, targeted services, and fault types).
}
\label{tab:ase24-real-data}
\resizebox{0.8\textwidth}{!}{%
\setlength\tabcolsep{3pt}
\begin{tabular}{l r r r r r c}
\hline
\textbf{Name} & \textbf{\#metrics}  & \textbf{\#svc}   & \textbf{\#t\_svc}  & \textbf{\#fault} & \textbf{\#cases} & \textbf{\#type}  \\ \hline \hline
Sock Shop 1 & 38 & 13 & 5 & 2 & 50 & cts \\ \hline
Sock Shop 2 & 46 & 15 & 5 & 5 & 125 & cts  \\ \hline
Online Boutique & 49 & 12 & 5  & 5 & 125 & cts \\ \hline
Train Ticket & 212 & 64 & 5 & 5 & 125 & cts \\ \hline
\end{tabular}%
}

{\footnotesize (*) The abbreviation convention is the same as Table \ref{tab:synthetic-data}.}
\end{table}

\subsubsection{Synthetic Datasets} \label{sec:synthetic-data}

We use three different synthetic data generators from three previous RCA studies~\citep{Li2022Circa, Azam2022rcd, Chakraborty2023CausIL} to create the synthetic datasets: CIRCA, RCD, and CausIL data generators. These data generators are used in various research works to evaluate RCA methods~\citep{Azam2022rcd, Li2022Circa, Chakraborty2023CausIL, liu2023pyrca}. Their mechanisms are as follows:

CIRCA data generator~\citep{Li2022Circa} generates a random causal directed acyclic graph (DAG) based on a given number of nodes and edges. From this DAG, time series data for each node is generated using a vector auto-regression (VAR) model. A fault is injected into a node by altering the noise term in the VAR model for two timestamps. RCD data generator~\citep{Azam2022rcd} uses the pyAgrum package~\citep{pyagrum} to generate a random DAG based on a given number of nodes, subsequently generating discrete time series data for each node, with values ranging from 0 to 5. A fault is introduced into a node by changing its conditional probability distribution. Meanwhile, CausIL data generator~\citep{Chakraborty2023CausIL} generates causal graphs and time series data that simulate the behaviour of microservice systems. It first constructs a DAG of services and metrics based on domain knowledge, then generates metric data for each node of the DAG using regressors trained on real metrics data. Unlike the CIRCA and RCD data generators, the CausIL data generator does not have the capability to inject faults.

To create our synthetic datasets, we first generate 10 DAGs whose nodes range from 10 to 50 for each of the synthetic data generators. Next, we generate fault-free datasets using these DAGs with different seedings, resulting in 100 cases for the CIRCA and RCD generators and 10 cases for the CausIL generator. We then create faulty datasets by introducing ten faults into each DAG and generating the corresponding faulty data, yielding 100 cases for the CIRCA and RCD data generators. The fault-free datasets are used to evaluate causal discovery methods, while the faulty datasets are used to assess RCA methods. We use all three dataset generators to alleviate each other's weaknesses, such as diverse causal graph structures and different types of metrics data (continuous and discrete), enabling a more comprehensive assessment of the performance of the studied causal inference-based RCA methods. Table \ref{tab:synthetic-data} shows the dataset characteristics .

\begin{figure}
\centering
\includegraphics[width=0.8\textwidth]{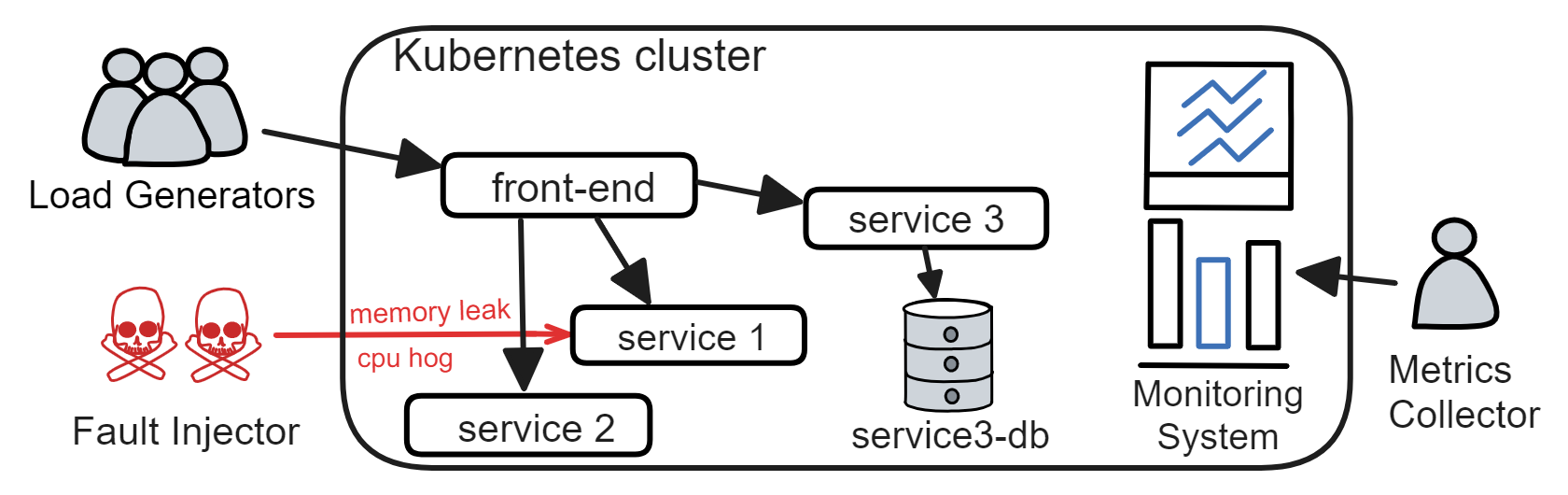}
\caption{Overview of our setup for microservice systems.} \label{fig:ase24-system_setup}
\end{figure}

\subsubsection{Benchmark Microservice Systems} \label{sec:realworld-data}

We deploy three popular benchmark microservice systems: Sock Shop~\citep{sockshop}, Online Boutique~\citep{ob}, and Train Ticket~\citep{tt}, which are widely used for evaluating RCA methods~\citep{Jinjin2018Microscope, Azam2022rcd, wu2021microdiag, xin2023causalrca, wu2022automatic, he2022graph, yu2021microrank, Wang2021evalcausal, pham2024baro}. Sock Shop~\citep{sockshop} is a sock-selling e-commerce application that consists of 15 services communicating with each other through HTTP requests. Online Boutique~\citep{ob}, with its 12 services, is an e-commerce application where users can browse items, add them to the cart, and purchase them. Train Ticket~\citep{tt} is one of the largest microservice systems, simulating a train ticket booking system with 64 services. Compared to Sock Shop and Online Boutique systems, Train Ticket system has longer and more complex failure propagation paths. 

To generate metrics data, we first deploy these microservice systems on a four-node Kubernetes cluster hosted by AWS. Next, we use the Istio service mesh~\citep{istio} with Prometheus~\citep{prometheus} and cAdvisor~\citep{cadvisor} to monitor and collect resource-level and service-level metrics of all services, as in previous works~\citep{Azam2022rcd, xin2023causalrca, pham2024baro}. To generate traffic, we use the load generators provided by these systems and customise them to explore all services with 100 to 200 users concurrently. We then introduce five common faults (CPU hog, memory leak, disk IO stress, network delay, and packet loss) into five different services within each system. Finally, we collect metrics data before and after the fault injection operation. An overview of our setup is presented in Figure~\ref{fig:ase24-system_setup}. Furthermore, we diversify our datasets by using the available Sock Shop data from a previous study~\citep{Azam2022rcd}, which we refer to as Sock Shop 1. We refer to our Sock Shop data as Sock Shop 2. The statistics of the collected datasets are shown in Table \ref{tab:ase24-real-data}.

\subsection{Evaluation Metrics}

\subsubsection{Causal Graph Construction.} We assess the accuracy of the estimated causal graph and its skeleton using F1-score, 
\begin{equation} \nonumber
F1 = \frac{2 \times Pre \times Rec}{Pre + Rec}, \quad Pre = \frac{TP}{TP + FP}, \quad Rec = \frac{TP}{TP + FN},
\end{equation}
where $TP$, $FP$, and $FN$ correspond to true positives, false positives, and false negatives, respectively. $TP$ is the number of correctly identified actual edges, $FP$ is the number of incorrectly identified edges, and $FN$ is the number of missing edges. The F1-score on the skeleton graph, denoted as F1-S, evaluates the accuracy of the edges without considering their directions. The F1-score on the full directed graph, denoted as F1, takes into account the orientations of the edges and penalises for incorrectly estimated directions of correctly identified adjacencies. Following previous works~\citep{Chakraborty2023CausIL, liu2023pyrca}, we also use Structural Hamming Distance (SHD)~\citep{raghu2018evaluation} to assess the estimated causal graph. The SHD score is obtained by summing the missing edges, extra edges, and incorrectly directed edges.

\subsubsection{Root Cause Analysis.} In this work, we evaluate the accuracy in identifying the root cause services as this is a standard practice in related work~\citep{Jinjin2018Microscope, Azam2022rcd, xin2023causalrca, Wang2018cloudranger, Chen2014Causeinfer, Ma2019Msrank, Ma2020Automap}. The root cause service is the service associated with the identified metric from the causal graph~\citep{Azam2022rcd, Ma2020Automap, xin2023causalrca}. Following existing works~\citep{Jinjin2018Microscope, Azam2022rcd, xin2023causalrca, Meng2020Microcause, yu2021microrank, Li2022Circa}, we use two standard metrics, $AC@k$ and $Avg@k$, to assess the performance of the RCA methods. 
$AC@k$ represents the probability the top $k$ results given by a method include the root cause. $AC@k$ scores range from $0$ to 1, and the higher the value, the better the method. Given a set of failure cases A, $AC@k$ is calculated as follows,
\begin{equation} \nonumber
    AC@k = \frac{1}{|A|} \sum\nolimits_{a\in A}\frac{\sum_{i<k}R^a[i]\in V^a_{rc}}{min(k, |V^a_{rc}|)},
\end{equation}
where $R^a[i]$ is the ranking result for the failure case a. $V^a_{rc}$ is the root cause set of case a. 
$Avg@k$, which measures the overall performance of RCA methods, is calculated as $Avg@k = \frac{1}{k}\sum\nolimits_{1\le j\le k} AC@j$.

\subsection{Experimental Settings}

For PC, FCI, LiNGAM, and CausalRCA methods, we use the implementation published in~\citep{xin2023causalrca}. For Granger, we use the standard implementation from the statsmodels package~\citep{statsmodels}. For PCMCI, we use the implementation in~\citep{Li2022Circa}. For fGES and NTLR, we use the source code in~\citep{Chakraborty2023CausIL} and~\citep{zhang2021gcastle}, respectively. For $\epsilon$-Diagnosis, we use the implementation in~\citep{liu2023pyrca}. For RCD, CIRCA, MicroCause, RUN, CausalAI, and BARO, we use their available implementation in~\citep{Azam2022rcd, Li2022Circa, run2024aaai, causalai23salesforce, pham2024baro}. Note that for CIRCA, since the call graph is unavailable, thus following~\citep{Azam2022rcd}, we use the PC algorithm to construct this graph. For the hyperparameter settings of the methods, we use the default values suggested by their respective papers. We confirmed the correctness of the source code by reproducing the presented results in the original and related papers. We conduct all experiments on Linux servers equipped with 8 CPUs, 16GB RAM.

%% file: papers/ase24/4.results.tex
\subsection[RQ1: Effectiveness in Constructing Causal Graphs?]{How Effective are Causal Discovery Algorithms in Constructing Causal Graphs?} \label{sec:rq1-results}

\begin{sidewaystable}[p]
\caption[Performance of six causal discovery methods with default settings.]{Performance of nine causal discovery methods on synthetic datasets with default settings. Best results are \textbf{bold}.}
\label{tab:q1-causal-default}
\resizebox{\textheight}{!}{%
\begin{tabular}{l|rrr|rrr|rrr|rrr|rrr|rrr}
\hline
 & \multicolumn{3}{c|}{CIRCA10} & \multicolumn{3}{c|}{CIRCA50} & \multicolumn{3}{c|}{RCD10} & \multicolumn{3}{c|}{RCD50} & \multicolumn{3}{c|}{CausIL10} & \multicolumn{3}{c}{CausIL50} \\ \hline
\textbf{} & \multicolumn{1}{l}{\textbf{F1}} & \multicolumn{1}{l}{\textbf{F1-S}} & \multicolumn{1}{l|}{\textbf{SHD}} & \multicolumn{1}{l}{\textbf{F1}} & \multicolumn{1}{l}{\textbf{F1-S}} & \multicolumn{1}{l|}{\textbf{SHD}} & \multicolumn{1}{l}{\textbf{F1}} & \multicolumn{1}{l}{\textbf{F1-S}} & \multicolumn{1}{l|}{\textbf{SHD}} & \multicolumn{1}{l}{\textbf{F1}} & \multicolumn{1}{l}{\textbf{F1-S}} & \multicolumn{1}{l|}{\textbf{SHD}} & \multicolumn{1}{l}{\textbf{F1}} & \multicolumn{1}{l}{\textbf{F1-S}} & \multicolumn{1}{l|}{\textbf{SHD}} & \multicolumn{1}{l}{\textbf{F1}} & \multicolumn{1}{l}{\textbf{F1-S}} & \multicolumn{1}{l}{\textbf{SHD}} \\ \hline \hline
PC & \textbf{0.49} & 0.65 & \textbf{16} & \textbf{0.38} & 0.47 & \textbf{104} & 0.3 & \textbf{0.59} & \textbf{14} & 0.24 & \textbf{0.46} & 120 & 0.45 & 0.75 & 16 & 0.3 & 0.46 & 145 \\ \hline
FCI & 0.43 & 0.63 & 19 & 0.33 & \textbf{0.48} & 115 & \textbf{0.36} & \textbf{0.59} & 16 & \textbf{0.3} & \textbf{0.46} & 137 & 0.5 & \textbf{0.76} & 16 & \textbf{0.31} & \textbf{0.47} & \textbf{144} \\ \hline
Granger & 0.46 & 0.6 & 26 & 0.18 & 0.23 & 463 & 0.1 & 0.21 & 19 & 0.19 & 0.42 & \textbf{86} & 0.44 & 0.62 & 28 & 0.13 & 0.22 & 650 \\ \hline
ICALiNGAM & 0.18 & 0.66 & 28 & 0.08 & 0.35 & 283 & 0.19 & 0.46 & \textbf{14} & 0.19 & 0.42 & \textbf{86} & 0.22 & 0.72 & 20 & 0.09 & 0.36 & 281 \\ \hline
DirectLiNGAM & 0.4 & 0.66 & 22 & 0.22 & 0.37 & 249 & 0.19 & 0.45 & \textbf{14} & 0.2 & 0.42 & \textbf{86} & \textbf{0.54} & \textbf{0.76} & \textbf{13} & 0.21 & 0.36 & 263 \\ \hline
GES & 0.42 & 0.66 & 20 & 0.34 & 0.44 & 160 & 0.23 & 0.32 & 15 & 0.23 & 0.32 & 92 & 0.47 & 0.67 & 18 & 0.22 & 0.41 & 205 \\ \hline
fGES & 0.3 & \textbf{0.67} & 22 & 0.18 & 0.44 & 165 & 0.25 & 0.32 & 15 & 0.24 & 0.31 & 92 & 0.36 & \textbf{0.76} & 19 & 0.23 & 0.41 & 195 \\ \hline
PCMCI & 0.12 & 0.18 & 32 & 0.04 & 0.07 & 986 & 0.22 & 0.38 & 44 & 0.06 & 0.11 & 1223 & 0.16 & 0.25 & 35 & 0.06 & 0.11 & 1101 \\ \hline
NTLR & 0.32 & 0.52 & 21 & 0.14 & 0.27 & 131 & 0.19 & 0.34 & 20 & - & - & - & 0.43 & 0.66 & 25 & 0.06 & 0.11 & 570 \\ \hline
\end{tabular}%
}
\end{sidewaystable}

\begin{sidewaystable}[p]
\caption[Performance of six causal discovery methods with hyperparameter tuning.]{Performance of six causal discovery methods on synthetic datasets with hyperparameter tuning. Best results are \textbf{bold}.}
\label{tab:q1-causal-tuning}
\resizebox{\textheight}{!}{%
\begin{tabular}{l|rrr|rrr|rrr|rrr|rrr|rrr}
\hline
 & \multicolumn{3}{c|}{CIRCA10} & \multicolumn{3}{c|}{CIRCA50} & \multicolumn{3}{c|}{RCD10} & \multicolumn{3}{c|}{RCD50} & \multicolumn{3}{c|}{CausIL10} & \multicolumn{3}{c}{CausIL50} \\ \hline
\textbf{} & \multicolumn{1}{l}{\textbf{F1}} & \multicolumn{1}{l}{\textbf{F1-S}} & \multicolumn{1}{l|}{\textbf{SHD}} & \multicolumn{1}{l}{\textbf{F1}} & \multicolumn{1}{l}{\textbf{F1-S}} & \multicolumn{1}{l|}{\textbf{SHD}} & \multicolumn{1}{l}{\textbf{F1}} & \multicolumn{1}{l}{\textbf{F1-S}} & \multicolumn{1}{l|}{\textbf{SHD}} & \multicolumn{1}{l}{\textbf{F1}} & \multicolumn{1}{l}{\textbf{F1-S}} & \multicolumn{1}{l|}{\textbf{SHD}} & \multicolumn{1}{l}{\textbf{F1}} & \multicolumn{1}{l}{\textbf{F1-S}} & \multicolumn{1}{l|}{\textbf{SHD}} & \multicolumn{1}{l}{\textbf{F1}} & \multicolumn{1}{l}{\textbf{F1-S}} & \multicolumn{1}{l}{\textbf{SHD}} \\ \hline \hline
PC & \textbf{0.5} & \textbf{0.66} & \textbf{17} & \textbf{0.31} & 0.36 & 170 & 0.31 & 0.64 & 21 & - & \textbf{-} & - & 0.47 & 0.76 & 17 & 0.22 & 0.38 & 192 \\ \hline
FCI & 0.42 & 0.65 & 20 & 0.28 & 0.43 & \textbf{143} & \textbf{0.49} & \textbf{0.87} & 15 & \textbf{-} & \textbf{-} & - & \textbf{0.53} & \textbf{0.78} & \textbf{16} & \textbf{0.25} & 0.42 & 177 \\ \hline
Granger & 0.38 & 0.58 & 32 & 0.1 & 0.16 & 829 & 0.25 & 0.45 & 30 & 0.07 & 0.14 & 780 & 0.45 & 0.68 & 27 & 0.18 & \textbf{0.44} & \textbf{165} \\ \hline
ICALiNGAM & 0.18 & \textbf{0.66} & 28 & 0.08 & 0.35 & 282 & 0.19 & 0.47 & \textbf{14} & 0.19 & 0.42 & 86 & 0.21 & 0.72 & 20 & 0.1 & 0.37 & 255 \\ \hline
fGES & 0.31 & \textbf{0.66} & 22 & 0.18 & \textbf{0.44} & 164 & 0.31 & 0.39 & 15 & 0.24 & 0.31 & 92 & 0.34 & 0.77 & 19 & 0.23 & 0.41 & 192 \\ \hline
PCMCI & 0.12 & 0.18 & 32 & 0.04 & 0.07 & 925 & 0.22 & 0.38 & 44 & 0.05 & 0.1 & 1225 & 0.16 & 0.25 & 35 & 0.06 & 0.11 & 1119 \\ \hline
\end{tabular}%
}
\\
{\footnotesize \textit{(*)  PC and FCI results on the RCD50 dataset were not obtained due to OOM errors during execution. GES, DirectLiNGAM, and NTLR were excluded for exceeding a 1-hour/case time-out.} }
\end{sidewaystable}

The performance of causal graph construction directly impacts the performance of causal inference-based RCA methods, yet previous works have overlooked this evaluation \citep{xin2023causalrca, Azam2022rcd, Jinjin2018Microscope, Chen2014Causeinfer, Meng2020Microcause, Wang2018cloudranger}. In this section, we evaluate a comprehensive list of common causal discovery methods: Granger, PC, PCMCI, FCI, DirectLiNGAM, ICALiNGAM, GES, fGES, and NTLR on six synthetic datasets: CIRCA10, RCD10, CausIL10, CIRCA50, RCD50, and CausIL50, to understand their effectiveness. All experiments are repeated 10 times, and we report the average of F1, F1-S, and SHD scores.

In Table \ref{tab:q1-causal-default}, we present the performance of these methods using the default hyperparameters. We have the following findings:

(1) \textbf{PC and FCI yield the best performance.} PC achieves the best score in 7 out of 18 cases, and FCI in 9 cases.

(2) \textbf{All methods struggle with estimating edge directions with Granger, NTLR, LiNGAM/GES-based methods being especially bad at identifying the edge directions.} The F1 scores of all methods are systematically lower than their F1-S scores. This reveals that \textit{we need to consider the capacity of causal discovery methods in estimating the edge directions when developing causal inference-based RCA methods for microservices in future work}.

(3) \textbf{The performance of all methods decrease significantly when the graph size increases.} This highlights that \textit{directly applying common causal discovery methods for RCA may be ineffective in large-scale microservice systems}.

(4) \textbf{There is still ample room for improvement in developing methods to construct a causal graph from metrics data.} The F1 scores of all methods are only within the range from 0.1 to 0.54, which are far from being ideal (the highest score is 1).

Furthermore, we also evaluate the studied methods under hyperparameter tuning setting using the Bayesian Information Criterion score~\citep{Schwarz1978bic, biza2020bictuning}, as described in our supplementary material, Sec. C. The experimental results are presented in Table \ref{tab:q1-causal-tuning}. We observe that: 

(5) \textbf{Hyperparameter tuning could improve the construction results of smaller graphs.} The performance of causal discovery methods with tuned hyperparameters on CIRCA10, RCD10, and CausIL10 is mostly better than when using default values.

(6) \textbf{Hyperparameter tuning does not perform well on larger graphs.} In CIRCA50, the tuned PC and FCI drop 23\% and 10\% respectively compared to the default settings. This issue could be due to the complexity of large graphs, which causes challenges for the hyperparameter tuning method to find optimal hyperparameter values. More work needs to improve the effectiveness of these methods for large and complex graphs.

\begin{tcolorbox}[left=2pt,right=2pt,top=0pt,bottom=0pt]
\textbf{Summary.} 
All common causal discovery methods have difficulties in dealing with large graphs and estimating edge directions. This suggests that directly applying causal discovery methods for 
RCA in large-scale microservice systems may not be effective. Hyperparameter tuning could improve the RCA performance on small graphs but still could not improve the performance on large graphs. {There is still ample room for further research}.
\end{tcolorbox}

\subsection[RQ2: Effectiveness in Locating Root Causes?]{How Effective are Causal Inference-based RCA Methods in Locating Root Causes?} \label{sec:rca-effectiveness}

In this section, we extensively evaluate whether state-of-the-art causal inference-based RCA methods can accurately identify the root causes of microservice system failures. The methods we evaluate are: PC-based, FCI-based, Granger-based, LiNGAM-based, fGES-based, NTLR-based, CausalRCA, MicroCause, CIRCA, RCD, NSigma, $\epsilon$-Diagnosis, BARO, CausalAI, and RUN. For the PC / FCI / Granger / fGES / LiNGAM / NTLR-based methods, we use either the popular PageRank or random walk scoring method to locate the root cause from the causal graphs. To assess whether these RCA methods perform better than random, we also include Dummy, a method that randomly chooses a node within the causal graph as the root cause. We thoroughly evaluate these methods using four synthetic datasets (RCD10, RCD50, CIRCA10, CIRCA50) and four datasets collected from three benchmark systems (Online Boutique, Sock Shop 1 \& 2, and Train Ticket), each with various types of failures. All experiments are repeated 10 times and averaged. 

Notably, CIRCA, RCD, NSigma, and $\epsilon$-Diagnosis require knowing the exact failure occurrence time $t_F$, which may not be practical as, in practice, this information is often unavailable. Meanwhile, BARO uses a customised anomaly detection technique to estimate the time $t_F$; however, this estimate may not always be accurate. To better understand the performance of these RCA methods, we include the experiments when we misspecify the failure occurrence time $t_F$. We experiment with the following variants of the RCA methods: RCD [$t_\Delta=0$], CIRCA [$t_\Delta=0$], NSigma [$t_\Delta=0$], $\epsilon$-Diagnosis [$t_\Delta=0$], BARO [$t_\Delta=0$] given the perfect $t_F$; and RCD [$t_\Delta=60$], CIRCA [$t_\Delta=60$], NSigma [$t_\Delta=60$], $\epsilon$-Diagnosis [$t_\Delta=60$], BARO [$t_\Delta=60$] given the specified time to be $t_F$ + 60 seconds. The choice of setting $t_\Delta$ to be 60 seconds is inspired by a previous RCA work that uses one-minute sampling intervals~\citep{Li2022Circa}, meaning that the monitoring system collects time series metrics data every 60 seconds.

In Table \ref{tab:q2-rca}, we report the overall performance of all methods in Avg@5. Our major findings are as follows: 

(1) \textbf{CausalRCA, RCD, CIRCA, NSigma, and BARO are generally better than other baselines in identifying root causes of microservice systems' failures.} Meanwhile, PC / FCI / Granger / LiNGAM / fGES / NTLR-PageRank/random walk, CausalAI, RUN, and MicroCause mostly perform similarly to Dummy, meaning they are no better or only slightly better than random selection in identifying the root cause of the datasets and benchmark systems used in our study. To the best of our knowledge, this is the first study to use Dummy as a baseline and surface this problem. Examining reported results in previous works \citep{Azam2022rcd, xin2023causalrca} yields the same conclusion regarding the effectiveness of the PC / FCI / Granger / LiNGAM / fGES / NTLR-PageRank/random walk approach. \textit{This insight again emphasises that directly applying these causal discovery methods for RCA in microservice systems may not be effective.}

(2) \textbf{CIRCA and NSigma are sensitive to the specification of the failure occurrence time, especially for large systems like Train Ticket, whilst RCD, $\epsilon$-Diagnosis, and BARO are more robust to this time.} While previous works assume the availability of this information when performing RCA on microservice systems, this finding suggests that \textit{the estimated failure occurrence time might significantly affect the robustness of interventional recognition-based RCA methods}, an aspect that is overlooked in previous works \citep{Azam2022rcd, Li2022Circa}. This finding also implies that \textit{future work should evaluate these RCA methods with different anomaly detectors.}

\begin{sidewaystable}[p]
\caption[Performance of twenty-one RCA methods on eight datasets.]{Performance of twenty-one RCA methods in terms of Avg@5 on eight datasets.  The fault types CPU, MEM, DISK, DELAY, LOSS, and SIM denote CPU hog, memory leak, disk IO stress, network delay, packet loss, and simulation faults, respectively.}
\label{tab:q2-rca}
\resizebox{\textwidth}{!}{%
\setlength\tabcolsep{3pt}
\begin{tabular}{l|r|r|r|r|rrrrr|rr|rrrrr|rrrrr}
\hline
 & RCD10 & RCD50 & CIRCA10 & CIRCA50 & \multicolumn{5}{c|}{Online Boutique} & \multicolumn{2}{c|}{Sock Shop 1} & \multicolumn{5}{c|}{Sock Shop 2} & \multicolumn{5}{c}{Train Ticket} \\ \hline
 
 & SIM & SIM & SIM & SIM & \multicolumn{1}{c}{CPU} & \multicolumn{1}{c}{MEM} & \multicolumn{1}{c}{DISK} & \multicolumn{1}{c}{DELAY} & \multicolumn{1}{c|}{LOSS} & \multicolumn{1}{c}{CPU} & \multicolumn{1}{c|}{MEM} & \multicolumn{1}{c}{CPU} & \multicolumn{1}{c}{MEM} & \multicolumn{1}{c}{DISK} & \multicolumn{1}{c}{DELAY} & \multicolumn{1}{c|}{LOSS} & \multicolumn{1}{c}{CPU} & \multicolumn{1}{c}{MEM} & \multicolumn{1}{c}{DISK} & \multicolumn{1}{c}{DELAY} & \multicolumn{1}{c}{LOSS} \\ \hline \hline
Dummy & 0.3 & 0.06 & 0.3 & 0.06 & 0.25 & 0.24 & 0.26 & 0.26 & 0.25 & 0.36 & 0.33 & 0.37 & 0.38 & 0.33 & 0.38 & 0.37 & 0.07 & 0.08 & 0.06 & 0.07 & 0.07 \\
PC-PR & 0.24 & 0.11 & 0.25 & 0.07 & 0.34 & 0.15 & 0.32 & 0.33 & 0.38 & 0.38 & 0.27 & 0.46 & 0.47 & 0.4 & 0.22 & 0.42 & 0.07 & 0.15 & 0.09 & 0.10 & 0.07  \\
PC-RW & 0.13 & 0 & 0.31 & 0.01 & 0.34 & 0.31 & 0.39 & 0.18 & 0.32 & 0.44 & 0.44 & 0.48 & 0.46 & 0.49 & 0.45 & 0.36 & 0.16 & 0.05 & 0.11 & 0.03 & 0.09 \\
FCI-PR & 0.29 & 0.09 & 0.31 & 0.06 & 0.22 & 0.22 & 0.38 & 0.29 & 0.4 & 0.44 & 0.2 & 0.39 & 0.44 & 0.55 & 0.26 & 0.42 & 0 & 0.06 & 0.22 & 0.13 & 0.12 \\
FCI-RW & 0.1 & 0.04 & 0.31 & 0.04 & 0.56 & 0.56 & 	0.56 & 0.36 & 0.36 & 0.32 & 0.32 & 0.48 & 0.48 & 0.48 & 0.48 & 0.48 & 0 & 0 & 0 & 0 & 0 \\
Granger-PR & 0.31 & 0.06 & 0.4 & 0.06 & 0.22 & 0.45 & 0.41 & 0.31 & 0.25 & 0.34 & 0.31 & 0.46 & 0.81 & 0.58 & 0.38 & 0.38 & 0.04 & 0.04 & 0.14 & 0.04 & 0.04 \\
Granger-RW & 0.23 & 0 & 0.3 & 0.01 & 0.34 & 0.31 & 0.34 & 0.14 & 0.2 & 0.56 & 0.54 & 0.48 & 0.46 & 0.49 & 0.45 & 0.36 & 0.16 & 0.05 & 0.11 & 0.02 & 0.13 \\
ICALiNGAM-PR & 0.17 & 0.01 & 0.26 & 0.02 & 0.22 & 0.08 & 0.09 & 0.12 & 0.3 & 0.31 & 0.26 & 0.06 & 0.01 & 0.01 & 0.00 & 0.18 & 0.03 & 0 & 0.16 & 0.03 & 0 \\
ICALiNGAM-RW & 0.13 & 0 & 0.31 & 0.01 & 0.4 & 0.37 & 0.34 & 0.14 & 0.2 & 0.56 & 0.54 & 0.48 & 0.46 & 0.49 & 0.45 & 0.36 & 0.16 & 0.05 & 0.11 & 0.03 & 0.13 \\
fGES-PR & 0.15 & 0.02 & 0.39 & 0.04 & 0.26 & 0.27 & 0.28 & 0.26 & 0.26 & 0.27 & 0.33 & 0.26 & 0.21 & 0.37	& 0.33 & 0.40 & 0.06 & 0.04 & 0.11  & 0.05 & 0.06 \\
fGES-RW & 0.13 & 0 & 0.31 & 0.01 & 0.4 & 0.38 & 0.34 & 0.14 & 0.22 & 0.36 & 0.38 & 0.48 & 0.46 & 0.49 & 0.45 & 0.36 & 0.16 & 0.05 & 	0.04  & 0.03 & 0.13 \\
NTLR-PR & 0.26 & 0.04 & 0.19 & 0.01 & 0.2 & 0.1 & 0.22 & 0.1 & 0.42 & 0.25 & 0.25 & 0.08 & 0.05 & 0.04 & 0.03 & 0.11 & - & - & - & - & - \\
NTLR-RW & 0.14 & 0.05 & 0.3 & 0.02 & 0.34 & 0.31 & 0.34 & 0.14 & 0.2 & 0.56 & 0.54 & 0.48 & 0.46 & 0.49 & 0.45 & 0.36 & - & - & - & - & - \\ 
CausalRCA & 0.1 & 0.04 & 0.05 & 0 & 0.97 & 0.98 & 0.71 & 0.92 & 0.52 & 0.84 & 0.84 & 0.49 & 0.82 & 0.74  & 0.61 & 0.47 & 0.53 & 0.3 & 0.13 & 0.17 & 0.11 \\
CausalAI & 0.08 & 0.06 & 0 & 0 & 0.62 & 0.30 & 0.45 & 0.18 & 0.24 & 0.42 & 0.42 & 0.38 & 0.18 & 0.14 & 0.41 & 0.51 & 0 & 0 & 0.04 & 0.07 & 0.06 \\
RUN & 0.1 & - & 0.3 & - & 0.57 & 0.56 & 0.60 & 0.35 & 0.33 & 0.48 & 0.42 & 0.46 & 0.47 & 0.48 & 0.33 & 0.48 & - & - & - & - & - \\
MicroCause & 0.13 & 0.04 & 0.37 & 0.09 & 0.34 & 0.45 & 0.37 & 0.57 & 0.42 & 0.4 & 0.55 & 0.47 & 0.43 & 0.33 & 0.22 & 0.46 & - & - & - & - & - \\ \hline 
$\epsilon$-Diagnosis {[}$t_\Delta=60${]} & 0 & 0 & 0 & 0 & 0.23 & 0.03 & 0.12 & 0.16 & 0.16 & 0.5 & 0.66 & 0.51 & 0.38 & 0.49 & 0.46 & 0.46 & 0 & 0 & 0 & 0 & 0 \\
BARO {[}$t_\Delta=60${]} & 0.18 & 0.03 & 0.24 & 0.04 & 0.94 & 0.99 & 0.87 & 0.99 & 0.6 & 0.98 & 0.98 & 0.99 & 0.98 & 0.94  & 1 & 0.88 & 0.81 & 0.99 & 0.77 & 0.82 & 0.72 \\
RCD {[}$t_\Delta=60${]} & 0.48 & 0.21 & 0.18 & 0.03 & 0.74 & 0.59 & 0.67 & 0.67 & 0.5 & 0.43 & 0.67 & 0.43 & 0.35 & 0.63 & 0.5 & 0.44 & 0.11 & 0.09 & 0.1 & 0.12 & 0.19 \\
CIRCA {[}$t_\Delta=60${]} & 0.19 & 0.09 & 0.19 & 0.03 & 0.24 & 0.36 & 0.4 & 0.58 & 0.39 & 0.52 & 0.62 & 0.19	& 0.16 & 0.37 & 0.26 & 0.39 & 0 & 0 & 0.07 & 0.03 & 0.1  \\
NSigma {[}$t_\Delta=60${]} & 0.23 & 0.03 & 0.05 & 0 & 0.16 & 0.24 & 0.43 & 0.55 & 0.38 & 0.15 & 0.33 & 0.27 & 0.22 & 0.23 & 0.37 & 0.5 & 0.03 & 0 & 0.03 & 0.05 & 0.12 \\ \hline
$\epsilon$-Diagnosis {[}$t_\Delta=0${]} & 0 & 0 & 0 & 0 & 0.26 & 0.02 & 0.22 & 0.12 & 0.13 & 0.5 & 0.55 & 0.49 & 0.34 & 0.49 & 0.51 & 0.5 & 0 & 0.02 & 0 & 0 & 0.02 \\
BARO {[}$t_\Delta=0${]} & 0.22 & 0.05 & 0.92 & 0.87 & 0.97 & 1 & 0.91 & 0.98 & 0.67 & 0.99 & 0.99 & 1 & 0.98 & 0.94  & 0.98 & 0.87 & 0.90 & 0.96 & 0.84 & 0.77 & 0.66 \\
RCD {[}$t_\Delta=0${]} & 0.89 & 0.62 & 0.28 & 0.05 & 0.91 & 0.74 & 0.67 & 0.38 & 0.47 & 0.63 & 0.66 & 0.62 & 0.46 & 0.62	& 0.48 & 0.37 & 0.08 & 0.01 & 0.31 & 0.09 & 0.12 \\
CIRCA {[}$t_\Delta=0${]} & 0.34 & 0.08 & 0.34 & 0.06 & 0.94 & 0.97 & 0.7 & 0.92 & 0.55 & 0.7 & 0.8 & 0.97 & 0.98 & 0.92 & 0.98 & 0.88 & 0.66 & 0.93 & 0.64 & 0.64 & 0.57 \\
NSigma {[}$t_\Delta=0${]} & 0.21 & 0.06 & 0.88 & 0.85 & 0.94 & 1 & 0.9 & 0.98 & 0.67 & 0.98 & 0.99 & 0.98 & 0.98 & 0.94 & 0.98 & 0.9 & 0.81 & 0.96 & 0.85 & 0.61 & 0.7 \\ \hline
\end{tabular}%
}
{\footnotesize \textit{(*) 'RW' stands for 'random walk', 'PR' stands for 'PageRank'. (-) MicroCause, RUN, and NTLR-based RCA have missing results because they exceed the limit of 2 hours per case.}} \\ 
\end{sidewaystable}

(3) \textbf{The performance of RCA methods on synthetic datasets may not accurately reflect their performance in real systems.} For example, RCD performs well on its synthetic and other datasets but poorly on CIRCA datasets. NSigma and BARO performs well on CIRCA synthetic datasets and other datasets but poorly on RCD synthetic datasets. Likewise, CausalRCA performs well on Online Boutique, Sock Shop 1 \& 2, but badly on synthetic datasets. These findings suggest \textbf{the generation of synthetic datasets may not be consistent across different works and may not accurately represent metrics data of microservices.} For example, RCD data generator introduces faults by altering the conditional probability of a node whereas real-world anomalies are often characterised by surges in metrics such as high CPU usage, workload drops \citep{lee2023eadro, Soldani2022rcasurvey}. \textit{It is thus important to reconsider the process of generating synthetic data when developing and evaluating RCA methods in future work}.

\begin{tcolorbox}[left=2pt,right=2pt,top=0pt,bottom=0pt]
\textbf{Summary.}
CausalRCA, RCD, CIRCA, NSigma, and BARO are generally the best RCA methods for microservices with metrics data. However, the performance of RCD, CIRCA, and NSigma are sensitive to the estimated failure occurrence time. Future works should evaluate these methods with different anomaly detectors to understand better their effectiveness when using actual estimated failure occurrence time. Existing synthetic datasets used in previous research \citep{Azam2022rcd, Li2022Circa} may not accurately reflect the performance of the RCA methods for real-world microservice systems. Finally, large-scale microservice systems still pose a great challenge for causal inference-based RCA methods.
\end{tcolorbox}

\subsection[RQ3: Efficiency in Causal Discovery and Root cause Analysis?]{How Efficient are Causal Discovery and Causal Inference-based RCA Methods?} \label{sec:efficency}

To promptly detect and troubleshoot failures so as to ensure minimal system downtime, it is crucial to develop efficient and effective RCA methods. In this section, we conduct an efficiency analysis of the studied causal discovery and RCA methods using all the available datasets. We record the running time per case of all methods on Linux machines, each with 8 CPU and 16GB memory. 

\subsubsection{Causal Graph Construction}
To evaluate the efficiency of causal discovery methods, we report the runtimes of nine representative methods on six synthetic datasets. D
etailed experimental results are provided in our supplementary material, Table S1. Our findings are:

(1) \textbf{The runtime of all causal discovery methods increases significantly with the graph complexity.} When the size of causal graphs increases from 10 to 50 nodes, these methods become seven to thousands of times slower, taking up to hours to build a causal graph. This observation complements our finding in Section \ref{sec:rq1-results} that \textit{the causal discovery algorithms is neither effective nor efficient on large microservice systems, presenting a challenge of how to select appropriate input metrics for effective RCA in such systems}.

(2) \textbf{The runtime of kernel-based conditional independence testing (PC with KCI \citep{wu2022automatic}, Kernel-based DirectLiNGAM) is prohibitively long.} Running PC with the KCI independence test as described in \citep{wu2022automatic} takes an average of over 1 hour/case to estimate a 10-node graph by CIRCA and RCD synthetic data generators. We can see that \textit{there is a large room for improvement in efficiency if one wants to employ a causal discovery algorithm in their RCA method for diagnosing failures of microservices.}

\subsubsection{Root Cause Analysis} \label{sec:speed-rca}
To evaluate the efficiency of RCA methods, in Table \ref{tab:q3-rca-speed}, we provide the running time of twenty-one RCA methods on synthetic and microservices datasets. We found that:

(1) \textbf{NSigma and BARO are consistently faster, followed by RCD whilst CausalRCA, MicroCause, RUN, and NTLR-based methods tend to be the slowest}.  
Based on our observations, NSigma and BARO have a small running time since they do not learn any causal graph, instead, they just compare the normal and abnormal metrics data. RCD also runs fast since it performs hierarchical learning and thus only needs to learn small causal graphs from subsets of metrics data while most other RCA methods learn full causal graphs from all metrics data. CausalRCA, MicroCause, RUN, and NTLR-based methods are significantly slower than others, due to their reliance on advanced causal discovery algorithms like DAG-GNN, PCMCI, NGCD, and NTLR.

(2) \textbf{The running time of most RCA methods increases significantly with the complexity of microservice systems}. Most causal inference-based RCA methods can handle Sock Shop (38-46 metrics) and Online Boutique (49 metrics) within seconds. However, Train Ticket (212 metrics) causes the methods to be much slower; the running time of the methods are from few minutes to one hour.

(3) \textbf{The running time of PC / FCI / Granger / LiNGAM / GES / NTLR-based RCA methods is proportional to the runtime of their corresponding causal discovery methods} (Table S1, supplementary material). Among these RCA methods, the PC-based methods are generally the fastest and NTLR-based methods are the slowest. This is similar to the behaviours of their corresponding causal discovery methods: PC is the fastest and NTLR is the slowest. This shows that \textit{for these RCA methods, their running time depends greatly on the running time of the causal graph construction step}. 

\begin{tcolorbox}[left=2pt,right=2pt,top=0pt,bottom=0pt]
\textbf{Summary.} The running time of causal inference-based RCA methods increases significantly as the size of microservice systems grows. Most causal inference-based RCA methods can handle a small set of metrics (<50) within seconds but slow down remarkably when dealing with larger microservice systems that involve a larger set of metrics. NSigma and BARO are always faster than others whilst CausalRCA, MicroCause, RUN, and NTLR-based methods are usually the slowest.
\end{tcolorbox}

\subsection[RQ4: Performance with Different Input Data Length?]{How do Different Methods Perform with Different Data Lengths?} \label{sec:eval-input-data}

In this section, we assess whether the performance of causal discovery and causal inference-based RCA methods depend on the input data length. This evaluation is to address the question that \textit{once a failure is detected, whether we can run the RCA methods immediately and thus use a smaller amount of input data or wait to collect more data so as to increase the accuracy of the RCA methods}. For the causal discovery methods, we evaluate their performance with the input data lengths varying from 125 to 4000 data points, corresponding to approximately 2 to 60 minutes of metrics data. For the causal inference-based RCA methods, we evaluate with the input data lengths varying from 60 to 600 data points, corresponding to 1 to 10 minutes of metrics data. It is worth noting that prior works on RCA only evaluate the methods using a smaller number of data points. For example, the works in \citep{wu2022automatic} and \citep{xin2023causalrca} use only 60 data points, while the work in \citep{Azam2022rcd} uses from 535 to 593 data points. All experiments are repeated five times, and we report the average results. Note here we only repeat the experiments 5 times instead of 10 due to the prohibitive running time of this task, i.e., it takes over 400 hours on 8 CPU and 16GB RAM machines for each repeat, making it prohibitive to run with 10 repeats.

\begin{table}[]
\centering
\caption{The running time (in seconds) of twenty-one RCA methods on eight datasets.}
\label{tab:q3-rca-speed}
\resizebox{0.8\textwidth}{!}{%
\setlength\tabcolsep{2pt}
\begin{tabular}{l|r|r|r|r|r|r|r|r}
\hline
 & CIRCA10 & RCD10 & CIRCA50 & RCD50 & \multicolumn{1}{c|}{SS1} & \multicolumn{1}{c|}{SS2} & \multicolumn{1}{c|}{OB} & \multicolumn{1}{c}{TT} \\ \hline \hline
PC-PR & 0.18 & 0.05 & 1.78 & 0.39 & 1.53 & 2.16 & 3.39 & 129.65 \\ \hline
PC-RW & 0.17 & 0.05 & 1.77 & 0.39 & 1.63 & 2.24 & 3.53 & 131.27 \\ \hline
FCI-PR & 0.19 & 0.04 & 1.85 & 0.37 & 2.62 & 2.52 & 4.9 & 154.95 \\ \hline
FCI-RW & 0.21 & 0.06 & 1.93 & 0.38 & 2.62 & 2.51 & 5.21 & 152.59 \\ \hline
Granger-PR & 1.05 & 1.3 & 36.32 & 27.55 & 5.72 & 13.28 & 12.9 & 196.3 \\ \hline
Granger-RW & 0.98 & 1.25 & 35.67 & 25.92 & 5.62 & 13.21 & 13.53 & 245.4 \\ \hline
ICA-PR & 0.47 & 0.07 & 6.68 & 6.07 & 0.51 & 2.76 & 1.51 & 16.07 \\ \hline
ICA-RW & 0.44 & 0.07 & 6.11 & 5.60 & 0.53 & 2.77 & 1.55 & 16.39 \\ \hline
fGES-PR & 0.63 & 0.16 & 10.76 & 2.04 & 7.36 & 5.40 & 11.49 & 372.8 \\ \hline
fGES-RW & 0.63 & 0.17 & 10.73 & 2.06 & 7.31 & 5.46 & 11.47 & 381.12 \\ \hline
NTLR-PR & 12.97 & 39.71 & 663.07 & 6179.77 & 487.88 & 471.11 & 448.94 & - \\ \hline
NTLR-RW & 12.7 & 39.26 & 672.11 & 6181.34 &  454.18 & 462.36 & 437.16 & - \\ \hline
CausalRCA & 53.79 & 51.3 & 197.6 & 165.89 & 79.33 & 143.97 & 146.67 & 1326.34 \\ \hline
CausalAI & 0.12 & 0.13 & 6.49 & 1.12 & 7.86 & 17.7 & 16.37 & 643.29 \\ \hline
RUN & 1078.65 & 1095 & - & - & 2051.12 & 4938.75 & 1548.28 & - \\ \hline
MicroCause & 27.6 & 24.69 & 1900.77 & 1430.89 & 75.78 & 206.64 & 257.73 & - \\ \hline
$\epsilon$-Diagnosis & 1 & 1 & 6.3 & 6.62 & 3.54 & 5.42 & 5.28 & 21.92 \\ \hline
RCD & 0.1 & 0.06 & 0.37 & 0.13 & 2.1 & 3.08 & 3.05 & 12.44 \\ \hline
CIRCA & 0.18 & 0.06 & 2.01 & 0.35 & 1.76 & 7 & 4.19 & 3792.29 \\ \hline
NSigma & 0.01 & 0.01 & 0.01 & 0.01 & 0.01 & 0.01 & 0.01 & 0.01 \\ \hline
BARO & 0.01 & 0.01 & 0.01 & 0.01 & 0.01 & 0.01 & 0.01 & 0.01 \\ \hline
\end{tabular}%
}

{\footnotesize \textit{(*) SS1, SS2, OB, TT denote Sock Shop 1, Sock Shop 2, Online Boutique, Train Ticket.}}
\end{table}

\subsubsection{Graph Construction} 
\begin{sidewaysfigure}[p]
\includegraphics[width=\textheight]{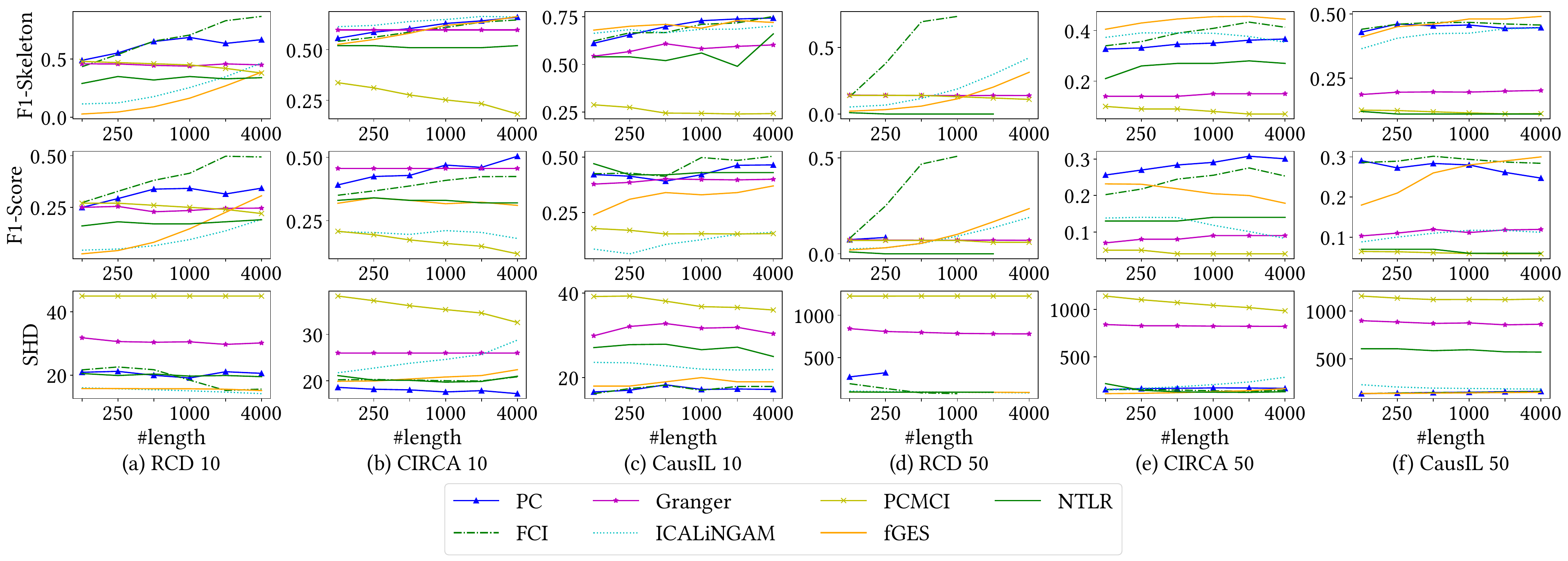}

{\footnotesize \textit{(*) PC, FCI, and NTLR results on RCD50 were partially obtained due to OOM errors during execution and exceeding the time limit.}} 

\caption[Performance of seven causal discovery methods with different data lengths.]{Performance of seven causal discovery methods on six synthetic datasets with different data lengths.}  \label{fig:rq4-causal-graph}
\end{sidewaysfigure}

We plot the performance of seven causal discovery methods (PC, FCI, Granger, ICALiNGAM, PCMCI, fGES, NTLR) on six synthetic datasets 
in Figure~\ref{fig:rq4-causal-graph}. Our observations are:

(1) \textbf{For most datasets, most methods (PC, FCI, ICALiNGAM, fGES) improve their accuracy when being given more 
data.} For example, on RCD10, the F1 score of FCI increases from 0.25 to 0.5 when the 
data length increases from 125 to 4000.

(2) \textbf{The accuracy of Granger and NTLR appears to plateau with increasing input metrics data whilst PCMCI's accuracy experiences a downward trend.} In all datasets, both Granger and NTLR maintain similar F1, F1-S, and SHD scores across all the input data lengths whilst the F1, F1-S, and SHD of PCMCI often decline with more input data.

(3) \textbf{On CausIL50 dataset, increasing the input data length does not significantly affect the performance of the causal discovery methods.} Most causal discovery methods maintain similar F1, F1-S, and SHD scores when the input metrics data length increases from 125 to 4000.

\begin{sidewaysfigure}[p]
\includegraphics[width=\textheight]{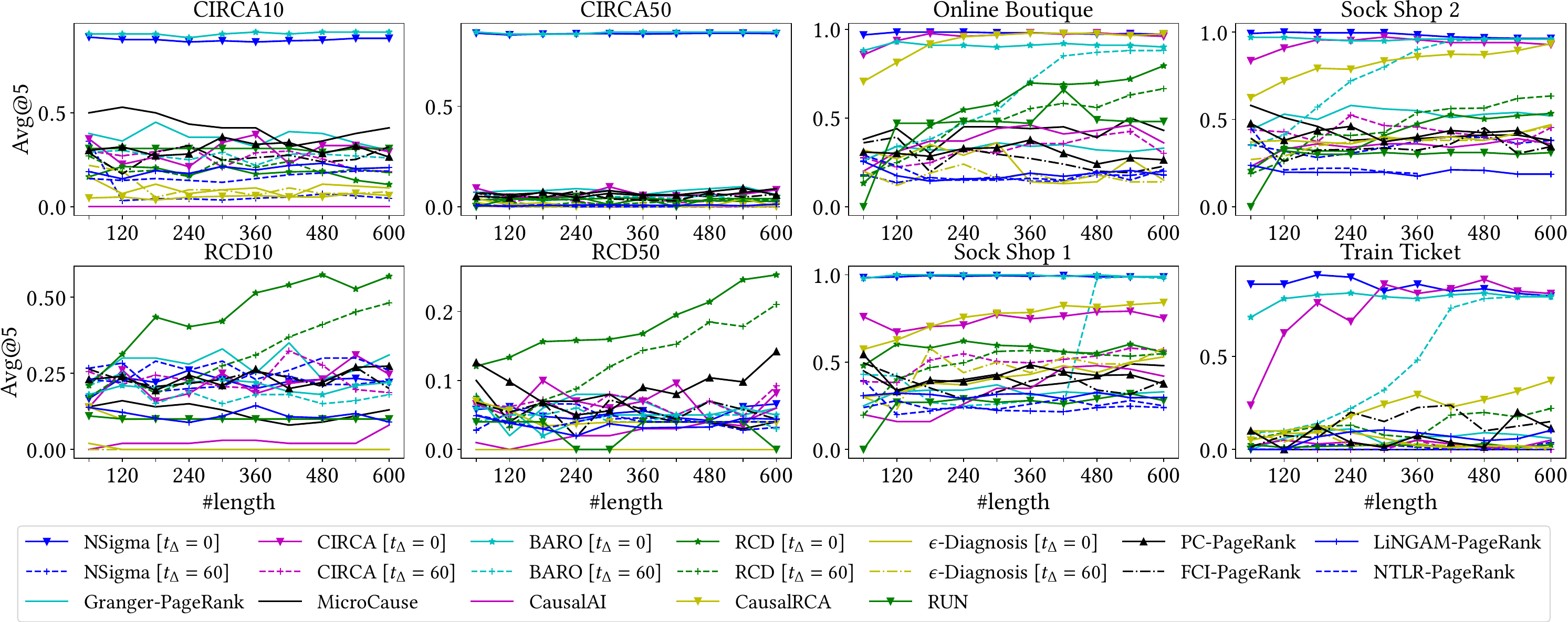}

\caption[Performance of fourteen RCA methods with different data lengths.]{Performance of fourteen RCA methods on eight datasets with different data lengths.}  \label{fig:rq4-rca}
\end{sidewaysfigure}

\subsubsection{Root Cause Analysis}
We plot the performance of studied RCA methods on eight different datasets with varying input data lengths in Figure~\ref{fig:rq4-rca}. Our major findings are:

(1) \textbf{With an accurate specification of the failure occurrence time, NSigma, BARO, and CIRCA perform notably well even with less input metrics data.} However, as we emphasised in Section \ref{sec:rca-effectiveness}, their performance drops significantly if there is a misspecification of the failure occurrence time. Notably, increasing the input data lengths, in this case, does not help NSigma and CIRCA, but can help BARO to significantly alleviate the performance degradation. The robustness of BARO in this context can be attributed to its use of median and IQR for hypothesis testing instead of mean and standard deviation as in NSigma and CIRCA.

(2) \textbf{RCD shows remarkable improvements with increased input data lengths.} For example, in Online Boutique, the Avg@5 score of RCD {[}$t_\Delta=0${]} increases from 0.13 to 0.79 when increasing the data length from 60 to 600 data points. This property could be attributed to its usage of $\Psi$-PC \citep{Jaber2020PsiFCI, Spirtes1993Causal}, a novel algorithm that learns a causal graph based on normal data and soft-abnormal data.

(3) \textbf{CausalRCA steadily improves performance when the input data length increases in majority of datasets.} For instance, in Sock Shop 1, its Avg@5 score increases from 0.62 to 0.93 with increased input data length from 120 to 600 data points. One possible reason is that it relies on a deep learning model which typically achieves better performance with more input data.

Based on these findings (1), (2), and (3), we can see that \textit{increasing the data length usually leads to improved performance of RCD, CIRCA, and CausalRCA methods}. However, \textit{this improvement could come with a trade-off that the root cause may not be promptly identified as it generally takes time to collect more abnormal data}. 

(4) \textbf{PC / FCI / Granger / LiNGAM / NTLR-PageRank, RUN, MicroCause, $\epsilon$-Diagnosis, and CausalAI do not perform well on the studied datasets and adjusting the input data length does not significantly affect their performance.} 
Their performance is consistently low across all datasets and having more metrics data does not seem to help. 

\begin{tcolorbox}[left=2pt,right=2pt,top=0pt,bottom=0pt]
\textbf{Summary.}
Given an accurate failure occurrence time, NSigma, BARO, and CIRCA perform notably well even with a short metric data lengths, making them ideal candidates for quickly diagnosing failures of microservice systems. Longer input data lengths could significantly improve the performance of both causal discovery and RCA methods. CausalRCA, BARO, CIRCA (given precise failure occurrence time), and RCD exhibit improvements when the input data length increases.
\end{tcolorbox}

%% file: papers/ase24/5.discussion.tex
\subsection{The Advantages and Disadvantages of the Studied RCA Methods} \label{sec:advantage-disadvantage} 

Based on our findings, we can conclude that the existing causal inference-based RCA methods for microservice systems need further improvements to work well in practice. We point out the advantages and disadvantages of each method as follows.

\textbf{PC/FCI/Granger/LiNGAM/NTLR-based, RUN, CausalAI, and MicroCause}. These methods learn a complete causal graph from all the metrics data. \textit{With default hyperparameter settings, their performance is not consistent across different datasets; hyperparameter tuning may help to ensure good performance on small-scale microservice systems, but does not help much with large-scale microservice systems}. Finally, for the datasets and microservice benchmark systems used in our studies (10 to 212 nodes), \textit{the performance of these methods is no better or only slightly better than random selection}. This issue might arise from their poor performance in the causal graph construction, as indicated by our findings in Section \ref{sec:rq1-results}. 

\textbf{CIRCA}. CIRCA relies on a user-constructed causal graph, demanding knowledge about microservice structure and metrics, rather than using causal discovery methods. It employs a hypothesis testing scoring method that requires precise failure occurrence time specification for optimal accuracy. However, \textit{CIRCA is highly sensitive to this value; it performs much worse when this value deviates slightly, e.g., 1 minute}. Another drawback of CIRCA is that \textit{it requires the call graph of microservices and manual work in mapping the metrics to construct the causal graph. This leads to difficulties in the adoption of this technique in real-world microservice systems, given their inherent dynamic nature.} Though note that using the PC algorithm to construct the causal graph can still result in a reasonable performance of the algorithm even though it may come with some efficiency trade-offs, particularly in large-scale microservices. 

\textbf{RCD}. The main advantage of RCD is that it does not learn a complete causal graph from all the metrics, but it employs a divide-and-conquer strategy to learn smaller causal graphs from metrics subsets. \textit{This approach makes it highly efficient, especially on large-scale microservices}, i.e., its running time is significantly lower than a majority of RCA methods. \textit{A drawback of RCD is that it requires a sufficient amount of metrics data to perform well.} In practice, this could lead to delays in troubleshooting the failures of microservice systems as it takes time to collect the abnormal/failure data. 

\textbf{NSigma}. NSigma also requires the specification of the failure occurrence time. Similar to CIRCA, its performance is also sensitive to this value. \textit{Given an accurate specification of the failure occurrence time, NSigma can achieve very high accuracy. However, with a slight deviation of the failure occurrence time, e.g., 1 minute, its performance significantly worsens}. Another advantage of NSigma is that \textit{its runtime is very small, making it the fastest method among our studied RCA methods, even on large-scale microservices.}

\textbf{BARO}. BARO can achieve very high accuracy and offer better resistance to the specified failure occurrence time when given enough metrics data. 
This robustness stems from the use of median-based hypothesis testing.
An advantage of BARO is that \textit{its runtime is very small, making it one of the fastest methods among our studied RCA methods.}

\textbf{CausalRCA.} An advantage of CausalRCA is that \textit{it does not require the specification of a failure occurrence time in order to diagnose the root cause.} By using DAG-GNN, a gradient-based causal discovery method, it can diagnose the root cause better than other methods that use PC or FCI. However, due to this characteristic, \textit{CausalRCA is less efficient than other methods.} Its running time is much higher compared to other RCA methods, especially on large-scale microservices with many metrics (>200).

\subsection{Future Research Directions} \label{sec:future-works}

We identify several challenges of causal inference-based RCA methods in the context of microservices. We outline these challenges 
and propose possible future research work:

(1) \textbf{Working with Large-scale Microservice Systems.}
Our findings show that most causal discovery methods (to construct the causal graphs) and RCA methods (to identify the failure's root cause) perform well on a small set (<50 metrics) of metrics data and perform badly on a larger set (>200 metrics).
Hence, \textit{selecting an optimal subset of metrics~\citep{thalheim2017sieve}, implementing divide-and-conquer strategies~\citep{Azam2022rcd}, or adding blocked edge sets as domain knowledge~\citep{Chakraborty2023CausIL} may help to reduce the need to construct a full causal graph from all the metrics data}, and thus, improve the performance of root cause identification. Additionally, \textit{using call graphs~\citep{Li2022Circa} or hand-crafted graphs from engineers~\citep{li2022actionable} may also benefit causal inference-based RCA methods}, albeit at a higher cost and risk of error~\citep{li2022actionable}.

(2) \textbf{Efficiency of the RCA Methods.} Existing works mainly focus on demonstrating the accuracy of the RCA methods without evaluating their efficiency (the running time). Our study suggests that the running time of some RCA methods could be very long, especially when dealing with large-scale microservice systems, and this can lead to a delay in the troubleshooting of the failure. \textit{Future research work could aim to develop RCA techniques that can achieve high accuracy whilst maintaining a reasonable running time.}

(3) \textbf{Sensitiveness to the Failure Occurrence Time.} Our study reveals that \textit{some RCA methods~\citep{Azam2022rcd, Li2022Circa} are sensitive to the failure occurrence time with varying degrees, especially in larger microservice systems}. Future works should extensively evaluate RCA methods within integrated anomaly detection and RCA pipelines. Such evaluations can provide valuable insights into the actual effectiveness of these methods with actual anomaly detectors. 

(4) \textbf{Using a Variety of Datasets from Different Microservice Systems.} Our findings indicate the importance of \textit{using multiple different datasets and microservice benchmark systems to evaluate causal inference-based RCA methods comprehensively}. High performance on one dataset or microservice system does not necessarily reflect high performance on other datasets/systems, as various factors, such as the number of services, metrics, and their inherent dynamic relationships, can significantly affect the RCA performance.

(5) \textbf{Synthetic Dataset Generation}. Since it is not always possible to deploy real microservice systems and simulate a variety of failure scenarios, it is beneficial for the research community to develop methods to generate synthetic data that closely mimics the behaviour of microservice systems. Our results suggest that synthetic datasets used in previous work may not accurately reflect the performance of RCA methods in the real world. \textit{Better methods for generating synthetic data are needed to enhance the development of future causal inference-based RCA methods for microservice systems.}

(6) \textbf{Systematic Hyperparameter Tuning.} In this work, we perform a hyperparameter tuning process via the BIC score~\citep{Schwarz1978bic, biza2020bictuning} and evaluate the performance of common causal discovery methods. Previous works usually choose these hyperparameters empirically, and these chosen values might not be the most optimal choice to achieve the maximal performance. Therefore, \textit{future work can also develop more advanced hyperparameter tuning approaches to improve the performance of 
RCA methods.}

(7) \textbf{Develop End-to-end Anomaly Detection and RCA}. Future research in RCA for microservices should consider developing more accurate anomaly detection modules. The performance of the RCA pipeline, when considered as a unified system, should be thoroughly evaluated. \textit{This approach would eliminate the need for provided information like the failure occurrence time~\citep{Li2022Circa, Azam2022rcd, pham2024baro}.}

%% file: papers/ase24/6.threats_to_validity.tex
We now discuss threats to the validity of our study, along with the means we undertook to mitigate these threats.

\subsection{Construct Validity}

The construct validity threat of our evaluation primarily concerns the hyperparameter setting and the evaluation metrics. To address this, we conduct a hyperparameter tuning for studied causal discovery methods. For the studied RCA methods, we use the default values suggested in their papers. We also employ well-established evaluation metrics used in previous works~\citep{Azam2022rcd, xin2023causalrca, Meng2020Microcause, yu2021microrank, Li2022Circa} to compare the performance of causal discovery and RCA methods.

\subsection{Internal Validity}

Regarding the studied methods, we re-use the code from various published RCA works, and we have also performed experiments to replicate the results of these source codes to ensure their correctness. To increase the internal validity of our experiment results and avoid the randomness factors in the causal discovery and RCA methods, we repeat the experiments multiple times for each dataset and method and report the average results. We use standard evaluation metrics extensively used in the literature to evaluate the performance of causal graphs and RCA methods. There may be other threats related to the underlying tools, our extracted data, that we have not considered here. To enable exploration of these potential threats and to facilitate replication 
of our work, we make available our tools and data.

\subsection{External Validity}

We evaluate the methods using various synthetic datasets and benchmark systems, along with four common faults. These systems and faults are used in multiple published works on RCA with different characteristics and domains. We acknowledge that different software applications and faults could have different properties and failure propagation mechanisms, which could impact the conclusions in this paper. However, we believe that the datasets and systems we use are representative since they have been used in many previous studies~\citep{Jinjin2018Microscope, Azam2022rcd, wu2021microdiag, xin2023causalrca, wu2022automatic, he2022graph, yu2021microrank, zhou2018trainticket, Wang2021evalcausal} and could help us to derive various important insights for causal inference-based RCA methods for microservice systems.

\subsection{Conclusion Validity}

The conclusion validity threat of our evaluation is related to the fault types used in our experiments. Microservice systems can experience different faults that affect the RCA results. To address this, we use five different fault types in our study (CPU, MEM, DISK, DELAY, and LOSS), covering a wide range of different failure scenarios in microservice systems. This allows us to properly evaluate the performance of the studied RCA methods. This is a significant improvement over previous studies, which typically involved only 2-3 fault types~\citep{Azam2022rcd, xin2023causalrca, wu2021microdiag, lee2023eadro}.

%% file: papers/ase24/7.related_work.tex
Studying causal inference-based RCA methods and evaluating their performance are important research topics. The work described in~\citep{Soldani2022rcasurvey} conducts a comprehensive survey on anomaly detection and RCA methods for (micro) service-based cloud systems. The studied methods include both causal inference-based and other methods. This survey, however, does not include any evaluation of the RCA methods. 
The works in~\citep{Arya2021evalcausalai} and~\citep{Wang2021evalcausal} evaluate Granger algorithms for AIOps on the Train Ticket microservice system using time series data constructed from logs. Our work, on the other hand, evaluates a wide range of causal discovery and causal inference-based RCA methods on various synthetic datasets and microservice systems. 

The work described in~\citep{Wu2021evalcausal} evaluates six popular causal discovery methods and combines them with PageRank to identify the root causes from metrics data. These methods are evaluated using the Sock Shop and Train Ticket systems. This work, however, does not evaluate the effectiveness of the constructed causal graph, the scoring methods, recent state-of-the-art causal inference-based methods (RCD, CIRCA, CausalRCA, RUN), and other important aspects such as the impact of hyperparameter tuning, input data length, among others. More recently, there is the work in~\citep{Siebert2023causal} that surveys different causal inference-based methods with the applications in software engineering. This survey, however, does not include any evaluation of these causal inference-based techniques. There are also research works that study and/or evaluate the performance of causal discovery methods for time series data, such as~\citep{Assaad2022causaltimesurvey, Moraffah2021causalsurvey, Glymour2019causal}. However, these works only focus on generic time series data, not metrics data from microservice systems.

%% file: papers/ase24/8.conclusion.tex

In this chapter, we presented a comprehensive evaluation and in-depth analysis of nine causal discovery methods and twenty-one causal inference-based RCA methods for microservice systems using metrics data. We derived many valuable insights from our evaluation and conclude that the performance of existing causal inference-based RCA methods can be further improved before being efficiently and effectively applied in practice. We also released new datasets to facilitate further research in this area and suggested possible future research directions. These contributions advance the understanding of causal inference-based RCA methods for microservice systems and pave the way for more robust and impactful solutions.

%% file: chapters/8.conclusion.tex
\chapter{Conclusion} \label{chap:conclusion}

\section{Summary of Contributions}

This thesis addresses the central research question: \textit{Can failures in microservice systems be detected and diagnosed automatically from multimodal observability data using machine learning and statistical techniques?} Through systematic investigation across multiple data modalities, novel methods, and comprehensive benchmarking, this thesis answers this question. The contributions span two complementary dimensions: novel methods that advance the state of the art in anomaly detection and root cause analysis (RCA), and benchmarking infrastructure that enables reproducible evaluation of these methods.

Along the \textbf{methods dimension}, this thesis develops a progression of approaches that address increasingly challenging aspects of automated failure diagnosis. The progression begins with metric-based analysis, where \barotool{} (Chapter~\ref{chap:fse24}) introduces an end-to-end framework that jointly performs anomaly detection and RCA on multivariate time-series metrics. A key insight behind \barotool{} is that existing RCA methods assume perfect anomaly detection, yet in practice, imprecise detection significantly degrades downstream diagnosis. \barotool{} addresses this through Multivariate Bayesian Online Change Point Detection and a novel nonparametric hypothesis testing technique called RobustScorer, achieving robust root cause localisation even under imprecise detection.

Recognising that metrics alone cannot capture all failure manifestations, the thesis expands to event-based analysis. \eventadl{} (Chapter~\ref{chap:eventadl}) is the first open-box anomaly detection and localisation framework designed for event data, a modality largely unexplored for RCA despite its richness in capturing system activities such as API calls, infrastructure changes, and security actions. Grounded in an empirical analysis of 520 real-world incidents from Amazon Web Services, \eventadl{} learns Event Semantic Patterns and Event Frequency Patterns during training, detects deviations online, and localises root causes through an Intervention Graph with time-aware random walks. Its open-box design provides interpretable explanations, achieving F1-scores of at least 90\% for detection and 100\% top-3 accuracy for localisation on real cloud-based systems.

The progression culminates in multimodal analysis, where \torai{} (Chapter~\ref{chap:torai}) combines metrics, logs, and traces to perform RCA without relying on service call graphs. Existing multimodal methods assume complete trace coverage to construct such graphs, an assumption that fails in real-world systems containing black-box services without traces. \torai{} addresses these blind spots through severity-based clustering and causal analysis, outperforming baselines in systems with incomplete instrumentation. Together, these three methods demonstrate that moving from single-modality to multimodal analysis progressively addresses the blind spots and limitations inherent in any individual data source.

Along the \textbf{benchmarking dimension}, this thesis contributes standardised resources and systematic evaluation efforts. \rcaeval{} (Chapter~\ref{chap:www25}) provides the first comprehensive open-source benchmark for RCA in microservice systems, comprising three large-scale datasets with 735 failure cases and fifteen reproducible baselines. By standardising datasets, evaluation protocols, and baselines, \rcaeval{} addresses a critical gap that previously made fair comparison of RCA methods difficult. Complementing this infrastructure, the systematic evaluation in Chapter~\ref{chap:ase24} examines nine causal discovery algorithms and twenty-one causal inference-based RCA methods, revealing that no single method consistently performs best, that performance on synthetic data does not reliably predict real-system performance, and that most methods struggle with large systems exceeding 200 metrics.

The two dimensions are coupled rather than merely parallel, and the results reported in the preceding chapters bear out the relationships outlined in Chapter~\ref{chap:introduction}. The three methods advance along two axes, widening the range of telemetry that can be diagnosed, from metrics to events and then to metrics, logs, and traces combined, while successively removing the assumptions that constrained the preceding method, namely precise anomaly segmentation, the manifestation of anomalies in numeric telemetry, and the availability of a complete service call graph. Because all three are unsupervised, efficient enough for online use, and produce a ranked list of candidate root causes with supporting evidence, they compose into a single diagnostic capability rather than competing for the same deployment. An operator instrumented only for metrics obtains diagnosis from \barotool{}, an operator whose failures surface as infrastructure and control-plane activity obtains diagnosis from \eventadl{}, and an operator with partial tracing coverage obtains diagnosis from \torai{}, all through the same form of output. The benchmarking dimension closes the loop around them, since the evaluation in Chapter~\ref{chap:ase24} produced the RE1 dataset and the evidence about where existing causal inference-based techniques break down, \rcaeval{} generalises that evaluation into a reusable benchmark, and the proposed methods are released through \rcaeval{} itself, with \barotool{} distributed as one of its baselines and \torai{} integrated into the same codebase. What this thesis delivers is therefore not a collection of related methods but a cohesive framework for automated anomaly detection and RCA across multiple data modalities.

These two dimensions are complementary: the novel methods push the boundaries of what automated RCA can achieve, while the benchmarking infrastructure provides the foundation for evaluation and comparison. They advance the scientific understanding and practical applicability of automated failure diagnosis in microservice systems.

\section{Thesis Impact and Significance}

The contributions of this thesis advance both the scientific understanding and practical applicability of automated anomaly detection and RCA in microservice systems. From a \textbf{scientific perspective}, this thesis makes several notable advances:

\textbf{Methodological Innovation.} The thesis introduces novel techniques across multiple dimensions: Multivariate Bayesian Online Change Point Detection for metrics, Event Semantic and Frequency Patterns for events, severity-based clustering for multimodal RCA without service call graphs, and robust hypothesis testing for root cause localisation. These techniques collectively expand the methodological toolkit available to researchers and demonstrate the value of combining statistical rigor with domain-specific insights.

\textbf{Empirical Understanding.} Through the systematic analysis of 520 real-world incidents and comprehensive evaluation of existing causal inference-based methods, this thesis provides unprecedented empirical insights into how failures manifest in observability data and which techniques work (or fail) in practice. These insights inform not only the design of new methods but also the realistic assessment of existing approaches.

\textbf{Multimodal Integration.} While prior work has largely focused on individual modalities, this thesis demonstrates the necessity and feasibility of multimodal approaches. By exploring metrics, events, logs, and traces both independently and collectively, it establishes a framework for understanding system behaviour and diagnosing failures comprehensively.

\textbf{Benchmark and Reproducibility.} RCAEval provides the research community with standardised datasets, reproducible baselines, and evaluation protocols, addressing a critical gap that has hindered progress. This contribution facilitates fair comparisons, accelerates research development, and lowers barriers to entry for new researchers.

From a \textbf{practical perspective}, this thesis addresses real-world operational challenges:

\textbf{Operational Feasibility.} By developing unsupervised methods that do not require extensive labelled data, this thesis acknowledges the reality that failure patterns evolve rapidly and labelled data become quickly outdated. The proposed methods (BARO, EventADL, and TORAI) are designed to operate with minimal human intervention.

\textbf{Interpretability and Trust.} EventADL's open-box design, along with BARO's hypothesis testing approach, provide interpretable explanations for detected anomalies and identified root causes. This interpretability is essential for operator trust and enables engineers to validate automated recommendations before taking corrective actions.

\textbf{Robustness to Incomplete Data.} TORAI's ability to perform RCA in presence of blind spots addresses a pervasive challenge in real-world systems where instrumentation is incomplete, third-party services lack traces, or legacy components cannot be easily modified. This robustness makes automated RCA practical in heterogeneous, evolving environments.

\textbf{Efficiency.} Methods like BARO and EventADL are designed for computational efficiency, enabling near-real-time anomaly detection and root cause localisation. This efficiency is critical for minimizing mean time to recovery (MTTR) and reducing the operational and financial impact of failures.

\textbf{Industrial Relevance and Path to Adoption.} The problems addressed in this thesis are drawn from operational practice rather than posed in the abstract. \eventadl{} was designed around an empirical analysis of 520 real-world incidents at Amazon Web Services and evaluated on real cloud-based systems, and the blind spot problem motivating \torai{} reflects the incomplete instrumentation routinely encountered in production deployments, where third-party and legacy components emit no traces. Each method is released as an open-source artifact (Appendix~B), with \barotool{} receiving the ACM SIGSOFT Best Artifact Award at FSE 2024, and \rcaeval{} packages the datasets, baselines, and evaluation protocols needed to reproduce the reported results. Releasing usable tooling alongside the methods lowers the cost for practitioners to trial these techniques on their own systems, which is a precondition for industrial uptake. Converting that potential into measured operational benefit, however, requires evaluation inside production environments and with the engineers who would use these tools, a direction taken up in Section~\ref{sec:future-directions}.

\section{Limitations and Reflections}

While this thesis makes substantial contributions, it is important to acknowledge its limitations and reflect on areas where further research is needed.

The evaluation in this thesis relies on three benchmark microservice systems (Train Ticket, Sock Shop, and Online Boutique) and selected real-world incidents. Although these systems are well-established in the research community, they may not capture the full complexity of large-scale commercial deployments at companies like Netflix, Amazon, or Google, which often operate thousands of services with custom infrastructure. Expanding evaluation to a broader range of systems, including commercial platforms where possible, would strengthen claims of generalisability.

A related challenge is concept drift. Software systems evolve continuously through code updates, configuration changes, and workload variations, causing the patterns characterising normal behaviour to shift over time. While \barotool{}'s online change point detection and \eventadl{}'s pattern learning are designed to adapt, they require periodic retraining or recalibration. More sophisticated adaptive learning techniques that detect and respond to concept drift automatically, without manual intervention, remain an open problem.

Beyond detection and localisation, a gap persists between diagnosis and remediation. This thesis evaluates root cause localisation at both service-level (coarse-grained) and indicator-level (fine-grained) granularity. However, operators often require even more specific information, such as the exact code module, configuration parameter, or external dependency responsible for a failure. While fine-grained root cause metrics or events provide valuable diagnostic clues, translating them into actionable remediation steps remains a largely manual process. Automated mapping from identified root causes to specific remediation actions, potentially leveraging knowledge bases, historical incident data, or large language models, represents an important direction.

Finally, scalability remains a concern. The evaluation study in Chapter~\ref{chap:ase24} reveals that most causal inference-based RCA methods struggle with large systems exceeding 200 metrics. While \torai{} and \barotool{} demonstrate better scalability through clustering and hypothesis testing respectively, systems with thousands of services may still pose challenges. Investigating whether existing automated anomaly detection and RCA methods can scale to such systems without sacrificing accuracy is an important area for future work.

\section{Future Research Directions}
\label{sec:future-directions}

Building on the contributions and limitations identified in this thesis, several promising avenues for future research emerge, organised below into six thematic clusters.

A natural and immediate next step is to \textbf{evaluate the proposed methods on the proposed benchmark}. This thesis contributes both novel RCA methods (\barotool{}, \eventadl{}, \torai{}) and a comprehensive benchmark (\rcaeval{} with 735 failure cases and fifteen baselines). A systematic evaluation of the proposed methods within the \rcaeval{} framework would enable direct, fair comparison against existing baselines under standardised conditions. Such an evaluation would reveal the complementary strengths and weaknesses of each method across different fault types, system scales, and data modalities. The results could inform the design of ensemble approaches that combine the strengths of multiple methods, for example leveraging \barotool{}'s robustness to imprecise detection, \eventadl{}'s event-based analysis, and \torai{}'s multimodal integration. 

A second direction concerns the move \textbf{from root cause analysis to explainable RCA}. The methods proposed in this thesis identify \textit{where} a failure originates, and \eventadl{}'s open-box design goes some way towards explaining \textit{why} an anomaly was flagged. Neither, however, reconstructs the full account an operator needs, which is how the root cause produced the symptoms that were observed. \citet{Soldani2022rcasurvey} identify this as an open challenge. Addressing it means recovering the propagation of a failure across the microservice system and the causal chain that links the root cause to each observed symptom, then rendering that chain in a form an operator can act on. Such explanations would change what operators can do with a diagnosis. Rather than resolving the incident alone, an operator who knows the propagation path can act on the architecture, for example by placing bulkheads along that path to contain the blast radius of similar failures and prevent the anomalies from recurring. Explainability of this kind cannot be assessed by accuracy metrics alone. A user-centred evaluation with professional site reliability engineers, measuring whether the generated explanations reduce diagnosis time and increase trust in the automated recommendation, would complement the accuracy-oriented evaluation adopted throughout this thesis and provide stronger evidence of practical value.

A third direction concerns \textbf{validation in industrial environments}. The evaluation in this thesis is conducted on benchmark microservice systems and on real incident data from a large cloud provider, which establishes technical validity but leaves the operational impact of the proposed methods on industrial practice largely unmeasured. Industry-driven case studies, in which \barotool{}, \eventadl{}, or \torai{} are deployed alongside the monitoring stack of a production system and assessed over an extended period, would provide that evidence. Such studies would report the measures that matter to practitioners, including the effect on mean time to detection and mean time to recovery, the rate of false alarms under realistic workload variation, the cost of instrumentation and retraining, and the willingness of on-call engineers to act on automated diagnoses. They would also expose constraints that benchmark evaluation cannot surface, such as multi-tenant deployments, continuous release cycles, and the organisational processes surrounding incident response. Closing this loop between research prototypes and industrial practice is essential for the contributions of this thesis to reach the systems that motivated them.

A fourth direction concerns the transition \textbf{from diagnosis to proactive mitigation}. This thesis focuses on detection and diagnosis, but the natural next step is automated remediation. Future research could develop self-healing systems that not only detect and diagnose failures but also automatically execute corrective actions, such as restarting services, scaling resources, rolling back deployments, or rerouting traffic. Such systems would need to reason about the safety and effectiveness of potential actions, possibly using reinforcement learning to learn optimal remediation policies from historical incidents. Closely related is the integration of RCA into broader incident management workflows involving alerting, escalation, communication, and post-mortem analysis. Intelligent triage systems that prioritize incidents based on severity, collaborative platforms that facilitate knowledge sharing among operators, and automated documentation systems that generate post-mortem reports from detected anomalies and identified root causes all represent important avenues.

A fifth direction involves \textbf{foundation models and continuous learning} for RCA. Recent advances in foundation models and large language models have demonstrated capabilities in reasoning over heterogeneous data. Multimodal foundation models pre-trained on large corpora of metrics, logs, traces, and events from diverse microservice systems could learn generalizable representations of normal and anomalous behaviour, enabling few-shot or zero-shot RCA on new systems without extensive retraining. Such models are also a plausible vehicle for the explainable RCA discussed above, since they could render a reconstructed propagation path as a natural language narrative, a visualisation, or a counterfactual account of what would have prevented the failure. Additionally, because microservice systems are dynamic with continuous deployments and evolving workloads, RCA systems must adapt without manual retraining. Techniques from continual learning, online learning, and transfer learning could enable models to incrementally update their understanding of system behaviour while avoiding catastrophic forgetting.

A sixth direction concerns \textbf{generalisation, domain knowledge, and privacy}. Most RCA methods are trained and evaluated on individual systems, requiring separate models for each deployment. Future research could investigate cross-system generalisation through system-agnostic features or meta-learning to quickly adapt to new systems with minimal data. Causal discovery from observational data alone is also challenging in complex systems with confounders and feedback loops. Hybrid approaches that combine data-driven causal discovery with domain knowledge, such as service dependencies, deployment configurations, and known failure modes, could significantly improve causal graph construction and root cause inference. Finally, as RCA systems become more automated and potentially cloud-based, ensuring the security and privacy of observability data becomes paramount. Privacy-preserving techniques such as federated learning, differential privacy, and homomorphic encryption could enable RCA without exposing sensitive telemetry data.

\section{Closing Remarks}

The journey of this thesis began with a fundamental question: \textit{Can we automate the detection and diagnosis of failures in increasingly complex microservice systems?} Through systematic investigation spanning multiple data modalities, developing novel techniques grounded in statistical rigor and domain insights, and conducting comprehensive evaluations on real systems and incidents, this thesis demonstrates that automated anomaly detection and root cause analysis are not only feasible but also practical and effective.

The contributions of this thesis, \barotool{}, \eventadl{}, \torai{}, \rcaeval{}, and the systematic evaluation of causal inference-based methods, collectively establish a comprehensive framework for automated failure troubleshooting in microservice systems. These contributions provide tools and insights that can reduce mean time to recovery, minimize the impact of failures, and empower operators to manage complex systems confidently.

Yet, as with any scientific endeavor, each answer raises new questions. The limitations and future research directions outlined in this chapter highlight that while significant progress has been made, much work remains. The most immediate next step is evaluating the proposed methods within the proposed benchmark to unify the two contribution dimensions. Beyond that, the move from root cause localisation to explainable RCA that reconstructs how a failure propagated, the validation of these methods through industry-driven case studies in production environments, the transition from reactive diagnosis to proactive mitigation, the integration of foundation models, the challenges of continuous learning in evolving systems, and the imperative of security and privacy all represent frontiers for future exploration.

As microservice architectures continue to evolve and the scale and complexity of cloud systems grow, the importance of automated anomaly detection and root cause analysis will only increase. This thesis lays a solid foundation for this critical research area, providing methodologies, benchmarks, and insights that will guide future investigations. We hope that the contributions presented here will inspire and enable researchers and practitioners to build more reliable, resilient, and self-managing microservice systems that can detect failures swiftly, diagnose their causes accurately, and ultimately, heal themselves autonomously.

The pursuit of truly autonomous, self-healing systems is a challenge that will require sustained effort from the research and industrial communities. This thesis represents one step forward on that journey. The road ahead is long, but the destination, microservice systems that can operate reliably at scale with minimal human intervention, is well worth the effort.

%% file: appendices/a.tex
\chapter*{Appendix A: List of Publications}
\addcontentsline{toc}{chapter}{Appendix A: List of Publications}

This thesis is a compilation of the following peer-reviewed papers. I declare that I am the main contributor to these works within the period of my PhD candidature. I carried out studies [1--4] in collaboration with my PhD advisors: Dr. Huong Ha (initial principal supervisor), Prof. Hongyu Zhang, and Prof. Xiuzhen Zhang. I carried out study [5] as lead author during my internship at Amazon Web Services. I declare that I take full responsibility for the inclusion of all the following works in my PhD thesis: \\

\noindent [1] \textbf{Luan Pham}, Huong Ha, and Hongyu Zhang. ``BARO: Robust Root Cause Analysis for Microservices via Multivariate Bayesian Online Change Point Detection.'' In Proceedings of the ACM on Software Engineering (FSE), 2024, pp.~2214--2237. [\textbf{CORE~A*}] (\textbf{Chapter~\ref{chap:fse24}}) \\

\noindent [2] \textbf{Luan Pham}, Huong Ha, and Hongyu Zhang. ``Root Cause Analysis for Microservice System Based on Causal Inference: How Far Are We?'' In Proceedings of the 39th IEEE/ACM International Conference on Automated Software Engineering (ASE), 2024, pp.~706--715. [\textbf{CORE~A*}] (\textbf{Chapter~\ref{chap:ase24}}) \\

\noindent [3] \textbf{Luan Pham}, Hongyu Zhang, Huong Ha, Flora Salim, and Xiuzhen Zhang. ``RCAEval: A Benchmark for Root Cause Analysis of Microservice Systems with Telemetry Data.'' In Companion Proceedings of the ACM on Web Conference (WWW), 2025, pp.~777--780. [\textbf{CORE~A*}] (\textbf{Chapter~\ref{chap:www25}}) \\

\noindent [4] \textbf{Luan Pham}, Huong Ha, Xiuzhen Zhang, and Hongyu Zhang. ``TORAI: Multi-Source Root Cause Analysis for Blind Spots in Microservice Service Call Graph.'' In Proceedings of the ACM on Software Engineering (FSE), 2026, Article FSE130, 23 pages. [\textbf{CORE~A*}] (\textbf{Chapter~\ref{chap:torai}}) \\

\noindent [5] \textbf{Luan Pham}, Victor Nicolet, Joey Dodds, Hui Guan, and Daniel Kroening. ``EventADL: Open-Box Anomaly Detection and Localization Framework for Events in Cloud-Based Service Systems.'' In Proceedings of the ACM on Software Engineering (FSE), 2026, Article FSE179. [\textbf{CORE~A*}] (\textbf{Chapter~\ref{chap:eventadl}})

%% file: appendices/b.tex
\chapter*{Appendix B: List of Contributed Artifacts}
\addcontentsline{toc}{chapter}{Appendix B: List of Contributed Artifacts}

This thesis contributes the following tools: \\

\begin{enumerate}[leftmargin=*]
    \item \textbf{BARO}: This tool won the \textbf{ACM SIGSOFT Best Artifact Award} at the ACM FSE 2024 Conference. The source code is available at \url{https://github.com/phamquiluan/baro}.
    \item \textbf{EventADL}: An open-box anomaly detection and root cause localisation framework for event data. The artifact is available at \url{https://doi.org/10.5281/zenodo.19433492}.
    \item \textbf{RCAEval}: An open-source benchmark for root cause analysis in microservice systems. RCAEval integrates the artifacts of both the ASE'24 causal-inference evaluation (Chapter~\ref{chap:ase24}) and the WWW'25 RCAEval short paper (Chapter~\ref{chap:www25}). The source code is available at \url{https://github.com/phamquiluan/rcaeval}.
    \item \textbf{TORAI}: A multi-source RCA method for blind spots in microservice systems. TORAI is integrated into the RCAEval framework and is available at \url{https://github.com/phamquiluan/rcaeval}.
\end{enumerate}

%% file: appendices/c.tex
\chapter*{Appendix C: List of Services}
\addcontentsline{toc}{chapter}{Appendix C: List of Services}

During my PhD candidature, I have served as a reviewer for the following journals:
\begin{itemize}[leftmargin=*,nolistsep]
    \item \textbf{ACM Transactions on Software Engineering and Methodology}.
    \item \textbf{IEEE Transactions on Dependable and Secure Computing}.
    \item \textbf{IEEE Transactions on Software Engineering}.
    \item \textbf{IEEE Signal Processing Letters}.
    \item \textbf{Springer Nature, Empirical Software Engineering}.
    \item \textbf{Springer Nature, Applied Intelligence}.
\end{itemize}

\noindent I have served as a Program Committee Member for the following conferences:
\begin{itemize}[leftmargin=*,nolistsep]
    \item IEEE/ACM International Conference on Software Engineering\\\textbf{ICSE 2026, ICSE~2027} (Demonstration Track).
    \item ACM International Conference on the Foundations of Software Engineering\\\textbf{FSE 2025, FSE 2026} (Artifact Track).
    \item International Conference on Evaluation and Assessment in Software Engineering\\\textbf{EASE 2025, EASE 2026} (AI Models / Data Track).
    \item IEEE/ACM International Conference on Automated Software Engineering\\\textbf{ASE 2025, ASE 2026} (Tool Track).
    \item IEEE/ACM International Conference on Mining Software Repositories\\\textbf{MSR 2025} (\textbf{Junior Program Committee Member}).
    \item IEEE International Symposium on Software Reliability Engineering\\\textbf{ISSRE 2025} (Artifact Track).
    \item ACM Conference on AI and Agentic Systems\\\textbf{ACM CAIS 2026} (Main Track).
\end{itemize}

\noindent I have served as a Reviewer for the following conferences:
\begin{itemize}[leftmargin=*,nolistsep]
    \item International Conference on Machine Learning\\\textbf{ICML 2026} (Main Track; \textbf{Silver Reviewer Award}).
    \item The Conference on Neural Information Processing Systems\\\textbf{NeurIPS 2026} (Main Track).
    \item International Joint Conference on Artificial Intelligence / European Conference on Artificial Intelligence\\\textbf{IJCAI-ECAI 2026} (Main Track \& Survey Track; \textbf{Gold Tier Reviewer Award}).
    \item International Joint Conference on Artificial Intelligence\\\textbf{IJCAI 2025} (Main Track \& Survey Track).
    \item European Conference on Artificial Intelligence\\\textbf{ECAI 2025} (Main Track \& Emergency).
    \item International Conference on Pattern Recognition\\\textbf{ICPR 2024, ICPR 2026} (Main Track).
    \item ACM SIGKDD International Conference on Knowledge Discovery and Data Mining\\\textbf{KDD 2025} (Artifact Track).
    \item International Conference on Learning Representations\\\textbf{ICLR 2025} (SCI-FM Workshop).
\end{itemize}

%% file: bib/references.bib
@article{Soldani2018microservice,
title = {The pains and gains of microservices: A Systematic grey literature review},
journal = {Journal of Systems and Software},
volume = {146},
pages = {215-232},
year = {2018},
issn = {0164-1212},
author = {Jacopo Soldani and Damian Andrew Tamburri and Willem-Jan {Van Den Heuvel}},
}

@article{Gregory2025Optus,
  author       = {Mark A. Gregory},
  title        = {Optus Triple Zero outage has left multiple people dead. A telecommunications expert explains what went wrong – and how to fix it},
  journal      = {Law Society Journal},
  year         = {2025},
  month        = {Sep 22},
  url          = {https://lsj.com.au/articles/optus-triple-zero-outage-has-left-multiple-people-dead-a-telecommunications-expert-explains-what-went-wrong-and-how-to-fix-it/},
  note         = {Accessed: 2025-12-22}
}

@inproceedings{
  orchard2025root,
  title={Root Cause Analysis of Outliers with Missing Structural Knowledge},
  author={William Roy Orchard and Nastaran Okati and Sergio Hernan Garrido Mejia and Patrick Bl{\"o}baum and Dominik Janzing},
  booktitle={The Thirty-ninth Annual Conference on Neural Information Processing Systems},
  year={2025},
  url={https://openreview.net/forum?id=7Nxq4RQApu}
}

@article{zhang2025adaptive,
  title={Adaptive root cause localization for microservice systems with multi-agent recursion-of-thought},
  author={Zhang, Lingzhe and Jia, Tong and Wang, Kangjin and Hong, Weijie and Duan, Chiming and He, Minghua and Li, Ying},
  journal={arXiv preprint arXiv:2508.20370},
  year={2025}
}

@inproceedings{budhathoki2022causal,
  title={Causal structure-based root cause analysis of outliers},
  author={Budhathoki, Kailash and Minorics, Lenon and Bl{\"o}baum, Patrick and Janzing, Dominik},
  booktitle={International conference on machine learning},
  pages={2357--2369},
  year={2022},
  organization={PMLR}
}

@inproceedings{lee2023eadro,
  title={Eadro: An end-to-end troubleshooting framework for microservices on multi-source data},
  author={Lee, Cheryl and Yang, Tianyi and Chen, Zhuangbin and Su, Yuxin and Lyu, Michael R},
  booktitle={2023 IEEE/ACM 45th International Conference on Software Engineering (ICSE)},
  pages={1750--1762},
  year={2023},
  organization={IEEE}
}

@misc{ocsf,
  author       = {{Open Cybersecurity Schema Framework}},
  title        = {Open Cybersecurity Schema Framework (OCSF)},
  year         = {2022},
  url          = {https://github.com/ocsf},
  note         = {Accessed: 2025-08-04}
}

@article{avizienis2004basic,
  title={Basic concepts and taxonomy of dependable and secure computing},
  author={Avizienis, Algirdas and Laprie, J-C and Randell, Brian and Landwehr, Carl},
  journal={IEEE transactions on dependable and secure computing},
  volume={1},
  number={1},
  pages={11--33},
  year={2004},
  publisher={IEEE}
}

@inproceedings{wu2020performance,
  title={Performance diagnosis in cloud microservices using deep learning},
  author={Wu, Li and Bogatinovski, Jasmin and Nedelkoski, Sasho and Tordsson, Johan and Kao, Odej},
  booktitle={International Conference on Service-Oriented Computing},
  pages={85--96},
  year={2020},
  organization={Springer}
}

@inproceedings{wang2020root,
  title={Root-cause metric location for microservice systems via log anomaly detection},
  author={Wang, Lingzhi and Zhao, Nengwen and Chen, Junjie and Li, Pinnong and Zhang, Wenchi and Sui, Kaixin},
  booktitle={2020 IEEE international conference on web services (ICWS)},
  pages={142--150},
  year={2020},
  organization={IEEE}
}

@inproceedings{wu2020microrca,
  title={Microrca: Root cause localization of performance issues in microservices},
  author={Wu, Li and Tordsson, Johan and Elmroth, Erik and Kao, Odej},
  booktitle={2020 IEEE/IFIP Network Operations and Management Symposium},
  pages={1--9},
  year={2020},
}

@inproceedings{wu2021microdiag,
  title={Microdiag: Fine-grained performance diagnosis for microservice systems},
  author={Wu, Li and Tordsson, Johan and Bogatinovski, Jasmin and Elmroth, Erik and Kao, Odej},
  booktitle={2021 IEEE/ACM International Workshop on Cloud Intelligence (CloudIntelligence)},
  pages={31--36},
  year={2021},
  organization={IEEE}
}

@inproceedings{yu2021microrank,
  title={Microrank: End-to-end latency issue localization with extended spectrum analysis in microservice environments},
  author={Yu, Guangba and Chen, Pengfei and Chen, Hongyang and Guan, Zijie and Huang, Zicheng and Jing, Linxiao and Weng, Tianjun and Sun, Xinmeng and Li, Xiaoyun},
  booktitle={Proceedings of the Web Conference (WWW'21)},
  pages={3087--3098},
  year={2021}
}

@inproceedings{Chakraborty2023CausIL,
author = {Chakraborty, Sarthak and Garg, Shaddy and Agarwal, Shubham and Chauhan, Ayush and Saini, Shiv Kumar},
title = {CausIL: Causal Graph for Instance Level Microservice Data},
year = {2023},
booktitle = {Proceedings of the ACM Web Conference},
pages = {2905–2915},
series = {WWW '23}
}

@inproceedings{he2022graph,
  title={Graph based Incident Extraction and Diagnosis in Large-Scale Online Systems},
  author={He, Zilong and Chen, Pengfei and Luo, Yu and Yan, Qiuyu and Chen, Hongyang and Yu, Guangba and Li, Fangyuan},
  booktitle={Proceedings of the 37th IEEE/ACM International Conference on Automated Software Engineering (ASE'22)},
  pages={1-13},
  year={2022}
}

@article{Soldani2022rcasurvey,
author = {Soldani, Jacopo and Brogi, Antonio},
title = {Anomaly Detection and Failure Root Cause Analysis in (Micro) Service-Based Cloud Applications: A Survey},
year = {2022},
volume = {55},
number = {3},
journal = {ACM Computing Surveys},
articleno = {59},
numpages = {39},
doi = {10.1145/3501297},
}

@inproceedings{Azam2022rcd,
 author = {Ikram, Azam and Chakraborty, Sarthak and Mitra, Subrata and Saini, Shiv and Bagchi, Saurabh and Kocaoglu, Murat},
 booktitle = {Advances in Neural Information Processing Systems (NeurIPS'22)},
 pages = {31158--31170},
 title = {Root Cause Analysis of Failures in Microservices through Causal Discovery},
 volume = {35},
 year = {2022}
}

@inproceedings{Wang2018cloudranger,
  author={Wang, Ping and Xu, Jingmin and Ma, Meng and Lin, Weilan and Pan, Disheng and Wang, Yuan and Chen, Pengfei},
  booktitle={18th IEEE/ACM International Symposium on Cluster, Cloud and Grid Computing (CCGRID'18)}, 
  title={CloudRanger: Root Cause Identification for Cloud Native Systems}, 
  year={2018},
  volume={},
  number={},
  pages={492-502}
}

@inproceedings{Meng2020Microcause,
  author={Meng, Yuan and Zhang, Shenglin and Sun, Yongqian and Zhang, Ruru and Hu, Zhilong and Zhang, Yiyin and Jia, Chenyang and Wang, Zhaogang and Pei, Dan},
  booktitle={IEEE/ACM 28th International Symposium on Quality of Service (IWQoS'20)}, 
  title={Localizing Failure Root Causes in a Microservice through Causality Inference}, 
  year={2020},
  volume={},
  number={},
  pages={1-10}
}

@article{Runge2019PCMCI,
author = {Jakob Runge  and Peer Nowack  and Marlene Kretschmer and Seth Flaxman and Dino Sejdinovic },
title = {Detecting and quantifying causal associations in large nonlinear time series datasets},
journal = {Science Advances},
volume = {5},
number = {11},
year = {2019},
}

@book{Spirtes1993Causal,
  title={Causation, prediction, and search},
  author={Spirtes, Peter and Glymour, Clark N and Scheines, Richard},
  year={2000},
  publisher={MIT press}
}

@inproceedings{Jinjin2018Microscope,
author="Lin, Jinjin
and Chen, Pengfei
and Zheng, Zibin",
title="Microscope: Pinpoint Performance Issues with Causal Graphs in Micro-service Environments",
booktitle="Service-Oriented Computing",
year="2018",
pages="3--20"
}

@inproceedings{Chen2014Causeinfer,
  author={Chen, Pengfei and Qi, Yong and Zheng, Pengfei and Hou, Di},
  booktitle={IEEE Conference on Computer Communications (INFOCOM'14)}, 
  title={CauseInfer: Automatic and distributed performance diagnosis with hierarchical causality graph in large distributed systems}, 
  year={2014},
  volume={},
  number={},
  pages={1887-1895}
}

@inproceedings{aggarwal2021causal,
  title={Causal modeling based fault localization in cloud systems using golden signals},
  author={Aggarwal, Pooja and Nagar, Seema and Gupta, Ajay and Shwartz, Larisa and Mohapatra, Prateeti and Wang, Qing and Paradkar, Amit and Mandal, Atri},
  booktitle={2021 IEEE 14th International Conference on Cloud Computing (CLOUD)},
  pages={124--135},
  year={2021},
  organization={IEEE}
}

@inproceedings{aggarwal2020localization,
  title={Localization of operational faults in cloud applications by mining causal dependencies in logs using golden signals},
  author={Aggarwal, Pooja and Gupta, Ajay and Mohapatra, Prateeti and Nagar, Seema and Mandal, Atri and Wang, Qing and Paradkar, Amit},
  booktitle={International Conference on Service-Oriented Computing},
  pages={137--149},
  year={2020},
  organization={Springer}
}

@article{chen2016causeinfer,
  title={CauseInfer: Automated end-to-end performance diagnosis with hierarchical causality graph in cloud environment},
  author={Chen, Pengfei and Qi, Yong and Hou, Di},
  journal={IEEE transactions on services computing},
  volume={12},
  number={2},
  pages={214--230},
  year={2016},
  publisher={IEEE}
}

@inproceedings{Ma2020Automap,
author = {Ma, Meng and Xu, Jingmin and Wang, Yuan and Chen, Pengfei and Zhang, Zonghua and Wang, Ping},
title = {AutoMAP: Diagnose Your Microservice-Based Web Applications Automatically},
year = {2020},
booktitle = {Proceedings of The Web Conference},
pages = {246–258},
series = {WWW'20}
}

@inproceedings{liu2021microhecl,
  title={Microhecl: High-efficient root cause localization in large-scale microservice systems},
  author={Liu, Dewei and He, Chuan and Peng, Xin and Lin, Fan and Zhang, Chenxi and Gong, Shengfang and Li, Ziang and Ou, Jiayu and Wu, Zheshun},
  booktitle={2021 IEEE/ACM 43rd International Conference on Software Engineering: Software Engineering in Practice (ICSE-SEIP)},
  pages={338--347},
  year={2021},
  organization={IEEE}
}

@inproceedings{chen2022adaptive,
  title={Adaptive performance anomaly detection for online service systems via pattern sketching},
  author={Chen, Zhuangbin and Liu, Jinyang and Su, Yuxin and Zhang, Hongyu and Ling, Xiao and Yang, Yongqiang and Lyu, Michael R},
  booktitle={Proceedings of the 44th International Conference on Software Engineering},
  pages={61--72},
  year={2022}
}

@inproceedings{yu2023cmdiagnostor,
  title={CMDiagnostor: An Ambiguity-Aware Root Cause Localization Approach Based on Call Metric Data},
  author={Yu, Qingyang and Pei, Changhua and Hao, Bowen and Li, Mingjie and Li, Zeyan and Zhang, Shenglin and Lu, Xianglin and Wang, Rui and Li, Jiaqi and Wu, Zhenyu and others},
  booktitle={Proceedings of the ACM Web Conference 2023},
  pages={2937--2947},
  year={2023}
}

@inproceedings{Ma2019Msrank,
  author={Ma, Meng and Lin, Weilan and Pan, Disheng and Wang, Ping},
  booktitle={IEEE International Conference on Web Services (ICWS'19)}, 
  title={MS-Rank: Multi-Metric and Self-Adaptive Root Cause Diagnosis for Microservice Applications}, 
  year={2019},
  volume={},
  number={},
  pages={60-67},
}

@article{zhang1996birch,
  title={BIRCH: an efficient data clustering method for very large databases},
  author={Zhang, Tian and Ramakrishnan, Raghu and Livny, Miron},
  journal={ACM sigmod record},
  volume={25},
  number={2},
  pages={103--114},
  year={1996},
  publisher={ACM New York, NY, USA}
}

@inproceedings{siffer2017anomaly,
  title={Anomaly detection in streams with extreme value theory},
  author={Siffer, Alban and Fouque, Pierre-Alain and Termier, Alexandre and Largouet, Christine},
  booktitle={Proceedings of the 23rd ACM SIGKDD international conference on knowledge discovery and data mining},
  pages={1067--1075},
  year={2017}
}

@article{Brin1998Pagerank,
title = {The anatomy of a large-scale hypertextual Web search engine},
journal = {Computer Networks and ISDN Systems},
volume = {30},
number = {1},
pages = {107-117},
year = {1998},
author = {Sergey Brin and Lawrence Page},
}

@inproceedings{Li2022Circa,
author = {Li, Mingjie and Li, Zeyan and Yin, Kanglin and Nie, Xiaohui and Zhang, Wenchi and Sui, Kaixin and Pei, Dan},
title = {Causal Inference-Based Root Cause Analysis for Online Service Systems with Intervention Recognition},
year = {2022},
booktitle = {Proceedings of the 28th ACM SIGKDD Conference on Knowledge Discovery and Data Mining (KDD'22)},
pages = {3230–3240},
}

@inproceedings{Jaber2020PsiFCI,
 author = {Jaber, Amin and Kocaoglu, Murat and Shanmugam, Karthikeyan and Bareinboim, Elias},
 booktitle = {Advances in Neural Information Processing Systems (NeurIPS'20)},
 pages = {9551--9561},
 title = {Causal Discovery from Soft Interventions with Unknown Targets: Characterization and Learning},
 volume = {33},
 year = {2020}
}

@book{wu2022automatic,
  title={Automatic performance diagnosis and recovery in cloud microservices},
  author={Wu, Li},
  year={2022},
  publisher={Technische Universitaet Berlin (Germany)}
}

@article{xin2023causalrca,
title = {CausalRCA: Causal inference based precise fine-grained root cause localization for microservice applications},
journal = {Journal of Systems and Software},
volume = {203},
pages = {111724},
year = {2023},
author = {Ruyue Xin and Peng Chen and Zhiming Zhao},
}

@InProceedings{Yu2019DagGNN,
  title = 	 {{DAG}-{GNN}: {DAG} Structure Learning with Graph Neural Networks},
  author =       {Yu, Yue and Chen, Jie and Gao, Tian and Yu, Mo},
  booktitle = 	 {Proceedings of the 36th International Conference on Machine Learning (ICML'19)},
  pages = 	 {7154--7163},
  year = 	 {2019},
  volume = 	 {97},
  month = 	 {09--15 Jun}
}

@inproceedings{biza2020bictuning,
  title={Tuning causal discovery algorithms},
  author={Biza, Konstantina and Tsamardinos, Ioannis and Triantafillou, Sofia},
  booktitle={International Conference on Probabilistic Graphical Models},
  pages={17--28},
  year={2020},
  organization={PMLR}
}

@article{zhou2018trainticket,
  title={Fault analysis and debugging of microservice systems: Industrial survey, benchmark system, and empirical study},
  author={Zhou, Xiang and Peng, Xin and Xie, Tao and Sun, Jun and Ji, Chao and Li, Wenhai and Ding, Dan},
  journal={IEEE Transactions on Software Engineering},
  volume={47},
  number={2},
  pages={243--260},
  year={2018},
  publisher={IEEE}
}

@inproceedings{mariani2018localizing,
  title={Localizing faults in cloud systems},
  author={Mariani, Leonardo and Monni, Cristina and Pezz{\'e}, Mauro and Riganelli, Oliviero and Xin, Rui},
  booktitle={2018 IEEE 11th International Conference on Software Testing, Verification and Validation (ICST)},
  pages={262--273},
  year={2018},
  organization={IEEE}
}

@book{wohlin2012experimentation,
  title={Experimentation in software engineering},
  author={Wohlin, Claes and Runeson, Per and H{\"o}st, Martin and Ohlsson, Magnus C and Regnell, Bj{\"o}rn and Wessl{\'e}n, Anders},
  year={2012},
  publisher={Springer Science \& Business Media}
}

@inproceedings{pan2021dycauserca,
  title={Faster, deeper, easier: crowdsourcing diagnosis of microservice kernel failure from user space},
  author={Pan, Yicheng and Ma, Meng and Jiang, Xinrui and Wang, Ping},
  booktitle={Proceedings of the 30th ACM SIGSOFT International Symposium on Software Testing and Analysis},
  pages={646--657},
  year={2021}
}

@INPROCEEDINGS{dan2021tracerca,
  author={Li, Zeyan and Chen, Junjie and Jiao, Rui and Zhao, Nengwen and Wang, Zhijun and Zhang, Shuwei and Wu, Yanjun and Jiang, Long and Yan, Leiqin and Wang, Zikai and Chen, Zhekang and Zhang, Wenchi and Nie, Xiaohui and Sui, Kaixin and Pei, Dan},
  booktitle={2021 IEEE/ACM 29th International Symposium on Quality of Service (IWQOS'21)}, 
  title={Practical Root Cause Localization for Microservice Systems via Trace Analysis}, 
  year={2021},
  volume={},
  number={},
  pages={1-10}
}

@article{Granger1980causal,
title = {Testing for causality: A personal viewpoint},
journal = {Journal of Economic Dynamics and Control},
volume = {2},
pages = {329-352},
year = {1980},
author = {C.W.J. Granger},
}

@inproceedings{Spirtes1995fci,
author = {Spirtes, Peter and Meek, Christopher and Richardson, Thomas},
title = {Causal Inference in the Presence of Latent Variables and Selection Bias},
year = {1995},
booktitle = {Proceedings of the Eleventh Conference on Uncertainty in Artificial Intelligence},
pages = {499–506},
numpages = {8},
series = {UAI'95}
}

@article{Shimizu2006Lingam,
  author  = {Shohei Shimizu and Patrik O. Hoyer and Aapo Hyvarinen and Antti Kerminen},
  title   = {A Linear Non-Gaussian Acyclic Model for Causal Discovery},
  journal = {Journal of Machine Learning Research},
  year    = {2006},
  volume  = {7},
  number  = {72},
  pages   = {2003--2030}
}

@article{Chickering2002Ges,
author = {Chickering, David Maxwell},
title = {Learning Equivalence Classes of Bayesian-Network Structures},
year = {2002},
publisher = {JMLR.org},
volume = {2},
pages = {445–498},
journal = {Journal of Machine Learning Research}
}

@article{Schwarz1978bic,
author = {Gideon Schwarz},
title = {{Estimating the Dimension of a Model}},
volume = {6},
journal = {The Annals of Statistics},
number = {2},
publisher = {Institute of Mathematical Statistics},
pages = {461 -- 464},
year = {1978}
}

@article{Ramsey2017fges,
  author       = {Joseph D. Ramsey and
                  Madelyn Glymour and
                  Ruben Sanchez{-}Romero and
                  Clark Glymour},
  title        = {A million variables and more: the Fast Greedy Equivalence Search algorithm for learning high-dimensional graphical causal models, with an application
                  to functional magnetic resonance images},
  journal      = {International Journal of Data Science and Analytics},
  volume       = {3},
  number       = {2},
  pages        = {121--129},
  year         = {2017}
}

@book{Spitzer1976randomwalk,
author = {Spitzer, Frank Ludvig},
title = { Principles of random walk},
publisher = { Springer-Verlag New York },
pages = {408},
year = { 1976 }
}

@inproceedings{Arya2021evalcausalai,
author = {Arya, Vijay and Shanmugam, Karthikeyan and Aggarwal, Pooja and Wang, Qing and Mohapatra, Prateeti and Nagar, Seema},
title = {Evaluation of Causal Inference Techniques for AIOps},
year = {2021},
booktitle = {Proceedings of the 3rd ACM India Joint International Conference on Data Science and Management of Data},
pages = {188–192},
series = {CODS-COMAD '21}
}

@article{friedman1983graph,
  title={Graph-theoretic measures of multivariate association and prediction},
  author={Friedman, Jerome H and Rafsky, Lawrence C},
  journal={The Annals of Statistics},
  pages={377--391},
  year={1983},
  publisher={JSTOR}
}

@article{friedman1979multivariate,
  title={Multivariate generalizations of the Wald-Wolfowitz and Smirnov two-sample tests},
  author={Friedman, Jerome H and Rafsky, Lawrence C},
  journal={The Annals of Statistics},
  pages={697--717},
  year={1979},
  publisher={JSTOR}
}

@inproceedings{luo2014correlating,
  title={Correlating events with time series for incident diagnosis},
  author={Luo, Chen and Lou, Jian-Guang and Lin, Qingwei and Fu, Qiang and Ding, Rui and Zhang, Dongmei and Wang, Zhe},
  booktitle={Proceedings of the 20th ACM SIGKDD international conference on Knowledge discovery and data mining},
  pages={1583--1592},
  year={2014}
}

@inproceedings{Wang2021evalcausal,
  author={Wang, Qing and Shwartz, Larisa and Grabarnik, Genady Ya. and Arya, Vijay and Shanmugam, Karthikeyan},
  booktitle={IEEE 14th International Conference on Cloud Computing (CLOUD'21)}, 
  title={Detecting Causal Structure on Cloud Application Microservices Using Granger Causality Models}, 
  year={2021},
  volume={},
  number={},
  pages={558-565}
}

@inproceedings{samir2019dla,
  title={Dla: Detecting and localizing anomalies in containerized microservice architectures using markov models},
  author={Samir, Areeg and Pahl, Claus},
  booktitle={2019 7th International Conference on Future Internet of Things and Cloud (FiCloud)},
  pages={205--213},
  year={2019},
  organization={IEEE}
}

@article{brandon2020graph,
  title={Graph-based root cause analysis for service-oriented and microservice architectures},
  author={Brand{\'o}n, {\'A}lvaro and Sol{\'e}, Marc and Hu{\'e}lamo, Alberto and Solans, David and P{\'e}rez, Mar{\'\i}a S and Munt{\'e}s-Mulero, Victor},
  journal={Journal of Systems and Software},
  volume={159},
  pages={110432},
  year={2020},
  publisher={Elsevier}
}

@article{zheng2018dags,
  title={Dags with no tears: Continuous optimization for structure learning},
  author={Zheng, Xun and Aragam, Bryon and Ravikumar, Pradeep K and Xing, Eric P},
  journal={Advances in neural information processing systems},
  volume={31},
  year={2018}
}

@article{causalai23salesforce,
  title={Salesforce CausalAI Library: A Fast and Scalable Framework for Causal Analysis of Time Series and Tabular Data},
  author={Arpit, Devansh and Fernandez, Matthew and Feigenbaum, Itai and Yao, Weiran and Liu, Chenghao and Yang, Wenzhuo and Josel, Paul and Heinecke, Shelby and Hu, Eric and Wang, Huan and others},
  journal={arXiv preprint arXiv:2301.10859},
  year={2023}
}

@article{van2020evaluation,
  title={An evaluation of change point detection algorithms},
  author={Van den Burg, Gerrit JJ and Williams, Christopher KI},
  journal={arXiv preprint arXiv:2003.06222},
  year={2020}
}

@inproceedings{Wu2021evalcausal,
  author={Wu, Li and Tordsson, Johan and Elmroth, Erik and Kao, Odej},
  booktitle={2021 IEEE International Conference on Autonomic Computing and Self-Organizing Systems (ACSOS'21)}, 
  title={Causal Inference Techniques for Microservice Performance Diagnosis: Evaluation and Guiding Recommendations}, 
  year={2021},
  volume={},
  number={},
  pages={21-30}
}

@inproceedings{thalheim2017sieve,
  title={Sieve: Actionable insights from monitored metrics in distributed systems},
  author={Thalheim, J{\"o}rg and Rodrigues, Antonio and Akkus, Istemi Ekin and Bhatotia, Pramod and Chen, Ruichuan and Viswanath, Bimal and Jiao, Lei and Fetzer, Christof},
  booktitle={Proceedings of the 18th ACM/IFIP/USENIX Middleware Conference (Middleware'17)},
  pages={14--27},
  year={2017}
}

@article{liu2023pyrca,
  title={PyRCA: A Library for Metric-based Root Cause Analysis},
  author={Liu, Chenghao and Yang, Wenzhuo and Mittal, Himanshu and Singh, Manpreet and Sahoo, Doyen and Hoi, Steven CH},
  journal={arXiv preprint arXiv:2306.11417},
  year={2023}
}

@inproceedings{li2022actionable,
  title = {Actionable and Interpretable Fault Localization for Recurring Failures in Online Service Systems},
  booktitle = {Proceedings of the 30th {{ACM Joint Meeting}} on {{European Software Engineering Conference}} and {{Symposium}} on the {{Foundations}} of {{Software Engineering}}},
  author = {Li, Zeyan and Zhao, Nengwen and Li, Mingjie and Lu, Xianglin and Wang, Lixin and Chang, Dongdong and Nie, Xiaohui and Cao, Li and Zhang, Wenchi and Sui, Kaixin and Wang, Yanhua and Du, Xu and Duan, Guoqing and Pei, Dan},
  year = {2022},
  month = nov,
  series = {{{ESEC}}/{{FSE}}'22}
}

@article{Siebert2023causal,
title = {Applications of statistical causal inference in software engineering},
journal = {Information and Software Technology},
volume = {159},
year = {2023},
author = {Siebert, Julien},
}

@inproceedings{Chen2020incidentmanagement,
author = {Chen, Zhuangbin and Kang, Yu and Li, Liqun and Zhang, Xu and Zhang, Hongyu and Xu, Hui and Zhou, Yangfan and Yang, Li and Sun, Jeffrey and Xu, Zhangwei and Dang, Yingnong and Gao, Feng and Zhao, Pu and Qiao, Bo and Lin, Qingwei and Zhang, Dongmei and Lyu, Michael R.},
title = {Towards Intelligent Incident Management: Why We Need It and How We Make It},
year = {2020},
booktitle = {Proceedings of the 28th ACM Joint Meeting on European Software Engineering Conference and Symposium on the Foundations of Software Engineering},
pages = {1487–1497},
numpages = {11},
location = {Virtual Event, USA} 
}

@INPROCEEDINGS{Chen2019incidenttriage,
  author={Chen, Junjie and He, Xiaoting and Lin, Qingwei and Xu, Yong and Zhang, Hongyu and Hao, Dan and Gao, Feng and Xu, Zhangwei and Dang, Yingnong and Zhang, Dongmei},
  booktitle={2019 IEEE/ACM 41st International Conference on Software Engineering: Software Engineering in Practice (ICSE-SEIP)}, 
  title={An Empirical Investigation of Incident Triage for Online Service Systems}, 
  year={2019},
  volume={},
  number={},
  pages={111-120}
}

@article{Vowels2022causaldiscoverysurvey,
author = {Vowels, Matthew J. and Camgoz, Necati Cihan and Bowden, Richard},
title = {D’Ya Like DAGs? A Survey on Structure Learning and Causal Discovery},
year = {2022},
address = {New York, NY, USA},
volume = {55},
number = {4},
journal = {ACM Computing Surveys},
articleno = {82}
}

@inproceedings{raghu2018evaluation,
  title={Evaluation of causal structure learning methods on mixed data types},
  author={Raghu, Vineet K and Poon, Allen and Benos, Panayiotis V},
  booktitle={Proceedings of 2018 ACM SIGKDD Workshop on Causal Discovery},
  pages={48--65},
  year={2018},
  organization={PMLR}
}

@article{Assaad2022causaltimesurvey,
author = {Assaad, Charles K. and Devijver, Emilie and Gaussier, Eric},
title = {Survey and Evaluation of Causal Discovery Methods for Time Series},
year = {2022},
issue_date = {May 2022},
publisher = {AI Access Foundation},
volume = {73},
journal = {Journal of Artificial Intelligence Research},
}

@inproceedings{Chen2019airalert,
author = {Chen, Yujun and Yang, Xian and Lin, Qingwei and Zhang, Hongyu and Gao, Feng and Xu, Zhangwei and Dang, Yingnong and Zhang, Dongmei and Dong, Hang and Xu, Yong and Li, Hao and Kang, Yu},
title = {Outage Prediction and Diagnosis for Cloud Service Systems},
year = {2019},
publisher = {Association for Computing Machinery},
address = {New York, NY, USA},
booktitle = {The World Wide Web Conference},
pages = {2659–2665},
location = {San Francisco, CA, USA},
series = {WWW '19}
}

@inproceedings{Shan2019Ediagnosis,
author = {Shan, Huasong and Chen, Yuan and Liu, Haifeng and Zhang, Yunpeng and Xiao, Xiao and He, Xiaofeng and Li, Min and Ding, Wei},
title = {{$\epsilon$}-Diagnosis: Unsupervised and Real-Time Diagnosis of Small-Window Long-Tail Latency in Large-Scale Microservice Platforms},
year = {2019},
booktitle = {The World Wide Web Conference (WWW'19)},
pages = {3215–3222}
}

@article{Moraffah2021causalsurvey,
author = {Moraffah, Raha and Sheth, Paras and Karami, Mansooreh and Bhattacharya, Anchit and Wang, Qianru and Tahir, Anique and Raglin, Adrienne and Liu, Huan},
title = {Causal Inference for Time Series Analysis: Problems, Methods and Evaluation},
year = {2021},
publisher = {Springer-Verlag},
address = {Berlin, Heidelberg},
volume = {63},
number = {12},
journal = {Knowledge and Information Systems},
pages = {3041–3085}
}

@article{Glymour2019causal,
author={Glymour, Clark and Zhang, Kun and Spirtes, Peter},
title={Review of Causal Discovery Methods Based on Graphical Models},      
journal={Frontiers in Genetics},      
volume={10},           
year={2019},
}

@misc{statsmodels,
  author       = {Skipper Seabold and Josef Perktold},
  title        = {Statsmodels: statistical modelling and econometrics in Python},
  howpublished = {\url{https://github.com/statsmodels/statsmodels/}},
  year         = {2023},
  note         = {Accessed: Mar 16, 2024}
}

@misc{cadvisor,
author = {Bobby Page and Ivan Wan-Geh (“iwankgb”) and Joe Davis (“dims”) and others},
title = {Container Advisor - an open-source tool to monitor containers},
url={https://github.com/google/cadvisor},
year={2024},
lastaccessed = {Sep 13, 2024},
}

@misc{pyagrum,
  author       = {Gaspard Ducamp and Christophe Gonzales and Pierre-Henri Wuillemin},
  title        = {A library dedicated to Bayesian networks and Probabilistic Graphical Models (pyAgrum)},
  howpublished = {\url{https://pyagrum.readthedocs.io/en/1.0.0/}},
  year         = {2023},
  note         = {Accessed: Mar 16, 2024}
}

@misc{istio,
  author       = {Istio Project},
  title        = {The Istio service mesh},
  howpublished = {\url{https://istio.io/}},
  year         = {2023},
  note         = {Accessed: Sep 12, 2024}
}

@misc{prometheus,
  author       = {Matt T. Proud and Julius Volz},
  title        = {An open-source monitoring and alerting toolkit (Prometheus)},
  howpublished = {\url{https://prometheus.io/}},
  year         = {2023},
  note         = {Accessed: Sep 12, 2024}
}

@misc{elasticsearch,
  author       = {Shay Banon},
  title        = {Elasticsearch -- Log Monitoring \& Centralized Log Management},
  howpublished = {\url{https://www.elastic.co/}},
  year         = {2024},
  note         = {Accessed: Sep 12, 2024}
}

@misc{vector,
  author       = {Luke Steensen and Datadog (vectordotdev)},
  title        = {A lightweight, ultra-fast tool for building observability pipelines (Vector)},
  howpublished = {\url{https://vector.dev/}},
  year         = {2024},
  note         = {Accessed: Sep 12, 2024}
}

@misc{loki,
  author       = {Tom Wilkie and Grafana Labs},
  title        = {Grafana Loki -- A horizontally scalable, highly available, multi-tenant log aggregation system},
  howpublished = {\url{https://grafana.com/oss/loki/}},
  year         = {2024},
  note         = {Accessed: Sep 12, 2024}
}

@misc{jaeger,
  author       = {Yuri Shkuro (original author) and Uber Technologies, Inc.},
  title        = {Jaeger: open source, distributed tracing platform},
  howpublished = {\url{https://www.jaegertracing.io/}},
  year         = {2024},
  note         = {Accessed: Sep 12, 2024}
}

@article{fang2023low,
  title={On Low-Rank Directed Acyclic Graphs and Causal Structure Learning},
  author={Fang, Zhuangyan and Zhu, Shengyu and Zhang, Jiji and Liu, Yue and Chen, Zhitang and He, Yangbo},
  journal={IEEE Transactions on Neural Networks and Learning Systems},
  year={2023},
  publisher={IEEE}
}

@misc{odigos2025pitfalls,
  author       = {{Odigos}},
  title        = {Solving the Pitfalls of Distributed Tracing in Real-World Microservices},
  howpublished = {\url{https://odigos.io/blog/solving-pitfalls-of-distributed-tracing-in-real-world-microservices}},
  year         = {2025},
  note         = {Accessed on September 2, 2025}
}

@inproceedings{shen2023deepflow,
  title={Network-centric distributed tracing with DeepFlow: Troubleshooting your microservices in zero code},
  author={Shen, Junxian and Zhang, Han and Xiang, Yang and Shi, Xingang and Li, Xinrui and Shen, Yunxi and Zhang, Zijian and Wu, Yongxiang and Yin, Xia and Wang, Jilong and others},
  booktitle={Proceedings of the ACM SIGCOMM 2023 Conference},
  pages={420--437},
  year={2023}
}

@inproceedings{cotroneo2019bad,
  title={How bad can a bug get? an empirical analysis of software failures in the openstack cloud computing platform},
  author={Cotroneo, Domenico and De Simone, Luigi and Liguori, Pietro and Natella, Roberto and Bidokhti, Nematollah},
  booktitle={Proceedings of the 2019 27th ACM Joint Meeting on European Software Engineering Conference and Symposium on the Foundations of Software Engineering},
  pages={200--211},
  year={2019}
}

@misc{dynatrace,
author = {Dynatrace, Inc.},
title = {Dynatrace: Unified observability and security},
url={https://www.dynatrace.com},
year={2024},
lastaccessed = {Sep 12, 2024},
}

@misc{datadog,
author = {Olivier Pomel and Alexis Lê-Quôc},
title = {Datadog: Modern monitoring and security},
url={https://www.datadoghq.com},
year={2024},
lastaccessed = {Sep 12, 2024},
}

@inproceedings{altenbernd2025amocrca,
  title={Amocrca: at most one change segmentation and relative correlation ranking for root cause analysis},
  author={Altenbernd, Anton and Wu, Zhiyuan and Kao, Odej},
  booktitle={Proceedings of the 33rd ACM International Conference on the Foundations of Software Engineering},
  pages={1386--1393},
  year={2025}
}

@misc{dynatrace1,
author = {Dynatrace, Inc.},
title = {Set up anomaly detection based on your business needs},
url={https://www.dynatrace.com/news/blog/metric-events-set-up-anomaly-detection-based-on-your-business-needs/},
year={2024},
lastaccessed = {Sep 12, 2024},
}

@misc{dynatrace3,
author = {Dynatrace, Inc.},
title = {Dynatrace: Root cause analysis with example},
url={https://docs.dynatrace.com/docs/platform/davis-ai/problem-and-root-cause/root-cause-analysis#expand--details-and-example},
year={2024},
lastaccessed = {Sep 12, 2024},
}

@misc{uber,
  author       = {Schwarz, Mathias and Neverov, Andrew},
  title        = {{Up: Portable Microservices Ready for the Cloud}},
  howpublished = {Uber Engineering Blog, \url{https://www.uber.com/en-AU/blog/up-portable-microservices-ready-for-the-cloud}},
  month        = {September},
  day          = {7},
  year         = {2023},
  note         = {Accessed: 2025-09-24},
}

@misc{datadog2,
author = {Datadog},
title = {Automated root cause analysis with Watchdog RCA},
url={https://www.datadoghq.com/blog/datadog-watchdog-automated-root-cause-analysis/},
year={2024},
lastaccessed = {Sep 12, 2024},
}

@misc{datadog1,
author = {Datadog},
title = {Introducing anomaly detection in Datadog},
url={https://www.datadoghq.com/blog/introducing-anomaly-detection-datadog/},
year={2024},
lastaccessed = {Sep 12, 2024},
}

@misc{googlesre,
author = {Google Inc.},
title = {Google - Site Reliability Engineering},
url={https://sre.google/sre-book/monitoring-distributed-systems/},
year={2024},
lastaccessed = {Sep 12, 2024},
}

@misc{gmm,
author = {scikit-learn},
title = {Gaussian Mixture Model},
url={https://scikit-learn.org/stable/modules/generated/sklearn.mixture.GaussianMixture.html},
year={2024},
lastaccessed = {Sep 12, 2024},
}

@misc{sockshop,
  author = {Weaveworks},
  title = {{microservices-demo}: Deployment scripts \& config for Sock Shop},
  howpublished = {\url{https://github.com/microservices-demo/microservices-demo}},
  year         = {2023},
  note         = {Accessed: 2025-09-24},
}

@misc{ob,
  author       = {GoogleCloudPlatform},
  title        = {{microservices-demo}: Sample cloud-first application with 10 microservices},
  howpublished = {\url{https://github.com/GoogleCloudPlatform/microservices-demo}},
  year         = {2025},
  note         = {Accessed: 2025-09-24},
}

@misc{tt,
  author       = {FudanSELab},
  title        = {{train-ticket}: Train Ticket — A Benchmark Microservice System},
  howpublished = {\url{https://github.com/FudanSELab/train-ticket}},
  year         = {2025},
  note         = {Accessed: 2025-09-24},
}

@misc{unibcp,
  title = {Bayesian Change Point Detection},
  author = {Johannes Kulick},
  url ={https://github.com/hildensia/bayesian_changepoint_detection},
  year={2023},
  lastaccessed = {Sep 28, 2023},
}

@misc{stressng,
  author = {Colin King},
  title = {Stress test for Computer system (stress-ng)},
  howpublished = {\url{https://wiki.ubuntu.com/Kernel/Reference/stress-ng}},
  year = {2024},
  note = {Accessed: Feb 21, 2024}
}

@misc{tc,
  author       = {Alexey N. Kuznetsov},
  title        = {Traffic Control},
  howpublished = {\url{https://man7.org/linux/man-pages/man8/tc.8.html}},
  year         = {2024},
  note         = {Accessed: Feb 21, 2024}
}

@misc{modifiedzscore,
  title = {Modified z score},
  author = {{IBM Corporation}},
  url={https://www.ibm.com/docs/en/cognos-analytics/11.1.0?topic=terms-modified-z-score},
  year={2023},
  lastaccessed = {Sep 28, 2023},
}

@misc{robustzscore,
  author = {{Asia Pacific Laboratory Accreditation Cooperation (APLAC)}},
  title = {Statistical Procedures, Calculations and Formulae, Appendix D},
  howpublished = {\url{https://www.apac-accreditation.org/app/uploads/2017/08/aplac_t017_appendix_d.pdf}},
  year = {2023},
  note = {Accessed: Sep 28, 2023}
}

@article{adams2007bayesian,
  title={Bayesian online changepoint detection},
  author={Adams, Ryan Prescott and MacKay, David JC},
  journal={arXiv preprint arXiv:0710.3742},
  year={2007}
}

@inproceedings{xuan2007modeling,
  title={Modeling changing dependency structure in multivariate time series},
  author={Xuan, Xiang and Murphy, Kevin},
  booktitle={Proceedings of the 24th international conference on Machine learning},
  pages={1055--1062},
  year={2007}
}

@book{Coles2001EVT,
  address = {London},
  author = {Coles, Stuart},
  publisher = {Springer-Verlag},
  series = {Springer Series in Statistics},
  title = {An introduction to statistical modeling of extreme values},
  year = 2001
}

@article{BlazquezGarc2021ADreview,
author = {Bl\'{a}zquez-Garc\'{\i}a, Ane and Conde, Angel and Mori, Usue and Lozano, Jose A.},
title = {A Review on Outlier/Anomaly Detection in Time Series Data},
year = {2021},
issue_date = {April 2022},
address = {New York, NY, USA},
volume = {54},
number = {3},
journal = {ACM Computing Surveys},
articleno = {56},
}

@inproceedings{du2017deeplog,
  title={Deeplog: Anomaly detection and diagnosis from system logs through deep learning},
  author={Du, Min and Li, Feifei and Zheng, Guineng and Srikumar, Vivek},
  booktitle={Proceedings of the 2017 ACM SIGSAC conference on computer and communications security},
  pages={1285--1298},
  year={2017}
}

@inproceedings{pham2024root,
  title={Root Cause Analysis for Microservice System based on Causal Inference: How Far Are We?},
  author={Pham, Luan and Ha, Huong and Zhang, Hongyu},
  booktitle={The 39th IEEE/ACM International Conference on Automated Software Engineering (ASE 2024)},
  year={2024}
}

@inproceedings{yu2023nezha,
  title={Nezha: Interpretable Fine-Grained Root Causes Analysis for Microservices on Multi-modal Observability Data},
  author={Yu, Guangba and Chen, Pengfei and Li, Yufeng and Chen, Hongyang and Li, Xiaoyun and Zheng, Zibin},
  booktitle={Proceedings of the 31st ACM Joint European Software Engineering Conference and Symposium on the Foundations of Software Engineering},
  pages={553--565},
  year={2023}
}

@article{zhang2023diagfusion,
  title={Robust failure diagnosis of microservice system through multimodal data},
  author={Zhang, Shenglin and Jin, Pengxiang and Lin, Zihan and Sun, Yongqian and Zhang, Bicheng and Xia, Sibo and Li, Zhengdan and Zhong, Zhenyu and Ma, Minghua and Jin, Wa and others},
  journal={IEEE Transactions on Services Computing},
  year={2023},
  publisher={IEEE}
}

@inproceedings{hou2021pdiagnose,
  title={Diagnosing performance issues in microservices with heterogeneous data source},
  author={Hou, Chuanjia and Jia, Tong and Wu, Yifan and Li, Ying and Han, Jing},
  booktitle={2021 IEEE Intl Conf on Parallel \& Distributed Processing with Applications, Big Data \& Cloud Computing, Sustainable Computing \& Communications, Social Computing \& Networking (ISPA/BDCloud/SocialCom/SustainCom)},
  pages={493--500},
  year={2021},
  organization={IEEE}
}

@inproceedings{he2017drain,
  title={Drain: An online log parsing approach with fixed depth tree},
  author={He, Pinjia and Zhu, Jieming and Zheng, Zibin and Lyu, Michael R},
  booktitle={2017 IEEE international conference on web services (ICWS)},
  pages={33--40},
  year={2017},
  organization={IEEE}
}

@article{he2021survey,
  title={A survey on automated log analysis for reliability engineering},
  author={He, Shilin and He, Pinjia and Chen, Zhuangbin and Yang, Tianyi and Su, Yuxin and Lyu, Michael R},
  journal={ACM computing surveys (CSUR)},
  volume={54},
  number={6},
  pages={1--37},
  year={2021},
  publisher={ACM New York, NY, USA}
}

@inproceedings{liu2020unsupervised,
  title={Unsupervised detection of microservice trace anomalies through service-level deep bayesian networks},
  author={Liu, Ping and Xu, Haowen and Ouyang, Qianyu and Jiao, Rui and Chen, Zhekang and Zhang, Shenglin and Yang, Jiahai and Mo, Linlin and Zeng, Jice and Xue, Wenman and others},
  booktitle={2020 IEEE 31st International Symposium on Software Reliability Engineering (ISSRE)},
  pages={48--58},
  year={2020},
  organization={IEEE}
}

@article{patel2020clustering,
  title={Clustering cloud workloads: K-means vs gaussian mixture model},
  author={Patel, Eva and Kushwaha, Dharmender Singh},
  journal={Procedia computer science},
  volume={171},
  pages={158--167},
  year={2020},
  publisher={Elsevier}
}

@article{luo2022depth,
  title={An in-depth study of microservice call graph and runtime performance},
  author={Luo, Shutian and Xu, Huanle and Lu, Chengzhi and Ye, Kejiang and Xu, Guoyao and Zhang, Liping and He, Jian and Xu, Chengzhong},
  journal={IEEE Transactions on Parallel and Distributed Systems},
  volume={33},
  number={12},
  pages={3901--3914},
  year={2022},
  publisher={IEEE}
}

@inproceedings{zhang2023benefit,
  title={The Benefit of Hindsight: Tracing $\{$Edge-Cases$\}$ in Distributed Systems},
  author={Zhang, Lei and Xie, Zhiqiang and Anand, Vaastav and Vigfusson, Ymir and Mace, Jonathan},
  booktitle={20th USENIX Symposium on Networked Systems Design and Implementation (NSDI 23)},
  pages={321--339},
  year={2023}
}

@article{giamattei2023monitoring,
  title={Monitoring tools for DevOps and microservices: A systematic grey literature review},
  author={Giamattei, Luca and Guerriero, Antonio and Pietrantuono, Roberto and Russo, Stefano and Malavolta, Ivano and Islam, Tanjina and Dinga, Madalina and Koziolek, Anne and Singh, Snigdha and Armbruster, Martin and others},
  journal={Journal of Systems and Software},
  pages={111906},
  year={2023},
  publisher={Elsevier}
}

@article{janes2023open,
  title={Open tracing tools: Overview and critical comparison},
  author={Janes, Andrea and Li, Xiaozhou and Lenarduzzi, Valentina},
  journal={Journal of Systems and Software},
  pages={111793},
  year={2023},
  publisher={Elsevier}
}

@article{pham2024baro,
  title={BARO: Robust Root Cause Analysis for Microservices via Multivariate Bayesian Online Change Point Detection},
  author={Pham, Luan and Ha, Huong and Zhang, Hongyu},
  journal={Proceedings of the ACM on Software Engineering},
  volume={1},
  number={FSE},
  pages={2214--2237},
  year={2024},
  publisher={ACM New York, NY, USA}
}

@inproceedings{zhang2021cloudrca,
  title={CloudRCA: A root cause analysis framework for cloud computing platforms},
  author={Zhang, Yingying and Guan, Zhengxiong and Qian, Huajie and Xu, Leili and Liu, Hengbo and Wen, Qingsong and Sun, Liang and Jiang, Junwei and Fan, Lunting and Ke, Min},
  booktitle={Proceedings of the 30th ACM International Conference on Information \& Knowledge Management},
  pages={4373--4382},
  year={2021}
}

@article{zhu2024hemirca,
  title={HeMiRCA: Fine-Grained Root Cause Analysis for Microservices with Heterogeneous Data Sources},
  author={Zhu, Zhouruixing and Lee, Cheryl and Tang, Xiaoying and He, Pinjia},
  journal={ACM Transactions on Software Engineering and Methodology},
  year={2024},
  publisher={ACM New York, NY}
}

@inproceedings{rouf2024instantops,
  title={InstantOps: A Joint Approach to System Failure Prediction and Root Cause Identification in Microserivces Cloud-Native Applications},
  author={Rouf, Raphael and Rasolroveicy, Mohammadreza and Litoiu, Marin and Nagar, Seema and Mohapatra, Prateeti and Gupta, Pranjal and Watts, Ian},
  booktitle={Proceedings of the 15th ACM/SPEC International Conference on Performance Engineering},
  pages={119--129},
  year={2024}
}

@article{xie2024tvdiag,
  title={TVDiag: A Task-oriented and View-invariant Failure Diagnosis Framework with Multimodal Data},
  author={Xie, Shuaiyu and Wang, Jian and He, Hanbin and Wang, Zhihao and Zhao, Yuqi and Zhang, Neng and Li, Bing},
  journal={arXiv preprint arXiv:2407.19711},
  year={2024}
}

@article{gu2023trinityrcl,
  title={TrinityRCL: Multi-Granular and Code-Level Root Cause Localization Using Multiple Types of Telemetry Data in Microservice Systems},
  author={Gu, Shenghui and Rong, Guoping and Ren, Tian and Zhang, He and Shen, Haifeng and Yu, Yongda and Li, Xian and Ouyang, Jian and Chen, Chunan},
  journal={IEEE Transactions on Software Engineering},
  volume={49},
  number={5},
  pages={3071--3088},
  year={2023},
  publisher={IEEE}
}

@inproceedings{liu2022microcbr,
  title={MicroCBR: Case-Based Reasoning on Spatio-temporal Fault Knowledge Graph for Microservices Troubleshooting},
  author={Liu, Fengrui and Wang, Yang and Li, Zhenyu and Ren, Rui and Guan, Hongtao and Yu, Xian and Chen, Xiaofan and Xie, Gaogang},
  booktitle={International Conference on Case-Based Reasoning},
  pages={224--239},
  year={2022},
  organization={Springer}
}

@inproceedings{wang2021groot,
  title={Groot: An event-graph-based approach for root cause analysis in industrial settings},
  author={Wang, Hanzhang and Wu, Zhengkai and Jiang, Huai and Huang, Yichao and Wang, Jiamu and Kopru, Selcuk and Xie, Tao},
  booktitle={2021 36th IEEE/ACM International Conference on Automated Software Engineering (ASE)},
  pages={419--429},
  year={2021},
  organization={IEEE}
}

@inproceedings{zheng2024mulan,
  title={MULAN: Multi-modal Causal Structure Learning and Root Cause Analysis for Microservice Systems},
  author={Zheng, Lecheng and Chen, Zhengzhang and He, Jingrui and Chen, Haifeng},
  booktitle={Proceedings of the ACM on Web Conference 2024},
  pages={4107--4116},
  year={2024}
}

@inproceedings{ashok2024traceweaver,
  title={TraceWeaver: Distributed Request Tracing for Microservices Without Application Modification},
  author={Ashok, Sachin and Harsh, Vipul and Godfrey, Brighten and Mittal, Radhika and Parthasarathy, Srinivasan and Shwartz, Larisa},
  booktitle={Proceedings of the ACM SIGCOMM 2024 Conference},
  pages={828--842},
  year={2024}
}

@inproceedings{run2024aaai,
  author = {Cheng{-}Ming Lin and Ching Chang and Wei{-}Yao Wang and Kuang{-}Da Wang and Wen{-}Chih Peng},
  title = {Root Cause Analysis in Microservice Using Neural Granger Causal Discovery},
  booktitle = {Thirty-Eighth {AAAI} Conference on Artificial Intelligence},
  pages = {206--213},
  year = {2024},
}

@misc{jsonlogic,
  title        = {{JSONLogic: A lightweight, safe way to share logic between systems}},
  author       = {Jeremy Wadhams},
  howpublished = {\url{https://jsonlogic.com/}},
  note         = {Accessed: 2025-07-02},
  year         = {n.d.},
}

@misc{zhang2021gcastle,
  title={gCastle: A Python Toolbox for Causal Discovery}, 
  author={Keli Zhang and Shengyu Zhu and Marcus Kalander and Ignavier Ng and Junjian Ye and Zhitang Chen and Lujia Pan},
  year={2021},
}

@inproceedings{le2023log,
  title={Log parsing with prompt-based few-shot learning},
  author={Le, Van-Hoang and Zhang, Hongyu},
  booktitle={2023 IEEE/ACM 45th International Conference on Software Engineering (ICSE)},
  pages={2438--2449},
  year={2023},
  organization={IEEE}
}

@misc{barogithub,
  title = {BARO: Root Cause Analysis for Microservices},
  author={Pham, Luan and Ha, Huong and Zhang, Hongyu},
  url={https://github.com/phamquiluan/baro},
  year={2024},
  lastaccessed = {May 14, 2024},
}

@misc{barogithubzenodo,
  title = {BARO Software Artifacts at Zenodo},
  author={Pham, Luan and Ha, Huong and Zhang, Hongyu},
  url={https://doi.org/10.5281/zenodo.11094092},
  year={2024},
  lastaccessed = {May 14, 2024},
}

@misc{aws_cloudtrail_events,
  author       = {{Amazon Web Services}},
  title        = {Understanding CloudTrail Events},
  year         = {2024},
  howpublished = {\url{https://docs.aws.amazon.com/awscloudtrail/latest/userguide/cloudtrail-events.html}},
  note         = {Accessed: 2025-06-06}
}

@misc{azure_monitor_eventhub_ingestion,
  author       = {{Microsoft Corporation}},
  title        = {Ingest Events from Azure Event Hubs into Azure Monitor Logs},
  year         = {2024},
  howpublished = {\url{https://learn.microsoft.com/en-us/azure/azure-monitor/logs/ingest-logs-event-hub}},
  note         = {Accessed: 2025-06-06}
}

@misc{google_cloud_audit_logs,
  author       = {{Google Cloud}},
  title        = {Cloud Audit Logs Overview},
  year         = {2024},
  howpublished = {\url{https://cloud.google.com/logging/docs/audit}},
  note         = {Accessed: 2025-06-06}
}

@article{lee2024explainable,
  title={Explainable time series anomaly detection using masked latent generative modeling},
  author={Lee, Daesoo and Malacarne, Sara and Aune, Erlend},
  journal={Pattern Recognition},
  volume={156},
  pages={110826},
  year={2024},
  publisher={Elsevier}
}

@inproceedings{lu2022matrix,
  title={Matrix profile XXIV: scaling time series anomaly detection to trillions of datapoints and ultra-fast arriving data streams},
  author={Lu, Yue and Wu, Renjie and Mueen, Abdullah and Zuluaga, Maria A and Keogh, Eamonn},
  booktitle={Proceedings of the 28th ACM SIGKDD Conference on Knowledge Discovery and Data Mining},
  pages={1173--1182},
  year={2022}
}

@inproceedings{zengy2022shadewatcher,
  title={Shadewatcher: Recommendation-guided cyber threat analysis using system audit records},
  author={Zeng, Jun and Wang, Xiang and Liu, Jiahao and Chen, Yinfang and Liang, Zhenkai and Chua, Tat-Seng and Chua, Zheng Leong},
  booktitle={2022 IEEE symposium on security and privacy (SP)},
  pages={489--506},
  year={2022},
  organization={IEEE}
}

@article{landauer2024critical,
  title={A critical review of common log data sets used for evaluation of sequence-based anomaly detection techniques},
  author={Landauer, Max and Skopik, Florian and Wurzenberger, Markus},
  journal={Proceedings of the ACM on Software Engineering},
  volume={1},
  number={FSE},
  pages={1354--1375},
  year={2024},
  publisher={ACM New York, NY, USA}
}

@inproceedings{meng2019loganomaly,
  title={LogAnomaly: Unsupervised detection of sequential and quantitative anomalies in unstructured logs},
  author={Meng, Weibin and Liu, Ying and Zhu, Yichen and Zhang, Shenglin and Pei, Dan and Liu, Yuqing and Chen, Yihao and Zhang, Ruizhi and Tao, Shimin and     Sun, Pei and others},
  booktitle={IJCAI},
  pages={4739--4745},
  year={2019}
}

@inproceedings{zhang2019robust,
  title={Robust log-based anomaly detection on unstable log data},
  author={Zhang, Xu and Xu, Yong and Lin, Qingwei and Qiao, Bo and Zhang, Hongyu and Dang, Yingnong and Xie, Chunyu and Yang, Xinsheng and Cheng, Qian and Li,  Ze and others},
  booktitle={Proceedings of the 2019 27th ACM joint meeting on European software engineering conference and symposium on the foundations of software            engineering},
  pages={807--817},
  year={2019}
}

@inproceedings{deepsvdd,
  title={Deep one-class classification},
  author={Ruff, Lukas and Vandermeulen, Robert and Goernitz, Nico and Deecke, Lucas and Siddiqui, Shoaib Ahmed and Binder, Alexander and M{\"u}ller, Emmanuel and Kloft, Marius},
  booktitle={International conference on machine learning},
  pages={4393--4402},
  year={2018},
  organization={PMLR}
}

@inproceedings{rdp,
  author       = {Hu Wang and
                  Guansong Pang and
                  Chunhua Shen and
                  Congbo Ma},
  title        = {Unsupervised Representation Learning by Predicting Random Distances},
  booktitle    = {{IJCAI}},
  pages        = {2950--2956},
  publisher    = {ijcai.org},
  year         = {2020}
}

@inproceedings{rca,
  author       = {Boyang Liu and
                  Ding Wang and
                  Kaixiang Lin and
                  Pang{-}Ning Tan and
                  Jiayu Zhou},
  title        = {{RCA:} {A} Deep Collaborative Autoencoder Approach for Anomaly Detection},
  booktitle    = {{IJCAI}},
  pages        = {1505--1511},
  publisher    = {ijcai.org},
  year         = {2021}
}

@inproceedings{neutral,
  author       = {Chen Qiu and
                  Timo Pfrommer and
                  Marius Kloft and
                  Stephan Mandt and
                  Maja Rudolph},
  title        = {Neural Transformation Learning for Deep Anomaly Detection Beyond Images},
  booktitle    = {Proceedings of Machine Learning Research},
  volume       = {139},
  pages        = {8703--8714},
  year         = {2021}
}

@inproceedings{icl,
  title={Anomaly detection for tabular data with internal contrastive learning},
  author={Shenkar, Tom and Wolf, Lior},
  booktitle={International Conference on Learning Representations},
  year={2022}
}

@article{dif,
  author       = {Hongzuo Xu and
                  Guansong Pang and
                  Yijie Wang and
                  Yongjun Wang},
  title        = {Deep Isolation Forest for Anomaly Detection},
  journal      = {{IEEE} Trans. Knowl. Data Eng.},
  volume       = {35},
  number       = {12},
  pages        = {12591--12604},
  year         = {2023}
}

@misc{hyglad,
      title={Hypergraph-Guided Regex Filter Synthesis for Event-Based Anomaly Detection},
      author={Margarida Ferreira and Victor Nicolet and Luan Pham and Joey Dodds and Daniel Kroening and Ines Lynce and Ruben Martins},
      year={2025},
      eprint={2509.06911},
      archivePrefix={arXiv},
      primaryClass={cs.SE},
      url={https://arxiv.org/abs/2509.06911},
}

@misc{yahoo_amazon_downtime_2018,
  author       = "{Yahoo Finance}",
  title        = "{Amazon's One‑Hour Downtime on Prime Day May Have Cost It \$72 Million to \$99 Million}",
  howpublished = "\url{https://finance.yahoo.com/news/amazon-apos-one-hour-downtime-145350120.html}",
  year         = "2018",
  month        = "Jul",
  note         = "Accessed: 2025‑06‑23"
}

@inproceedings{pham2025rcaeval,
  title={RCAEval: A Benchmark for Root Cause Analysis of Microservice Systems with Telemetry Data},
  author={Pham, Luan and Zhang, Hongyu and Ha, Huong and Salim, Flora and Zhang, Xiuzhen},
  booktitle={Companion Proceedings of the ACM on Web Conference 2025},
  pages={777--780},
  year={2025}
}

@manual{alibaba-actiontrail,
  title        = {Alibaba Cloud ActionTrail},
  author       = {{Alibaba Cloud}},
  year         = {2025},
  note         = {ActionTrail monitors and records your Alibaba Cloud account activities; latest documentation updated February 12, 2025},
  url          = {https://www.alibabacloud.com/help/en/actiontrail/}
}

@misc{aydore2022detecting,
  title={Detecting anomalous events from categorical data using autoencoders},
  author={Aydore, Sergul and Coskun, Baris and Melis, Luca},
  year={2022},
  month=dec # "~27",
  publisher={Google Patents},
  note={US Patent 11,537,902}
}

@inproceedings{event_ijcai16_ape,
  author       = {Ting Chen and
                  Lu{-}An Tang and
                  Yizhou Sun and
                  Zhengzhang Chen and
                  Kai Zhang},
  title        = {Entity Embedding-Based Anomaly Detection for Heterogeneous Categorical Events},
  booktitle    = {Proceedings of the Twenty-Fifth International Joint Conference on
                  Artificial Intelligence},
  pages        = {1396--1403},
  year         = {2016},
}

@article{cusum,
  author    = {E. S. Page},
  title     = {Continuous Inspection Schemes},
  journal   = {Biometrika},
  year      = {1954},
  volume    = {41},
  number    = {1–2},
  pages     = {100--115},
  doi       = {10.1093/biomet/41.1-2.100}
}

@article{brianlogsurvey2025,
   title={A Comprehensive Study of Machine Learning Techniques for Log-Based Anomaly Detection},
   volume={30},
   ISSN={1573-7616},
   url={http://dx.doi.org/10.1007/s10664-025-10669-3},
   DOI={10.1007/s10664-025-10669-3},
   number={5},
   journal={Empirical Software Engineering},
   publisher={Springer Science and Business Media LLC},
   author={Ali, Shan and Boufaied, Chaima and Bianculli, Domenico and Branco, Paula and Briand, Lionel},
   year={2025},
   month=jun 
}

@article{cheng2023ai,
  title={{AI} for {IT} Operations ({AIOps}) on Cloud Platforms: Reviews, Opportunities and Challenges},
  author={Cheng, Qian and Sahoo, Doyen and Saha, Amrita and Yang, Wenzhuo and Liu, Chenghao and Woo, Gerald and Singh, Manpreet and Saverese, Silvio and Hoi,   Steven CH},
  journal={arXiv preprint arXiv:2304.04661},
  year={2023}
}

@inproceedings{le2021neurallog,
  title={Log-based anomaly detection without log parsing},
  author={Le, Van-Hoang and Zhang, Hongyu},
  booktitle={2021 36th IEEE/ACM International Conference on Automated Software Engineering (ASE)},
  pages={492--504},
  year={2021},
  organization={IEEE}
}

@article{li2022constructing,
      title={Constructing Large-Scale Real-World Benchmark Datasets for AIOps}, 
      author={Zeyan Li and Nengwen Zhao and Shenglin Zhang and Yongqian Sun and Pengfei Chen and Xidao Wen and Minghua Ma and Dan Pei},
      year={2022},
      eprint={2208.03938},
      archivePrefix={arXiv},
      primaryClass={cs.SE},
      url={https://arxiv.org/abs/2208.03938}, 
}

@inproceedings{amin2019cadence,
  title={Cadence: Conditional anomaly detection for events using noise-contrastive estimation},
  author={Amin, Mohammad Ruhul and Garg, Pranav and Coskun, Baris},
  booktitle={Proceedings of the 12th ACM Workshop on Artificial Intelligence and Security},
  pages={71--82},
  year={2019}
}

@misc{coskun2022detecting,
  title={Detecting anomalous events using autoencoders},
  author={Coskun, Baris and Ding, Wei and Melis, Luca},
  year={2022},
  month=jun # "~28",
  publisher={Google Patents},
  note={US Patent 11,374,952}
}

@misc{opentelemetry_events_2025,
  author       = {OpenTelemetry},
  title        = {Semantic Conventions for Events},
  year         = {2025},
  howpublished = {\url{https://opentelemetry.io/docs/specs/semconv/general/events/}},
  note         = {Accessed: 2025-06-06}
}

@article{van2004workflow,
  title={Workflow Mining: Discovering Process Models from Event Logs},
  author={Van der Aalst, Wil and Weijters, Ton and Maruster, Laura},
  journal={IEEE Transactions on Knowledge and Data Engineering},
  volume={16},
  number={9},
  pages={1128--1142},
  year={2004},
  publisher={IEEE}
}

@article{pham2026torai,
      title={TORAI: Multi-Source Root Cause Analysis for Blind Spots in Microservice Service Call Graph},
      author={Pham, Luan and Ha, Huong and Zhang, Xiuzhen and Zhang, Hongyu},
      journal={Proceedings of the ACM on Software Engineering},
      volume={3},
      number={FSE},
      articleno={FSE130},
      numpages={23},
      year={2026},
      publisher={ACM New York, NY, USA},
      doi={10.1145/3808137}
}

@article{pham2026eventadl,
      title={EventADL: Open-Box Anomaly Detection and Localization Framework for Events in Cloud-Based Service Systems},
      author={Pham, Luan and Nicolet, Victor and Dodds, Joey and Guan, Hui and Kroening, Daniel},
      journal={Proceedings of the ACM on Software Engineering},
      volume={3},
      number={FSE},
      year={2026},
      publisher={ACM New York, NY, USA}
}

@article{pham2026graph,
  title={Graph-Free Root Cause Analysis},
  author={Pham, Luan},
  journal={arXiv preprint arXiv:2601.21359},
  year={2026}
}

@misc{gaia,
  title={{GAIA}: Generic AIOps Atlas},
  author={CloudWise-OpenSource},
  year={2025},
  howpublished={\url{https://github.com/CloudWise-OpenSource/GAIA-DataSet}},
  note={CloudWise GAIA Dataset for AIOps}
}

@misc{aiops21,
  title={{AIOps} 2021 Challenge Dataset},
  author={AIOps Nankai},
  year={2021},
  howpublished={\url{https://www.aiops.cn/gitlab/aiops-nankai/data/trace/aiops2021/}}
}

@inproceedings{gu2025adamas,
  title={ADAMAS: Adaptive Domain-Aware Performance Anomaly Detection in Cloud Service Systems},
  author={Gu, Wenwei and Gu, Jiazhen and Liu, Jinyang and Chen, Zhuangbin and Zhang, Jianping and Kuang, Jinxi and Feng, Cong and Yang, Yongqiang and Lyu, Michael R},
  booktitle={Proceedings of the IEEE/ACM 47th International Conference on Software Engineering},
  pages={911--923},
  year={2025}
}

@inproceedings{gu2024kpiroot,
  title={Kpiroot: Efficient monitoring metric-based root cause localization in large-scale cloud systems},
  author={Gu, Wenwei and Sun, Xinying and Liu, Jinyang and Huo, Yintong and Chen, Zhuangbin and Zhang, Jianping and Gu, Jiazhen and Yang, Yongqiang and Lyu, Michael R},
  booktitle={2024 IEEE 35th International Symposium on Software Reliability Engineering (ISSRE)},
  pages={403--414},
  year={2024},
  organization={IEEE}
}

@inproceedings{guo2024logformer,
  title={Logformer: A pre-train and tuning pipeline for log anomaly detection},
  author={Guo, Hongcheng and Yang, Jian and Liu, Jiaheng and Bai, Jiaqi and Wang, Boyang and Li, Zhoujun and Zheng, Tieqiao and Zhang, Bo and Peng, Junran and Tian, Qi},
  booktitle={Proceedings of the AAAI conference on artificial intelligence},
  volume={38},
  pages={135--143},
  year={2024}
}

@inproceedings{li2020swisslog,
  title={Swisslog: Robust and unified deep learning based log anomaly detection for diverse faults},
  author={Li, Xiaoyun and Chen, Pengfei and Jing, Linxiao and He, Zilong and Yu, Guangba},
  booktitle={2020 IEEE 31st International Symposium on Software Reliability Engineering (ISSRE)},
  pages={92--103},
  year={2020},
  organization={IEEE}
}

@inproceedings{li2023conan,
  title={Conan: Diagnosing batch failures for cloud systems},
  author={Li, Liqun and Zhang, Xu and He, Shilin and Kang, Yu and Zhang, Hongyu and Ma, Minghua and Dang, Yingnong and Xu, Zhangwei and Rajmohan, Saravan and Lin, Qingwei and others},
  booktitle={2023 IEEE/ACM 45th International Conference on Software Engineering: Software Engineering in Practice (ICSE-SEIP)},
  pages={138--149},
  year={2023},
  organization={IEEE}
}

@inproceedings{lee2023heterogeneous,
  title={Heterogeneous anomaly detection for software systems via semi-supervised cross-modal attention},
  author={Lee, Cheryl and Yang, Tianyi and Chen, Zhuangbin and Su, Yuxin and Yang, Yongqiang and Lyu, Michael R},
  booktitle={2023 IEEE/ACM 45th International Conference on Software Engineering (ICSE)},
  pages={1724--1736},
  year={2023},
  organization={IEEE}
}

@inproceedings{nedelkoski2019anomaly,
  title={Anomaly detection from system tracing data using multimodal deep learning},
  author={Nedelkoski, Sasho and Cardoso, Jorge and Kao, Odej},
  booktitle={2019 IEEE 12th International Conference on Cloud Computing (CLOUD)},
  pages={179--186},
  year={2019},
  organization={IEEE}
}

@inproceedings{ren2024slim,
  title={SLIM: a Scalable Light-weight Root Cause Analysis for Imbalanced Data in Microservice},
  author={Ren, Rui and Yang, Jingbang and Yang, Linxiao and Gu, Xinyue and Sun, Liang},
  booktitle={Proceedings of the 2024 IEEE/ACM 46th International Conference on Software Engineering: Companion Proceedings},
  pages={328--330},
  year={2024}
}

@inproceedings{roy2024exploring,
  title={Exploring llm-based agents for root cause analysis},
  author={Roy, Devjeet and Zhang, Xuchao and Bhave, Rashi and Bansal, Chetan and Las-Casas, Pedro and Fonseca, Rodrigo and Rajmohan, Saravan},
  booktitle={Companion proceedings of the 32nd ACM international conference on the foundations of software engineering},
  pages={208--219},
  year={2024}
}

@inproceedings{sun2024art,
  title={Art: A unified unsupervised framework for incident management in microservice systems},
  author={Sun, Yongqian and Shi, Binpeng and Mao, Mingyu and Ma, Minghua and Xia, Sibo and Zhang, Shenglin and Pei, Dan},
  booktitle={Proceedings of the 39th IEEE/ACM International Conference on Automated Software Engineering},
  pages={1183--1194},
  year={2024}
}

@article{sun2025interpretable,
  title={Interpretable failure localization for microservice systems based on graph autoencoder},
  author={Sun, Yongqian and Lin, Zihan and Shi, Binpeng and Zhang, Shenglin and Ma, Shiyu and Jin, Pengxiang and Zhong, Zhenyu and Pan, Lemeng and Guo, Yicheng and Pei, Dan},
  journal={ACM Transactions on Software Engineering and Methodology},
  volume={34},
  number={2},
  pages={1--28},
  year={2025},
  publisher={ACM New York, NY}
}

@inproceedings{wang2023incremental,
  title={Incremental causal graph learning for online root cause analysis},
  author={Wang, Dongjie and Chen, Zhengzhang and Fu, Yanjie and Liu, Yanchi and Chen, Haifeng},
  booktitle={Proceedings of the 29th ACM SIGKDD conference on knowledge discovery and data mining},
  pages={2269--2278},
  year={2023}
}

@inproceedings{wang2024mrca,
  title={MRCA: Metric-level root cause analysis for microservices via multi-modal data},
  author={Wang, Yidan and Zhu, Zhouruixing and Fu, Qiuai and Ma, Yuchi and He, Pinjia},
  booktitle={Proceedings of the 39th IEEE/ACM International Conference on Automated Software Engineering},
  pages={1057--1068},
  year={2024}
}

@inproceedings{yang2021semi,
  title={Semi-supervised log-based anomaly detection via probabilistic label estimation},
  author={Yang, Lin and Chen, Junjie and Wang, Zan and Wang, Weijing and Jiang, Jiajun and Dong, Xuyuan and Zhang, Wenbin},
  booktitle={2021 IEEE/ACM 43rd International Conference on Software Engineering (ICSE)},
  pages={1448--1460},
  year={2021},
  organization={IEEE}
}

@inproceedings{zhang2024trace,
  title={Trace-based multi-dimensional root cause localization of performance issues in microservice systems},
  author={Zhang, Chenxi and Dong, Zhen and Peng, Xin and Zhang, Bicheng and Chen, Miao},
  booktitle={Proceedings of the IEEE/ACM 46th International Conference on Software Engineering},
  pages={1--12},
  year={2024}
}

@misc{eventadl_artifact,
  author       = {Pham, Luan},
  title        = {Artifacts of "EventADL: Open-Box Anomaly Detection and Localization Framework for Events in Cloud-Based Service Systems"},
  month        = apr,
  year         = 2026,
  publisher    = {Zenodo},
  doi          = {10.5281/zenodo.19433493},
  url          = {https://doi.org/10.5281/zenodo.19433493}
}

@article{torai_sourcecode,
  author = {Luan Pham},
  title = {{Source code of "TORAI: Multi-Source Root Cause Analysis for Blind Spots in Microservice Service Call Graph"}},
  year = {2026},
  month = {4},
  url = {https://figshare.com/articles/software/Source_code_of_TORAI_Multi-Source_Root_Cause_Analysis_for_Blind_Spots_in_Microservice_Service_Call_Graph_/31938495},
  doi = {10.6084/m9.figshare.31938495.v1}
}

@article{torai_datasets,
  author = {Luan Pham},
  title = {{Datasets for "TORAI: Multi-Source Root Cause Analysis for Blind Spots in Microservice Service Call Graph"}},
  year = {2026},
  month = {4},
  url = {https://figshare.com/articles/dataset/Datasets_for_TORAI_Multi-Source_Root_Cause_Analysis_for_Blind_Spots_in_Microservice_Service_Call_Graph_/31925976},
  doi = {10.6084/m9.figshare.31925976.v1}
}


%% file: bib/strings.bib
@STRING{ ICWS = {Proceedings of the IEEE International Conference on Web Services (ICWS)} }

@STRING{ WWW = {Proceedings of the International Conference on World Wide Web (WWW)} }

@STRING{ SERVICES = {Proceedings of the IEEE Congress on Services (SERVICES)} }

@STRING{ IJCAI = {Proceedings of the International Joint Conference on Artificial Intelligence (IJCAI)} }

@STRING{ AAAI = {Proceedings of the National Conference on Artificial Intelligence (AAAI)} }

@STRING{ AGENTS = {Proceedings of the Annual Conference on Autonomous Agents (AGENTS)} }

@STRING{ UAI = {Proceedings of the International Conference on Uncertainty in Artificial Intelligence (UAI)} }

@STRING{ SPRINGER = {Springer} }

@STRING{ MIT = {The MIT Press} }

@STRING{ ACM = {ACM Press} }

@STRING{ ICML = {Proceedings of the International Conference on Machine Learning (ICML)} }

@STRING{ SIGMOD = {Proceedings of the ACM International Conference on Management of Data (SIGMOD)} }
